\documentclass[11pt]{article}
\usepackage[utf8]{inputenc}
\usepackage[letterpaper]{geometry}
\usepackage{color}
\usepackage{booktabs}
\usepackage{pmboxdraw}
\usepackage{multirow}
\usepackage{amsmath}
\usepackage{amssymb}
\usepackage{setspace}
\usepackage{wasysym}
\usepackage[authoryear,longnamesfirst]{natbib}
\usepackage[unicode=true,
 bookmarks=true,bookmarksnumbered=false,bookmarksopen=true,bookmarksopenlevel=1,
 breaklinks=true,pdfborder={0 0 1},backref=false,colorlinks=true]
 {hyperref}
\hypersetup{
citecolor=blue,
urlcolor=blue, 
linkcolor=blue,
colorlinks=true}
\usepackage{url}

\makeatletter

\usepackage{amsfonts}\usepackage{mathrsfs}\usepackage{bm}\usepackage{verbatim}\usepackage{dsfont}\usepackage{tabularx}
\usepackage{lipsum}\usepackage{ltablex}\usepackage{array}\usepackage{makecell}\usepackage{caption}\usepackage{multirow}\usepackage{float}\usepackage{wasysym}\usepackage{graphicx}\usepackage{soul}\usepackage{rotating}
\usepackage[hang,flushmargin]{footmisc}
\usepackage[longnamesfirst,authoryear]{natbib}
\normalbaselines 

\providecommand{\U}[1]{\protect\rule{.1in}{.1in}}
\newtheorem{theorem}{Theorem}[section]

\newtheorem{assumption}{Assumption}

\newtheorem{lemma}[theorem]{Lemma}
\newtheorem{proposition}[theorem]{Proposition}

\newenvironment{proof}[1][Proof]{\noindent\textbf{#1.} }{\ \rule{0.5em}{0.5em}}

\@addtoreset{lemma}{Appendix B}

\numberwithin{equation}{section}

\newcommand{\Keywords}[1]{\par\noindent {\small{\em Keywords\/}: #1}} 
\newcommand{\JELclass}[1]{\par\noindent {\small{\em JEL classification\/}: #1}} 

\newtheorem{assumptionalt}{Assumption}[assumption]

\newenvironment{assumptionp}[1]{
  \renewcommand\theassumptionalt{#1}
  \assumptionalt
}{\endassumptionalt}

\makeatother

\graphicspath{{./figures/}}

\begin{document}
\title{High Dimensional Binary Choice Model with Unknown Heteroskedasticity
or Instrumental Variables}
\author{{Fu Ouyang}\thanks{School of Economics, The University of Queensland, St Lucia, QLD 4072,
Australia. Email: \protect\protect\protect\protect\href{http://f.ouyang@uq.edu.au}{f.ouyang@uq.edu.au}.} \\
 {\normalsize{}{}{}{}University of Queensland} \and {Thomas T.
Yang}\thanks{Corresponding author. Research School of Economics, The Australian
National University, Canberra, ACT 0200, Australia. Email: \protect\protect\protect\protect\href{http://tao.yang@anu.edu.au}{tao.yang@anu.edu.au}.} \\
 {\normalsize{}{}{}{}Australian National University}}
\date{\today}

\maketitle
\onehalfspacing 
\begin{abstract}
This paper proposes a new method for estimating high-dimensional binary choice models. We consider a semiparametric model that places no distributional assumptions on the error term, allows for heteroskedastic errors, and permits endogenous regressors. Our approaches extend the special regressor estimator originally proposed by \citet{Lewbel2000}. This estimator becomes impractical in high-dimensional settings due to the curse of dimensionality associated with high-dimensional conditional density estimation. To overcome this challenge, we introduce an innovative data-driven dimension reduction method for nonparametric kernel estimators, which constitutes the main contribution of this work. The method combines distance covariance-based screening with cross-validation (CV) procedures, making special regressor estimation feasible in high dimensions. Using this new feasible conditional density estimator, we address variable and moment (instrumental variable) selection problems for these models. We apply penalized least squares (LS) and generalized method of moments (GMM) estimators with an $L_1$ penalty. A comprehensive analysis of the oracle and asymptotic properties of these estimators is provided. Finally, through Monte Carlo simulations and an empirical study on the migration intentions of rural Chinese residents, we demonstrate the effectiveness of our proposed methods in finite sample settings.

\vspace{4mm}

\JELclass{C12, C14, C21, C52}

\medskip{}
\vspace{2cm}

\end{abstract}

\onehalfspacing
\section{Introduction}

In this paper, we study the estimation of semiparametric binary choice
models in high-dimensional settings. We adopt a standard model specification
in which the 0-1 valued dependent variable $Y$ is modeled as 
\[
Y=\boldsymbol{1}(\boldsymbol{X}_{1}^{s^{\ast}\prime}\boldsymbol{\beta}^{\ast}+\varepsilon>0)=\boldsymbol{1}\left(X_{1}\beta_{1}^{\ast}+X_{2}\beta_{2}^{\ast}+...+X_{s^{\ast}}\beta_{s^{\ast}}^{\ast}+\varepsilon>0\right),
\]
where $\boldsymbol{X}_{1}^{s^{*}}\equiv(X_{1},X_{2},...,X_{s^{\ast}})'$
represent the explanatory variables (signals), $\boldsymbol{\beta}^{*}\equiv(\beta_{1}^{\ast},\beta_{2}^{\ast},...,\beta_{s^{\ast}}^{\ast})'$
are the unknown true coefficients to estimate, $\varepsilon$ represents
an unobserved disturbance (error term), and $\boldsymbol{1}\left(\cdot\right)$
is the indicator function that equals one if $\cdot$ is true and 0
otherwise. In the context of high-dimensional settings, we allow for
the possibility that either the number of candidate explanatory variables
$\boldsymbol{X}_{1}^{p_{n}}\equiv\left(X_{1},X_{2},...,X_{p_{n}}\right)'$,
denoted by $p_{n}$, diverges as the sample size $n$ increases, or
the number of available instrumental variables (IV) $\boldsymbol{Z}_{1}^{p_{n}}\equiv\left(Z_{1},Z_{2},...,Z_{p_{n}}\right)'$,
also denoted by $p_{n}$, diverges depending on the specific application.

Estimating binary choice models is far more challenging than estimating
linear models due to the non-linearity of the indicator function.
Researchers often resort to imposing distributional assumptions on
the error term $\varepsilon$, such as assuming it follows a normal
or logistic distribution, to enable feasible estimation. However,
the likelihood of $\varepsilon$ precisely conforming to these specific
distributions is very low. Complicating things further, heteroskedasticity
in $\varepsilon$, common in economic data, can invalidate these distributional
assumptions. Furthermore, when endogeneity issues are in play, which
is also frequently encountered in empirical studies, these distributional
assumptions provide no help in identifying $\boldsymbol{\beta}^{*}$,
even if a sufficient number of valid IVs are available in $\boldsymbol{Z}_{1}^{p_{n}}$.
One may address the endogeneity issue using the control function
method detailed in standard textbooks like \citet{Wooldridge2010}.
The key idea is to include the residuals $\boldsymbol{u}$ obtained
from regressing $\boldsymbol{X}_{1}^{s^{*}}$ on the valid IVs in
$\boldsymbol{Z}_{1}^{p_{n}}$ in the regression, with the hope of
controlling the endogeneity through additional control variables $\boldsymbol{u}$.
However, it is essential to acknowledge an important limitation of
the control function approach--it demands not just any set of valid
IVs in $\boldsymbol{Z}_{1}^{p_{n}}$ but rather the exact right set
of valid IVs. For example, missing any valid IVs of $\boldsymbol{Z}_{1}^{p_{n}}$
can lead to a violation of the requisite assumptions for this approach.
We refer interested readers to Section 5 of \citet{LewbelDongYang}
for more details.

Dealing with many candidate explanatory variables or IVs adds another
layer of difficulty. The processes of recovering the model's sparsity
structure (e.g., \citet{Tibshirani1996}), choosing the strongest
instruments (e.g., \citet{Belloni_et_al2012}), selecting the appropriate
moment conditions (e.g., \citet{Liao2013} and \citet{ChengLiao2015}),
and conducting estimation and inferences in high-dimensional GMM (e.g.,
\citet{BelloniEtal2018}, \citet{GautierTsybakov}, and \citet{GoldEtal2020})
become even more challenging in the context of the binary choice model.
To our knowledge, existing literature on high-dimensional binary choice
models often assumes independence between the error term $\varepsilon$
and the explanatory variables $\boldsymbol{X}$. Classic works like
\citet{FanLi2001} and more recent studies such as \citet{chetverikov2023}
and \citet{KhanEtal2023} rely on this assumption. However, these
papers exclude the possibility of heteroskedasticity in the error
term. Furthermore, we are not aware of any research on high-dimensional
binary choice models that allow for endogeneity and instrumental variables
in the usual sense.\footnote{Here, the usual sense means the following: A valid IV
$Z$ for an endogenous variable $X$ ($\textrm{Cov}\left(X,\varepsilon\right)\neq0$)
in linear models satisfies both the relevance and the exogeneity conditions,
that is, $\textrm{Cov}\left(X,Z\right)\neq0$ and $\textrm{Cov}\left(\varepsilon,Z\right)=0$.} Another strand of literature considers binary prediction, possibly
in high-dimensional settings. These works do not assume distributional
assumptions and consider a framework similar to that in \citet{Manski1975}.
Important contributions in this direction include \citet{ElliottLieli2013},
\citet{ChenLee2018}, and \citet{babiiEtal2021}. An interesting observation
in \citet{babiiEtal2021} is that logistic regression can provide
accurate predictions even if the independence between $\boldsymbol{X}$
and $\varepsilon$ fails to hold or $\varepsilon$ does not obey logistic
distribution.

In this paper, we propose a feasible special regressor method to address
these challenges. Our approach is semiparametric, as it does not rely
on any distributional assumptions regarding $\varepsilon$ and allows
for general heteroskedasticity. When certain explanatory variables
are endogenous, our method only requires the availability of IVs that
are valid in the conventional sense. The special regressor method,
initially proposed in \citet{Lewbel2000}, is based on the assumption
that there exists a special regressor $V$ in the model 
\begin{equation}
Y=\boldsymbol{1}(V+\boldsymbol{X}_{1}^{s^{\ast}\prime}\boldsymbol{\beta}^{\ast}+\varepsilon>0)=\boldsymbol{1}\left(V+X_{1}\beta_{1}^{\ast}+X_{2}\beta_{2}^{\ast}+...+X_{s^{\ast}}\beta_{s^{\ast}}^{\ast}+\varepsilon>0\right),\label{EQ:model}
\end{equation}
where the coefficient before $V$ is normalized to $1$. $V$ is assumed
to be a continuous regressor with support larger than that of $-(\boldsymbol{X}_{1}^{s^{\ast}\prime}\boldsymbol{\beta}^{\ast}+\varepsilon)$
and satisfy $V\perp\left(\varepsilon,\boldsymbol{X}_{1}^{s^{\ast}}\right)|\boldsymbol{Z}_{1}^{p_{n}}$.\footnote{When all candidate explanatory variables are exogenous, $\boldsymbol{Z}_{1}^{p_{n}}=\boldsymbol{X}_{1}^{p_{n}}$
and the exclusion restriction on $V$ can be expressed as $V\perp\varepsilon|\boldsymbol{X}_{1}^{p_{n}}$.} Then, under the classic exogeneity condition, $\mathbb{E}\left(\boldsymbol{Z}_{1}^{p_{n}}\varepsilon\right)=0$,
\citet{Lewbel2000} showed that 
\begin{equation}
\mathbb{E}[\boldsymbol{Z}_{1}^{p_{n}}(\tilde{Y}-X_{1}\beta_{1}^{\ast}-X_{2}\beta_{2}^{\ast}-...-X_{s^{\ast}}\beta_{s^{\ast}}^{\ast})]=0,\label{EQ:momentCond}
\end{equation}
where $\tilde{Y}$ is defined as 
\begin{equation}
\tilde{Y}=\frac{Y-\boldsymbol{1}\left(V>0\right)}{f\left(V|\boldsymbol{Z}_{1}^{p_{n}}\right)}.\label{EQ:Ytide}
\end{equation}
Using the moment conditions in equation (\ref{EQ:momentCond}), we
can construct the regular GMM estimator to estimate $\boldsymbol{\beta}^{*}$.
The main advantage of the special regressor approach is its ability
to estimate the coefficients in a manner akin to linear models.

A drawback of the special regressor approach is its requirement to
estimate the conditional density $f\left(V|\boldsymbol{Z}_{1}^{p_{n}}\right)$.
Even when dealing with a moderate number of elements (e.g., $p_{n}=10$)
in $\boldsymbol{Z}_{1}^{p_{n}}$, nonparametrically estimating $f\left(V|\boldsymbol{Z}_{1}^{p_{n}}\right)$
is challenging due to the curse of dimensionality. As a result, the
special regressor estimation becomes infeasible if we allow the number
of elements in $\boldsymbol{Z}_{1}^{p_{n}}$ to diverge. This paper
is motivated by the observation that often only a few elements in
$\boldsymbol{Z}_{1}^{p_{n}}$ are ``relevant'' to $V$. Then, the
estimator becomes feasible, provided we can distinguish these relevant
$Z$s from the irrelevant ones with high probability. 

Interestingly, the foundational study
by \citet{HallRacineLi} revealed a surprising prevalence of ``irrelevant''
components in estimating conditional densities. Applying the special
regressor approach, \citet{DongLewbel2015} adopted ``negative age''
as the special regressor to study inter-state migration. Their sample
consists of 23 to 59 years old individuals who had completed
education and were not retired in 1990. Of all the regressors, ``negative
age'' is expected to be independent of factors such as ``education''
(since all individuals had completed their education), ``gender'',
``race'', and potentially ``government benefit''. This can be
further seen from the application of \citet{XueYangZhou}, which studied
the migration intention of rural residents in China using the special
regressor method. The special regressor adopted in their paper is
the ``average daily precipitation''. Since precipitation is strongly
exogenous and unlikely to be affected by individual characteristics,
they assumed $V$ to be independent of all other regressors and only
calculate $\hat{f}\left(V\right)$ for (\ref{EQ:Ytide}). Notably,
this study did not formally test the validity of this assumption.
Our paper proposes a data-driven procedure to address this challenge
rigorously.

Dimension reduction for conditional density estimation poses a significant
challenge. The cross-validation (CV) approach, as proposed in \citet{HallRacineLi},
is computationally demanding, making it difficult to apply in high-dimensional
scenarios. \citet{Efromovich2010} required tensor products of basis
over each dimension; this makes his approach infeasible in high-dimensional
settings. The approach in \citet{ZhangJMP2023} suffers from the same
issue because it needs to compute conditional expectations at each
dimension. In this paper, we present an innovative dimension reduction
technique inspired by the ``sure independence screening'' method
introduced in \citet{FanLv2008} for linear models. To mitigate the
high computational burden associated with high-dimensional linear
models, \citet{FanLv2008} suggested regressing the dependent variable
on each explanatory variable individually and retaining only those
with the strongest correlation to the dependent variable. However,
using correlation to measure dependence is inappropriate for our goals,
as it captures only linear relationships. Consequently, we adopt the
concept of ``distance covariance'' (DC), first detailed and studied
in \citet{Szekely_et_al}. As a metric of general dependence, DC is
zero if and only if the two random vectors under examination are independent.

Leveraging this desirable property, we propose a DC-based screening
procedure. When estimating $f\left(V|\boldsymbol{Z}_{1}^{p_{n}}\right)$,
the primary strategy is to employ the screening method to reduce the
dimension of the conditioning set. Subsequently, the CV procedure
from \citet{HallRacineLi} is applied post-screening to refine the
screening outcomes further. Integrating these two methods bypasses
the often complicated task of determining optimal tuning parameters.
This procedure also reduces the necessity for ``perfect'' variable
selection during the screening phase, offering a more dependable and
practical approach to estimate $f\left(V|\boldsymbol{Z}_{1}^{p_{n}}\right)$
in the context of high-dimensional data.\footnote{This two-step selection approach can be extended to other nonparametric
estimation scenarios, making it of independent interest.} We extensively explore the theoretical attributes of our proposed
selection method and the subsequent conditional density estimation.
Our findings demonstrate that our selection technique can identify
the components in $\boldsymbol{Z}_{1}^{p_{n}}$ that are relevant to $V$
with a high probability. Furthermore, the post-selection conditional
density estimator achieves the ``oracle'' rate of convergence.

With the feasible estimator for $f\left(V|\boldsymbol{Z}_{1}^{p_{n}}\right)$
in hand, we proceed to address the variable selection and moment (IV)
selection problems for high-dimensional binary choice models. Specifically,
for the case where there are many exogenous explanatory variables
$\boldsymbol{X}_{1}^{p_{n}}$, we introduce two estimators obtained
from combining linear regression with either weighted or unweighted
$L_{1}$ penalties. These estimators can estimate $\boldsymbol{\beta}^{\ast}$
while selecting true signals within $\boldsymbol{X}_{1}^{p_{n}}$
simultaneously. When there are many candidate IVs $\boldsymbol{Z}_{1}^{p_{n}}$,
a mixture of valid and invalid, we propose a GMM with $L_{1}$ penalty
procedure, which can simultaneously estimate $\boldsymbol{\beta}^{\ast}$
and detect invalid IVs within $\boldsymbol{Z}_{1}^{p_{n}}$. We establish
the asymptotic properties of these approaches, providing a solid theoretical
foundation for their application. Furthermore, we demonstrate the practical effectiveness of these methods in finite samples using both simulated and real-world data.

The rest of this paper is organized as follows. In Section \ref{SEC:f(v,x)reduce},
we propose a novel dimension reduction method for the conditional
density estimation. With the feasible conditional density estimator,
we address the classic variable and moment (IV) selection problems
in the context of special regressor estimation in Section \ref{SEC:theoryapply}.
We investigate the small sample properties of our methods through Monte Carlo experiments
in Section \ref{SEC:simu}, where we also provide a practical guide for choosing tuning parameters and kernel functions to implement our proposed procedures. In Section \ref{SEC:applications}, we further illustrate our approaches by examining migration decisions of rural residents in China. Section \ref{SEC:conclusion} concludes this
paper. 

Proofs and tables are presented in the online supplementary appendix. Specifically, Appendix \ref{APP:introduction} provides a gentle introduction
to the DC and CV procedures. The proofs of all the main theorems are collected
in Appendices \ref{APP:A}--\ref{APP:C}, while the proofs of all technical
lemmas are deferred to Appendix \ref{APP:lemmas}. An investigation
of variable selections with unweighted $L_{1}$ penalty is presented
in Appendix \ref{APP:lasso}. Lastly, Appendix \ref{APP:tables} contains
tables for simulation and application results.

For ease of reference, we list the notations maintained throughout this paper here.

\noindent \textbf{Notation.} All vectors are column vectors. Unless stated otherwise, we follow the convention of using capital letters for random variables and their corresponding lowercase letters for sample realizations. We use
$\rho$ to denote the eigenvalues of a matrix. For instance, $\rho_{\min}$
and $\rho_{\max}$ denote a matrix's minimum and maximum eigenvalues,
respectively. The notation $\left\Vert \boldsymbol{x}\right\Vert $
represents the Euclidean norm of a vector $\boldsymbol{x}$. For a
matrix $\mathbf{A},$ we define $\left\Vert \mathbf{A}\right\Vert \equiv\sqrt{\text{trace}\left(\mathbf{AA}^{\prime}\right)}$
and $\left\Vert \mathbf{A}\right\Vert _{\infty}\equiv\max_{jl}\mathbf{A}_{jl}$,
where $\mathbf{A}_{jl}$ is the $\left(j,l\right)$-th element of
$\mathbf{A}$. The symbol $\left\Vert \cdot\right\Vert _{0}$ represents
the $L_{0}$-norm, which counts the total number of nonzero elements
in a vector. $\left\Vert \cdot\right\Vert _{1}$ denotes the $L_{1}$-norm. For deterministic series $\left\{ a_{n}\right\} _{n=1}^{\infty}$
and $\left\{ b_{n}\right\} _{n=1}^{\infty}$, the notation $a_{n}\propto b_{n}$
means that $0<C_{1}\leq\lim\inf_{n\rightarrow\infty}\left\vert a_{n}/b_{n}\right\vert \leq\lim\sup_{n\rightarrow\infty}\left\vert a_{n}/b_{n}\right\vert \leq C_{2}<\infty$
for some constants $C_{1}$ and $C_{2}$, $a_{n}\lesssim b_{n}$ means
$\lim\sup_{n\rightarrow\infty}\left\vert a_{n}/b_{n}\right\vert \leq C<\infty$ for some constant $C$, $a_{n}\gtrsim b_{n}$ if $b_{n}\lesssim a_{n},$
and $a_{n}\ll b_{n}$ if $a_{n}=o\left(b_{n}\right),$ and $a_{n}\gg b_{n}$
if $b_{n}\ll a_{n}$. The term $\mathcal{A}^{c}$ stands for the complement
of the set $\mathcal{A}$, and $\left\vert \mathcal{A}\right\vert $
is the number of elements in $\mathcal{A}$. As $n$ tends to infinity,
the notations $\overset{P}{\rightarrow}$ and $\overset{d}{\rightarrow}$
indicate convergence in probability and distribution, respectively.
The symbol $C$ denotes various positive constants, which may change
from one instance to the next.

\section{Dimension Reduction for the Conditional Density\label{SEC:f(v,x)reduce}}

As previously mentioned, an essential prerequisite for implementing
the special regressor estimator is the ability to estimate $f\left(V|\boldsymbol{Z}_{1}^{p_{n}}\right)$,
which is used to construct $\tilde{Y}$ as defined in equation (\ref{EQ:Ytide}).
To achieve this, we presuppose a certain sparsity in the conditional
set of $f\left(V|\boldsymbol{Z}_{1}^{p_{n}}\right)$. Specifically,
we assume that only a subset of the $Z$s are included in the conditional
set and refer to them as ``relevant to $V$''. This assumption is
detailed in Assumption \ref{A:main}. To measure the dependence between
$V$ and the $Z$s, we use the \textit{distance covariance} (DC), first
introduced and examined in \citet{Szekely_et_al}. Section \ref{SEC:distance_cov}
provides a brief overview of this concept. In Section \ref{SEC:screening},
we describe our screening process and outline its theoretical properties.
We propose a practical dimension reduction technique in Section \ref{SEC:threshold}
by combining our screening method with the approach from \citet{HallRacineLi}.
We further extend the theory from \citet{HallRacineLi} in Section
\ref{SEC:further}, allowing for applying the dimension
reduction approach to a broader range of scenarios. Section \ref{SEC:proceduref}
summarizes the practical procedure for easier reference.

We emphasize that this novel approach to dimension reduction for conditional
density is the main innovation of this paper. This method has the
potential to be adapted to various scenarios involving high-dimensional
conditional density estimation, including the studies discussed in
Sections \ref{SEC:VariableSelection} and \ref{SEC:MSelection}.

\subsection{Review of the Distance Covariance}\label{SEC:distance_cov}

We measure the dependence between $V$ and another regressor or IV
$Z_{l}$ using DC introduced and studied in \citet{Szekely_et_al}.
We briefly review this statistic in this section and refer interested
readers to Appendix \ref{app:DC} and \citet{Szekely_et_al} for more
details.

Let $\phi_{V}\left(t\right)$, $\phi_{Z_{l}}\left(s\right)$ and $\phi_{V,Z_{l}}\left(t,s\right)$
denote the characteristic functions of $V,Z_{l},$ and $\left(V,Z_{l}\right),$
respectively. Specifically, 
\[
\phi_{V,Z_{l}}\left(t,s\right)=\mathbb{E}\left(e^{\mathrm{i}tV+\mathrm{i}sZ_{l}}\right)=\int\int e^{\mathrm{i}tv+\mathrm{i}sz_{l}}f\left(v,z_{l}\right)dvdz_{l},
\]
where $\mathrm{i}$ denotes the imaginary unit, and $\phi_{V}\left(t\right)\textrm{ and }\phi_{Z_{l}}\left(s\right)$
are similarly defined. The DC between univariate random variables
$V$ and $Z_{l}$ is defined as 
\[
\mathcal{V}^{2}\left(V,Z_{l}\right)=\int_{\mathbb{R}^{2}}\left\Vert \phi_{V,Z_{l}}\left(t,s\right)-\phi_{V}\left(t\right)\phi_{Z_{l}}\left(s\right)\right\Vert \omega\left(t,s\right)dtds,
\]
where $\left\Vert \phi\right\Vert ^{2}=\phi\bar{\phi}$ for a complex-valued
function $\phi$, with $\bar{\phi}$ being the conjugate of $\phi$,
and 
\[
\omega\left(t,s\right)=\frac{1}{\pi^{2}t^{2}s^{2}}.
\]
One nice property of DC is that $\mathcal{V}\left(V,Z_{l}\right)=0$
if and only if $V$ and $Z_{l}$ are independent. While other
positive weighting functions can ensure this property as well,
this particular choice of $\omega(t,s)$ (also recommended in \citet{Szekely_et_al})
is notable for yielding a very simple sample analog, as follows.

To estimate $\mathcal{V}^{2}\left(V,Z_{l}\right)$, the sample DC
is
\[
\mathcal{V}_{n}^{2}\left(V,Z_{l}\right)=S_{n1}\left(V,Z_{l}\right)+S_{n2}\left(V,Z_{l}\right)-2S_{n3}\left(V,Z_{l}\right),
\]
where 
\begin{align*}
S_{n1}\left(V,Z_{l}\right) & =\frac{1}{n^{2}}\sum_{j,k=1}^{n}\left\vert v_{j}-v_{k}\right\vert \left\vert z_{lj}-z_{lk}\right\vert ,\\
S_{n2}\left(V,Z_{l}\right) & =\left(\frac{1}{n^{2}}\sum_{j,k=1}^{n}\left\vert v_{j}-v_{k}\right\vert \right)\left(\frac{1}{n^{2}}\sum_{j,k=1}^{n}\left\vert z_{lj}-z_{lk}\right\vert \right), \\
S_{n3}\left(V,Z_{l}\right) & =\frac{1}{n^{3}}\sum_{j=1}^{n}\sum_{k,m=1}^{n}\left\vert v_{j}-v_{k}\right\vert \left\vert z_{lj}-z_{lm}\right\vert ,
\end{align*}
and $(v_{j},z_{lj}),j=1,2,...,n,$ are i.i.d. realizations
of $(V,Z_{l})$. \citet{Szekely_et_al} showed that 
\[
\mathcal{V}_{n}^{2}\left(V,Z_{l}\right)\rightarrow\mathcal{V}^{2}\left(V,Z_{l}\right)\text{ almost surely.}
\]
and proposed the test-statistic for null hypothesis $H_{0}:V\perp Z_{l}$
defined as $n\mathcal{V}_{n}^{2}(V,Z_{l})/S_{n2}(V,Z_{l})$, where
the denominator makes the test statistic scale-free (e.g., increasing
$V$ or $Z_{l}$ by ten times does not alter the test statistic).

As will be detailed in Section \ref{SEC:screening}, we propose a
DC screening procedure in this paper based on the following statistic:
\begin{equation}
\hat{\mathcal{T}}_{nl}=\frac{n^{1/2}\mathcal{V}_{n}^{2}\left(V,Z_{l}\right)}{S_{n2}\left(V,Z_{l}\right)}.\label{EQ:statistic}
\end{equation}

\subsection{Dimension Reduction for Conditional Density Estimation via Screening\label{SEC:screening}}

Estimating conditional density in high-dimensional settings is a challenging
problem. The well-known paper, \citet{HallRacineLi}, proposed a data-driven
CV procedure to automatically reduce the dimension of conditional
density estimation. However, it is infeasible to handle the high dimensional
case due to the high computation cost and the curse of dimensionality.
To tackle this problem, we use the \textit{screening} technique initially
proposed by \citet{FanLv2008} for ultra-high dimensional linear models.
The idea of the screening in the linear model case is to calculate
the correlation between the dependent variable and one independent
variable at a time and keep the independent variables with the highest
correlation with the dependent variable for the model, e.g., 5\% of
the candidate independent variables.

In the context of nonparametric conditional density estimation, the
work most closely related to this paper is \citet{LiZhongZhu}, who
extended the screening method to explore more general nonlinear relationships
between the dependent variable and independent variables, leveraging
the concept of distance correlation.\footnote{The relationship between distance correlation and distance covariance
is analogous to that of ordinary correlation and covariance. In \citet{LiZhongZhu},
the empirical distance correlation is defined as $\mathcal{V}_{n}(V,Z_{l})/\sqrt{\mathcal{V}_{n}(V,V)\cdot\mathcal{V}_{n}(Z_{l},Z_{l})}$.} Using a similar idea, we measure the dependence between $V$ and
another regressor or IV $Z_{l}$ one at a time, employing the statistic
defined in (\ref{EQ:statistic}). It is worth noting that \citet{LiZhongZhu}
imposed restrictive sub-Gaussian tail assumption. Our work differs
from theirs by allowing for much heavier-tailed distributions, as
stated in the second part of Assumption \ref{A:main}, presented below.
This flexibility significantly broadens the applicability of the DC-based
screening approach, particularly in economics, where thin-tail conditions
like sub-Gaussian distributions can be rather restrictive. 

Before presenting the key assumptions and theoretical results, we
introduce some technical terms. Let $(\tilde{V},\tilde{Z}_{l})$ denote
an independent copy of $(V,Z_{l})$. We define the quantity 
\[
S_{2}\left(V,Z_{l}\right)=\mathbb{E}(\vert V-\tilde{V}\vert)\mathbb{E}(\vert Z_{l}-\tilde{Z}_{l}\vert).
\]
Additionally, we introduce the term $\kappa_{n}$, which represents
a slowly diverging sequence with the rate of divergence no faster
than $\log n$. 

\setlength{\parskip}{0pt}
\begin{assumptionp}{1}\label{A:main} For all $l=1,2,...,p_{n}$,
the following conditions hold: 
\begin{enumerate}
\item[(1)] $\{V_{i},Z_{il}\}_{i=1}^{n}$ are i.i.d. across $i$. 
\item[(2)] There exist only a small number of $Z$s, without loss of generality,
say $\left(Z_{1},Z_{2},...,Z_{p^{\ast}}\right)$, relevant to $V$
such that 
\begin{align}
V & \perp\left(Z_{p^{\ast}+1},Z_{p^{\ast}+2},...,Z_{p_{n}}\right),\text{ }\label{EQ:Indep}\\
\text{and }V & \perp\left(Z_{p^{\ast}+1},Z_{p^{\ast}+2},...,Z_{p_{n}}\right)|\left(Z_{1},Z_{2},...,Z_{p^{\ast}}\right).\label{EQ:condIndep1}
\end{align}
\item[(3)] $V$ and $Z_{l}$ satisfy 
\[
\max_{l=1,...,p_{n}}\{\mathbb{E}(\vert VZ_{l}\vert^{2+\delta}),\mathbb{E}(\vert V\vert^{2+\delta}),\mathbb{E}(\vert Z_{l}\vert^{2+\delta})\}<\infty,
\]
for some positive $\delta\geq1,$\ and 
\[
p_{n}\lesssim\min\{n^{1+\delta/2},n^{\delta}\}.
\]
\item[(4)] 
\[
\min_{l=1,...,p^{\ast}}\frac{\mathcal{V}^{2}\left(V,Z_{l}\right)}{S_{2}\left(V,Z_{l}\right)}\gtrsim n^{-1/2}\kappa_{n}\sqrt{\log n},\text{ and}
\]
\[
0<\sqrt{D_{1}}\leq\min\{\mathbb{E}(|V-\tilde{V}|),\min_{l=1,...,p_{n}}\{\mathbb{E}(|Z_{l}-\tilde{Z}_{l}|)\}\}\leq\sqrt{D_{2}}<\infty,
\]
for some positive $D_{1}$ and $D_{2}$. 
\end{enumerate}
\end{assumptionp}
\setlength{\parskip}{1em}

Equation (\ref{EQ:Indep}) is essential for the screening procedure. Equation  (\ref{EQ:condIndep1}) is to ensure we can apply the method in \citet{HallRacineLi}.
Both equations (\ref{EQ:Indep}) and (\ref{EQ:condIndep1}) can be
implied by condition (19) in \citet{HallRacineLi}. Alternatively,
we can express the sparsity of the conditional density of $V$ given $\boldsymbol{Z}_{1}^{p_{n}}$ solely through conditional independence,
specifically using equation (\ref{EQ:condIndep1}) only. However,
as discussed intensively in \citet{HallRacineLi}, the definition
of conditional independence per se can be ambiguous.\footnote{That is the following: $V\perp Z_{1}|Z_{2},$ $V\perp Z_{2}|Z_{1},$
but $V\not\perp$ $\left(Z_{1},Z_{2}\right)$ may hold at the same
time.} We refer readers to their paper's example of linear combinations
of standard normal random variables (on page 1016). To avoid such
ambiguity, \citet{HallRacineLi} focus on conventional independence,
which can imply both equations (\ref{EQ:Indep}) and (\ref{EQ:condIndep1}).

The DC screening procedure proposed below can only guarantee to identify
$(Z_{p^{*}+1},Z_{p^{*}+2}...,Z_{p_{n}})$ that satisfy both (\ref{EQ:Indep})
and (\ref{EQ:condIndep1}) in Assumption \ref{A:main}(2) as irrelevant.
There are situations where certain covariates are dependent on $V$,
but only indirectly through other covariates. For example, $V$ is
a function of $Z_{1}$, and $Z_{2}$ is correlated with $Z_{1}$ but
independent of all other determinants of $V$. This violates Assumption
\ref{A:main}(2), and as a result, the DC screening is likely to identify
$Z_{2}$ as relevant. To address this issue, we generalize the findings
in \citet{HallRacineLi} in Section \ref{SEC:further}, showing that
applying their CV-based bandwidth selection algorithm post-screening
can eliminate the influence of such $Z_{2}$ on $V$ when estimating
the conditional density of $V$ on $(Z_{1},Z_{2})$.

Assumption \ref{A:main}(3) imposes mild moment conditions for $V$
and $\boldsymbol{Z}_{1}^{p_{n}}$. It is apparent that higher dimension
$p_{n}$ requires more restrictive moment conditions, i.e., larger
$\delta$. We restrict $\delta\geq1$ so that the selection error
can be negligible when establishing Theorem \ref{TH:post-selection}.
The \emph{false discovery
rate} (FDR) defined in equation (\ref{EQ:FDR}) below can also be better
controlled for larger $\delta$ due to sharper error bounds. We allow
$V$ and $Z_{l}$ to have as low as $\left(2+\delta\right)$-th finite
moment, but jointly $\mathbb{E}(\left\vert VZ_{l}\right\vert ^{2+\delta})$
should be finite. To accommodate ultra-high dimensional data (i.e.,
$p_{n}\propto\exp(n^{C})$ for some positive $C$), we do need to
impose thin tail restrictions, such as sub-Gaussian, on $V$ and $\boldsymbol{Z}_{1}^{p_{n}}$ (see, e.g., Assumption C1 in \citet{LiZhongZhu}). However, ultra-high
dimensional data are rare in economics. In this sense, we consider
our assumption quite general. Assumption \ref{A:main}(4), which distinguishes
relevant variables from irrelevant ones, is also made in \citet{LiZhongZhu}

Let $\varsigma_{n}$ denote the threshold for the test statistic $\mathcal{\hat{T}}_{nl}$
defined in equation (\ref{EQ:statistic}). We determine $Z_{l}$ to
be relevant to $V$ if 
\[
\widehat{\mathcal{J}}_{l}\equiv\mathbf{1}(\mathcal{\hat{T}}_{nl}\geq\varsigma_{n})=1.
\]
We define the set of indices of the truly relevant variables and the
set of indices of selected relevant variables, respectively, as 
\begin{equation}
\mathcal{A}^{\ast}=\left\{ 1,2,...,p^{\ast}\right\} \text{ and }\widehat{\mathcal{A}}^{\ast}=\{l:\widehat{\mathcal{J}}_{l}=1,l=1,2,...,p_n\}.\label{EQ:Astarhat}
\end{equation}
Following the multiple test literature (see, e.g., \citet{BenjaminiHochberg}),
we adopt the \emph{true positive rate} (TPR) and the FDR defined as 
\begin{align}
\text{TPR}_{n} & =\frac{\sum_{l=1}^{p_{n}}\mathbf{1}(\widehat{\mathcal{J}}_{l}=1\text{ and }\mathcal{V}^{2}\left(V,Z_{l}\right)\neq0)}{\sum_{l=1}^{p_{n}}\mathbf{1}\left(\mathcal{V}^{2}\left(V,Z_{l}\right)\neq0\right)}\text{ and}\nonumber \\
\text{ FDR}_{n} & =\frac{\sum_{l=1}^{p_{n}}\mathbf{1}(\widehat{\mathcal{J}}_{l}=1\text{ and }\mathcal{V}^{2}\left(V,Z_{l}\right)=0)}{\sum_{l=1}^{p_{n}}\widehat{\mathcal{J}}_{l}+1}\label{EQ:FDR}
\end{align}
to measure the performance of the screening procedure proposed above.
The following results are on the Type-I and Type-II errors, which
are useful for showing the properties of TPR$_{n}$ and FDR$_{n}$
summarized in Theorem \ref{TH:PDR}.

\setlength{\parskip}{0pt}
\begin{theorem} \label{TH:main1} Suppose Assumption \ref{A:main}
holds. Let $\varsigma_{n}=C_{\varsigma}\sqrt{\kappa_{n}\log n}$ for
some positive $C_{\varsigma}$. 
\begin{enumerate}
\item[(1)] If $\mathcal{V}^{2}\left(V,Z_{l}\right)/S_{2}\left(V,Z_{l}\right)\lesssim n^{-1/2}\sqrt{\log n}$,
then 
\[
\Pr(\widehat{\mathcal{J}}_{l}=1)\leq C_{1}n^{-1-\delta/2}\varsigma_{n}^{-2+\delta}+C_{2}n^{-\delta}\varsigma_{n}^{-2},
\]
for some positive $C_{1}$ and $C_{2}$. 
\item[(2)] If $\mathcal{V}^{2}\left(V,Z_{l}\right)/S_{2}\left(V,Z_{l}\right)\gtrsim n^{-1/2}\kappa_{n}\sqrt{\log n}$,
then 
\[
\Pr(\widehat{\mathcal{J}}_{l}=1)\geq1-C_{3}n^{-1-\delta/2}\varsigma_{n}^{-2+\delta}-C_{4}n^{-\delta}\varsigma_{n}^{-2},
\]
for some positive $C_{3}$ and $C_{4}$. 
\end{enumerate}
\end{theorem}
\setlength{\parskip}{1em}

Theorem \ref{TH:main1} shows that with some suitably chosen $\varsigma_{n}$,
the proposed test statistic $\mathcal{\hat{T}}_{nl}$ can separate
the relevant variables from irrelevant variables with high probability.
However, there is some power loss up to $\sqrt{\kappa_{n}\log n}$.
The error bound is directly linked to $\delta$. The larger $\delta$
or more restrictive moment conditions are, the sharper the bound becomes.
There is some middle ground where our procedure is silent. Built on
Theorem \ref{TH:main1}, the theorem below shows the properties of
our screening procedure in terms of TPR$_{n}$ and FDR$_{n}$.

\setlength{\parskip}{0pt}
\begin{theorem} \label{TH:PDR} Suppose Assumption \ref{A:main}
holds. Let $\varsigma_{n}=C_{\varsigma}\sqrt{\kappa_{n}\log n}$ for
some positive $C_{\varsigma}$. Then 
\begin{align*}
 & \mathbb{E}\left(\text{\emph{TPR}}_{n}\right)=1-C_{1}n^{-1-\delta/2}\varsigma_{n}^{-2+\delta}-C_{2}n^{-\delta}\varsigma_{n}^{-2},\\
 & \text{\emph{FDR}}_{n}\overset{P}{\rightarrow}0,
\end{align*}
and 
\[
\text{\emph{Pr}}(\widehat{\mathcal{A}}^{\ast}=\mathcal{A}^{\ast})\rightarrow1,\newline\ 
\]
for some positive $C_{1}$ and $C_{2}$.
\end{theorem}
\setlength{\parskip}{1em}

In addition to the properties of TPR$_{n}$ and FDR$_{n}$, Theorem
\ref{TH:PDR} also establishes the ``oracle'' property that our
screening procedure can select the truly relevant variables with probability
approaching one. However, the theoretical rate of $\varsigma_{n}$ only
provides a limited guide in practice. For this reason, we propose
a practical procedure for choosing the threshold in the following
subsection, utilizing the CV approach proposed in \citet{HallRacineLi}.
It is worth noting that this procedure does not rely on the perfect
selection.

\subsection{Choice of $\varsigma_{n}$ and Post Screening Density Estimation\label{SEC:threshold}}

In this section, we first discuss how to choose $\varsigma_{n}$ based
on the results stated in Theorem \ref{TH:PDR}. Then, we propose a
more practical way of handling variable selection, which utilizes
the method by \citet{HallRacineLi}. After briefly reviewing their
method, we present our procedure and derive the convergence rate of
the post-selection conditional density estimator. This result is summarized
in Theorem \ref{TH:post-selection}.

Our procedure involves keeping as many as $Z_{l}$ that the method
of \citet{HallRacineLi} can handle from the initial screening stage,
after which we apply the method of \citet{HallRacineLi} to reduce
the impact of included irrelevant variables. This is similar to the
approach proposed by \citet{FanLv2008}, who suggested keeping 5\%
of the most correlated covariates and then using Lasso to refine the
variable selection further. More importantly, we will show in Section
\ref{SEC:further} that the method of \citet{HallRacineLi} can eliminate
the impact of certain covariates that may depend on $V$ but only
through some other relevant covariates.

\noindent \textbf{Choosing $\varsigma_{n}^{\ast}$ based on theoretical
results}

\noindent Based on the rates derived in Theorems \ref{TH:main1} and
\ref{TH:PDR}, one can choose 
\begin{equation}
\varsigma_{n}^{\ast}=Cn^{-1/2}\left(\log n\right)^{3/4}\label{EQ:xi_n_prac}
\end{equation}
for some positive constant $C$, which is equivalent to setting $\kappa_{n}=\sqrt{\log n}$.
By doing so, we can identify relevant $Z_{l}$ satisfying $\mathcal{V}^{2}\left(V,Z_{l}\right)/S_{2}\left(V,Z_{l}\right)\gtrsim n^{-1/2}\kappa_{n}\sqrt{\log n}=n^{-1/2}\log n$
with high probability. Note that this choice results in a loss of
power up to the order of $\log n$.

\noindent \textbf{A practical way of choosing $\varsigma_{n}^{\ast}$}

\noindent We propose choosing the threshold following the idea of
\citet{FanLv2008}. Specifically, we calculate $\mathcal{\hat{T}}_{nl}$
for all $l=1,2,...,p_{n}$ and then rank them from highest to lowest.
We then retain the $Z$s with the highest $\mathcal{\hat{T}}_{nl}$
values while ensuring that we do not include too many variables to
the extent that the method proposed in \citet{HallRacineLi} becomes
impractical. Denote the number of retained $Z$s as $\tilde{p}$.
Note that $\tilde{p}$ is a fixed integer that does not change with
$n$. Without loss of generality, assume that $Z_{1},Z_{2},...,Z_{\tilde{p}}$
are the $\tilde{p}$ variables with the highest $\mathcal{\hat{T}}_{nl}$
values. We then proceed to estimate the conditional density using
the method presented in \citet{HallRacineLi}, which effectively reduces
the dimension of the conditional set and automatically eliminates
the impact of irrelevant variables, as demonstrated in their work.

Here, we briefly overview the procedure proposed in \citet{HallRacineLi} with more details in Appendix \ref{app:CV}.
We recommend referring to the original paper for a more comprehensive
understanding of the details and theoretical properties. Throughout
our presentation, we will assume that all $Z$s are continuously distributed
for brevity. First, we define some technical terms: 
\begin{align*}
\hat{f}\left(z_{1},...,z_{\tilde{p}},v\right)= & n^{-1}\sum_{i=1}^{n}\left[\Pi_{l=1}^{\tilde{p}}K_{h_{l}}\left(z_{li}-z_{l}\right)\right]K_{h_{V}}\left(v_{i}-v\right),\\
\hat{f}\left(z_{1},...,z_{\tilde{p}}\right)= & n^{-1}\sum_{i=1}^{n}\Pi_{l=1}^{\tilde{p}}K_{h_{l}}\left(z_{li}-z_{l}\right),\text{ and}\\
\hat{G}\left(z_{1},...,z_{\tilde{p}}\right)= & n^{-2}\sum_{i_{1}=1}^{n}\sum_{i_{2}=1}^{n}\left\{ \left[\Pi_{l=1}^{\tilde{p}}K_{h_{l}}\left(z_{li_{1}}-z_{l}\right)\right]\left[\Pi_{l=1}^{\tilde{p}}K_{h_{l}}\left(z_{li_{2}}-z_{l}\right)\right]\right.\\
 & \left.\times\int K_{h_{V}}\left(v_{i_{1}}-v\right)K_{h_{V}}\left(v_{i_{2}}-v\right)dv\right\} ,
\end{align*}
where $K_{h}\left(x\right)=h^{-1}K\left(x/h\right)$ and $K(\cdot)$ is a
standard kernel function. The conditional density of $V$ on $\left(Z_{1},Z_{2},...,Z_{\tilde{p}}\right)$
is estimated as 
\begin{equation}
\hat{f}\left(v|z_{1},...,z_{\tilde{p}}\right)=\frac{\hat{f}\left(v,z_{1},...,z_{\tilde{p}}\right)}{\hat{f}\left(z_{1},...,z_{\tilde{p}}\right)}.\label{EQ:fv|z}
\end{equation}
The CV criterion is defined as 
\begin{equation}
\text{CV}\left(h_{V},h_{1},...,h_{\tilde{p}}\right)=\hat{I}_{n1}\left(h_{V},h_{1},...,h_{\tilde{p}}\right)-2\hat{I}_{n2}\left(h_{V},h_{1},...,h_{\tilde{p}}\right),\label{EQ:CV}
\end{equation}
where $\hat{I}_{n1}$ and $\hat{I}_{n2}$ are computed as 
\[
\hat{I}_{n1}\left(h_{1},...,h_{\tilde{p}}\right)=\frac{1}{n}\sum_{i=1}^{n}\frac{\hat{G}_{-i}\left(z_{1i},...,z_{\tilde{p}i}\right)}{\hat{f}_{-i}\left(z_{1i},...,z_{\tilde{p}i}\right)^{2}}\text{ and }\hat{I}_{n2}\left(h_{V},h_{1},...,h_{\tilde{p}}\right)=\frac{1}{n}\sum_{i=1}^{n}\frac{\hat{f}_{-i}\left(v_{i},z_{1i},...,z_{\tilde{p}i}\right)}{\hat{f}_{-i}\left(z_{1i},...,z_{\tilde{p}i}\right)}
\]
with $\hat{G}_{-i}$ and $\hat{f}_{-i}$ being the leave-one-out (leave
the $i$-th observation out) estimator. Note that we omit the weighting
function used in \citet{HallRacineLi} for conciseness.

We can compute the optimal bandwidth 
\begin{equation}
\hat{\boldsymbol{h}}\equiv(\hat{h}_{V},\hat{h}_{1},...,\hat{h}_{\tilde{p}})=\arg\min_{(h_{V},h_{1},...,h_{\tilde{p}})}\text{CV}\left(h_{V},h_{1},...,h_{\tilde{p}}\right),\label{EQ:bandwidth}
\end{equation}
and then plug $\hat{\boldsymbol{h}}$ into equation (\ref{EQ:fv|z})
to obtain $\hat{f}\left(v|z_{1},...,z_{\tilde{p}}\right)$.

As demonstrated in Theorem 2 of \citet{HallRacineLi}, the CV procedure
(\ref{EQ:bandwidth}) can select asymptotically optimal bandwidths
for relevant variables and push the bandwidths for irrelevant variables
to their upper limits, effectively making them appear uniform in the
conditional set and thus less informative to $V$, with the probability
approaching one as $n\rightarrow\infty$.

We provide the convergence rate of the post-selection conditional
density estimation in Theorem \ref{TH:post-selection}, which is an
immediate result from Theorem 2 of \citet{HallRacineLi} and our Theorem
\ref{TH:PDR}. The following technical conditions are necessary, akin to those in \citet{HallRacineLi}. One note is that we need
to assume $\tilde{p}\geq p^{\ast}$ to ensure we do not miss any relevant
$Z$s. Although we assume all covariates are continuous below,
our procedure can also accommodate discrete variables. Demonstrating
the validity of this extension is straightforward, given \citet{HallRacineLi}.

\begin{assumptionp}{2}\label{A:continuousf_kernelK} 
\begin{enumerate}
\item[(1)] $V$ and $\boldsymbol{Z}_{1}^{p_{n}}$ are continuous random variables.$\ $The
densities $f\left(v,z_{1},...,z_{p^{\ast}}\right)$, $f\left(z_{1},...,z_{p^{\ast}}\right),$
and $f\left(v|z_{1},...,z_{p^{\ast}}\right)$ are bounded and $r$-th
order continuously differentiable for an even positive integer $r$. 
\item[(2)] $K\left(\cdot\right)$ is nonnegative, symmetric about 0, and compactly
supported. $K\left(0\right)\neq0$, $\int K\left(u\right)du=1,$ $\int K\left(u\right)^{2}du<\infty,$
$\int u^{j}K\left(u\right)du=0$ for $j=1,2,...,r-1,$ and $\int u^{r}K\left(u\right)du\neq0$. 
\item[(3)] $\tilde{p}\geq p^{\ast}$. 
\end{enumerate}
\end{assumptionp}

\begin{theorem} \label{TH:post-selection}Suppose Assumptions \ref{A:main}
and \ref{A:continuousf_kernelK} hold. Then the $\hat{f}\left(v|z_{1},...,z_{\tilde{p}}\right)$
estimated with the bandwidths $\hat{\boldsymbol{h}}$ obtained from
equation (\ref{EQ:bandwidth}) satisfies 
\[
\hat{f}\left(v|z_{1},...z_{\tilde{p}}\right)-f\left(v|z_{1},...,z_{p^{\ast}}\right)=O_{P}\left(n^{-\frac{r}{p^{\ast}+1+2r}}\right).
\]
Furthermore, if $r>\left(p^{\ast}+1\right)/2,$ $\hat{f}\left(v|z_{1},...,z_{\tilde{p}}\right)-f\left(v|z_{1},...,z_{p^{\ast}}\right)=o_{P}\left(n^{-1/4}\right).$
\end{theorem}

Note that we omit some minor technical details required for the above
theorem, specifically, conditions (22)--(24) in \citet{HallRacineLi}.
The main implication of Theorem \ref{TH:post-selection} is that irrelevant
variables have no impact on the convergence rate of $\hat{f}\left(v|z_{1},...,z_{\tilde{p}}\right)$
since the fastest convergence rate achievable for $\hat{f}\left(v|z_{1},...,z_{\tilde{p}}\right)$
would be $n^{-\frac{r}{p^{\ast}+1+2r}}$ if we only included the $p^{\ast}$
truly relevant variables in the estimation. When $r>\left(p^{\ast}+1\right)/2$,
the nonparametric estimation has an error on the order of $o_{P}\left(n^{-1/4}\right)$,
which meets the minimum requirement for parametric estimation with
nuisance nonparametric estimation as the first step. The $o_{P}\left(n^{-1/4}\right)$
holds uniformly (with some loss of power of $\log n$), which can
be demonstrated easily as in \citet{LiRacine}. Since $p^{\ast}$
is unknown, practitioners may want to choose a high-order kernel with
order $r>\left(\tilde{p}+1\right)/2.$

By utilizing the method proposed in \citet{HallRacineLi}, we do not
need a perfect variable selection during the initial screening for
conditional density estimation. We may be reasonably conservative
and include some potentially irrelevant variables in the first-step
screening, but they will have no impact on the asymptotic convergence
rate. It is worth noting that \citet{HallRacineLi} does not eliminate
these irrelevant variables from the estimation. Instead, their method
reduces the influence of irrelevant variables by assigning them large
bandwidths selected through the CV procedure (\ref{EQ:bandwidth}).

\subsection{Some Further Results without the Conventional Independence\label{SEC:further}}

Consider the following illustrative scenario: 
\begin{equation}
V=g\left(Z_{1},e\right),\text{ }Z_{1}=U^{\ast}+U_{1},\text{ }\left(Z_{2},Z_{3},...,Z_{s}\right)\perp\left(U_{1},e\right),\text{ for some }s>1,\label{EQ:V_Z_Z}
\end{equation}
where $g(\cdot,\cdot)$ is an unknown continuous function. Apparently,
$V\perp\left(Z_{2},Z_{3},...,Z_{s}\right)|Z_{1}$, but $V\not\perp Z_{1}|(Z_{2},...,Z_{s})$
due to the presence of $U_{1}$. Thus, the conditional independence
is unambiguous in this case. However, it is possible that $\left(Z_{2},Z_{3},...,Z_{s}\right)$
and $V$ are dependent through $U^{\ast}$. If such dependence exists,
then (\ref{EQ:V_Z_Z}) violates Assumption \ref{A:main}(2) (and hence
condition (19) in \citet{HallRacineLi}). Proposition \ref{CORO:Post-Estimation}
presented below demonstrates that the procedure proposed in Section
\ref{SEC:threshold} can be applied to scenario (\ref{EQ:V_Z_Z}),
yielding the same convergence rate as obtained in Theorem \ref{TH:post-selection}.
This broadens the applicability of the theory of \citet{HallRacineLi}.

Note that in our exposition, we use (\ref{EQ:V_Z_Z}) as an illustrating
example, where only a scalar $Z_{1}$ enters $g(\cdot,\cdot)$. We
intentionally adopt this simplified setup to facilitate understanding
and avoid tedious discussions on the potential ambiguity of conditional
(in)dependence that can arise with different sub-vectors of $(Z_{1},...,Z_{s})$.
In practice, when the conditional independence is ambiguous, the CV
criterion may have multiple local minima, and we may be unable to
identify the global minimum. Extending our approach to cases with
multiple relevant $Z$s in $g(\cdot,\cdot)$ is straightforward, albeit
more technically involved. Due to space constraints, we do not pursue
this direction formally in this paper.

\begin{proposition} \label{CORO:Post-Estimation}Suppose we have i.i.d.
data $\left(v_{i},z_{1i},...,z_{si}\right),$ $i=1,2,...,n$, from
the model (\ref{EQ:V_Z_Z}), and Assumption \ref{A:continuousf_kernelK}
holds. Then the $\hat{f}\left(v|z_{1},...,z_{s}\right)$ estimated
with the bandwidths obtained in equation (\ref{EQ:bandwidth}) satisfies 
\[
\hat{f}\left(v|z_{1},...,z_{s}\right)-f\left(v|z_{1}\right)=O_{P}\left(n^{-\frac{r}{2+2r}}\right).
\]
\end{proposition}

To understand the significance of the above corollary, note that it
obtains the same rate as Theorem \ref{TH:post-selection} when $p^{\ast}=1$.

However, our method is unsuitable in scenarios where all noise variables
are highly dependent on $V$ (e.g., through $Z_{1}$), as our screening
procedure fails to exclude these noise variables.

\subsection{The Practical Procedure of the Conditional Density Estimation\label{SEC:proceduref}}

\setlength{\parskip}{0pt}
We conclude this section by summarizing our proposed procedure for conditional density estimation with dimension reduction: 
\begin{itemize}
\item[-] \textbf{Step 1}: Calculate $\mathcal{\hat{T}}_{nl}$ defined as in
(\ref{EQ:statistic}) for all $l=1,2,...,p_{n}$. 
\item[-] \textbf{Step 2}: Rank $\mathcal{\hat{T}}_{nl}$ values from largest
to smallest. Keep $\tilde{p}$ elements (as many as possible while
ensuring the method in \citet{HallRacineLi} can handle them) of $Z$s
with the highest $\mathcal{\hat{T}}_{nl},$ say, $Z_{1},Z_2,...,Z_{\tilde{p}}$. 
\item[-] \textbf{Step 3}: Find the asymptotically optimal bandwidths $\hat{\boldsymbol{h}}$
that minimize the CV criterion (\ref{EQ:CV}). 
\item[-] \textbf{Step 4}: Substitute $\hat{\boldsymbol{h}}$ into equation
(\ref{EQ:fv|z}) to obtain the conditional density estimate: 
\begin{equation}
\hat{f}(v|\boldsymbol{z}_{1}^{\tilde{p}})\equiv\hat{f}(v|z_{1},...,z_{\tilde{p}}).\label{EQ:ftidef}
\end{equation}
\end{itemize}
\setlength{\parskip}{1em}


\section{High Dimensional Binary Choice Model\label{SEC:theoryapply}}

In this section, we explore high-dimensional binary choice models.
We start by using the feasible estimator $\hat{f}(v|\boldsymbol{z}_{1}^{\tilde{p}})$,
as derived in Section \ref{SEC:threshold}, to approximate $f\left(v|\boldsymbol{z}_{1}^{p_{n}}\right)$.
This leads to the feasible $\tilde{Y}$ in equation (\ref{EQ:Ytide}).
We then apply the special regressor approach in high-dimensional settings.
The analysis in this section follows existing methods but incorporates
an estimated $\tilde{Y}$.

To facilitate understanding in this section, consider viewing $\hat{f}(v|\boldsymbol{z}_{1}^{\tilde{p}})$
from the previous section as a ``regular'' nonparametric estimator
and ignore that it is a post-selection estimator. Then, the
estimators introduced in this section are treated as building upon existing methods
but incorporating a nonparametric first-step plug-in. We emphasize
that we have accounted for the selection issue related to the estimation
of $f(v|\boldsymbol{z}_{1}^{p_{n}})$ when deriving all the theoretical
results. This treatment is demonstrated in Appendices
\ref{APP:B1} and \ref{APP:C}.

We begin by reviewing the special regressor estimator and presenting
the conditions required to validate the moment conditions
defined in equation (\ref{EQ:momentCond}). With a feasible estimator
for the conditional density, these moment
conditions enable the estimation of the model parameters. We can then
apply either an LS or a GMM estimator, incorporating appropriate regularization techniques
commonly used for linear models. To illustrate this, we explore the
classic variable selection problem in Section \ref{SEC:VariableSelection} and discuss how to choose valid moment
conditions from a large set of candidate IVs in Section \ref{SEC:MSelection}.
Other scenarios, such as selecting optimal instruments (see, e.g.,
\citet{Belloni_et_al2012}), can be similarly studied but may involve
more complex technical details.

It is important to note that the problems discussed in Sections \ref{SEC:VariableSelection}
and \ref{SEC:MSelection} are fundamentally distinct: The notations
used in each section are specific to their context and may have different
meanings, even though they might appear identical.

\subsection{Review of the Special Regressor Estimator}

The following technical conditions placed in \citet{Lewbel2000} are
required for the validity of the special regressor estimator. \begin{assumptionp}
{3}\label{A:specialR} 
\begin{enumerate}
\item[(1)] Equation (\ref{EQ:model}) holds. 
\item[(2)] The error $\varepsilon$ and covariates in equation (\ref{EQ:model})
satisfy $V\perp\left(\varepsilon,\boldsymbol{X}_{1}^{s^{\ast}}\right)|\boldsymbol{Z}_{1}^{p_{n}}$. 
\item[(3)] The conditional distribution of $V$ given $\boldsymbol{Z}_{1}^{p_{n}}$
is absolutely continuous, and has support $\left[L,K\right]$ for
some constants $L$ and $K,$ $-\infty\leq L<0\leq K\leq\infty.$
The support of $-X_{1}\beta_{1}^{\ast}-X_{2}\beta_{2}^{\ast}-...-X_{s^{\ast}}\beta_{s^{\ast}}^{\ast}-\varepsilon$
is a subset of the interval $\left[L,K\right]$. 
\item[(4)] $f\left(V|\boldsymbol{Z}_{1}^{p_{n}}\right)$ is bounded and bounded
away from zero. 
\end{enumerate}
\end{assumptionp}

Note that in the case where all candidate covariates $\boldsymbol{X}_{1}^{p_{n}}$
are exogenous, $\boldsymbol{Z}_{1}^{p_{n}}$ is simply $\boldsymbol{X}_{1}^{p_{n}}$,
and hence Assumption \ref{A:specialR}(2) can be expressed as $V\perp\varepsilon|\boldsymbol{X}_{1}^{p_{n}}$.
Assumption \ref{A:specialR}(3) is the ``large support'' condition
for $V$. This condition is similar to the ``overlap'' condition for the average
treatment effects estimator, which assumes that the propensity score
is bounded away from zero and one. The ``large support'' condition
can be restrictive. For instance, the special regressor ``age''
in \citet{DongLewbel2015} and ``precipitation'' in \citet{XueYangZhou} have bounded support, which excludes the inclusion of $\varepsilon$
with larger support. Lastly, Assumption \ref{A:specialR}(4) is a technical condition
that is necessary to prevent the ``irregular slower-than-$\sqrt{n}$
convergence'' property, as explained in \citet{KhanTamer}. 

The special regressor approach comes at the cost of imposing strong
assumptions on $V$. Specifically, these assumptions require $V$
to have strong exogeneity and large support. However, despite these
constraints, these assumptions yield the following useful identity:
\[
\mathbb{E}[\tilde{Y}-X_{1}\beta_{1}^{\ast}-X_{2}\beta_{2}^{\ast}-...-X_{s^{\ast}}\beta_{s^{\ast}}^{\ast}\left\vert \boldsymbol{Z}_{1}^{p_{n}}\right.]=\mathbb{E}\left[\varepsilon\left\vert \boldsymbol{Z}_{1}^{p_{n}}\right.\right]
\]
with 
\begin{equation}
\tilde{Y}=\frac{Y-\boldsymbol{1}\left(V>0\right)}{f\left(V|\boldsymbol{Z}_{1}^{p_{n}}\right)},\label{EQ:Ytide_repeat}
\end{equation}
which implies the moment conditions that resemble those for the linear
IV models: For $j\in\{1,...,p_{n}\}$, 
\[
\mathbb{E}[Z_{j}(\tilde{Y}-X_{1}\beta_{1}^{\ast}-X_{2}\beta_{2}^{\ast}-...-X_{s^{\ast}}\beta_{s^{\ast}}^{\ast})]=0,
\]
if $\mathbb{E}\left[\varepsilon|Z_{j}\right]=0$ is satisfied.\footnote{To see this, consider 
\begin{align*}
\mathbb{E}[Z_{j}(\tilde{Y}-X_{1}\beta_{1}^{\ast}-X_{2}\beta_{2}^{\ast}-...-X_{s^{\ast}}\beta_{s^{\ast}}^{\ast})] & =\mathbb{E}[Z_{j}\mathbb{E}(\tilde{Y}-X_{1}\beta_{1}^{\ast}-X_{2}\beta_{2}^{\ast}-...-X_{s^{\ast}}\beta_{s^{\ast}}^{\ast}\left\vert \boldsymbol{Z}_{1}^{p_{n}}\right.)]\\
 & =\mathbb{E}\left[Z_{j}\mathbb{E}\left(\varepsilon\left\vert \boldsymbol{Z}_{1}^{p_{n}}\right.\right)\right]=\mathbb{E}\left[Z_{j}\mathbb{E}\left(\varepsilon\left\vert Z_{j}\right.\right)\right]=0.
\end{align*}
} We refer interested readers to \citet{Lewbel2000} for a detailed
discussion on the technical conditions, the proofs, and the estimators.
In subsequent sections, we explore the applications of the special
regressor estimators in high-dimensional settings.

\subsection{Adaptive Lasso for the Binary Choice Model\label{SEC:VariableSelection}}

\subsubsection{Model and Results}

In this section, we attempt to replicate some findings regarding the
adaptive Lasso in a high-dimensional setting. The adaptive Lasso,
introduced by \citet{Zou2006}, is a widely used penalized regression
method in econometrics and statistics. This approach was extended
to high-dimensional settings in subsequent studies (\citet{HuangEtal2008}
and \citet{LinEtal2009}). \citet{HuangEtal2008} examined scenarios
with an extremely high dimension, where the number of covariates increases
exponentially. In these cases, it was assumed that both the distributions
of covariates and error terms decay at an exponential rate to manage
the extremely high-dimensional covariates. On the other hand, \citet{LinEtal2009}
eased the tail (or moment) conditions set by \citet{HuangEtal2008}
and concentrated on moderately high-dimensional situations, where
the number of covariates increases at polynomial rates.

In typical economic applications, while economists might handle numerous covariates, their number is generally far less than that of observations. Additionally, many economic variables, such as wages, exhibit heavy-tailed distributions, making the assumption of light tails rather restrictive. Given these considerations, we focus on replicating results for the scenario described in \citet{LinEtal2009}, as it presents a more practical case for economic applications.

Suppose the true model is given by 
\begin{equation}
Y=\mathbf{1}\left(V+X_{1}\beta_{1}^{*}+X_{2}\beta_{2}^{*}+\cdots+X_{s^{*}}\beta_{s^{*}}^{*}+\varepsilon>0\right),\label{EQ:model_Y-1}
\end{equation}
where $(X_{1},\dots,X_{s^{*}})$ are the true signals among all candidate
covariates $\boldsymbol{X}_{1}^{p_{n}}$. Researchers do not know which
components of $\boldsymbol{X}_{1}^{p_{n}}$ correspond to $(X_{1},\dots,X_{s^{*}})$
until the data reveal this. We assume $\boldsymbol{X}_{1}^{p_{n}}$ to
be exogenous, and thus it plays the role of $\boldsymbol{Z}_{1}^{p_{n}}$
in Assumption~\ref{A:specialR}. We use the notation $X$ instead
of $Z$ in this section to avoid confusion. The true parameter values
are collected into a $p_{n}\times1$ vector $\boldsymbol{\beta}^{*}=(\beta_{1}^{*},\beta_{2}^{*},\dots,\beta_{s^{*}}^{*},0,0,\dots,0)^{\prime}$,
corresponding to the true signal variables, followed by zeros for
the remaining candidate covariates. It is worth noting that the true
signals, $(X_{1},\dots,X_{s^{*}})$, may or may not overlap with the
$X$s relevant to $V$, because the two issues (namely, ``true signals
in the binary choice model'' and ``relevant $X$s to $V$'') are
generally independent. However, we do not reorder $X$ to distinguish
these two issues in this section for notational convenience.

We denote the feasible estimator of $\tilde{y}_{i}$ by 
\begin{equation}
\widehat{\tilde{y}}_{i}=\frac{y_{i}-\mathbf{1}(v_{i}>0)}{\hat{f}(v_{i}|\boldsymbol{x}_{1i}^{\tilde{p}})},\label{eq:feasibley}
\end{equation}
where $\hat{f}(v_{i}|\boldsymbol{x}_{1i}^{\tilde{p}})$ is obtained
using the procedure outlined in Section \ref{SEC:proceduref}, but
with a $2r$-th order kernel function (refer to Theorem \ref{TH:AdaptiveLasso}
and the subsequent discussion for a detailed explanation). 

The adaptive
Lasso estimator for our case is then obtained by: 
\begin{equation}
\boldsymbol{\hat{\beta}}=\arg\min_{\boldsymbol{\beta}}\sum_{i=1}^{n}(\widehat{\tilde{y}}_{i}-\boldsymbol{x}_{i}^{\prime}\boldsymbol{\beta})^{2}+\lambda_{n}\sum_{j=1}^{p_{n}}\varpi_{j}|\beta_{j}|,\label{eq:adaptiveLasso}
\end{equation}
where $\boldsymbol{x}_{i}\equiv(x_{1i},\dots,x_{s^{*}i},\dots,x_{p_{n}i})^{\prime}$,
and 
\[
\varpi_{j}=|\tilde{\beta}_{j}|^{-\gamma},\quad j=1,\dots,p_{n},
\]
are adaptive weights used for penalizing different coefficients in the $L_1$ penalty, with $\gamma>0$, and $\tilde{\beta}_{j}$ being an initial estimator, as discussed in
Section \ref{sec:Initial-Estimator}.

We make the following assumptions, which are similar to those placed
in \citet{LinEtal2009}.
\begin{assumptionp}{4}\label{A:adaptiveLasso} 
\begin{enumerate}
\item[(1)] There exists a positive constant $D_{3}$ such that $\rho_{\min}\left(\mathbb{E}\left(\boldsymbol{X}_{1}^{s^{\ast}}\boldsymbol{X}_{1}^{s^{\ast}\prime}\right)\right)\geq D_{3}$.
Furthermore, $\max_{l=1,2,\ldots,p_{n}}\{\mathbb{E}(\left\vert X_{l}\right\vert ^{2+\delta})\}<\infty$
for some $\delta>4.$ 
\item[(2)] The initial estimator $\tilde{\boldsymbol{\beta}}$ satisfies 
\[
\max_{j=1,\ldots,p_{n}}\left|\tilde{\beta}_{j}-\beta_{j}\right|=O_{P}\left(n^{-\varrho}\right),\varrho>0,
\]
and there exist positive constants $\alpha$ and $C$ such that 
\begin{equation}
\min_{j=1,\ldots,s^{\ast}}\left\vert \beta_{j}\right\vert \geq Cn^{-\alpha},\quad\alpha<\frac{1}{2}.\label{eq:beta-min}
\end{equation}
Moreover, $\varrho>\alpha.$ 
\item[(3)] $s^{\ast}\propto n^{C_{s}}$, $p_{n}\propto n^{C_{p}}$, $\lambda_{n}\propto n^{C_{\lambda}}$,
for some non-negative $C_s$, $C_p$, and $C_\lambda$. Further, $C_{s}<\min\left\{ 1-2\alpha,\frac{1}{2}\right\} $,
$C_{p}<1$, 
\begin{equation}
\frac{1}{2}-\varrho\gamma+C_{s}<C_{\lambda}<\frac{1}{2}-\alpha\gamma-\frac{C_{s}}{2},\quad\gamma>\max\left\{ \frac{C_{s}}{\varrho-\alpha},\frac{3C_{s}/2}{\varrho-\alpha}\right\} .\label{eq:tuningrestric}
\end{equation}
\item[(4)] $\Omega_{n}$ defined in (\ref{EQ:omegan-1}) is positive definite
with finite eigenvalues. 
\end{enumerate}
\end{assumptionp}

While \citet{LinEtal2009} assumed $\mathbf{X}_{n} \equiv (\boldsymbol{x}_{1},\ldots,\boldsymbol{x}_{n})'$ to be non-random, we extend their work to accommodate random $\mathbf{X}_{n}$. Assumption \ref{A:adaptiveLasso}(1) is the full rank condition, the same as in \citet{LinEtal2009}. Assumption \ref{A:adaptiveLasso}(2) assumes the existence of an initial estimator converging to the true parameters at a polynomial rate and imposes the so-called ``beta-min" condition, which is crucial for deriving the ``oracle" property and conducting inference. Otherwise, well-known impossibility results for the uniform validity of post-selection estimators, such as those obtained in \citet{LeebPotscher2006,LeebPotscher2008}, apply. Assumption \ref{A:adaptiveLasso}(4) is standard considering $\Omega_{n}$ is a variance matrix. We now discuss Assumption \ref{A:adaptiveLasso}(3) in detail. We impose a slightly stricter condition than \citet{LinEtal2009} due to the relaxation of their setting to one with random $\mathbf{X}_{n}$. The trade-off is that the rate of $s^{*}$ imposes some restrictions on the rates of tuning parameters as described in (\ref{eq:tuningrestric}). For intuition, by setting $C_{s}=0$, we need $1/2-\varrho\gamma<C_{\lambda}<1/2-\alpha\gamma$, and this set is not empty because $\alpha<\varrho$. For $\gamma$, we only require $\gamma>0$. It is clear that the larger the $\gamma$, the wider the valid range of $C_{\lambda}$. When $C_{s}$ is positive, the range of valid $C_{\lambda}$ is narrower, and we need $\gamma$ to be sufficiently large to ensure the range of valid $C_{\lambda}$ is not empty.

With these technical conditions in place, we establish theoretical
results for the adaptive Lasso estimator, akin to Theorem 1 of \citet{LinEtal2009}.
It is important to note that our goal is not to improve existing results
on Lasso estimation, which is beyond the scope of this paper. We define
\[
\Sigma_{n}\equiv\left(\left[\mathbb{E}\left(\boldsymbol{X}_{1}^{s^{*}}\boldsymbol{X}_{1}^{s^{*}\prime}\right)\right]^{-1}\Omega_{n}\left[\mathbb{E}\left(\boldsymbol{X}_{1}^{s^{*}}\boldsymbol{X}_{1}^{s^{*}\prime}\right)\right]^{-1}\right),
\]
which is useful for the asymptotics of $\hat{\boldsymbol{\beta}}$. $\Omega_n$ is defined in (\ref{EQ:omegan-1}).

\begin{theorem} \label{TH:AdaptiveLasso} Suppose Assumptions \ref{A:main},
\ref{A:continuousf_kernelK}, \ref{A:specialR}, and \ref{A:adaptiveLasso}
hold, and $\delta\geq\frac{2r}{p^{*}+1+2r}$ with $r>\frac{p^{*}+1}{2}$.
Additionally, we adopt a $2r$-th order kernel to construct $\hat{f}(v|\boldsymbol{x}_{1}^{\tilde{p}})$
for $\widehat{\tilde{y}}_{i}$. Then $\hat{\boldsymbol{\beta}}^{(2)}=\boldsymbol{0}$
with probability approaching one, and for any $s^{*}\times1$ vector
$\boldsymbol{e}$ with $\|\boldsymbol{e}\|=1$, 
\[
\sqrt{n}\boldsymbol{e}'\Sigma_{n}^{-1/2}\left(\hat{\boldsymbol{\beta}}^{(1)}-\boldsymbol{\beta}^{*(1)}\right)\stackrel{d}{\rightarrow}N(0,1),
\]
where $\hat{\boldsymbol{\beta}}^{(1)}$ denotes the first $s^{*}$
elements of $\hat{\boldsymbol{\beta}}$, $\hat{\boldsymbol{\beta}}^{(2)}$
denotes the remaining $p_{n}-s^{*}$ elements of $\hat{\boldsymbol{\beta}}$,
and $\boldsymbol{\beta}^{*(1)}$ is defined similarly. \end{theorem}

In this theorem, we require that $r>\frac{p^{*}+1}{2}$ so that 
\[
\hat{f}(v|\boldsymbol{x}_{1}^{\tilde{p}})-f(v|\boldsymbol{x}_{1}^{p^{*}})=o_{P}\left(n^{-1/4}\right)
\]
as shown in Theorem \ref{TH:post-selection}. Furthermore, we adopt
a $2r$-th order kernel for $\hat{f}(v|\boldsymbol{x}_{1}^{\tilde{p}})$
to ensure that the bias term is of order $o\left(n^{-1/2}\right)$,
which is required for the $\sqrt{n}$-convergence of the parameters,
as explained in \citet{Lewbel2000}.

The above theorem presents the ``oracle'' property, implying the
``super efficiency'' that the procedure can effectively eliminate
irrelevant regressors with probability approaching one, and the resulting
estimator has the same asymptotic properties as one obtained using
only the true signals. We can obtain this result thanks to the ``beta-min''
condition. A relaxation of this condition and the related error bounds
for the Lasso estimator can be found in Appendix \ref{APP:lasso}.

We recommend that practitioners select $\lambda_n$ by 10-fold CV and experiment with a range of $\gamma$ values, choosing the value at which the results stabilize. For further guidance and illustrating examples, we refer readers to Sections \ref{SEC:simu} and \ref{SEC:applications}.

A final note is that we use the U-statistics technique to derive the
asymptotic property of our estimator $\hat{\boldsymbol{\beta}}$. As an anonymous referee pointed out, an
interesting alternative way is to find the Neyman-orthogonal score
with respect to $\hat{f}(v|x)$ and apply empirical processes techniques
(e.g., \citet{Andrews1994}) afterwards. We note that finding the Neyman-orthogonal score in our case is challenging and requires considerable effort. Unlike in  \citet{ChernozhukovEtal2022} and many related papers, where the first-step estimation involves $X$, our first step is to obtain the dependent variable $\widehat{\tilde{Y}}$ for the final estimation. This distinction inhibits us from directly applying the results in existing literature. To apply empirical processes techniques,
we need to verify that $\hat{f}(v|x)$ belongs to a Donsker class.
Note that $\hat{f}(v|x)$ involves bandwidths as tuning parameters.
Typically, different technical conditions are required to handle $\hat{f}(v|x)$
in various scenarios. For example, see \citet{MammenEtal2011}, \citet{Escanciano2014},
and \citet{SeoOtsu2018}. We leave this direction for future work.

\subsubsection{Initial Estimator\label{sec:Initial-Estimator}}
We recommend two options for the initial estimator: Ordinary Least Squares (OLS) and the Lasso estimator, with the latter detailed in Appendix \ref{APP:lasso}. If the number of covariates is considerably smaller than the number of observations and the Gram matrix of covariates ($n^{-1}\mathbf{X}_{n}'\mathbf{X}_{n}$) is full rank, we recommend using the OLS estimator for its simplicity. However, if the number of covariates is large or the Gram matrix is nearly singular, we suggest the Lasso estimator, as outlined in Appendix \ref{APP:lasso}.

In this section, we cite a result from \citet{LinEtal2009}, which demonstrates that the OLS estimator can yield consistent estimates that may serve as initial estimators. For details on the Lasso estimator, we refer readers to Appendix \ref{APP:lasso}. Let $\widehat{\mathbf{\tilde{Y}}} \equiv (\widehat{\tilde{y}}_{1}, \ldots, \widehat{\tilde{y}}_{n})'$. By definition, the OLS estimator is given by 
\[
\tilde{\boldsymbol{\beta}} = \left(\mathbf{X}_{n}'\mathbf{X}_{n}\right)^{-1}\mathbf{X}_{n}'\widehat{\mathbf{\tilde{Y}}}.
\]
The following result is Proposition 2.1 of \citet{LinEtal2009}, if we adopt $\mathbf{\tilde{Y}}$ in $\tilde{\boldsymbol{\beta}}$. The generalization of using $\widehat{\mathbf{\tilde{Y}}}$ instead of $\mathbf{\tilde{Y}}$ is straightforward given the proof of Theorem \ref{TH:AdaptiveLasso}. We omit the proof for brevity. 
\begin{proposition} Suppose $n^{-1}\mathbf{X}_{n}'\mathbf{X}_{n}$
is full rank, and let $\rho_{n}=\rho_{\min}(n^{-1}\mathbf{X}_{n}'\mathbf{X}_{n})$.
Then, under Assumption \ref{A:adaptiveLasso}, 
\[
\max_{j=1,\ldots,p_{n}}\left|\tilde{\beta}_{j}-\beta_{j}^{*}\right|=O_{P}\left(\rho_{n}^{-1}n^{-\frac{1-C_{p}}{2}}\right).
\]
\end{proposition}

Clearly, if we can find a $\varrho>0$ such that 
\[
\rho_{n}^{-1}n^{-\frac{1-C_{p}}{2}}\apprle n^{-\varrho},
\]
then $\tilde{\boldsymbol{\beta}}$ is a valid candidate for the initial
estimator.

\subsection{Moment Selection}\label{SEC:MSelection}
Assuming certain moment conditions hold, \citet{Liao2013} examined the selection of valid moment conditions from a fixed set of candidates for GMM estimation. \citet{ChengLiao2015} expanded this approach by selecting moment conditions that are both valid and relevant while allowing for an increasing number of potential IVs. For simplicity, we focus solely on validity in this section, assuming all IVs are relevant. While it is possible to consider both validity and relevance, doing so introduces additional complexity. In what follows, we adopt the approach of \citeauthor{ChengLiao2015} (\citeyear{ChengLiao2015}), allowing for an increasing number of potential IVs, and aim to replicate their results under similar technical conditions.

The model remains unchanged: 
\[
Y=\boldsymbol{1}\left(V+X_{1}\beta_{1}^{\ast}+X_{2}\beta_{2}^{\ast}+...+X_{s^{\ast}}\beta_{s^{\ast}}^{\ast}+\varepsilon>0\right).
\]
Following \citet{Liao2013} and \citet{ChengLiao2015}, we assume
that we have prior knowledge of the first $k^{\ast}$ ($k^{\ast}\geq s^{\ast}$)
IVs, denoted as $\boldsymbol{Z}^{\ast}\equiv\left(Z_{1},...,Z_{k^{\ast}}\right)^{\prime}$,
all of which are valid for identifying the parameters $\boldsymbol{\beta}^{\ast}\equiv\left(\beta_{1}^{\ast},\beta_{2}^{\ast},...,\beta_{s^{\ast}}^{\ast}\right)^{\prime}$.\footnote{Note that $\boldsymbol{Z}^{\ast}$ (known valid IVs) may or may not
overlap with $\boldsymbol{Z}_{1}^{p^{\ast}}$ ($Z$s relevant to $V$)
defined in Section \ref{SEC:f(v,x)reduce}.} We assume $k^{\ast}$ is fixed to simplify analysis. There are other
valid IVs, denoted by $\underline{\boldsymbol{Z}}_{A}=(Z_{k^{*}+1},Z_{k^{*}+2},...,Z_{k^{*}+d_{A}})^{\prime}$,
which are mixed with invalid IVs, denoted by 
\[
\underline{\boldsymbol{Z}}_{B}=(Z_{k^{*}+d_{A}+1},Z_{k^{*}+d_{A}+1},...,Z_{k^{*}+d_{A}+d_{B}})^{\prime}.
\]
We do not know which IVs are valid or invalid. We use $A$ to denote
the indices of the valid instruments ($\underline{\boldsymbol{Z}}_{A}$),
$B$ to denote the indices of the invalid instruments ($\underline{\boldsymbol{Z}}_{B}$),
and $D=A\cup B$ to denote the indices for all candidate IVs to be
selected. The objective is to choose the valid IVs from $\underline{\boldsymbol{Z}}_{D}=(\underline{\boldsymbol{Z}}_{A}^{\prime},\underline{\boldsymbol{Z}}_{B}^{\prime})'$.
We refer to \citet{Liao2013} and \citet{ChengLiao2015} for more
information and applications related to this setting. We denote 
\[
p_{n}\equiv k^{\ast}+d_{A}+d_{B}.
\]
With this notation, $\boldsymbol{Z}_{1}^{p_{n}}=\left(\boldsymbol{Z}^{\ast\prime},\underline{\boldsymbol{Z}}_{A}^{\prime},\underline{\boldsymbol{Z}}_{B}^{\prime}\right)^{\prime}$.
To further simplify notations, we denote $\boldsymbol{X}\equiv\left(X_{1},...,X_{s^{\ast}}\right)^{\prime}$
and $\boldsymbol{\beta}\equiv\left(\beta_{1},...,\beta_{s^{\ast}}\right)^{\prime}$.
For the same reason as in the last section, the two issues (namely,
``valid IV'' and ``relevant $Z$s to $V$'') are generally independent.
We do not reorder $Z$ to distinguish these two issues in this section for ease of notation.


To implement the moment selection, we introduce the auxiliary parameter
\underline{$\boldsymbol{\eta}$} and its true value \underline{$\boldsymbol{\eta}$}$^{\ast}$
as 
\[
\mathbb{E}[\underline{\boldsymbol{Z}}_{D}(\tilde{Y}-\boldsymbol{X}^{\prime}\boldsymbol{\beta})]=\underline{\boldsymbol{\eta}}\text{ and }\mathbb{E}[\underline{\boldsymbol{Z}}_{D}(\tilde{Y}-\boldsymbol{X}^{\prime}\boldsymbol{\beta}^{\ast})]=\underline{\boldsymbol{\eta}}^{\ast}
\]
with $\tilde{Y}$ defined in (\ref{EQ:Ytide_repeat}). By definition,
\underline{$\eta$}$_{j}^{\ast}=0$ if $j\in A,$\ and \underline{$\eta$}$_{j}^{\ast}\neq0$
if $j\in B$. We denote the parameters to be estimated as $\boldsymbol{\theta}\equiv\left(\boldsymbol{\beta}^{\prime},\underline{\boldsymbol{\eta}}^{\prime}\right)^{\prime}$.
The moment conditions can now be expressed as 
\[
\boldsymbol{m}\left(\boldsymbol{\theta}\right)=\left[\begin{array}{c}
\boldsymbol{Z}^{\ast}\left(\tilde{Y}-\boldsymbol{X}^{\prime}\boldsymbol{\beta}\right)\\
\underline{\boldsymbol{Z}}_{D}\left(\tilde{Y}-\boldsymbol{X}^{\prime}\boldsymbol{\beta}\right)-\underline{\boldsymbol{\eta}}
\end{array}\right].
\]
The feasible sample analog is given by 
\[
\overline{\boldsymbol{\hat{m}}}_{n}\left(\boldsymbol{\theta}\right)=\left[\begin{array}{c}
n^{-1}\sum_{i=1}^{n}\boldsymbol{z}_{i}^{\ast}\left(\widehat{\tilde{y}}_{i}-\boldsymbol{x}_{i}^{\prime}\boldsymbol{\beta}\right)\\
n^{-1}\sum_{i=1}^{n}\underline{\boldsymbol{z}}_{Di}\left(\widehat{\tilde{y}}_{i}-\boldsymbol{x}_{i}^{\prime}\boldsymbol{\beta}\right)-\underline{\boldsymbol{\eta}}
\end{array}\right],
\]
where $\underline{\boldsymbol{z}}_{Di}$ is a $\left(d_{A}+d_{B}\right)\times1$
vector representing the realization of $\underline{\boldsymbol{Z}}_{D}$
for the observation $i$, $\boldsymbol{z}_{i}^{\ast}$ is a $k^{\ast}\times1$
vector, defined similarly to $\underline{\boldsymbol{z}}_{Di}$, and
\[
\widehat{\tilde{y}}_{i}=\frac{y_{i}-\boldsymbol{1}\left(v_{i}>0\right)}{\hat{f}(v_{i}|\boldsymbol{z}_{1i}^{\tilde{p}})}
\]
with $\hat{f}(v_{i}|\boldsymbol{z}_{1i}^{\tilde{p}})$ being obtained
using the procedure outlined in Section \ref{SEC:proceduref}, but
with a $2r$-th order kernel function (for the same reason as listed
in the last section).

We perform the selection and estimation using weighted $L_{1}$ penalty
proposed as in \citet{ChengLiao2015}. Alternative methods employing
different penalty functions can be handled similarly. Specifically,
we define a penalized GMM estimator for $\boldsymbol{\theta}$ as
\begin{equation}
\boldsymbol{\hat{\theta}}=\arg\min_{\boldsymbol{\theta}}\overline{\boldsymbol{\hat{m}}}_{n}\left(\boldsymbol{\theta}\right)^{\prime}\mathbf{W}_{n}\overline{\boldsymbol{\hat{m}}}_{n}\left(\boldsymbol{\theta}\right)+\lambda_{n}\sum_{j=1}^{d_{A}+d_{B}}\varpi_{j} |\underline{\eta}_{j} |,\label{EQ:SCAD}
\end{equation}
where $\mathbf{W}_{n}$ is a $p_{n}\times p_{n}$ positive definite
weighting matrix and 
\[
\varpi_{j}= |\underline{\tilde{\eta}}_{j} |{}^{-\gamma},\quad j=1,...,d_{A}+d_{B},
\]
where $\gamma>0,$ $\underline{\tilde{\eta}}_{j}$ is some initial
estimator, and one candidate is from (\ref{eq:initialGMM}). We impose
the following technical conditions, which are essentially the same
as those placed in \citet{ChengLiao2015}, including restrictions
on tuning parameters.

\begin{assumptionp}{5}\label{A:GMM} 
\begin{enumerate}
\item[(1)] Observations are i.i.d. across $i$. 
\item[(2)] The rank of $\mathbb{E}\left(\boldsymbol{Z}^{\ast}\boldsymbol{X}^{\prime}\right)$
is $s^{\ast}$. $\mathbb{E}\left(\boldsymbol{Z}^{\ast}\varepsilon\right)=\boldsymbol{0},$
$\mathbb{E}\left(Z_{j}\varepsilon\right)=0$ if $j\in A,$ and $\mathbb{E}\left(Z_{j}\varepsilon\right)=\underline{\eta}_{j}^{\ast}\neq0$
if $j\in B$. 
\item[(3)] $\Omega_{n}$ defined in (\ref{EQ:omegan}) satisfy $\rho_{\min}\left(\Omega_{n}\right)\geq C^{-1}$
for some $C>0,$ for all $n$. 
\item[(4)] $\mathbf{W}_{n}$ is a $p_{n}\times p_{n}$ positive definite matrix
with uniformly finite and bounded away from 0 eigenvalues. 
\item[(5)] $\max_{j=1,...,p_{n}}\mathbb{E}(Z_{j}^{4})$ and $\max_{j=1,...,s^{\ast}}\mathbb{E}(X_{j}^{4})$
are uniformly bounded for all $n$. 
\item[(6)] $k^{*}$ is fixed. $a_{n}\equiv\min_{j\in B}\{\vert\underline{\eta}_{j}^{\ast}\vert\}\gg\sqrt{\left.p_{n}\right/n}$. 
\item[(7)] $b_{n}\equiv\max_{j=1,...,d_{A}+d_{B}}|\underline{\tilde{\eta}}_{j}-\underline{\eta}_{j}^{*}|=O_{P}(\sqrt{\left.p_{n}\right/n}).$
$a_{n},b_{n}$, $p_{n}$, $\gamma$ and $\lambda_{n}$ jointly satisfy
\[
b_{n}^{\gamma}\lambda_{n}^{-1}=o_{P}\left(\sqrt{\left.n\right/p_{n}}\right),a_{n}^{-\gamma}\lambda_{n}\ll1/\sqrt{np_{n}},\text{ and }p_{n}\ll n^{1/3}/\log n.
\]
\end{enumerate}
\end{assumptionp}

Parts (1), (4), and (5) of Assumption \ref{A:GMM} are standard. We
place the classic rank condition in Assumption \ref{A:GMM}(2), which
implicitly assumes $k^{\ast}\geq s^{\ast}$ since $\mathbb{E}\left(\boldsymbol{Z}^{\ast}\boldsymbol{X}^{\prime}\right)$
is a $k^{\ast}\times s^{\ast}$ matrix. The rest of Assumption \ref{A:GMM}(2)
defines valid and invalid IVs. In Assumption \ref{A:GMM}(3), $\Omega_{n}$
is the variance of the moment conditions. Note that the dimension
of $\boldsymbol{\theta}_{B}^{\ast}$ (defined below in (\ref{eq:paramOfInterest}))
might be diverging, so we consider a linear combination of $\boldsymbol{\hat{\theta}}_{B}\boldsymbol{-\theta}_{B}^{\ast}$
for the asymptotics, as did in \citet{ChengLiao2015}. We require
$\Omega_{n}$ to be of full rank because we need $\Omega_{n}^{-1}$
to construct the weight of this linear combination. In Assumption
\ref{A:GMM}(6), we assume $k^{\ast}$ is fixed and we need $\min_{j\in B}\{|\underline{\eta}_{j}^{\ast}|\}$
to be big enough so that our procedure can effectively distinguish
valid IVs from invalid ones.

Assumption \ref{A:GMM}(7) places restrictions on the rate of initial estimator $\underline{\tilde{\eta}}_{j}$ that can be obtained as 
\begin{equation}
\tilde{\boldsymbol{\theta}}=\arg\min_{\boldsymbol{\theta}}\overline{\boldsymbol{\hat{m}}}_{n}\left(\boldsymbol{\theta}\right)^{\prime}\mathbf{W}_{n}\overline{\boldsymbol{\hat{m}}}_{n}\left(\boldsymbol{\theta}\right),\label{eq:initialGMM}
\end{equation}
which is the GMM estimator without the penalty term. \citet{ChengLiao2015}
has shown that $\max_{j=1,...,d_{A}+d_{B}} |\underline{\tilde{\eta}}_{j}-\underline{\eta}_{j}^{*} |=O_{P} (\sqrt{\left.p_{n}\right/n} )$
(see their discussion after Assumption 5.1). $p_{n}\ll n^{1/3}/\log n$ is
needed to verify the Lindeberg condition for asymptotic normality.
Note that \citet{ChengLiao2015} essentially assumes that $a_{n}=C>0$. 
When $a_{n}=C>0$, we can, for example, take $\lambda_{n}=1/\sqrt{np_{n}\log n}$
and any $\gamma\geq1,$ due to $p_{n}\ll n^{1/3}/\log n.$

Since $\underline{\boldsymbol{\hat{\eta}}}_{A}=\boldsymbol{0}$ will
be shown to occur with high probability, regarding the asymptotic
distribution, the parameters of our interest are 
\begin{equation}
\boldsymbol{\theta}_{B}\equiv(\boldsymbol{\beta}^{\prime},\underline{\boldsymbol{\eta}}_{B}^{\prime})^{\prime}.\label{eq:paramOfInterest}
\end{equation}
We denote the corresponding estimator as $\boldsymbol{\hat{\theta}}_{B}\equiv(\boldsymbol{\hat{\beta}}^{\prime},\underline{\boldsymbol{\hat{\eta}}}_{B}^{\prime})^{\prime}$
and the true value as $\boldsymbol{\theta}_{B}^{\ast}\equiv(\boldsymbol{\beta}^{\ast\prime},\underline{\boldsymbol{\eta}}_{B}^{\ast\prime})^{\prime}$.
The partial derivative of the moment conditions with respect to $\boldsymbol{\theta}_{B}\equiv(\boldsymbol{\beta}^{\prime},\underline{\boldsymbol{\eta}}_{B}^{\prime})^{\prime}$
is denoted as 
\begin{equation}
\Gamma_{\boldsymbol{\theta}_{B}}=\mathbb{E}\left(\begin{array}{cc}
-\boldsymbol{Z}^{\ast}\boldsymbol{X}^{\prime} & \mathbf{0}_{k^{\ast}\times d_{B}}\\
-\underline{\boldsymbol{Z}}_{A}\boldsymbol{X}^{\prime} & \mathbf{0}_{d_{A}\times d_{B}}\\
-\underline{\boldsymbol{Z}}_{B}\boldsymbol{X}^{\prime} & -\boldsymbol{I}_{d_{B}\times d_{B}}
\end{array}\right).\label{eq:GammaTheta}
\end{equation}
$\Gamma_{\boldsymbol{\theta}_{B}}^{\prime}\Gamma_{\boldsymbol{\theta}_{B}}$
is of rank $s^{\ast}+d_{B},$ due to the assumption that
$\mathbb{E}\left(\boldsymbol{Z}^{\ast}\boldsymbol{X}^{\prime}\right)$
is of rank $s^{\ast}.$ We define the following matrix: 
\[
\Sigma_{n}\equiv\left(\Gamma_{\boldsymbol{\theta}_{B}}^{\prime}\mathbf{W}_{n}\Gamma_{\boldsymbol{\theta}_{B}}\right)^{-1}\Gamma_{\boldsymbol{\theta}_{B}}^{\prime}\mathbf{W}_{n}\Omega_{n}\mathbf{W}_{n}\Gamma_{\boldsymbol{\theta}_{B}}\left(\Gamma_{\boldsymbol{\theta}_{B}}^{\prime}\mathbf{W}_{n}\Gamma_{\boldsymbol{\theta}_{B}}\right)^{-1},
\]
which is useful for the asymptotics.

We reproduce the results in \citet{Liao2013} and \citet{ChengLiao2015} in the following theorem.

\begin{theorem} \label{TH:GMM} Suppose Assumptions \ref{A:main},
\ref{A:continuousf_kernelK}, \ref{A:specialR}, and \ref{A:GMM}
hold. Further, $r>\left(p^{\ast}+1\right)/2$ and we adopt $2r$-th
order kernel and the $\hat{\boldsymbol{h}}$ obtained from (\ref{EQ:bandwidth})
to construct $\hat{f}(v|\boldsymbol{z}_{1}^{\tilde{p}})$ for $\widehat{\tilde{y}}_{i}$,
where the $r$ is defined in Assumption \ref{A:continuousf_kernelK}.
Then 
\begin{enumerate}
\item[(1)] $\Pr(\underline{\hat{\eta}}_{j}=0,\text{ }\forall\text{ }j\in A)\rightarrow1$
and $\Pr(\underline{\hat{\eta}}_{j}\neq0,\text{ }\forall\text{ }j\in B)\rightarrow1$. 
\item[(2)] For any $\left(s^{\ast}+d_{B}\right)\times1$ vector $\boldsymbol{e}$
such that $\left\Vert \boldsymbol{e}\right\Vert =1$, 
\[
\sqrt{n}\boldsymbol{e}^{\prime}\Sigma_{n}^{-1/2}(\boldsymbol{\hat{\theta}}_{B}\boldsymbol{-\theta}_{B}^{\ast})\overset{d}{\rightarrow}N\left(0,1\right).
\]
\end{enumerate}
\end{theorem}

For the same reason as for Theorem \ref{TH:AdaptiveLasso}, we impose restrictions on the adopted kernel function, as specified in the theorem. As in the last section, we recommend to choose $\lambda_n$ by 10-fold CV and try a range of $\gamma$. For practical examples and further guidance, see Sections \ref{SEC:simu} and \ref{SEC:applications}.

Finally, an interesting direction would be to permit $s^{*}$ to diverge
while simultaneously considering variable selection and moment selection.
We leave this topic for future research.

\section{Simulation\label{SEC:simu}}

This section assesses the finite sample performance of the procedures
proposed in Section \ref{SEC:theoryapply} through Monte Carlo experiments.
For all designs studied in this section, we consider sample sizes
$n=500,1000, 2000$ and base our findings on 500 independent replications
conducted using the R programming language and MATLAB. All simulation results are
reported in tables collected in Appendix \ref{APP:tables}.

\subsection{Monte Carlo for Section \ref{SEC:VariableSelection}}

We first investigate the adaptive Lasso method as discussed in Section
\ref{SEC:VariableSelection}. We explore three distinct simulation
designs based on the following data-generating process: 
\[
Y=\mathbf{1}(V+\beta_{0}+\beta_{1}X_{1}+\cdots+\beta_{p}X_{p_n}+\varepsilon>0).
\]
Here, we consider $p_n\in\{14,29,49\}$, corresponding to binary choice
models with $p_{n}=15,30$, and 50 covariates, respectively. The true
parameters are set as follows: $\beta_{0}=\beta_{1}=0$, $\beta_{2}=\beta_{3}=1$,
and $\beta_{l}=0$ for all $l=4,\ldots,p_n$. $V$ is defined as $V=X_{1}+X_{1}X_{2}+\mathbf{1}(X_{2}>0)+e_{v}$,
where $e_{v}\sim\text{Logistic}(0,2)$. The three designs differ in
the distributions of $\varepsilon$ and $\boldsymbol{X}_{1}^{p_n}=(X_{1},\ldots,X_{p_n})'$: 
\begin{itemize}
\item[-] \textbf{Design 1}: $\varepsilon\sim N(0,\pi^{2}/3)$ and $\boldsymbol{X}_{1}^{p_n}\sim\text{MVN}(\mathbf{0},\boldsymbol{I}_{p_n})$,
where $\text{MVN}$ stands for multivariate normal distribution and
$\boldsymbol{I}_{p_n}$ is the $p_n\times p_n$ identity matrix. 
\item[-] \textbf{Design 2}: $\varepsilon=e_{y}\cdot e^{\vert X_{3}\vert/1.8}$
with $e_{y}\sim N(0,1)$. All other aspects of Design 1 remain the
same. 
\item[-] \textbf{Design 3}: $\boldsymbol{X}_{1}^{p_n}$ is generated from a multivariate normal distribution whose marginal distributions are standard normal and the correlation between $X_j$ and $X_l$ is $0.5^{\vert j-l \vert}$. All other aspects are
consistent with Design 2. 
\end{itemize}

Design 1 is a benchmark design, where the Probit model is correctly
specified up to a scale of $\pi/\sqrt{3}$. We anticipate that both the standard Probit and our adaptive Lasso estimators are consistent,
with the former being more efficient. Design 2 is a heteroskedastic
Probit model, where $\varepsilon$ has the same standard deviation
as Design 1. We expect our adaptive Lasso regression to give consistent estimates
in this design, but the Probit estimator will be biased. Design 3
examines the effectiveness of our procedure in handling the more general
dependence structure as in (\ref{EQ:V_Z_Z}), where $V$ and $X_{3},...,X_{p_n}$
are dependent but only through their dependence on $(X_{1},X_{2})$.
It is worth noting that by construction, $V$ has a stronger dependence
on $X_{1}$ than on $X_{2}$. For instance, in Designs 1 and 2, $\mathrm{Corr}(V,X_{1})=0.25$,
while $\mathrm{Corr}(V,X_{2})=0.1$.

For each simulation design, we estimate the parameters $\boldsymbol{\beta}\equiv(\beta_{0},\beta_{1},\ldots,\beta_{p_n})'$
using two methods: adaptive Lasso regression and Probit regression. The former
is implemented with the following procedure: 
\begin{itemize}
\item[-] \textbf{Step 1} (DC screening): Compute $\hat{\mathcal{T}}_{nl}$
using (\ref{EQ:statistic}) for $X_{l}$, $l=1,...,p_n$, rank $\hat{\mathcal{T}}_{nl}$
from largest to smallest, and select the top $\tilde{p}=4$ $X_{l}$'s
with the largest $\hat{\mathcal{T}}_{nl}$, denoted by $\tilde{\mathcal{A}}$. 
\item[-] \textbf{Step 2} (Bandwidth selection): Apply the CV approach proposed
by \citet{HallRacineLi} to determine the optimal bandwidth $\hat{\boldsymbol{h}}$
for estimating $f(v|x_{1}^{\tilde{p}})$, as specified in (\ref{EQ:bandwidth}).
Here, we use the Gaussian kernel. 
\item[-] \textbf{Step 3} (Estimate $f(v|x_{1}^{\tilde{p}})$): Compute $\hat{f}(v|x_{1}^{\tilde{p}})$
using a 4th-order Gaussian kernel (as explained below Theorem \ref{TH:AdaptiveLasso})
and $\hat{\boldsymbol{h}}$ obtained in Step 2. Both Steps 2 and 3
are implemented using the R package $\mathtt{np}$ (\citet{hayfield2008nonparametric}). 
\item[-] \textbf{Step 4} (Adaptive Lasso regression): Obtain the estimate $\hat{\boldsymbol{\beta}}$
using the adaptive Lasso regression (\ref{eq:adaptiveLasso}) with the feasible $\hat{f}(v|x_{1}^{\tilde{p}})$
calculated in Step 3. Here, we select tuning parameters $\lambda_{n}$
through a 10-fold CV for each $\gamma\in\{2,3,4\}$. Both the CV tuning parameter selection and the
adaptive Lasso regression are implemented using the R package $\mathtt{glmnet}$
(\citet{Friedman2010glmnet}). 
\end{itemize}
Probit results are obtained using R's built-in function $\mathtt{glm}()$.
Note that $\mathtt{glm}()$ assumes a standard deviation of 1 for
$\varepsilon$ (default scale normalization) and estimates the coefficient
on $V$, denoted as $\rho$.\footnote{As a result, $\mathtt{glm}()$ fits the following Probit model: 
\[
Y=\mathbf{1}(\rho V+\beta_{0}^{Probit}+\beta_{1}^{Probit}X_{1}+\cdots+\beta_{p}^{Probit}X_{p_n}+e>0),
\]
where $e=\sqrt{3}\varepsilon/\pi\sim N(0,1)$, $\rho=\sqrt{3}/\pi$,
and $\beta^{Probit}=(\beta_{0}^{Probit},\beta_{1}^{Probit},\cdots,\beta_{p_n}^{Probit})'=\sqrt{3}\beta/\pi$.} To facilitate comparison, we compute the estimates of ratios $\beta_{l}/\rho$
for the Probit estimation.

The simulation results are displayed in tables presented in Appendix G.1, beginning with ``Table 1" for Design 1, ``Table 2" for Design 2, and so on. For each design, tables labeled ``A", ``B", and ``C" correspond to the variable selection results in Steps 1 and 4 of our proposed procedure, the outcomes of the adaptive Lasso regression, and the Probit estimation, respectively. For example, in Table \ref{tab:1A}, we report $\Pr(\{1\} \in \tilde{\mathcal{A}})$ and $\Pr(\{2\} \in \tilde{\mathcal{A}})$, representing the probabilities that $X_{1}$ and $X_{2}$ are selected as variables relevant to $V$. The column labeled ``Correct (\%)" in Table \ref{tab:1A} displays the average number (percentage) of the true zero coefficients correctly identified as zero, and the column labeled ``Incorrect" shows the average count of the two true nonzero coefficients incorrectly set to zero. Table \ref{tab:1B} summarizes the adaptive Lasso estimator's performance for Design 1, presenting mean bias (MEANB), root mean squared errors (RMSE), median bias (MEDB), and median absolute deviation (MAD) for $(\hat{\beta}_{2}, \hat{\beta}_{3})$. Table \ref{tab:1C} contains the same set of statistics for the Probit estimator of $(\beta_{2}/\rho,\beta_{3}/\rho)$ as in Table \ref{tab:1B}.

The variable selection outcomes are encouraging, as presented
in Tables \ref{tab:1A}--\ref{tab:3A}. The screening selects $(X_{1},X_{2})$
as relevant variables for $V$ with probability approaching one. Notably,
for $X_{1}$, which has moderate dependence with $V$, the $\Pr(\{1\}\in\tilde{\mathcal{A}})$
attains 1 even when $n\leq 500$ across all three designs. For $X_{2}$,
which has weak dependence on $V$, the $\Pr(\{2\}\in\tilde{\mathcal{A}})$
approaches 1 when $n\geq1000$ and is more sensitive to the number
of candidate variables ($p$) in smaller sample sizes.\footnote{It is worth highlighting that the screening performs slightly better
in Design 3 compared to Designs 1 and 2. This improvement can be attributed
to the enhanced correlation between $X_{1}$ and $X_{2}$ in Design
3, strengthening their association with $V$. In fact, in Design 3, $\text{Corr}(V,X_{1})\approx 0.30$
and $\text{Corr}(V,X_{2})\approx 0.21$.} Additionally, the adaptive Lasso regression can automatically exclude irrelevant regressors (e.g., $X_{1}$) from the regression with increasing probability as the sample size increases, which is evidenced in the ``Correct (\%)" columns of Tables \ref{tab:1A}--\ref{tab:3A}. 

As expected, when the Probit model is correctly specified (up to scale), the Probit estimator is consistent and performs better than our adaptive Lasso estimator, as shown in Tables \ref{tab:1B} and \ref{tab:1C}. However, in the presence of heteroskedasticity, observed in Tables \ref{tab:2C} and \ref{tab:3C}, the Probit estimator maintains a bias of approximately 10\%–20\% for $\beta_{3}/\rho$ that persists as the sample size increases. Our adaptive Lasso estimator, which uses an inverse density as a plug-in term, is prone to generating extreme estimates, resulting in comparatively larger RMSEs, especially in smaller sample sizes. However, focusing on the MAD, which is robust to outliers, we observe that our adaptive Lasso estimator retains its $\sqrt{n}$-consistency across all three designs, aligning with our asymptotic theory, as illustrated in Tables \ref{tab:1B}–\ref{tab:3B}. Finally, although the adaptive Lasso variable selection is influenced by the choice of the tuning parameter $\gamma$, the resulting estimates appear to be pretty robust to this choice.

\subsection{Monte Carlo for Section \ref{SEC:MSelection}}

In this section, we examine the penalized GMM approach proposed in Section
\ref{SEC:MSelection} through three simulation designs (Designs 4--6)
based on the following binary choice model: 
\[
Y=\boldsymbol{1}(V+\beta_{0}+\beta_{1}X+\varepsilon>0),
\]
where the true parameters $\boldsymbol{\beta}=(\beta_{0},\beta_{1})'=(0,1)$
and $X$ is an endogenous regressor. To estimate $\boldsymbol{\beta}$,
we use instrumental variables $\boldsymbol{Z}^{*}=(1,Z^{*})'$, $\underline{\boldsymbol{Z}}_{A}=(Z_{1},...,Z_{d_{A}})'$,
and $\underline{\boldsymbol{Z}}_{B}=(Z_{d_{A}+1},...,Z_{d_{A}+d_{B}})'$
with $(d_{A},d_{B})\in\{(6,7),(14,14),(24,24)\}$, corresponding to
cases with $p_{n}=15,30,$ and 50, respectively. Designs 4--6 differ
in how $V$ and $\boldsymbol{Z}_{1}^{p_{n}}=(\boldsymbol{Z}^{*\prime},\underline{\boldsymbol{Z}}_{A}^{\prime},\underline{\boldsymbol{Z}}_{B}^{\prime})'$
are dependent and how variables in $\underline{\boldsymbol{Z}}_{B}$
are correlated with $\varepsilon$. Specifically, letting $\boldsymbol{e}=(e_{1},...,e_{4})'\sim\text{MVN}(\boldsymbol{0},\boldsymbol{I}_{4})$,
$\boldsymbol{u}=(u_{1},...,u_{d_{A}+d_{B}+1})'\sim\text{MVN}(\boldsymbol{0},\boldsymbol{I}_{d_{A}+d_{B}+1})$,
and $\boldsymbol{e}\perp\boldsymbol{u}$, we set 
\begin{itemize}
\item[-] \textbf{Design 4}: $X=(e_{1}+e_{2}+e_{3})/\sqrt{3}$, $\varepsilon=(e_{1}+e_{4})/\sqrt{2}$,
$Z^{*}=(e_{2}+u_{1})/\sqrt{2}$, $Z_{1}=(e_{2}+u_{2})/\sqrt{2}$,
\[
Z_{j}=\frac{1}{2}e_{3}+\frac{\sqrt{3}}{2}u_{j+1}\text{ for all }j=2,...,d_{A},\text{ and }
\]
\[
Z_{j}=\frac{1}{2}e_{1}+\frac{\sqrt{3}}{2}u_{j+1}\text{ for all }j=d_{A}+1,...,d_{A}+d_{B}.
\]
The special regressor $V=Z^{*}+Z^{*}\cdot Z_{1}+\boldsymbol{1}(Z_{1}>0)+e_{v}$
with $e_{v}\sim\text{Logistic}(0,2)$ and $e_{v}\perp(\boldsymbol{e},\boldsymbol{u})$. 
\item[-] \textbf{Design 5}: $Z_{j}=e_{2}/2+\sqrt{3}u_{j+1}/2$ for $j=2,...,d_{A}$.
All other aspects are the same as Design 4. 
\item[-] \textbf{Design 6}: $Z_{j}=e_{1}/\sqrt{2}+u_{j+1}/\sqrt{2}\text{ for }j=d_{A}+1,...,d_{A}+d_{B}$.
All other aspects remain the same as Design 5. 
\end{itemize}

Design 4 is a benchmark design where $V$ is solely dependent on $(Z^{*},Z_{1})$
with different levels among all candidate IVs. In this design, it
is evident from the data-generating process that all candidate IVs
are relevant to the endogenous regressor $X$. Note that $Z^{*}$
and $\underline{\boldsymbol{Z}}_{A}$ are valid IVs for $X$, while
$\underline{\boldsymbol{Z}}_{B}$ are invalid IVs. In Design 5, we
modify Design 4 by replacing $e_{3}$ with $e_{2}$ to examine the
effectiveness of the DC screening procedure under a more general dependence
structure, as in (\ref{EQ:V_Z_Z}). Consequently, $V$ becomes dependent
on $Z_{2},...,Z_{d_{A}}$ as well, but only through their correlation
with $(Z^{*},Z_{1})$, so that $V\perp(Z_{2},...,Z_{d_{A}})|(Z^{*},Z_{1})$
and this conditional independence is not ambiguous.\footnote{This conditional independence doesn't alter the identification strength
of $\boldsymbol{Z}_{1}^{p_{n}}$.} Design 6 aims to strengthen the correlation between $\underline{\boldsymbol{Z}}_{B}$
and $\varepsilon$ to assess the sensitivity of the penalized GMM method
in identifying invalid IVs.

To implement the penalized GMM method, we first obtain a feasible estimate
$\hat{f}(v|z_{1}^{\tilde{p}})$ for $f(v|z_{1}^{p_{n}})$. This follows the same Steps 1–3 as in Designs 1–3, with $\tilde{p}=4$, but substituting $X$s with $Z$s in this context. After computing the feasible $\hat{f}(v|z_{1}^{\tilde{p}})$ estimates, we proceed to
obtain $\hat{\boldsymbol{\theta}}$ from (\ref{EQ:SCAD}). We solve the minimization problem in the penalized GMM estimation using the optimal projected gradient algorithm proposed by \citet{schmidt2010graphical}, setting the tuning parameters as $\lambda_{n} = 1/\sqrt{n p_n}$ and $\gamma\in\{1,2,3\}$. Throughout all designs, we use the identity matrix as the weighting matrix, i.e., \(\mathbf{W}_{n} = \boldsymbol{I}_{p_{n}}\). Our simulation studies reveal that the simple identity matrix often outperforms the theoretically more efficient optimal GMM weighting matrix. An intuitive explanation for this observation is that, in finite samples, the efficiency gains from using the optimal weighting matrix are often insufficient to outweigh the errors introduced by its estimation, particularly given that our estimator incorporates a nonparametrically estimated plug-in term. Thus, we recommend practitioners use the identity matrix as the default weighting matrix.

We present the simulation results for Designs 4–6 in Tables 4–6, respectively. The outcomes for variable screening and moment selection are shown in tables labeled ``A", and the performance of the penalized GMM estimation is illustrated in tables labeled ``B". Specifically, let $\tilde{\mathcal{Z}}$ denote the set of relevant $Z$s selected by DC screening. Table \ref{tab:4A} reports $\Pr(Z^{*} \in \tilde{\mathcal{Z}})$ and $\Pr(Z_{1} \in \tilde{\mathcal{Z}})$. Additionally, the ``Correct (\%)" column in Table \ref{tab:4A} presents the average count (percentage) of truly valid IVs (other than $Z^{*}$) that are successfully selected, and the ``Incorrect (\%)" column reports the average number (percentage) of invalid IVs mistakenly categorized as valid. Table \ref{tab:4B}--\ref{tab:6B} provides the same set of performance metrics as Tables \ref{tab:1B}–\ref{tab:3B} for $(\hat{\beta}_{0}, \hat{\beta}_{1})$ in Designs 4--6.

As shown in Tables \ref{tab:4A}--\ref{tab:6A}, the DC
screening procedure consistently performs well. When $n\geq500$, it demonstrates an
almost certain ability to select all variables relevant to $V$. For IV (moment) selection, the penalized GMM procedure effectively distinguishes valid IVs from invalid ones with increasing accuracy as the sample size grows, as indicated by the ``Correct (\%)" and ``Incorrect (\%)" columns. The choice of tuning parameter $\gamma$ presents a trade-off: for a given $\lambda_n$, a larger (smaller) $\gamma$ improves (reduces) the likelihood of correctly identifying valid IVs but also raises (lowers) the risk of mistakenly classifying some invalid IVs as valid. We recommend that practitioners try a range of $\gamma$ values and carefully examine their IV selection outcomes to avoid inconsistent estimates.

Tables \ref{tab:4B}--\ref{tab:6B} present the performance of the penalized GMM estimator. The estimator shows noticeable bias in small samples, but as the sample size increases, this bias quickly diminishes, consistent with our asymptotic theory. Interestingly, the RMSE of the penalized GMM estimator does not consistently decrease with sample size. This may be due to the use of the estimated inverse density function as a plug-in term; when the estimated density approaches zero, ``outliers'' can appear in $\widehat{\tilde{Y}}$, leading to estimates that deviate significantly from the true values. However, focusing on the MAD, which is less sensitive to extreme values, we observe that the estimator converges at approximately a parametric rate, aligning with our Theorem \ref{TH:GMM}. Finally, as the result of stronger IVs, the penalized GMM estimator performs better in Design 6 compared to Designs 4 and 5.

\section{Empirical Illustration}\label{SEC:applications} 

China's economic reforms since the late 1970s have significantly transformed the country’s economy. In the early 1980s, the government began to relax restrictions on population mobility. Over time, rural residents were granted more freedom to leave their villages and seek employment in larger cities for higher wages. In the following analysis, we explore factors influencing the migration intentions of rural residents by applying our method to a dataset drawn from the Rural-Urban Migration in China (RUMiC) project.\footnote{RUMiC consists of three components: the Urban Household Survey, the Rural Household Survey, and the Migrant Household Survey. It was initiated by researchers from the Australian National University, the University of Queensland, and Beijing Normal University, with support from the Institute for the Study of Labor (IZA). RUMiC received funding from the Australian Research Council, the Australian Agency for International Development, the Ford Foundation, IZA, and the Chinese Foundation of Social Sciences. Further details on the survey are available at \url{https://datasets.iza.org/dataset/58/longitudinal-survey-on-rural-urban-migration-in-china}.} The survey focuses on individuals moving from rural areas to major Chinese cities. Participants (both migrants and workers) answered a wide range of questions. For more detailed information on the survey design and variable construction, refer to the survey website and \citet{Gongetal2008}. Currently, data from the 2008 wave are publicly available.

Previous studies have investigated the factors influencing rural residents' migration intentions using various methodologies, yielding mixed results. \citet{Zhao1999}, \citet{Zhao2003}, and \citet{Mullan2011} applied logistic regression models to cross-sectional data, while \citet{XueYangZhou} utilized the special regressor approach within a panel data framework that included interactive fixed effects. 
 
Thanks to the distribution-free nature of our method, we relax the assumption that the error term follows a logistic distribution, which is imposed in \citet{Zhao1999}, \citet{Zhao2003}, and \citet{Mullan2011}. Additionally, we explicitly examine the dependence of the special regressor on other included covariates using our screening procedure, whereas \citet{XueYangZhou} relied on intuition for such assumption. However, unlike their work, we focus on cross-sectional data, as our methods are not designed for panel data settings.

Departing from aforementioned works that considered only a limited set of explanatory variables--thereby risking omitted variable bias--we significantly broaden the scope of variables included in our analysis and avoid specifying the regression model based solely on the researcher's subjective judgment. Instead, we employ our proposed data-driven methods to identify the key factors influencing migration intentions. However, a limitation of this approach is that the final estimation results may lack clear causal interpretation. We emphasize that the primary purpose of this application is to illustrate the use of our methodology and provide insights for future research that can adopt more focused and rigorously designed studies.

The migration intention is modeled as follows:
\begin{equation}
Y=\mathbf{1}\left(V+\beta_{0}+X_{1}\beta_{1}+X_{2}\beta_{2}+\cdots+X_{p}\beta_{p}+\varepsilon>0\right),
\end{equation}
where \( Y \) represents the binary migration intention. We use data from the 2008 wave of the RUMiC survey. Following \citet{XueYangZhou}, we employ the negative logarithm of the average daily precipitation between April and August from two and three years prior as the special regressor, $V$. The model includes 51 explanatory variables (i.e., \( p = 51 \)). By applying the procedure proposed in Section \ref{SEC:VariableSelection}, we aim to identify predictors relevant to migration decisions.  After excluding observations with missing values, the final sample consists of 3,787 observations. Tables \ref{tab:AP2} and \ref{tab:AP3} in Appendix G.2 provide the definitions and summary statistics for these variables, respectively.

We standardize all variables, including $V$, to have a mean of 0 and variance of 1 before estimation. We apply the adaptive Lasso estimator proposed in Section \ref{SEC:VariableSelection}, specifically the one defined in equation (\ref{eq:adaptiveLasso}). The initial estimator is obtained using OLS, as outlined in Section \ref{sec:Initial-Estimator}. Data-driven bandwidths are determined according to the procedure in Section \ref{SEC:proceduref}. 

For the screening procedure, we set $\tilde{p} = 4$, meaning we retain the four covariates with the highest values of $\mathcal{\hat{T}}_{n}$. After applying the CV method from \citet{HallRacineLi} on these covariates, we found that the bandwidth of one covariate reached its upper limit, indicating that it has no impact on the conditional density. This suggests that $\tilde{p} = 4$ is sufficiently large for this application. The remained covariates are  ``Height'', ``Income'', and ``Age''.  

Next, we discuss the tuning parameters for the adaptive Lasso. For a given \(\gamma\) in the weight \(\varpi_j\), the parameter \(\lambda_n\) in equation (\ref{eq:adaptiveLasso}) is selected via 10-fold CV. After experimenting with \(\gamma\) values ranging from 1 to 15, we found that setting \(\gamma\) between 2 and 12 resulted in selecting the same five covariates. Based on this, we conduct the variable selection by setting \(\gamma\) within this range.

As the last step, we perform post-selection OLS and make inferences using the asymptotics from Theorem \ref{TH:AdaptiveLasso}, constructing confidence intervals by estimating the sample counterpart of \(\Sigma_n\). In \(\Sigma_n\), \(\Omega_n\) represents the variance of the influence term, the same as in \citet{Lewbel2000} for a fixed dimension setting. A plug-in estimate of \(\Omega_n\) can be constructed as follows. First, estimate \(\hat{u}_i = \widehat{\tilde{y}}_i - \boldsymbol{x}_i' \boldsymbol{\hat{\beta}}\) from the post-selection OLS. Then, compute
\[
\hat{\boldsymbol{q}}_i = \hat{u}_i \boldsymbol{x}_i + \widehat{\mathbb{E} \left( \hat{u}_i \boldsymbol{x}_i \,\middle|\, \boldsymbol{x}_{1i}^{p^{*}} \right)} - \widehat{\mathbb{E} \left( \hat{u}_i \boldsymbol{x}_i \,\middle|\, \boldsymbol{x}_{1i}^{p^{*}}, v_i \right)},
\]
where \(\widehat{\mathbb{E}(\cdot|\cdot)}\) denotes standard nonparametric kernel regression. In this case, we include ``Height", ``Income", and ``Age" as \(\boldsymbol{x}_{1i}^{p^{*}}\). Finally, compute
\[
\hat{\Omega}_n = \frac{1}{n}\sum_{i=1}^{n} \hat{\boldsymbol{q}}_i \hat{\boldsymbol{q}}'_i.
\]
With \(\hat{\Omega}_n\), \(\hat{\Sigma}_n\) can be straightforwardly obtained. Note \(\hat{\Omega}_n\) can be similarly calculated for the scenario in Section \ref{SEC:MSelection}. 

To ensure a fair and meaningful comparison, we apply the adaptive Logistic Lasso, using the initial estimator from logistic regression. This estimation is implemented using the standard \texttt{R} package \texttt{glmnet}. The tuning parameter $\lambda_n$ is also selected via 10-fold CV, with a fixed value of $\gamma$ for the weight $\varpi_j$. However, we find that the selection results are sensitive to the choice of $\gamma$. Specifically, when $\gamma \leq 3$, more than 19 covariates are selected, for $\gamma = 4$, six covariates are selected, whereas with $\gamma = 5$, only one covariate is selected. Based on insights from our method, we set $\gamma = 4$, as both methods (our proposed approach and adaptive Logistic Lasso) select a comparable number of covariates. The final estimates are then obtained by applying post-selection logistic regression.

We summarize the key findings as follows:

First, our results indicate that $V$ (the average precipitation) is not independent of all the covariates considered. Using the CV method by \citet{HallRacineLi}, we find that $V$ is dependent on ``Height", ``Income", and ``Age". This dependence is reasonable given that northern China is typically drier but less economically developed than southern China, and individuals from the north tend to be taller on average, explaining the association between $V$ and both ``Height" and ``Income". However, we do not have a clear explanation for the relation between $V$ and ``Age". These findings emphasize the importance of the screening procedure, as a model-free data analysis can reveal significant (possibly nonlinear) relationships that may otherwise be overlooked. Second, $V$ is consistently selected by the adaptive Logistic Lasso procedure and is highly significant with the expected sign in the post-selection logistic regression. This supports the decision to include $V$ in the model and justifies the normalization of its coefficient to 1 in the special regressor approach.

We report the post-selection estimation results for both methods in Table \ref{tab:AP1}. Standard errors are computed based solely on the post-selection estimates.  Both methods select several common regressors, including \texttt{a10} (``Height''), ``Age'', and \texttt{a23} (``Medical Expenses Reimbursed'').   Post-selection results from both approaches suggest that \texttt{a23} is insignificant. However, there are some discrepancies. For example, the signs of the coefficients for ``Height'' differ between the two methods. Our approach uniquely selects \texttt{nold} (``No. of Elderly''), whereas the adaptive Logistic Lasso selects \texttt{deduc5}, \texttt{e35}, and \texttt{e36}. These discrepancies may arise from biased estimates in the Logistic estimation, possibly due to model mis-specifications or heteroskedasticity in the error term.

\begin{center}
\begin{table}[H]
\centering
\global\long\def\thetable{AP1}%
\caption[Post-Selection Estimation Results]{Post Selection Estimation}  
\label{tab:AP1} 
\begin{tabular}[c]{lcc}
    \hline\hline
              & Benchmark & Our Approach \\
    VARIABLES & Logistic & Special Regressor \\
    \hline
   \textit{Controls:} & & \\
   nold (No. of Elderly)       &                & $0.073^{***}$ \\
                               &                & $[0.027]$     \\
   a10 (Height)                & $0.329^{***}$  & $-0.140^{***}$ \\    
                               & $[0.089]$      & $[0.015]$      \\        
   Age                         &  $-0.702^{***}$& $-0.216^{***}$\\   
                               & $[0.077]$      & $[0.014]$     \\
   a23 (Medical Expenses Reimbursed) &  $-0.371$& $-0.050$      \\
                               &  $[0.475]$     & $[0.031]$     \\  
   c18\_2 (Monthly Bonus and Allowance) &       & $-0.112^{***}$ \\
                               &                & $[0.014]$     \\
   deduc5 (College Degree)     & $-1.810$       & \\
                               & $[60.7]$       & \\
   e35 (No. of people in major cities contacted & $-0.742^{**}$ & \\
   \ \ \ \ \   during the last spring festival) & $[0.325]$     & \\
   e36 (Among e35 No. of people have  hukou)    & $0.340^{*}$   & \\
   \ \ \ \ \                                    & $[0.181]$     & \\                                  
   \textit{Special Regressor:} &                & \\
    Minus Log lagged rainfall  & $0.550^{***}$ & \\
                               & $[0.100]$      & \\
    \hline
  No. of Observations & 3,787 & 3,787 \\
    \hline\hline
\multicolumn{3}{l}{Note: Standard errors in brackets.}   \\
\multicolumn{3}{l}{***, **, and * indicate significance at 1\%,  5\%, and 10\% levels, respectively.}
\end{tabular}
\end{table}
\end{center}

We conclude this section by discussing the implications of our empirical results. The positive effect of \texttt{nold} on migration intentions is likely driven by financial pressures, as elderly individuals in rural China generally lack access to pensions, pushing younger family members to migrate for better financial support. The negative effect of ``Age" can be explained by younger individuals being less settled and more willing to take risks, making them more inclined to migrate. The negative coefficient for \texttt{c18\_2} (``Monthly Bonus and Allowance") suggests that better working conditions discourage migration, as individuals receiving higher compensation may be less motivated to leave their current jobs. However, we do not have a clear causal interpretation of the effect of ``Height". As previously noted, one challenge with including many covariates in the model is that the results may lack clear causal interpretations.

\section{Conclusion}\label{SEC:conclusion} 
In this paper, we introduce a new estimation procedure for semiparametric binary choice models in high-dimensional settings. This is achieved by combining an innovative dimension reduction method for conditional density estimation with the special regressor approach. We study the classic variable and moment selection problems in the binary choice model context. Monte Carlo simulations illustrate the finite sample properties of our methods. We illustrate our proposed approaches by studying migration intentions of rural residents in China. Our empirical findings indicate that the special regressor used in \citet{XueYangZhou} (``Precipitation'') is related to several covariates and therefore may not satisfy the strong exogeneity assumption. This underscores the value of our data-driven approach in guiding practitioners through empirical model specification.

For future research, one promising avenue is to integrate our novel
conditional density estimator with other estimators that engage with
high-dimensional conditional density estimates. 
\par\bigskip

\begin{center}
\textbf{\large{}{}{}{}Acknowledgments }{\large{}{}{} }{\large\par}
\par\end{center}
We thank the editor, Xiaohong Chen, an associate editor, and three anonymous referees for their helpful comments, which have substantially improved the paper. We are grateful for the valuable feedback and discussions provided by seminar participants at the University of Melbourne, the University of Queensland, and the University of Sydney, as well as conference attendees at the 17th International Symposium on Econometric Theory and Applications (SETA 2023) and the 31st Australia New Zealand Econometric Study Group Meeting (ANZESG 2023). Any remaining errors are our responsibility.
\par\bigskip

\bibliography{references}

\begin{thebibliography}{52}
\newcommand{\enquote}[1]{``#1''}
\expandafter\ifx\csname natexlab\endcsname\relax\def\natexlab#1{#1}\fi

\bibitem[\protect\citeauthoryear{Andrews}{Andrews}{1994}]{Andrews1994}
\textsc{Andrews, D.~W.} (1994): \emph{Chapter 37 Empirical process methods in econometrics}, vol.~4 of \emph{Handbook of Econometrics}, Elsevier.

\bibitem[\protect\citeauthoryear{Babii, Chen, Ghysels, and Kumar}{Babii et~al.}{2021}]{babiiEtal2021}
\textsc{Babii, A., X.~Chen, E.~Ghysels, and R.~Kumar} (2021): \enquote{Binary choice with asymmetric loss in a data-rich environment: theory and an application to racial justice,} \emph{arXiv preprint arXiv:2010.08463}.

\bibitem[\protect\citeauthoryear{Belloni, Chen, Chernozhukov, and Hansen}{Belloni et~al.}{2012}]{Belloni_et_al2012}
\textsc{Belloni, A., D.~Chen, V.~Chernozhukov, and C.~Hansen} (2012): \enquote{Sparse models and methods for optimal instruments with an application to eminent domain,} \emph{Econometrica}, 80, 2369--2429.

\bibitem[\protect\citeauthoryear{Belloni, Chernozhukov, Chetverikov, Hansen, and Kato}{Belloni et~al.}{2018}]{BelloniEtal2018}
\textsc{Belloni, A., V.~Chernozhukov, D.~Chetverikov, C.~Hansen, and K.~Kato} (2018): \enquote{High-Dimensional Econometrics and Regularized GMM,} \emph{arXiv preprint arXiv:1806.01888}.

\bibitem[\protect\citeauthoryear{Benjamini and Hochberg}{Benjamini and Hochberg}{1995}]{BenjaminiHochberg}
\textsc{Benjamini, Y. and Y.~Hochberg} (1995): \enquote{Controlling the false discovery rate: a practical and powerful approach to multiple testing,} \emph{Journal of the Royal Statistical Society: Series B (Statistical Methodology)}, 57, 289--300.

\bibitem[\protect\citeauthoryear{Bickel, Ritov, and Tsybakov}{Bickel et~al.}{2009}]{BickelEtal2009}
\textsc{Bickel, P.~J., Y.~Ritov, and A.~B. Tsybakov} (2009): \enquote{Simultaneous analysis of Lasso and Dantzig selector,} \emph{The Annals of Statistics}, 37, 1705 -- 1732.

\bibitem[\protect\citeauthoryear{Boucheron, Lugosi, and Massart}{Boucheron et~al.}{2013}]{Boucheron_L_M}
\textsc{Boucheron, S., G.~Lugosi, and P.~Massart} (2013): \emph{{Concentration inequalities: A nonasymptotic theory of independence}}, Oxford University Press.

\bibitem[\protect\citeauthoryear{Chen and Lee}{Chen and Lee}{2018}]{ChenLee2018}
\textsc{Chen, L.-Y. and S.~Lee} (2018): \enquote{Best subset binary prediction,} \emph{Journal of Econometrics}, 206, 39--56.

\bibitem[\protect\citeauthoryear{Cheng and Liao}{Cheng and Liao}{2015}]{ChengLiao2015}
\textsc{Cheng, X. and Z.~Liao} (2015): \enquote{Select the valid and relevant moments: An information-based LASSO for GMM with many moments,} \emph{Journal of Econometrics}, 186, 443--464.

\bibitem[\protect\citeauthoryear{Chernozhukov, Escanciano, Ichimura, Newey, and Robins}{Chernozhukov et~al.}{2022}]{ChernozhukovEtal2022}
\textsc{Chernozhukov, V., J.~C. Escanciano, H.~Ichimura, W.~K. Newey, and J.~M. Robins} (2022): \enquote{Locally robust semiparametric estimation,} \emph{Econometrica}, 90, 1501--1535.

\bibitem[\protect\citeauthoryear{Chetverikov and Sørensen}{Chetverikov and Sørensen}{2023}]{chetverikov2023}
\textsc{Chetverikov, D. and J.~R.-V. Sørensen} (2023): \enquote{Selecting penalty parameters of high-dimensional M-estimators using bootstrapping after cross-validation,} \emph{arXiv preprint arXiv:2104.04716}.

\bibitem[\protect\citeauthoryear{Dong and Lewbel}{Dong and Lewbel}{2015}]{DongLewbel2015}
\textsc{Dong, Y. and A.~Lewbel} (2015): \enquote{A simple estimator for binary choice models with endogenous regressors,} \emph{Econometric Reviews}, 34, 82--105.

\bibitem[\protect\citeauthoryear{Efromovich}{Efromovich}{2010}]{Efromovich2010}
\textsc{Efromovich, S.} (2010): \enquote{Dimension reduction and adaptation in conditional density estimation,} \emph{Journal of the American Statistical Association}, 105, 761--774.

\bibitem[\protect\citeauthoryear{Elliott and Lieli}{Elliott and Lieli}{2013}]{ElliottLieli2013}
\textsc{Elliott, G. and R.~P. Lieli} (2013): \enquote{Predicting binary outcomes,} \emph{Journal of Econometrics}, 174, 15--26.

\bibitem[\protect\citeauthoryear{Escanciano, Jacho-Chávez, and Lewbel}{Escanciano et~al.}{2014}]{Escanciano2014}
\textsc{Escanciano, J.~C., D.~T. Jacho-Chávez, and A.~Lewbel} (2014): \enquote{Uniform convergence of weighted sums of non and semiparametric residuals for estimation and testing,} \emph{Journal of Econometrics}, 178, 426--443.

\bibitem[\protect\citeauthoryear{Fan and Li}{Fan and Li}{2001}]{FanLi2001}
\textsc{Fan, J. and R.~Li} (2001): \enquote{Variable selection via nonconcave penalized likelihood and its oracle properties,} \emph{Journal of the American Statistical Association}, 96, 1348--1360.

\bibitem[\protect\citeauthoryear{Fan and Lv}{Fan and Lv}{2008}]{FanLv2008}
\textsc{Fan, J. and J.~Lv} (2008): \enquote{Sure independence screening for ultrahigh dimensional feature space,} \emph{Journal of the Royal Statistical Society: Series B (Statistical Methodology)}, 70, 849--911.

\bibitem[\protect\citeauthoryear{Friedman, Hastie, and Tibshirani}{Friedman et~al.}{2010}]{Friedman2010glmnet}
\textsc{Friedman, J., T.~Hastie, and R.~Tibshirani} (2010): \enquote{Regularization paths for generalized linear models via coordinate descent,} \emph{Journal of Statistical Software}, 33, 1--22.

\bibitem[\protect\citeauthoryear{Gautier and Tsybakov}{Gautier and Tsybakov}{2019}]{GautierTsybakov}
\textsc{Gautier, E. and A.~B. Tsybakov} (2019): \enquote{High-dimensional instrumental variables regression and confidence sets,} \emph{arXiv preprint arXiv:1812.11330}.

\bibitem[\protect\citeauthoryear{Gold, Lederer, and Tao}{Gold et~al.}{2020}]{GoldEtal2020}
\textsc{Gold, D., J.~Lederer, and J.~Tao} (2020): \enquote{Inference for high-dimensional instrumental variables regression,} \emph{Journal of Econometrics}, 217, 79--111.

\bibitem[\protect\citeauthoryear{Gong, Kong, Li, and Meng}{Gong et~al.}{2008}]{Gongetal2008}
\textsc{Gong, X., S.~T. Kong, S.~Li, and X.~Meng} (2008): \emph{Rural-urban migrants: a driving force for growth}, Canberra: Asia Pacific Press.

\bibitem[\protect\citeauthoryear{Hall, Racine, and Li}{Hall et~al.}{2004}]{HallRacineLi}
\textsc{Hall, P., J.~Racine, and Q.~Li} (2004): \enquote{Cross-validation and the estimation of conditional probability densities,} \emph{Journal of the American Statistical Association}, 99, 1015--1026.

\bibitem[\protect\citeauthoryear{Hayfield and Racine}{Hayfield and Racine}{2008}]{hayfield2008nonparametric}
\textsc{Hayfield, T. and J.~S. Racine} (2008): \enquote{Nonparametric econometrics: The np package,} \emph{Journal of Statistical Software}, 27, 1--32.

\bibitem[\protect\citeauthoryear{Huang, Ma, and Zhang}{Huang et~al.}{2008}]{HuangEtal2008}
\textsc{Huang, J., S.~Ma, and C.-H. Zhang} (2008): \enquote{Adaptive Lasso for sparse high-dimensional regression models,} \emph{Statistica Sinica}, 18, 1603 -- 1618.

\bibitem[\protect\citeauthoryear{Jing, Shao, and Wang}{Jing et~al.}{2003}]{JingShaoWang}
\textsc{Jing, B.-Y., Q.-M. Shao, and Q.~Wang} (2003): \enquote{Self-normalized Cram{\'e}r-type large deviations for independent random variables,} \emph{The Annals of Probability}, 31, 2167--2215.

\bibitem[\protect\citeauthoryear{Khan, Lan, Tamer, and Yao}{Khan et~al.}{2023}]{KhanEtal2023}
\textsc{Khan, S., X.~Lan, E.~Tamer, and Q.~Yao} (2023): \enquote{Estimating high dimensional monotone index models by iterative convex optimization,} \emph{arXiv preprint arXiv:2110.04388}.

\bibitem[\protect\citeauthoryear{Khan and Tamer}{Khan and Tamer}{2010}]{KhanTamer}
\textsc{Khan, S. and E.~Tamer} (2010): \enquote{Irregular identification, support conditions, and inverse weight estimation,} \emph{Econometrica}, 78, 2021--2042.

\bibitem[\protect\citeauthoryear{Leeb and P{\"o}tscher}{Leeb and P{\"o}tscher}{2006}]{LeebPotscher2006}
\textsc{Leeb, H. and B.~M. P{\"o}tscher} (2006): \enquote{{Can one estimate the conditional distribution of post-model-selection estimators?}} \emph{The Annals of Statistics}, 34, 2554 -- 2591.

\bibitem[\protect\citeauthoryear{Leeb and P{\"o}tscher}{Leeb and P{\"o}tscher}{2008}]{LeebPotscher2008}
---\hspace{-.1pt}---\hspace{-.1pt}--- (2008): \enquote{Sparse estimators and the oracle property, or the return of Hodges’ estimator,} \emph{Journal of Econometrics}, 142, 201--211.

\bibitem[\protect\citeauthoryear{Lewbel}{Lewbel}{2000}]{Lewbel2000}
\textsc{Lewbel, A.} (2000): \enquote{Semiparametric qualitative response model estimation with unknown heteroscedasticity or instrumental variables,} \emph{Journal of Econometrics}, 97, 145--177.

\bibitem[\protect\citeauthoryear{Lewbel, Dong, and Yang}{Lewbel et~al.}{2012}]{LewbelDongYang}
\textsc{Lewbel, A., Y.~Dong, and T.~T. Yang} (2012): \enquote{Comparing features of convenient estimators for binary choice models with endogenous regressors,} \emph{Canadian Journal of Economics}, 45, 809--829.

\bibitem[\protect\citeauthoryear{Li and Racine}{Li and Racine}{2007}]{LiRacine}
\textsc{Li, Q. and J.~S. Racine} (2007): \emph{Nonparametric econometrics: theory and practice}, Princeton University Press.

\bibitem[\protect\citeauthoryear{Li, Zhong, and Zhu}{Li et~al.}{2012}]{LiZhongZhu}
\textsc{Li, R., W.~Zhong, and L.~Zhu} (2012): \enquote{Feature screening via distance correlation learning,} \emph{Journal of the American Statistical Association}, 107, 1129--1139.

\bibitem[\protect\citeauthoryear{Liao}{Liao}{2013}]{Liao2013}
\textsc{Liao, Z.} (2013): \enquote{Adaptive GMM shrinkage estimation with consistent moment selection,} \emph{Econometric Theory}, 29, 857--904.

\bibitem[\protect\citeauthoryear{Lin, Xiang, and Zhang}{Lin et~al.}{2009}]{LinEtal2009}
\textsc{Lin, Z., Y.~Xiang, and C.~Zhang} (2009): \enquote{Adaptive Lasso in high-dimensional settings,} \emph{Journal of Nonparametric Statistics}, 21, 683--696.

\bibitem[\protect\citeauthoryear{Mammen, Rothe, and Schienle}{Mammen et~al.}{2012}]{MammenEtal2011}
\textsc{Mammen, E., C.~Rothe, and M.~Schienle} (2012): \enquote{Nonparametric regression with nonparametrically generated covariates,} \emph{The Annals of Statistics}, 40, 1132 -- 1170.

\bibitem[\protect\citeauthoryear{Manski}{Manski}{1975}]{Manski1975}
\textsc{Manski, C.~F.} (1975): \enquote{Maximum score estimation of the stochastic utility model of choice,} \emph{Journal of Econometrics}, 3, 205--228.

\bibitem[\protect\citeauthoryear{Mullan, Grosjean, and Kontoleon}{Mullan et~al.}{2011}]{Mullan2011}
\textsc{Mullan, K., P.~Grosjean, and A.~Kontoleon} (2011): \enquote{Land tenure arrangements and rural–urban migration in China,} \emph{World Development}, 39, 123--133.

\bibitem[\protect\citeauthoryear{Nagaev}{Nagaev}{1979}]{Nagaev}
\textsc{Nagaev, S.~V.} (1979): \enquote{Large deviations of sums of independent random variables,} \emph{The Annals of Probability}, 745--789.

\bibitem[\protect\citeauthoryear{Powell, Stock, and Stoker}{Powell et~al.}{1989}]{PowellEtal1989}
\textsc{Powell, J.~L., J.~H. Stock, and T.~M. Stoker} (1989): \enquote{Semiparametric estimation of index coefficients,} \emph{Econometrica}, 57, 1403 -- 1430.

\bibitem[\protect\citeauthoryear{Schmidt}{Schmidt}{2010}]{schmidt2010graphical}
\textsc{Schmidt, M.} (2010): \enquote{Graphical model structure learning with l1-regularization,} \emph{Ph.D. Thesis, University of British Columbia}.

\bibitem[\protect\citeauthoryear{Seo and Otsu}{Seo and Otsu}{2018}]{SeoOtsu2018}
\textsc{Seo, M.~H. and T.~Otsu} (2018): \enquote{Local M-estimation with discontinuous criterion for dependent and limited observations,} \emph{The Annals of Statistics}, 46, 344--369.

\bibitem[\protect\citeauthoryear{Serfling}{Serfling}{1980}]{Serfling1980}
\textsc{Serfling, R.~J.} (1980): \emph{Approximation theorems of mathematical statistics}, vol. 162, John Wiley \& Sons.

\bibitem[\protect\citeauthoryear{Su, Yang, Zhang, and Zhou}{Su et~al.}{2022}]{Su_et_al2022}
\textsc{Su, L., T.~T. Yang, Y.~Zhang, and Q.~Zhou} (2022): \enquote{A one-covariate-at-a-time method for nonparametric additive models,} \emph{arXiv preprint arXiv:2204.12023}.

\bibitem[\protect\citeauthoryear{Sz{\'e}kely, Rizzo, and Bakirov}{Sz{\'e}kely et~al.}{2007}]{Szekely_et_al}
\textsc{Sz{\'e}kely, G.~J., M.~L. Rizzo, and N.~K. Bakirov} (2007): \enquote{{Measuring and testing dependence by correlation of distances},} \emph{The Annals of Statistics}, 35, 2769 -- 2794.

\bibitem[\protect\citeauthoryear{Tibshirani}{Tibshirani}{1996}]{Tibshirani1996}
\textsc{Tibshirani, R.} (1996): \enquote{Regression shrinkage and selection via the lasso,} \emph{Journal of the Royal Statistical Society: Series B (Statistical Methodology)}, 58, 267--288.

\bibitem[\protect\citeauthoryear{Wooldridge}{Wooldridge}{2010}]{Wooldridge2010}
\textsc{Wooldridge, J.~M.} (2010): \emph{Econometric analysis of cross section and panel data}, MIT press.

\bibitem[\protect\citeauthoryear{Xue, Yang, and Zhou}{Xue et~al.}{2018}]{XueYangZhou}
\textsc{Xue, S., T.~T. Yang, and Q.~Zhou} (2018): \enquote{Binary choice model with interactive effects,} \emph{Economic Modelling}, 70, 338--350.

\bibitem[\protect\citeauthoryear{Zhang}{Zhang}{2023}]{ZhangJMP2023}
\textsc{Zhang, L.} (2023): \enquote{High-dimensional conditional density estimation and continuous difference-in-differences,} \emph{UCLA Working Paper}.

\bibitem[\protect\citeauthoryear{Zhao}{Zhao}{1999}]{Zhao1999}
\textsc{Zhao, Y.} (1999): \enquote{Leaving the countryside: rural-to-urban migration decisions in China,} \emph{American Economic Review}, 89, 281--286.

\bibitem[\protect\citeauthoryear{Zhao}{Zhao}{2003}]{Zhao2003}
---\hspace{-.1pt}---\hspace{-.1pt}--- (2003): \enquote{The role of migrant networks in labor migration: the case of China,} \emph{Contemporary Economic Policy}, 21, 500--511.

\bibitem[\protect\citeauthoryear{Zou}{Zou}{2006}]{Zou2006}
\textsc{Zou, H.} (2006): \enquote{The adaptive Lasso and its oracle properties,} \emph{Journal of the American Statistical Association}, 101, 1418--1429.

\end{thebibliography}

\newpage
\appendix
 
\setcounter{footnote}{0} \setcounter{table}{0}
\setcounter{figure}{0} \setcounter{section}{0}
\numberwithin{equation}{section}
\setcounter{page}{1}%

\begin{center}
{\Large \textbf{Online Appendix to}}$\vspace{0.08in}$

{\Large \textbf{\textquotedblleft High Dimensional Binary Choice Model with Unknown Heteroskedasticity or Instrumental Variables\textquotedblright}}

(NOT for Publication)

$\medskip$

Fu Ouyang\footnote{School of Economics, University of Queensland},
Thomas T. Yang\footnote{Research School of Economics, Australian National
University}

{\small \ \ \ \ \ \ \ \ \ }
\end{center}

Appendix \ref{APP:introduction} provides a gentle introduction
to the DC and CV procedures. The proofs of all the main theorems are collected
in Appendices \ref{APP:A}--\ref{APP:C}, while the proofs of all technical
lemmas are deferred to Appendix \ref{APP:lemmas}. An investigation
of variable selections with unweighted $L_{1}$ penalty is presented
in Appendix \ref{APP:lasso}. Lastly, Appendix \ref{APP:tables} contains
tables for all simulation and application results.

\section{An Introduction of the DC and CV Procedures\label{APP:introduction}}

We briefly elaborate on the methods of distance covariance (DC) in \citet{Szekely_et_al} and cross-validation (CV) in \citet{HallRacineLi} in this section to facilitate reading.

\subsection{On the DC procedure \label{app:DC}}

\citet{Szekely_et_al} proposed to use the DC to measure dependence
between two random variables. Recall that $\phi_{V}\left(t\right)$,
$\phi_{Z}\left(s\right)$, and $\phi_{V,Z}\left(t,s\right)$ denote
the characteristic functions of $V$, $Z$, and $(V,Z)$, respectively.
To ease reading, we assume both $V$ and $Z$ are scalar random variables,
and we simply present results without any proof; we refer readers
to the original paper for the proof. Observing the test of the hypothesis
of independence 
\[
\mathbb{H}_{0}:\phi_{V,Z}\left(t,s\right)=\phi_{V}\left(t\right)\phi_{Z}\left(s\right)\quad\text{vs.}\quad\mathbb{H}_{1}:\phi_{V,Z}\left(t,s\right)\neq\phi_{V}\left(t\right)\phi_{Z}\left(s\right),
\]
\citet{Szekely_et_al} adopted DC to measure the distance between $\phi_{V,Z}\left(t,s\right)$
and $\phi_{V}\left(t\right)\phi_{Z}\left(s\right):$ 
\[
\mathcal{V}^{2}\left(V,Z;\omega\right)=\int_{\mathbb{R}^{2}}\left\Vert \phi_{V,Z}\left(t,s\right)-\phi_{V}\left(t\right)\phi_{Z}\left(s\right)\right\Vert \omega\left(t,s\right)\,dt\,ds,
\]
where we let $\mathcal{V}^{2}\left(V,Z_{l};\omega\right)$ depend
on the weight $\omega\left(t,s\right)>0$. One remarkable property
of DC is that $\mathcal{V}^{2}\left(V,Z;\omega\right)=0$ if and only
if $V\perp Z$.

The choice of $\omega\left(t,s\right)$ is not essential theoretically.
However, when we take $\omega^{*}\left(t,s\right)=\frac{1}{\pi^{2}t^{2}s^{2}}$
(the weighting function in the main body of the paper), 
\begin{align*}
\mathcal{V}^{2}\left(V,Z;\omega^{*}\right) & =\mathbb{E}\left(\left|V-V'\right|\left|Z-Z'\right|\right)+\mathbb{E}\left|V-V'\right|\mathbb{E}\left|Z-Z'\right| -2\mathbb{E}\left(\left|V-V'\right|\left|Z-Z''\right|\right)\\
 & \equiv S_{1}\left(V,Z\right)+S_{2}\left(V,Z\right)-2S_{3}\left(V,Z\right),
\end{align*}
where $V'$ is an independent copy of $V$, and both $Z'$ and $Z''$
are independent copies of $Z$. We refer readers to Remark 3 in \citet{Szekely_et_al}
for this result. This choice of $\omega^{*}\left(t,s\right)$ greatly
facilitates the computation of its sample analog. Due to this, $\omega^{*}\left(t,s\right)$
is widely adopted for practice. Clearly, $S_{n1}$, $S_{n2}$, and
$S_{n3}$ defined in the main body of the paper are sample analogs
of $S_{1}$, $S_{2}$, and $S_{3}$, respectively.

It is important to make a test statistic scale free. For example,
one can adopt the distance correlation 
\[
\rho\left(V,Z\right)=\frac{\mathcal{V}^{2}\left(V,Z;\omega^{*}\right)}{\sqrt{\mathcal{V}^{2}\left(V,V;\omega^{*}\right)\mathcal{V}^{2}\left(Z,Z;\omega^{*}\right)}},
\]
and clearly $\rho\left(\epsilon V,\epsilon Z\right)=\rho\left(V,Z\right)$ for any non-zero constant $\epsilon$.
Observing that $\mathcal{V}^{2}\left(V,Z;\omega^{*}\right)\leq S_{2}$
(see Section 2.1 in \citet{Szekely_et_al}), we adopt 
\[
\tilde{\rho}\left(V,Z\right)=\frac{\mathcal{V}^{2}\left(V,Z;\omega^{*}\right)}{S_{2}\left(V,Z\right)},
\]
which is also scale free. We chose this due to the following limiting
result, which is Theorem 6 in \citet{Szekely_et_al} on $\tilde{\rho}\left(V,Z\right)$: Let 
\[
\alpha_{n}\left(V,Z\right)=\Pr\left(\frac{n\mathcal{V}_{n}^{2}\left(V,Z\right)}{S_{n2}\left(V,Z\right)}>\left[\Phi^{-1}\left(1-\alpha/2\right)\right]^{2}\right).
\]
Then for all $0<\alpha\leq0.215$, 
\[
\lim_{n\rightarrow\infty}\alpha_{n}\left(V,Z\right)\leq\alpha\text{ and }\sup_{V,Z}\left\{ \lim_{n\rightarrow\infty}\alpha_{n}\left(V,Z\right)\middle|\mathcal{V}^{2}\left(V,Z;\omega^{*}\right)=0\right\} =\alpha.
\]
Finally, DC can measure the dependence between vector random variables,
and the idea and implementation are essentially the same as the uni-variate
case. We omit the discussion on it for briefness.

\subsection{On the CV procedure}\label{app:CV}

\citet{HallRacineLi} introduced an innovative method to mitigate the impacts of irrelevant covariates in the conditional set for estimating conditional densities. The underlying concept is that the theoretical optimal bandwidths for these irrelevant covariates are effectively infinite, such that the covariates provide no useful information. Consequently, when determining the ``optimal'' bandwidths to minimize the CV error for conditional densities, it should be possible to achieve the same rate of convergence as if the irrelevant covariates had been removed from the estimation process.

The performance criterion is the integrated square error (ISE),
\begin{equation}
\text{ISE} = \int \left[\hat{f}\left(v \mid \boldsymbol{z}\right) - f\left(v \mid \boldsymbol{z}\right)\right]^{2} f\left(\boldsymbol{z}\right) \, d\boldsymbol{z} \, dv, \label{eq:ISE}
\end{equation}
which serves as the basis for the CV objective function. In the presence of discrete variables, let $\boldsymbol{Z} = \left(\boldsymbol{Z}^{\mathrm{c}}, \boldsymbol{Z}^{\mathrm{d}}\right)$ represent a division of $\boldsymbol{Z}$ into continuous and discrete components. The ISE can then be expressed as
\[
\text{ISE} = \sum_{\boldsymbol{z}^{\mathrm{d}}} \int \left[\hat{f}\left(v \mid \boldsymbol{z}\right) - f\left(v \mid \boldsymbol{z}\right)\right]^{2} f\left(\boldsymbol{z}\right) \, d\boldsymbol{z} \, dv,
\]
where 
\[
f\left(\boldsymbol{z}\right) = f^{\mathrm{c}}\left(\boldsymbol{z}^{\mathrm{c}} \mid \boldsymbol{z}^{\mathrm{d}}\right) \Pr\left(\boldsymbol{Z}^{\mathrm{d}} = \boldsymbol{z}^{\mathrm{d}}\right),
\]
$f^{\mathrm{c}}\left(\boldsymbol{z}^{\mathrm{c}} \mid \boldsymbol{z}^{\mathrm{d}}\right)$ denotes the density of $\boldsymbol{z}^{\mathrm{c}}$ conditional on $\boldsymbol{Z}^{\mathrm{d}} = \boldsymbol{z}^{\mathrm{d}}$, and $\sum_{\boldsymbol{z}^{\mathrm{d}}}$ takes a summation over all configurations (atoms) of the distribution of $\boldsymbol{Z}^{\mathrm{d}}$.

Since the analysis with both continuous and discrete covariates is
the same as the case with only continuous covariates, we focus
on the latter case for simpler illustration.

To understand how the CV defined in (\ref{EQ:CV}) serves as a surrogate for the ISE, we first expand the right side of (\ref{eq:ISE}) as
\[
\text{ISE} = I_{1n} - 2I_{2n} + I_{3n},
\]
where 
\begin{align*}
I_{1n} & = \int \hat{f}\left(v \mid \boldsymbol{z}\right)^2 f\left(\boldsymbol{z}\right) d\boldsymbol{z} dv,\\
 & = \int \left[\int \hat{f}\left(v, \boldsymbol{z}\right)^2 dv\right] \frac{f\left(\boldsymbol{z}\right)}{\hat{f}\left(\boldsymbol{z}\right)^2} d\boldsymbol{z}\\
 & \equiv \int \frac{\hat{G}\left(\boldsymbol{z}\right)}{\hat{f}\left(\boldsymbol{z}\right)^2} f\left(\boldsymbol{z}\right) d\boldsymbol{z},\\
I_{2n} & = \int \hat{f}\left(v \mid \boldsymbol{z}\right) f\left(v, \boldsymbol{z}\right) d\boldsymbol{z} dv\\
 & = \int \frac{\hat{f}\left(v, \boldsymbol{z}\right)}{\hat{f}\left(\boldsymbol{z}\right)} f\left(v, \boldsymbol{z}\right) d\boldsymbol{z} dv,
\end{align*}
and $I_{3n}$ does not depend on the tuning parameters $\boldsymbol{h}.$

Clearly, $\hat{I}_{1n}$ and $\hat{I}_{2n}$ are leave-one-out sample
analogs of $I_{1n}$ and $I_{2n}$, respectively. Further, since
$I_{3n}$ does not contain the parameters of interest, we can
ignore this term when minimizing the objective CV error.
Thus, we conclude that the CV defined in (\ref{EQ:CV}) is an appropriate proxy
for the ISE.

The main conclusion of \citet{HallRacineLi} is presented in their Theorem 2: the CV procedure (\ref{EQ:bandwidth}) can select asymptotically optimal
bandwidths for relevant variables and push the bandwidths for irrelevant
variables to their upper limits, effectively rendering them uniform
in the conditional set and thus less informative to $V$, with the
probability approaching one as $n \to \infty$.

A final remark is that an implementation of the method in \citet{HallRacineLi}
can be found in the ``\texttt{np}'' package in R, which accommodates both
continuous and discrete random variables.

\section{Main Proofs and Technical Lemmas for Section \ref{SEC:f(v,x)reduce}\label{APP:A}}

We first list all the technical lemmas needed to prove the
main theorems in the paper. The proofs of these lemmas are relegated
to Appendix \ref{APP:lemmas}. In these lemmas, we assume that $\left\{ X_{i}\right\} _{i=1}^{n}$
is an i.i.d. series. $H_{jkm}\equiv H\left(X_{j},X_{k},X_{m}\right)$
is some generic 3-rd order U-statistics and is symmetric in $\left(X_{j},X_{k},X_{m}\right)$. Without loss of generality, we assume $H_{jkm}$ is \textit{nonnegative}.
In addition, we let $H_{jkm,1}\equiv H_{jkm}\boldsymbol{1}\left(H_{jkm}\leq M\right)$
and $H_{jkm,2}\equiv H_{jkm}\boldsymbol{1}\left(H_{jkm}>M\right)$
for a large positive $M.$ We may sometimes write $\left(n-1\right)^{-1}$
or $\left(n-2\right)^{-1}$ simply as $n^{-1}$ to avoid tedious but
straightforward discussion.

We briefly introduce the technical lemmas below.

We show the probability bounds for the deviation of the 3rd and 2nd-order
U-statistics from their means in Lemmas \ref{LE:3rdU} and \ref{LE:2ndU},
respectively. The outline of the proof is as follows. Lemma \ref{LE:3rdU_NoM}
and \ref{LE:3rdU_M}\ show the probability bounds for the moderate
deviation of $H_{jkm,1}$ and $H_{jkm,2}$, respectively, when $j\neq k\neq m$.
Lemma \ref{LE:3rdU_-1} deals with the case when $k=m,$ and Lemma
\ref{LE:3rdU_-2} handles the case when $j=k=m.$ Lemma \ref{LE:3rdU}
collects all the results from Lemmas \ref{LE:3rdU_NoM}, \ref{LE:3rdU_M},
\ref{LE:3rdU_-1}, and \ref{LE:3rdU_-2} and derive the bound. Lemma \ref{LE:2ndU} is a special case of Lemma \ref{LE:3rdU} for the 2nd-order U-statistics. Lemma \ref{LE:Vn_deviation} uses the results developed in Lemmas \ref{LE:3rdU}
and \ref{LE:2ndU}, and shows the probability bounds for the numerator
and denominator in the statistics $\mathcal{\hat{T}}_{nl}.$

Regarding notations, $R_{nj}$ and $T_{nj},$ $j=1,2,...,$ are defined
and used within specific lemmas only.

Note the following set relationships hold: 
\[
\left\{ X_{1}X_{2}\geq a\right\} \subseteq\left\{ X_{1}\geq b\right\} \cup\left\{ X_{2}\geq a/b\right\} ,\text{ for any }X_{1},X_{2}\geq0\text{ and }a,b>0,
\]
\[
\left\{ X\geq a\right\} ,\left\{ X\leq-a\right\} \subseteq\left\{ \left\vert X\right\vert \geq a\right\} \text{ for any }X\text{ and }a\geq0,
\]
\[
\text{and }\left\{ X\geq a\right\} \subseteq\left\{ X\geq b\right\} \text{ for any }X\text{ and }a\geq b.
\]
These relationships can imply certain probability inequalities due
to the fact that ``if $A\subseteq B,$ then $\Pr\left(A\right)\leq\Pr\left(B\right)$.''\ We
frequently use those inequalities to show our results.

\begin{lemma} \label{LE:3rdU_NoM}Suppose that $B_{H2}=\mathbb{E}(H_{jkm}^{2})<\infty$.
For any positive $\varsigma_{n},$ 
\begin{align*}
 & \Pr\left(\left\vert \frac{\sqrt{n}}{n\left(n-1\right)\left(n-2\right)}\sum_{j,k,m=1,j\neq k\neq m}^{n}\left[H_{jkm,1}-\mathbb{E}\left(H_{jkm,1}\right)\right]\right\vert \geq\varsigma_{n}\right)\\
\leq & 2\exp\left(-\frac{\left\lfloor n/3\right\rfloor \varsigma_{n}^{2}}{2n\left(B_{H2}+Mn^{-1/2}\varsigma_{n}/3\right)}\right),
\end{align*}
where $\left\lfloor n/3\right\rfloor $ denotes the integer part of
$n/3.$ \end{lemma}

\begin{lemma} \label{LE:3rdU_M} Suppose that 
\[
\mathbb{E}\left[H\left(X_{j},X_{k},X_{m}\right)^{2+\delta}\right]\leq B_{H\delta}<\infty,
\]
holds for a positive $\delta.$ For any positive $\varsigma_{n},$
\begin{align*}
 & \Pr\left(\left\vert \frac{\sqrt{n}}{n\left(n-1\right)\left(n-2\right)}\sum_{j,k,m=1,j\neq k\neq m}^{n}\left[H_{jkm,2}-\mathbb{E}\left(H_{jkm,2}\right)\right]\right\vert \geq\varsigma_{n}\right)\\
 & \leq C_{1}M^{-\delta}n^{-1}\varsigma_{n}^{-2}+C_{2}M^{-\delta}n^{-\delta/2}\varsigma_{n}^{-\left(2+\delta\right)}+2\exp\left(-C_{3}M^{\delta}\varsigma_{n}^{2}\right),
\end{align*}
for some positive $C_{1}$, $C_{2}$ and $C_{3}.$ \end{lemma}

\begin{lemma} \label{LE:3rdU_-1}Suppose that 
\[
\mathbb{E}\left[H\left(X_{j},X_{k},X_{k}\right)^{2+\delta}\right]\leq B_{H\delta}<\infty
\]
holds for a positive $\delta$, and we let $B_{H1}=\max\left\{ \mathbb{E}\left[H\left(X_{j},X_{k},X_{m}\right)\right],\mathbb{E}\left[H\left(X_{j},X_{k},X_{k}\right)\right]\right\} $.
Then, for any positive $\varsigma_{n}$ and $M$ such that $\varsigma_{n}>10n^{-1/2}B_{H1}$
and $M<n^{1/2}\varsigma_{n}/5,$ 
\[
\Pr\left(\left\vert \frac{\sqrt{n}}{n^{3}}\sum_{j,k=1}^{n}\left[H_{jkk}-\mathbb{E}\left(H_{jkk}\right)\right]\right\vert \geq\varsigma_{n}\right)\leq Cn^{-2}M^{-\delta}\varsigma_{n}^{-2}
\]
holds for some positive $C.$ \end{lemma}

\begin{lemma} \label{LE:3rdU_-2}Suppose that 
\[
\mathbb{E}\left[H\left(X_{j},X_{j},X_{j}\right)^{2+\delta}\right]\leq B_{H\delta}<\infty
\]
holds for a positive $\delta$. Then for any positive $\varsigma_{n}$,
\[
\Pr\left(\left\vert \frac{\sqrt{n}}{n^{3}}\sum_{j=1}^{n}\left[H_{jjj}-\mathbb{E}\left(H_{jjj}\right)\right]\right\vert \geq\varsigma_{n}\right)\leq C_{1}n^{-4-5\delta/2}\varsigma_{n}^{-2-\delta}+2\exp\left(-C_{2}n^{4}\varsigma_{n}^{2}\right)
\]
holds for some positive $C_{1}$ and $C_{2}.$ \end{lemma}

\begin{lemma} \label{LE:3rdU}Suppose that 
\[
\mathbb{E}\left[H\left(X_{j},X_{k},X_{m}\right)^{2+\delta}\right]\leq B_{H\delta}<\infty,
\]
holds for a positive $\delta.$ For any positive $\varsigma_{n}$
such that $\varsigma_{n}\rightarrow\infty$ and $\varsigma_{n}\gtrsim\sqrt{\kappa_{n}\log\left(n\right)}$
for a positive slowly diverging $\kappa_{n},$ 
\[
\Pr\left(\left\vert \frac{\sqrt{n}}{n^{3}}\sum_{j,k,m=1}^{n}\left[H_{jkm}-\mathbb{E}\left(H_{jkm}\right)\right]\right\vert \geq\varsigma_{n}\right)\leq C_{1}n^{-1-\delta/2}\varsigma_{n}^{-2+\delta}+C_{2}n^{-\delta}\varsigma_{n}^{-2}
\]
holds for some positive $C_{1}$ and $C_{2}.$ If in addition $\varsigma_{n}\propto n^{C}$
for some positive $C,$ 
\[
\Pr\left(\left\vert \frac{\sqrt{n}}{n^{3}}\sum_{j,k,m=1}^{n}\left[H_{jkm}-\mathbb{E}\left(H_{jkm}\right)\right]\right\vert \geq\varsigma_{n}\right)\leq C_{3}n^{-1-\delta/2}\varsigma_{n}^{-2-\delta/2}+C_{4}n^{-\delta}\varsigma_{n}^{-2-3\delta/2},
\]
holds for some positive $C_{3}$ and $C_{4}.$ \end{lemma}

\begin{lemma} \label{LE:2ndU}A second order U-statistic, $\frac{1}{n^{2}}\sum_{j,k=1}^{n}H\left(X_{j},X_{k}\right),$
where $H$ is symmetric in $\left(X_{j},X_{k}\right),$ satisfies
\[
\mathbb{E}\left[H\left(X_{j},X_{k}\right)^{2+\delta}\right]\leq B_{H\delta}<\infty,
\]
for a positive $\delta$. Then for any positive $\varsigma_{n}$ such
that $\varsigma_{n}\rightarrow\infty$ and $\varsigma_{n}\gtrsim\sqrt{\kappa_{n}\log\left(n\right)}$
for a positive slowly diverging $\kappa_{n},$ 
\[
\Pr\left(\left\vert \frac{\sqrt{n}}{n^{2}}\sum_{j,k=1}^{n}\left[H_{jk}-\mathbb{E}\left(H_{jk}\right)\right]\right\vert \geq\varsigma_{n}\right)\leq C_{1}n^{-1-\delta/2}\varsigma_{n}^{-2+\delta}+C_{2}n^{-\delta}\varsigma_{n}^{-2}
\]
holds for some positive $C_{1}$ and $C_{2}.$ If in addition $\varsigma_{n}\propto n^{C}$
for some positive $C,$ 
\[
\Pr\left(\left\vert \frac{\sqrt{n}}{n^{2}}\sum_{j,k=1}^{n}\left[H_{jk}-\mathbb{E}\left(H_{jk}\right)\right]\right\vert \geq\varsigma_{n}\right)\leq C_{3}n^{-1-\delta/2}\varsigma_{n}^{-2-\delta/2}+C_{4}n^{-\delta}\varsigma_{n}^{-2-3\delta/2},
\]
holds for some positive $C_{3}$ and $C_{4}.$ \end{lemma}

\begin{lemma} \label{LE:Vn_deviation}Suppose Assumption \ref{A:main}
holds. $\varsigma_{n}$ is a series that satisfies $\sqrt{\kappa_{n}\log\left(n\right)}\lesssim\varsigma_{n}\lesssim n^{1/2}$
for a slowly diverging $\kappa_{n}$. Then 
\[
\Pr\left(\sqrt{n}\left\vert \mathcal{V}_{n}^{2}\left(V,Z_{l}\right)-\mathcal{V}^{2}\left(V,Z_{l}\right)\right\vert \geq\varsigma_{n}\right)\leq C_{1}n^{-1-\delta/2}\varsigma_{n}^{-2+\delta}+C_{2}n^{-\delta}\varsigma_{n}^{-2}
\]
and 
\[
\Pr\left(\left\vert S_{n2}\left(V,Z_{l}\right)-S_{2}\left(V,Z_{l}\right)\right\vert \geq\frac{D_{1}}{2}\right)\leq C_{3}n^{-2-3\delta/4}+C_{4}n^{-1-7\delta/4}
\]
hold uniformly for $l=1,2,...,p_{n}$ for some positive $C_{1},C_{2},C_{3},$
and $C_{4}.$ \end{lemma}

\begin{proof}[Proof of Theorem \ref{TH:main1}] The proof of
the theorem is built on Lemma \ref{LE:Vn_deviation}. In the following,
we first deal with the case with $\frac{\mathcal{V}^{2}\left(V,Z_{l}\right)}{S_{2}\left(V,Z_{l}\right)}\lesssim n^{-1/2}\sqrt{\log n}.$
We then handle the case with $\frac{\mathcal{V}^{2}\left(V,Z_{l}\right)}{S_{2}\left(V,Z_{l}\right)}\gtrsim n^{-1/2}\kappa_{n}\sqrt{\log n}.$\smallskip{}

\noindent \textbf{Some preparations.} It is not hard to see from Assumption
\ref{A:main} that $S_{2}\left(V,Z_{l}\right)$ is uniformly bounded
and bounded away from zero. Mathematically,  
\[
0<D_{1}\leq S_{2}\left(V,Z_{l}\right)\leq D_{2}<\infty
\]
holds for positive $D_1$ and $D_{2}.$ As a result, $\frac{\mathcal{V}^{2}\left(V,Z_{l}\right)}{S_{2}\left(V,Z_{l}\right)}\lesssim n^{-1/2}\sqrt{\log n}$
and $\frac{\mathcal{V}^{2}\left(V,Z_{l}\right)}{S_{2}\left(V,Z_{l}\right)}\gtrsim n^{-1/2}\kappa_{n}\sqrt{\log n}$
imply $\mathcal{V}^{2}\left(V,Z_{l}\right)\lesssim n^{-1/2}\sqrt{\log n}$
and $\mathcal{V}^{2}\left(V,Z_{l}\right)\gtrsim n^{-1/2}\kappa_{n}\sqrt{\log n},$
respectively.\smallskip{}

\noindent \textbf{Part 1 Decomposition. }We deal with the first part
of the theorem. As shown above, $\mathcal{V}^{2}\left(V,Z_{l}\right)\lesssim n^{-1/2}\sqrt{\log n}.$
We conduct the following decomposition: 
\begin{align}
\Pr\left(\mathcal{\hat{T}}_{nl}=\frac{n^{1/2}\mathcal{V}_{n}^{2}\left(V,Z_{l}\right)}{S_{n2}\left(V,Z_{l}\right)}\geq\varsigma_{n}\right) & \leq\Pr\left(\frac{1}{S_{n2}\left(V,Z_{l}\right)}\geq\frac{2}{D_{1}}\right)+\Pr\left(n^{1/2}\mathcal{V}_{n}^{2}\left(V,Z_{l}\right)\geq\frac{D_{1}\varsigma_{n}}{2}\right)\nonumber \\
 & =\Pr\left(S_{n2}\left(V,Z_{l}\right)\leq\frac{D_{1}}{2}\right)+\Pr\left(n^{1/2}\mathcal{V}_{n}^{2}\left(V,Z_{l}\right)\geq\frac{D_{1}\varsigma_{n}}{2}\right)\nonumber \\
 & \equiv T_{n1}+T_{n2}.\label{EQ:stat_case1}
\end{align}

\noindent \textbf{Part 1 on }$T_{n1}$\textbf{. }First, due to $S_{2}\left(V,Z_{l}\right)\geq D_{1},$
$S_{2}\left(V,Z_{l}\right)-D_{1}/2\geq\frac{D_{1}}{2}.$ Using it,
\begin{align*}
T_{n1} & =\Pr\left(S_{n2}\left(V,Z_{l}\right)\leq\frac{D_{1}}{2}\right)\leq\Pr\left(S_{n2}\left(V,Z_{l}\right)\leq S_{2}\left(V,Z_{l}\right)-\frac{D_{1}}{2}\right)\\
 & =\Pr\left(S_{n2}\left(V,Z_{l}\right)-S_{2}\left(V,Z_{l}\right)\leq-\frac{D_{1}}{2}\right)\leq\Pr\left(\left\vert S_{n2}\left(V,Z_{l}\right)-S_{2}\left(V,Z_{l}\right)\right\vert \geq\frac{D_{1}}{2}\right).
\end{align*}
After applying Lemma \ref{LE:Vn_deviation} on the above term, we
obtain 
\begin{equation}
T_{n1}\leq C_{3}n^{-2-3\delta/4}+C_{4}n^{-1-7\delta/4},\label{EQ:stat_case1_p1}
\end{equation}
for some positive $C_{3}$ and $C_{4}.$\smallskip{}

\noindent \textbf{Part 1 on }$T_{n2}$\textbf{.} Since $\mathcal{V}^{2}\left(V,Z_{l}\right)\lesssim n^{-1/2}\sqrt{\log n},$
$\varsigma_{n}=C_{\varsigma}\sqrt{\kappa_{n}\log n},$ and $\kappa_{n}\rightarrow\infty,$
$\mathcal{V}^{2}\left(V,Z_{l}\right)\ll n^{-1/2}\varsigma_{n},\ $which
implies $\mathcal{V}^{2}\left(V,Z_{l}\right)\leq D_{1}n^{-1/2}\varsigma_{n}/4$
for $n$ large enough. Using this observation, 
\begin{align*}
T_{n2} & =\Pr\left(\mathcal{V}_{n}^{2}\left(V,Z_{l}\right)\geq\frac{D_{1}n^{-1/2}\varsigma_{n}}{2}\right)=\Pr\left(\mathcal{V}_{n}^{2}\left(V,Z_{l}\right)-\mathcal{V}^{2}\left(V,Z_{l}\right)\geq\frac{D_{1}n^{-1/2}\varsigma_{n}}{2}-\mathcal{V}^{2}\left(V,Z_{l}\right)\right)\\
 & \leq\Pr\left(\mathcal{V}_{n}^{2}\left(V,Z_{l}\right)-\mathcal{V}^{2}\left(V,Z_{l}\right)\geq\frac{D_{1}n^{-1/2}\varsigma_{n}}{4}\right)\leq\Pr\left(n^{1/2}\left\vert \mathcal{V}_{n}^{2}\left(V,Z_{l}\right)-\mathcal{V}^{2}\left(V,Z_{l}\right)\right\vert \geq\frac{D_{1}\varsigma_{n}}{4}\right).
\end{align*}
Applying Lemma \ref{LE:Vn_deviation} on the above term, we obtain
\begin{equation}
T_{n2}\leq C_{1}n^{-1-\delta/2}\varsigma_{n}^{-2+\delta}+C_{2}n^{-\delta}\varsigma_{n}^{-2},\label{EQ:stat_case1_p2}
\end{equation}
for some positive $C_{1}$ and $C_{2}.$\smallskip{}

\noindent \textbf{Part 1 Final result. }We\textbf{ }put the previous
analysis together. Specifically, substitute (\ref{EQ:stat_case1_p1})
and (\ref{EQ:stat_case1_p2}) into (\ref{EQ:stat_case1}) and keep
only the dominating terms, we obtain 
\[
\Pr\left(\mathcal{\hat{T}}_{nl}=\frac{n^{1/2}\mathcal{V}_{n}^{2}\left(V,Z_{l}\right)}{S_{n2}\left(V,Z_{l}\right)}\geq\varsigma_{n}\right)\leq C_{1}n^{-1-\delta/2}\varsigma_{n}^{-2+\delta}+C_{2}n^{-\delta}\varsigma_{n}^{-2},
\]
where without loss of generality, we re-use constants $C_{1}$ and
$C_{2},$ and we can make $C_{1}$ and $C_{2}$ big enough so that
the inequality holds.\smallskip{}

We turn to the second part of the theorem where we assume $\mathcal{V}^{2}\left(V,Z_{l}\right)/S_{2}\left(V,Z_{l}\right)\gtrsim n^{-1/2}\kappa_{n}\sqrt{\log n}$.

\noindent \textbf{Part 2 Decomposition. }As discussed before, $\mathcal{V}^{2}\left(V,Z_{l}\right)$
satisfies $\mathcal{V}^{2}\left(V,Z_{l}\right)\gtrsim n^{-1/2}\kappa_{n}\sqrt{\log n}$
due to the fact that $S_{2}\left(V,Z_{l}\right)$ is uniformly bounded
and bounded away from zero$.$ Note 
\begin{align}
 & \Pr\left(\mathcal{\hat{T}}_{nl}=\frac{n^{1/2}\mathcal{V}_{n}^{2}\left(V,Z_{l}\right)}{S_{n2}\left(V,Z_{l}\right)}\geq\varsigma_{n}\right)=1-\Pr\left(\frac{n^{1/2}\mathcal{V}_{n}^{2}\left(V,Z_{l}\right)}{S_{n2}\left(V,Z_{l}\right)}<\varsigma_{n}\right)\nonumber \\
 & \geq1-\Pr\left(\frac{1}{S_{n2}\left(V,Z_{l}\right)}<\frac{1}{D_{2}+D_{1}/2}\right)-\Pr\left(n^{1/2}\mathcal{V}_{n}^{2}\left(V,Z_{l}\right)<\left(D_{2}+D_{1}/2\right)\varsigma_{n}\right)\nonumber \\
 & \equiv1-T_{n3}-T_{n4}.\label{EQ:stat_case2}
\end{align}

\noindent \textbf{Part 2 on $T_{n3}$.} Due to $S_{2}\left(V,Z_{l}\right)\leq D_{2},$
$D_{2}+D_{1}/2-S_{2}\left(V,Z_{l}\right)\geq D_{1}/2.$ Then, 
\begin{align}
T_{n3} & =\Pr\left(S_{n2}\left(V,Z_{l}\right)>D_{2}+D_{1}/2\right)=\Pr\left(S_{n2}\left(V,Z_{l}\right)-S_{2}\left(V,Z_{l}\right)>D_{2}+D_{1}/2-S_{2}\left(V,Z_{l}\right)\right)\nonumber \\
 & \leq\Pr\left(S_{n2}\left(V,Z_{l}\right)-S_{2}\left(V,Z_{l}\right)>D_{1}/2\right)\leq\Pr\left(\left\vert S_{n2}\left(V,Z_{l}\right)-S_{2}\left(V,Z_{l}\right)\right\vert >D_{1}/2\right)\nonumber \\
 & \leq C_{5}n^{-2-3\delta/4}+C_{6}n^{-1-7\delta/4}\label{EQ:stat_case2_p1}
\end{align}
for some positive $C_{5}$ and $C_{6}$, where the last line holds
by applying Lemma \ref{LE:Vn_deviation}.\smallskip{}

\noindent \textbf{Part 2 on $T_{n4}$.} Since $\mathcal{V}^{2}\left(V,Z_{l}\right)\gtrsim n^{-1/2}\kappa_{n}\sqrt{\log n},$
$\varsigma_{n}=C_{\varsigma}\sqrt{\kappa_{n}\log n},$ and $\kappa_{n}\rightarrow\infty,$
$\mathcal{V}^{2}\left(V,Z_{l}\right)\gg n^{-1/2}\varsigma_{n},\ $which
implies $\mathcal{V}^{2}\left(V,Z_{l}\right)\geq2n^{-1/2}\left(D_{2}+D_{1}/2+1\right)\varsigma_{n}$
for $n$ large enough. Using this result, we can write 
\begin{align}
T_{n4}= & \Pr\left(n^{1/2}\mathcal{V}_{n}^{2}\left(V,Z_{l}\right)<\left(D_{2}+D_{1}/2\right)\varsigma_{n}\right)\nonumber \\
= & \Pr\left(\mathcal{V}_{n}^{2}\left(V,Z_{l}\right)-\mathcal{V}^{2}\left(V,Z_{l}\right)<n^{-1/2}\left(D_{2}+D_{1}/2\right)\varsigma_{n}-\mathcal{V}^{2}\left(V,Z_{l}\right)\right)\nonumber \\
\leq & \Pr\left(\mathcal{V}_{n}^{2}\left(V,Z_{l}\right)-\mathcal{V}^{2}\left(V,Z_{l}\right)<-n^{-1/2}\varsigma_{n}\right)\leq\Pr\left(n^{1/2}\left\vert \mathcal{V}_{n}^{2}\left(V,Z_{l}\right)-\mathcal{V}^{2}\left(V,Z_{l}\right)\right\vert >\varsigma_{n}\right)\nonumber \\
\leq & C_{7}n^{-1-\delta/2}\varsigma_{n}^{-2+\delta}+C_{8}n^{-\delta}\varsigma_{n}^{-2}\label{EQ:stat_case2_p2}
\end{align}
where the last inequality holds by applying Lemma \ref{LE:Vn_deviation}.
\smallskip{}

\noindent \textbf{Part 2, final result}$.$ Substitute (\ref{EQ:stat_case2_p1})
and (\ref{EQ:stat_case2_p2}) into (\ref{EQ:stat_case2}) and keep
only the dominating terms to obtain 
\[
\Pr\left(\mathcal{\hat{T}}_{nl}=\frac{n^{1/2}\mathcal{V}_{n}^{2}\left(V,Z_{l}\right)}{S_{n2}\left(V,Z_{l}\right)}\geq\varsigma_{n}\right)\geq1-C_{3}n^{-1-\delta/2}\varsigma_{n}^{-2+\delta}-C_{4}n^{-\delta}\varsigma_{n}^{-2},
\]
where without loss of generality, we re-use constants $C_{3}$ and
$C_{4},$ and we can make $C_{3}$ and $C_{4}$ big enough so that
the inequality holds. \end{proof}

\begin{proof}[Proof of Theorem \ref{TH:PDR}] First, we study
TPR$_{n}.$ Note that 
\begin{align*}
\mathbb{E}\left(\text{TPR}_{n}\right) & =p^{\ast-1}\sum_{l=1}^{p_{n}}\mathbb{E}\left[\mathbf{1}\left(\widehat{\mathcal{J}}_{l}=1\text{ and }\mathcal{V}^{2}\left(V,Z_{l}\right)\neq0\right)\right]\\
 & =p^{\ast-1}\sum_{l=1}^{p^{\ast}}\Pr\left(\mathcal{\hat{T}}_{nl}\geq\varsigma_{n}\right)\geq1-C_{1}n^{-1-\delta/2}\varsigma_{n}^{-2+\delta}-C_{2}n^{-\delta}\varsigma_{n}^{-2},
\end{align*}
for some positive constants $C_{1}$ and $C_{2},$ where the inequality
holds by the second part of Theorem \ref{TH:main1} and Assumption
\ref{A:main}.

Now, we turn to FDR$_{n}.$ Note that 
\begin{align*}
\mathbb{E}\left[\sum_{l=1}^{p_{n}}1\left(\widehat{\mathcal{J}}_{l}=1,\text{ and }\mathcal{V}^{2}\left(V,Z_{l}\right)=0\right)\right] & =\sum_{l=p^{\ast}+1}^{p_{n}}\Pr\left(\mathcal{\hat{T}}_{nl}\geq\varsigma_{n}\right)\\
 & \leq\left(p_{n}-p^{\ast}\right)\left(C_{1}n^{-1-\delta/2}\varsigma_{n}^{-2+\delta}+C_{2}n^{-\delta}\varsigma_{n}^{-2}\right),
\end{align*}
holds for some positive constants $C_{3}$ and $C_{4}$ by the first
part of Theorem \ref{TH:main1}$.$ Recall that Assumption \ref{A:main}
imposes 
\[
p_{n}\lesssim\min\left\{ n^{1+\delta/2},n^{\delta}\right\} .
\]
Thus, by $\varsigma_{n}\rightarrow\infty,$ 
\[
\mathbb{E}\left[\sum_{l=1}^{p_{n}}1\left(\widehat{\mathcal{J}}_{l}=1\text{ and }\mathcal{V}^{2}\left(V,Z_{l}\right)=0\right)\right]\leq\left(p_{n}-p^{\ast}\right)\left(C_{1}n^{-1-\delta/2}\varsigma_{n}^{-2+\delta}+C_{2}n^{-\delta}\varsigma_{n}^{-2}\right)\rightarrow0.
\]
Then 
\[
\text{FDR}_{n}=\frac{\sum_{l=1}^{p_{n}}\mathbf{1}\left(\widehat{\mathcal{J}}_{l}=1\text{ and }\mathcal{V}^{2}\left(V,Z_{l}\right)=0\right)}{\sum_{l=1}^{p_{n}}\widehat{\mathcal{J}}_{l}+1}\overset{P}{\rightarrow}0
\]
by Markov inequality and the fact that $\sum_{l=1}^{p_{n}}\widehat{\mathcal{J}}_{l}+1\geq1$.

The last part of the theorem can be seen from 
\begin{align*}
\text{Pr}\left(\widehat{\mathcal{A}}^{\ast}=\mathcal{A}^{\ast}\right) & =\text{Pr}\left(\left\{ \cap_{l=1}^{p^{\ast}}\left\{ \widehat{\mathcal{J}}_{l}=1\right\} \right\} \cap\left\{ \cap_{l=p^{\ast}+1}^{p_{n}}\left\{ \widehat{\mathcal{J}}_{l}=0\right\} \right\} \right)\\
 & =1-\text{Pr}\left(\left\{ \cup_{l=1}^{p^{\ast}}\left\{ \widehat{\mathcal{J}}_{l}=0\right\} \right\} \cup\left\{ \cup_{l=p^{\ast}+1}^{p_{n}}\left\{ \widehat{\mathcal{J}}_{l}=1\right\} \right\} \right)\\
 & \geq1-\sum_{l=1}^{p^{\ast}}\text{Pr}\left(\widehat{\mathcal{J}}_{l}=0\right)-\sum_{l=p^{\ast}+1}^{p_{n}}\text{Pr}\left(\widehat{\mathcal{J}}_{l}=1\right)\\
 & \geq1-p_{n}\left(C_{1}n^{-1-\delta/2}\varsigma_{n}^{-2+\delta}+C_{2}n^{-\delta}\varsigma_{n}^{-2}\right)\\
 & \rightarrow1,
\end{align*}
where the fourth line holds by applying Theorem \ref{TH:main1} to
each $l,$ and the last line holds by Assumption \ref{A:main} with
$p_{n}\lesssim\min\left\{ n^{1+\delta/2},n^{\delta}\right\} $ and
$\varsigma_{n}\rightarrow\infty.$ \end{proof}

\medskip{}

\begin{proof}[Proof of Theorem \ref{TH:post-selection}] \textbf{Some
preparations. }We define one desirable event. We let 
\[
\hat{E}\equiv\left\{ \min_{l=1,2,...,p^{\ast}}\mathcal{\hat{T}}_{nl}>\max_{l=p^{\ast}+1,p^{\ast}+2,...,p_{n}}\mathcal{\hat{T}}_{nl}\right\} ,
\]
the event that all relevant variables have higher $\mathcal{\hat{T}}_{nl}.$
Take the $\varsigma_{n}$ in Theorems \ref{TH:main1} and \ref{TH:PDR},
then 
\begin{align}
\Pr\left(\hat{E}\right) & \geq\Pr\left(\min_{l=1,2,...,p^{\ast}}\mathcal{\hat{T}}_{nl}\geq\varsigma_{n}>\max_{l=p^{\ast}+1,p^{\ast}+2,...,p_{n}}\mathcal{\hat{T}}_{nl}\right)\nonumber \\
 & =1-\Pr\left(\left(\cup_{l=1}^{p^{\ast}}\left\{ \mathcal{\hat{T}}_{nl}<\varsigma_{n}\right\} \right)\cup\left(\cup_{l=p^{\ast}+1}^{p_{n}}\left\{ \mathcal{\hat{T}}_{nl}\geq\varsigma_{n}\right\} \right)\right)\nonumber \\
 & \geq1-\sum_{l=1}^{p^{\ast}}\Pr\left(\mathcal{\hat{T}}_{nl}<\varsigma_{n}\right)-\sum_{l=p^{\ast}+1}^{p_{n}}\Pr\left(\mathcal{\hat{T}}_{nl}\geq\varsigma_{n}\right)\nonumber \\
 & \geq1-o\left(1\right),\label{EQ:pr(e^)}
\end{align}
where the last line is a direct result from Theorems \ref{TH:main1}
and \ref{TH:PDR}.

Denote 
\[
\widehat{\mathcal{A}}^{0}=\left\{ 1,2,...,\tilde{p}\right\} ,
\]
which is the set of indices of $\tilde{p}$ $Z$s with the largest
$\mathcal{\hat{T}}_{nl}.$ We let 
\[
\hat{E}^{\ast}\equiv\left\{ \mathcal{A}^{\ast}\subseteq\widehat{\mathcal{A}}^{0}\right\} 
\]
denote the event that $\mathcal{\hat{T}}_{nl}$ of all relevant variables
are selected.\ By the assumption that $p^{\ast}\leq\tilde{p},$\ we
have 
\[
\hat{E}^{\ast}\supseteq\hat{E}
\]
because the event $\hat{E}$ implies the event $\hat{E}^{\ast}$.
Together with equation (\ref{EQ:pr(e^)}), we have
\[
\Pr\left(\hat{E}^{\ast}\right)\geq\Pr\left(\hat{E}\right)\geq1-o\left(1\right).
\]

\noindent \textbf{The result. }The theorem holds if we can show that
for any $a_{n}\rightarrow\infty$ and small $\varepsilon>0,$ 
\begin{equation}
\Pr\left(\left\vert \hat{f}\left(v|z_{1},...z_{\tilde{p}}\right)-f\left(v|z_{1},...,z_{p^{\ast}}\right)\right\vert \geq a_{n}n^{-\frac{r}{p^{\ast}+1+2r}}\right)\leq\varepsilon\label{EQ:fhatdiff}
\end{equation}
for $n$ large enough. Note that 
\begin{align}
 & \Pr\left(\left\vert \hat{f}\left(v|z_{1},...z_{\tilde{p}}\right)-f\left(v|z_{1},...,z_{p^{\ast}}\right)\right\vert \geq a_{n}n^{-\frac{r}{p^{\ast}+1+2r}}\right)\nonumber \\
= & \Pr\left(\left.\left\vert \hat{f}\left(v|z_{1},...z_{\tilde{p}}\right)-f\left(v|z_{1},...,z_{p^{\ast}}\right)\right\vert \geq a_{n}n^{-\frac{r}{p^{\ast}+1+2r}}\right\vert \hat{E}^{\ast}\right)\Pr\left(\hat{E}^{\ast}\right)\nonumber \\
 & +\Pr\left(\left.\left\vert \hat{f}\left(v|z_{1},...z_{\tilde{p}}\right)-f\left(v|z_{1},...,z_{p^{\ast}}\right)\right\vert \geq a_{n}n^{-\frac{r}{p^{\ast}+1+2r}}\right\vert \hat{E}^{\ast c}\right)\Pr\left(\hat{E}^{\ast c}\right)\nonumber \\
\leq & \Pr\left(\left.\left\vert \hat{f}\left(v|z_{1},...z_{\tilde{p}}\right)-f\left(v|z_{1},...,z_{p^{\ast}}\right)\right\vert \geq a_{n}n^{-\frac{r}{p^{\ast}+1+2r}}\right\vert \hat{E}^{\ast}\right)+\Pr\left(\hat{E}^{\ast c}\right)\nonumber \\
\equiv & T_{n1}+T_{n2}.\label{EQ:fhatdiff_decomp}
\end{align}
Note that we adopt a high-order kernel, and $Y$ and $Z$ are continuous.
Thus, everything in Theorem 3 of \citet{HallRacineLi} can be went
through with bias being the order of $h^{r}$. This leads to the convergence
rate $n^{-\frac{r}{p^{\ast}+1+2r}}$ by some simple calculation. For
$T_{n1},$ conditioning $\hat{E}^{\ast}$ means all relevant variables
are included in $\widehat{\mathcal{A}}^{0},$ and we can apply Theorem
3 in \citet{HallRacineLi} using the updated result above and we obtain
\[
\hat{f}\left(v|z_{1}2,...z_{\tilde{p}}\right)-f\left(v|z_{1},...,z_{p^{\ast}}\right)=O_{P}\left(n^{-\frac{r}{p^{\ast}+1+2r}}\right).
\]
Therefore, for $n$ greater than some positive $N_{1},$ 
\begin{equation}
T_{n1}=\Pr\left(\left.\left\vert \hat{f}\left(v|z_{1},...z_{\tilde{p}}\right)-f\left(v|z_{1},...,z_{p^{\ast}}\right)\right\vert \geq a_{n}n^{-\frac{r}{p^{\ast}+1+2r}}\right\vert \hat{E}^{\ast}\right)\leq\frac{\varepsilon}{2}.\label{EQ:fhatdiff_decomp1}
\end{equation}
For $T_{n2},$ as shown at the beginning of the proof, for $n$ greater
than some positive $N_{2},$ 
\begin{equation}
T_{n2}=\Pr\left(\hat{E}^{\ast c}\right)=1-\Pr\left(\hat{E}^{\ast}\right)\leq\frac{\varepsilon}{2}.\label{EQ:fhatdiff_decomp2}
\end{equation}
Substituting (\ref{EQ:fhatdiff_decomp1}) and (\ref{EQ:fhatdiff_decomp2})
into (\ref{EQ:fhatdiff_decomp}) yields (\ref{EQ:fhatdiff}), for
$n\geq\max\left\{ N_{1},N_{2}\right\} ,$ as desired. \end{proof}

\begin{proof}[Proof of Proposition \ref{CORO:Post-Estimation}] Define
\[
\mu_{f}\left(z_{1},z_{2},...,z_{s},v\right)\equiv\mathbb{E}\left\{ \left[\Pi_{l=1}^{s}K_{h_{l}}\left(z_{li}-z_{l}\right)\right]K_{h_{V}}\left(v_{i}-v\right)\right\} ,
\]
\[
\mu_{m}\left(z_{1},z_{2},...,z_{s}\right)\equiv\mathbb{E}\left[\Pi_{l=1}^{s}K_{h_{l}}\left(z_{li}-z_{l}\right)\right],
\]
and 
\[
\mu_{g}\left(v|z_{1},z_{2},...,z_{s}\right)\equiv\frac{\mu_{f}\left(z_{1},z_{2},...,z_{s},v\right)}{\mu_{m}\left(z_{1},z_{2},...,z_{s}\right)},
\]
and similarly define $\mu_{f}\left(z_{1},v\right),$ $\mu_{m}\left(z_{1}\right),$
and $\mu_{g}\left(v|z_{1}\right)$. By inspecting the proof of Theorem 3 in \citet{HallRacineLi} (or Theorem 4.2 in the working paper version),
the key is to show that 
\begin{equation}
f\left(V|Z_{1},Z_{2},...,Z_{s}\right)=f\left(V|Z_{1}\right)\label{EQ:f_v_z}
\end{equation}
and 
\begin{equation}
\mu_{g}\left(v|z_{1},z_{2},...,z_{s}\right)-\mu_{g}\left(v|z_{1}\right)=O\left(\left(nh_{1}\right)^{-1}\right),\text{ for }h_{1}^{r}\lesssim\left(nh_{1}\right)^{-1},\label{EQ:mu_g}
\end{equation}
because the rest part of its proofs are just on uniform convergence
of certain terms and do not need the conventional independence.\footnote{Note that $\mu_{f2}\left(x,y\right)/\mu_{m}\left(x\right)^{2}$ defined
above equation (6.34) in the working paper version of \citet{HallRacineLi}
is asymptotically negligible. So we do not have to worry about it.}

For equation (\ref{EQ:f_v_z}), 
\begin{align*}
f\left(V|Z_{1},Z_{2},...,Z_{s}\right) & =\frac{f\left(Z_{1},Z_{2},...,Z_{s},V\right)}{f\left(Z_{1},Z_{2},...,Z_{s}\right)}=\frac{f\left(Z_{1},Z_{2},...,Z_{s},V\right)}{f\left(Z_{1}\right)}\frac{f\left(Z_{1}\right)}{f\left(Z_{1},Z_{2},...,Z_{s}\right)}\\
 & =\frac{f\left(Z_{2},...,Z_{s},V|Z_{1}\right)}{f\left(Z_{2},...,Z_{s}|Z_{1}\right)}=\frac{f\left(Z_{2},...,Z_{s}|Z_{1}\right)f\left(V|Z_{1}\right)}{f\left(Z_{2},...,Z_{s}|Z_{1}\right)}=f\left(V|Z_{1}\right),
\end{align*}
where the second equality in the second line uses $\left(Z_{2},...,Z_{s}\right)\perp V|Z_{1}.$

For equation (\ref{EQ:mu_g}), we first define $Q\left(z_{1}\right)\equiv\mathbb{E}\left[\Pi_{l=2}^{s}K_{h_{l}}\left(z_{li}-z_{l}\right)|Z_{1}=z_{1}\right].$
$Q\left(z_{1}\right)$ is $r$-th order differentiable by Assumption
\ref{A:continuousf_kernelK}. Then, 
\begin{align*}
 & \mu_{f}\left(z_{1},z_{2},...,z_{s},v\right)\\
= & \mathbb{E}\left\{ \left[\Pi_{l=1}^{s}K_{h_{l}}\left(z_{li}-z_{l}\right)\right]K_{h_{V}}\left(v_{i}-v\right)\right\} \\
= & \mathbb{E}\left\{ K_{h_{1}}\left(z_{1i}-z_{1}\right)\mathbb{E}\left[\Pi_{l=2}^{s}K_{h_{l}}\left(z_{li}-z_{l}\right)|Z_{1}=z_{1i}\right]\mathbb{E}\left[K_{h_{V}}\left(v_{i}-v\right)|Z_{1}=z_{1i}\right]\right\} \\
= & Q\left(z_{1}\right)\mathbb{E}\left[K_{h_{V}}\left(v_{i}-v\right)|Z_{1}=z_{1}\right]f\left(z_{1}\right)+O\left(h_{1}^{r}\right),
\end{align*}
where the third line holds by the law of iterated expectation, and
$\left(Z_{2},...,Z_{s}\right)\perp V|Z_{1},$ and the fourth line
holds by the Taylor expansion around $Z_{1}=z_{1}$.\footnote{The above holds due to the following observation. For any smooth enough
functions $\phi_{1}\left(z\right)$ and $\phi_{2}\left(z\right)$,
\begin{align*}
\mathbb{E}\left\{ K_{h_{1}}\left(z_{1i}-z_{1}\right)\phi_{1}\left(z_{1i}\right)\phi_{2}\left(z_{1i}\right)\right\}  & =\int K_{h_{1}}\left(s-z_{1}\right)\phi_{1}\left(s\right)\phi_{2}\left(s\right)f\left(s\right)ds\\
 & =\phi_{1}\left(z_{1}\right)\phi_{2}\left(z_{1}\right)f\left(z_{1}\right)+O\left(h_{1}^{r}\right).
\end{align*}

{}} Similarly, 
\[
\mu_{m}\left(z_{1},z_{2},...,z_{s}\right)=\mathbb{E}\left[\Pi_{l=1}^{s}K_{h_{l}}\left(z_{li}-z_{l}\right)\right]=Q\left(z_{1}\right)f\left(z_{1}\right)+O\left(h_{1}^{r}\right).
\]
Together, they imply 
\[
\mu_{g}\left(v|z_{1},z_{2},...,z_{s}\right)=\frac{\mu_{f}\left(z_{1},z_{2},...,z_{s},v\right)}{\mu_{m}\left(z_{1},z_{2},...,z_{s}\right)}=\mathbb{E}\left[K_{h_{V}}\left(v_{i}-v\right)|Z_{1}=z_{1}\right]+O\left(h_{1}^{r}\right).
\]
The Taylor expansion around $Z_{1}=z_{1}$ yields 
\[
\mu_{g}\left(v|z_{1}\right)=\frac{\mu_{f}\left(z_{1},v\right)}{\mu_{m}\left(z_{1}\right)}=\mathbb{E}\left[K_{h_{V}}\left(v_{i}-v\right)|Z_{1}=z_{1}\right]+O\left(h_{1}^{r}\right).
\]
Take a difference of the terms in the above two equations, 
\[
\mu_{g}\left(v|z_{1},z_{2},...,z_{s}\right)-\mu_{g}\left(v|z_{1}\right)=O\left(h_{1}^{r}\right)=O(\left(nh_{1}\right)^{-1})
\]
when $h_{1}^{r}\lesssim\left(nh_{1}\right)^{-1}$, as desired. \end{proof}

\section{Main Proofs and Technical Lemmas for Section \ref{SEC:VariableSelection}\label{APP:B1}}

Since we try to reproduce the results in \citet{LinEtal2009}, we
will use similar notations as theirs.

We denote $\mathbf{X}_{n}=\left(\mathbf{X}_{1n},...,\mathbf{X}_{p_{n}n}^ {}\right)$
with $\mathbf{X}_{jn}^ {}=\left(x_{j1},x_{j2},...,x_{jn}\right)^{\prime},$
and decompose $\mathbf{X}_{n}$ into $(\mathbf{X}_{n}^{\left(1\right)},\mathbf{X}_{n}^{\left(2\right)}),$
where $\mathbf{X}_{n}^{\left(1\right)}=\left(\mathbf{X}_{1n}^ {},...,\mathbf{X}_{s^{\ast}n}^ {}\right)$
is the first $n\times s^{\ast}$ submatrix, and $\mathbf{X}_{n}^{\left(2\right)}=(\mathbf{X}_{(s^{\ast}+1)n},...,\mathbf{X}_{p_{n}n})$
is the last $n\times\left(p_{n}-s^{\ast}\right)$ submatrix.

The ``Gram matrix''\ is denoted as $\mathbf{C}_{n}=\frac{1}{n}\mathbf{X}_{n}^{\prime}\mathbf{X}_{n}$
and $\mathbf{C}_{n}^{\left(i,j\right)}=\frac{1}{n}\mathbf{X}_{n}^{\left(i\right)\prime}\mathbf{X}_{n}^{\left(j\right)}$
for $i,j=1,2.$ We similarly denote $\boldsymbol{x}_{i}^{\left(1\right)}\equiv\left(x_{1i},...,x_{s^{\ast}i}\right)^{\prime}$,
$\boldsymbol{x}_{i}^{\left(2\right)}\equiv\left(x_{\left(s^{\ast}+1\right)i},...,x_{p_{n}i}\right)^{\prime}$
and $\boldsymbol{\beta}^{\ast\left(1\right)}\equiv\left(\beta_{1}^{\ast},\beta_{2}^{\ast},...,\beta_{s^{\ast}}^{\ast}\right)^{\prime}.$

$\text{sgn}\left(\cdot\right)$ denotes the sign function. It is defined
as 
\[
\text{sgn}\left(b\right)=\left\{ \begin{array}{c}
1\\
-1\\
0
\end{array}\right.\begin{array}{c}
\text{if }b>0,\\
\text{if }b<0,\\
\text{if }b=0.
\end{array}
\]
Using this notation, $\hat{\boldsymbol{\beta}}=_{s}\mathbf{\boldsymbol{\beta}^{*}}$
if $\text{sgn}\left(\hat{\beta}_{j}\right)=\text{sgn}\left(\beta_{j}^{*}\right),$
$j=1,...,p_{n}$. Let

\[
\boldsymbol{S}^{\left(1\right)}=\left(\varpi_{1}\text{sgn}\left(\beta_{1}^{*}\right),...,\varpi_{s^{*}}\text{sgn}\left(\beta_{s^{*}}^{*}\right)\right),
\]
\[
\boldsymbol{\varpi}^{\left(1\right)}=\left(\varpi_{1},...,\varpi_{s^{*}}\right)'\text{ and }\boldsymbol{\varpi}^{\left(2\right)}=\left(\varpi_{s^{*}+1},...,\varpi_{p_{n}}\right)',
\]
where $\varpi_{j},j=1,...,p_{n},$ is defined in (\ref{eq:adaptiveLasso}).
Also 
\[
\hat{\boldsymbol{Q}}=\frac{\mathbf{X}_{n}^{\prime}\left(\widehat{\mathbf{\tilde{Y}}}-\mathbf{X}_{n}\mathbf{\boldsymbol{\beta}^{*}}\right)}{\sqrt{n}},\;\hat{\boldsymbol{Q}}^{\left(1\right)}=\frac{\mathbf{X}_{n}^{\left(1\right)\prime}\left(\widehat{\mathbf{\tilde{Y}}}-\mathbf{X}_{n}\mathbf{\boldsymbol{\beta}^{*}}\right)}{\sqrt{n}},\;\text{and}\;\hat{\boldsymbol{Q}}^{\left(2\right)}=\frac{\mathbf{X}_{n}^{\left(2\right)\prime}\left(\widehat{\mathbf{\tilde{Y}}}-\mathbf{X}_{n}\mathbf{\boldsymbol{\beta}^{*}}\right)}{\sqrt{n}}.
\]
Further, denote $u=\tilde{Y}-\boldsymbol{X}^{\prime}\boldsymbol{\beta}^{\ast},$

\begin{equation}
\boldsymbol{q}=u\boldsymbol{X}_{1}^{s^{\ast}}+\mathbb{E}\left(u\boldsymbol{X}_{1}^{s^{\ast}}\left\vert \boldsymbol{X}_{1}^{p^{\ast}}\right.\right)-\mathbb{E}\left(u\boldsymbol{X}_{1}^{s^{\ast}}\left\vert \boldsymbol{X}_{1}^{p^{\ast}},V\right.\right),\label{eq:q_influence}
\end{equation}
and 
\begin{equation}
\Omega_{n}\equiv\mathbb{E}\left[\boldsymbol{q}\boldsymbol{q}'\right].\label{EQ:omegan-1}
\end{equation}
Note that $\boldsymbol{q}$ is the population of the influence function
of $n^{-1}\sum_{i=1}^{n}$ $\boldsymbol{x}_{i}^{\left(1\right)}\left(\widehat{\tilde{y}}_{i}-\boldsymbol{x}_{i}^{\prime}\boldsymbol{\beta}^{\ast}\right)$
with the plugged-in $\hat{f}.$ We refer readers to \citet{Lewbel2000}
for the details. If we know $f$ for some reason, the population influence
function is simply $u\boldsymbol{X}_{1}^{s^{\ast}}.$

We present a few lemmas. The first lemma is Lemma 3.1 from \citet{LinEtal2009}.

\begin{lemma}\label{LE:sign_consistent}$\Pr\left(\hat{\boldsymbol{\beta}}=_{s}\mathbf{\boldsymbol{\beta}^{*}}\right)\geq\Pr\left(A\cap B\right),$
\[
A=\left\{ \left|\left(\mathbf{C}_{n}^{\left(1,1\right)}\right)^{-1}\hat{\boldsymbol{Q}}^{\left(1\right)}\right|<\sqrt{n}\left[\left|\boldsymbol{\beta}^{*\left(1\right)}\right|-\frac{\lambda_{n}}{2n}\left|\left(\mathbf{C}_{n}^{\left(1,1\right)}\right)^{-1}\boldsymbol{S}^{\left(1\right)}\right|\right]\right\} 
\]
and 
\[
B=\left\{ \left|\mathbf{C}_{n}^{\left(2,1\right)}\left(\mathbf{C}_{n}^{\left(1,1\right)}\right)^{-1}\hat{\boldsymbol{Q}}^{\left(1\right)}-\hat{\boldsymbol{Q}}^{\left(2\right)}\right|+\frac{\lambda_{n}}{2\sqrt{n}}\left|\mathbf{C}_{n}^{\left(2,1\right)}\left(\mathbf{C}_{n}^{\left(1,1\right)}\right)^{-1}\boldsymbol{S}^{\left(1\right)}\right|\leq\frac{\lambda_{n}}{2\sqrt{n}}\boldsymbol{\varpi}^{\left(2\right)}\right\} .
\]

\end{lemma}

\begin{lemma} \label{LE:matrixXX'} Suppose i.i.d. and Assumption
\ref{A:adaptiveLasso} holds. Then 
\[
\Pr\left(\rho_{\min}\left(\mathbf{C}_{n}^{\left(1,1\right)}\right)\geq\frac{D_{3}}{2}\right)=1-o\left(1\right).
\]

\end{lemma}

\begin{lemma} \label{LE:matrixXX'C21} Suppose i.i.d. and Assumption
\ref{A:adaptiveLasso} holds. Denote $C_{X2}=\max_{j=1,...,p_{n}}\mathbb{E}\left(X_{j}^{2}\right)$,
then 
\[
\Pr\left(\max_{j=s^{*}+1,...,p_{n}}\left\Vert \frac{1}{n}\mathbf{X}_{jn}^{\prime}\mathbf{X}_{n}^{\left(1\right)}\right\Vert \geq2C_{X2}\sqrt{s^{*}}\right)=o\left(1\right).
\]

\end{lemma}

\begin{lemma} \label{LE:prepare} Suppose Assumption \ref{A:main},
\ref{A:continuousf_kernelK}, \ref{A:specialR} and \ref{A:adaptiveLasso}
hold. In additional, we adopt $2r$-th order kernel to construct $\hat{f}\left(V\left\vert \boldsymbol{X}_{1}^{\tilde{p}}\right.\right)$
for $\widehat{\tilde{y}}_{i}$. Then 
\begin{align*}
 & \frac{1}{\sqrt{n}}\sum_{i=1}^{n}x_{ji}\left(\widehat{\tilde{y}}_{i}-\boldsymbol{x}_{i}^{\prime}\boldsymbol{\beta}^{\ast}\right)=\frac{1}{\sqrt{n}}\sum_{i=1}^{n}x_{ji}\left(\tilde{y}_{i}-\boldsymbol{x}_{i}^{\prime}\boldsymbol{\beta}^{\ast}\right)\\
 & +\frac{1}{\sqrt{n}}\sum_{i=1}^{n}x_{ji}\left(y_{i}-\boldsymbol{1}\left(v_{i}>0\right)\right)\left[\frac{\hat{f}_{\boldsymbol{X}i}-\mathbb{E}\left(\hat{f}_{\boldsymbol{X}i}\right)}{\mathbb{E}\left(\hat{f}_{V\boldsymbol{X}i}\right)}+\frac{\mathbb{E}\left(\hat{f}_{\boldsymbol{X}i}\right)\left(\mathbb{E}\left(\hat{f}_{V\boldsymbol{X}i}\right)-\hat{f}_{V\boldsymbol{X}i}\right)}{\left[\mathbb{E}\left(\hat{f}_{V\boldsymbol{X}i}\right)\right]^{2}}\right]+o_{P}\left(1\right),
\end{align*}
and 
\[
\Pr\left(\max_{j=1,2.,,,p_{n}}\left\vert \frac{1}{\sqrt{n}}\sum_{i=1}^{n}x_{ji}\left(\widehat{\tilde{y}}_{i}-\boldsymbol{x}_{i}^{\prime}\boldsymbol{\beta}^{\ast}\right)\right\vert \geq\varsigma_{n}\text{ }\right)=o\left(1\right)
\]
for any $\varsigma_{n}\propto n^{C}$,\ where $C$ is a positive
constant. \end{lemma}

\begin{lemma} \label{LE:AB}Suppose i.i.d. and Assumption \ref{A:adaptiveLasso}
holds. The events $A$ and $B$ defined in Lemma \ref{LE:sign_consistent}
satisfy 
\[
\Pr\left(A^{c}\right)\rightarrow0\text{ and }\Pr\left(B^{c}\right)\rightarrow0
\]

\end{lemma}

\begin{proof}[Proof of Theorem \ref{TH:AdaptiveLasso}]Lemmas
\ref{LE:sign_consistent} and \ref{LE:AB} imply that 
\[
\Pr\left(\hat{\boldsymbol{\beta}}=_{s}\mathbf{\boldsymbol{\beta}^{*}}\right)\geq\Pr\left(A\cap B\right)\geq1-\Pr\left(A^{c}\right)-\Pr\left(B^{c}\right)=1-o\left(1\right).
\]
By the definition of sign consistency and the sparsity of $\mathbf{\boldsymbol{\beta}^{*}}$,
\[
\Pr\left(\hat{\boldsymbol{\beta}}^{\left(2\right)}=\boldsymbol{0}\right)=1-o\left(1\right).
\]

Moreover, the first order condition implies (by setting $\hat{\boldsymbol{\beta}}^{\left(2\right)}=\boldsymbol{0}$)
that 
\[
\Pr\left(\mathbf{X}_{n}^{\left(1\right)\prime}\left(\widehat{\mathbf{\tilde{Y}}}-\mathbf{X}_{n}^{\left(1\right)}\hat{\boldsymbol{\beta}}^{\left(1\right)}\right)=\frac{\lambda_{n}}{2}\boldsymbol{S}^{\left(1\right)}\right)=1-o\left(1\right),
\]
thus 
\[
\Pr\left(\sqrt{n}\boldsymbol{e}'\Sigma_{n}^{-1/2}\left(\hat{\boldsymbol{\beta}}^{\left(1\right)}-\boldsymbol{\beta}^{*\left(1\right)}\right)=\boldsymbol{e}'\Sigma_{n}^{-1/2}\left(\mathbf{C}_{n}^{\left(1,1\right)}\right)^{-1}\hat{\boldsymbol{Q}}^{\left(1\right)}-\boldsymbol{e}'\Sigma_{n}^{-1/2}\left(\mathbf{C}_{n}^{\left(1,1\right)}\right)^{-1}\frac{\lambda_{n}}{2\sqrt{n}}\boldsymbol{S}^{\left(1\right)}\right)=1-o\left(1\right).
\]
Note that 
\[
\left|\boldsymbol{e}'\Sigma_{n}^{-1/2}\left(\mathbf{C}_{n}^{\left(1,1\right)}\right)^{-1}\frac{\lambda_{n}}{2\sqrt{n}}\boldsymbol{S}^{\left(1\right)}\right|\apprle\sqrt{s^{*}}\frac{\lambda_{n}}{\sqrt{n}}\left(\min_{j=1,...,s^{*}}\left|\tilde{\beta}_{j}\right|\right)^{-\gamma}\apprle_{P}n^{C_{\lambda}+\alpha\gamma+C_{s}/2-1/2}=o_{P}\left(1\right),
\]
due to $C_{\lambda}<1/2-\alpha\gamma-C_{s}/2$ imposed in Assumption
\ref{A:adaptiveLasso}. Therefore, 
\begin{equation}
\sqrt{n}\boldsymbol{e}'\Sigma_{n}^{-1/2}\left(\hat{\boldsymbol{\beta}}^{\left(1\right)}-\boldsymbol{\beta}^{*\left(1\right)}\right)=\boldsymbol{e}'\Sigma_{n}^{-1/2}\left(\mathbf{C}_{n}^{\left(1,1\right)}\right)^{-1}\hat{\boldsymbol{Q}}^{\left(1\right)}+o_{P}\left(1\right).\label{eq:beta^}
\end{equation}
So the key is to show the asymptotics of $\boldsymbol{e}'\Sigma_{n}^{-1/2}\left(\mathbf{C}_{n}^{\left(1,1\right)}\right)^{-1}\hat{\boldsymbol{Q}}^{\left(1\right)}.$

To this end

\begin{align}
\boldsymbol{e}'\Sigma_{n}^{-1/2}\left(\mathbf{C}_{n}^{\left(1,1\right)}\right)^{-1}\hat{\boldsymbol{Q}}^{\left(1\right)} & =\sum_{i=1}^{n}\frac{\boldsymbol{e}'}{\sqrt{n}}\Sigma_{n}^{-1/2}\left(\mathbf{C}_{n}^{\left(1,1\right)}\right)^{-1}\boldsymbol{x}_{i}^{\left(1\right)}\left(\widehat{\tilde{y}}_{i}-\boldsymbol{x}_{i}^{\left(1\right)\prime}\boldsymbol{\beta}^{\left(1\right)\ast}\right)\nonumber \\
 & =\sum_{i=1}^{n}\frac{\boldsymbol{e}'}{\sqrt{n}}\Sigma_{n}^{-1/2}\left(\left[\mathbb{E}\left(\boldsymbol{X}_{1}^{s^{\ast}}\boldsymbol{X}_{1}^{s^{\ast}\prime}\right)\right]\right)^{-1}\boldsymbol{q}_{i}+o_{P}\left(1\right)\nonumber \\
 & \equiv\sum_{i=1}^{n}\phi_{ni}+o_{P}\left(1\right)\label{eq:phni}
\end{align}
where the last line applies the results ($\boldsymbol{q}_{i}$ defined
in (\ref{eq:q_influence}) being the influence term) in \citet{Lewbel2000}.
To get the asymptotic normality, one sufficient condition is the Lindeberg's
condition of the triangular array Central Limit Theorem: 
\[
\sum_{i=1}^{n}\mathbb{E}\left[\phi_{ni}^{2}\mathbf{1}\left(\phi_{ni}>\epsilon\right)\right]=o\left(1\right),\text{ for any positive }\epsilon.
\]
Note 
\begin{align*}
 & \sum_{i=1}^{n}\mathbb{E}\left[\phi_{ni}^{2}\mathbf{1}\left(\phi_{ni}>\epsilon\right)\right]\\
\leq & \frac{1}{\epsilon^{2}}\sum_{i=1}^{n}\mathbb{E}\left[\phi_{ni}^{4}\right]=\frac{1}{n\epsilon^{2}}\mathbb{E}\left[\left|\boldsymbol{e}'\Sigma_{n}^{-1/2}\left(\left[\mathbb{E}\left(\boldsymbol{X}_{1}^{s^{\ast}}\boldsymbol{X}_{1}^{s^{\ast}\prime}\right)\right]\right)^{-1}\boldsymbol{q}_{i}\boldsymbol{q}_{i}'\left(\left[\mathbb{E}\left(\boldsymbol{X}_{1}^{s^{\ast}}\boldsymbol{X}_{1}^{s^{\ast}\prime}\right)\right]\right)^{-1}\left(\Sigma_{n}^{-1/2}\right)'\boldsymbol{e}\right|^{2}\right]\\
\stackrel{(\textrm{i})}{\apprle} & \frac{1}{n\epsilon^{2}}\mathbb{E}\left[\left(\rho_{\max}\left\{ \boldsymbol{q}_{i}\boldsymbol{q}_{i}'\right\} \right)^{2}\right]\leq\frac{1}{n\epsilon^{2}}\mathbb{E}\left[\left(\text{trace}\left\{ \boldsymbol{q}_{i}\boldsymbol{q}_{i}'\right\} \right)^{2}\right]=\frac{1}{n\epsilon^{2}}\mathbb{E}\left[\left(\sum_{j=1}^{s^{*}}q_{ij}^{2}\right)^{2}\right]\\
\leq & \frac{s^{*2}}{n\epsilon^{2}}\max_{j=1,...,s^{*}}\mathbb{E}\left(q_{ij}^{4}\right)\rightarrow0,
\end{align*}
where ($\textrm{i}$) holds due to the bounded eigenvalues of all
the matrices involved, and last line holds due to the moment conditions
and $s^{*2}/n\rightarrow0.$ Thus, we have verified the Lindeberg's
condition for $\sum_{i=1}^{n}\phi_{ni}.$ Apply the Lindeberg Central
Limit Theorem on $\sum_{i=1}^{n}\phi_{ni}$ in (\ref{eq:phni}), and
substitute the result back to (\ref{eq:beta^}), we can 
\[
\sqrt{n}\boldsymbol{e}'\Sigma_{n}^{-1/2}\left(\hat{\boldsymbol{\beta}}^{\left(1\right)}-\boldsymbol{\beta}^{*\left(1\right)}\right)\stackrel{d}{\rightarrow}N\left(0,1\right),
\]
by Slutsky's Theorem.
\end{proof}

\section{Main Proofs and Technical Lemmas for Section \ref{SEC:MSelection}\label{APP:C}}

We introduce some notations in the below. For the moment conditions
and the asymptotics, 
\[
u=\tilde{Y}-\boldsymbol{X}^{\prime}\boldsymbol{\beta}^{\ast},
\]
\[
\boldsymbol{q}^{\ast}=u\boldsymbol{Z}^{\ast}+\mathbb{E}\left(u\boldsymbol{Z}^{\ast}\left\vert \boldsymbol{Z}_{1}^{p^{\ast}}\right.\right)-\mathbb{E}\left(u\boldsymbol{Z}^{\ast}\left\vert \boldsymbol{Z}_{1}^{p^{\ast}},V\right.\right),
\]
\[
\underline{\boldsymbol{q}}_{D}=u\underline{\boldsymbol{Z}}_{D}-\underline{\boldsymbol{\eta}}^{\ast}+\mathbb{E}\left(u\underline{\boldsymbol{Z}}_{D}\left\vert \boldsymbol{Z}_{1}^{p^{\ast}}\right.\right)-\mathbb{E}\left(u\underline{\boldsymbol{Z}}_{D}\left\vert \boldsymbol{Z}_{1}^{p^{\ast}},V\right.\right).
\]
By the definition of notations 
\[
\boldsymbol{q}_{1}^{p_{n}}=\left(\boldsymbol{q}^{\ast\prime},\underline{\boldsymbol{q}}_{D}^{\prime}\right)^{\prime}.
\]
Then we define 
\begin{equation}
\Omega_{n}\equiv\mathbb{E}\left[\left(\begin{array}{c}
\boldsymbol{q}^{\ast}\\
\underline{\boldsymbol{q}}_{D}
\end{array}\right)\left(\begin{array}{cc}
\boldsymbol{q}^{\ast\prime} & \underline{\boldsymbol{q}}_{D}^{\prime}\end{array}\right)\right].\label{EQ:omegan}
\end{equation}
Note that $\boldsymbol{q}^{\ast}$ is the population of the influence
function of $n^{-1}\sum_{i=1}^{n}$ $\boldsymbol{z}_{i}^{\ast}\left(\widehat{\tilde{y}}_{i}-\boldsymbol{x}_{i}^{\prime}\boldsymbol{\beta}^{\ast}\right)$
with the plugged-in $\hat{f}.$ We refer readers to \citet{Lewbel2000}
for the details. If we know $f$ for some reason, the population influence
function is simply $u\boldsymbol{Z}^{\ast},$ the same as in \citet{ChengLiao2015}.
This similarly applies to $\underline{\boldsymbol{q}}_{D}.$

$\Gamma_{\boldsymbol{\theta}_{B}}$ (defined in (\ref{eq:GammaTheta}))
is the limit of 
\begin{equation}
\frac{\partial\overline{\boldsymbol{\hat{m}}}_{n}\left(\boldsymbol{\theta}\right)}{\partial\boldsymbol{\theta}_{B}^{\prime}}=\left(\begin{array}{cc}
-n^{-1}\sum_{i=1}^{n}\boldsymbol{z}_{i}^{\ast}\boldsymbol{x}_{i}^{\prime} & \mathbf{0}_{k^{\ast}\times d_{B}}\\
-n^{-1}\sum_{i=1}^{n}\underline{\boldsymbol{z}}_{A}^{\ast}\boldsymbol{x}_{i}^{\prime} & \mathbf{0}_{d_{A}\times d_{B}}\\
-n^{-1}\sum_{i=1}^{n}\underline{\boldsymbol{z}}_{B}^{\ast}\boldsymbol{x}_{i}^{\prime} & -\boldsymbol{I}_{d_{B}\times d_{B}}
\end{array}\right)\equiv\frac{\partial\overline{\boldsymbol{\hat{m}}}_{n}}{\partial\boldsymbol{\theta}_{B}^{\prime}},\label{EQ:pd_linear}
\end{equation}
and it is not a function of $\boldsymbol{\theta}$ and denoted as
$\frac{\partial\overline{\boldsymbol{\hat{m}}}_{n}}{\partial\boldsymbol{\theta}_{B}^{\prime}}.$

We let $z_{ji}$ be the $i$-th realization of $Z_{j}$ and $Z_{j}$
is the $j$-th element of $\boldsymbol{Z}_{1}^{p_{n}}.$ Similarly,
$q_{ji}$ denote the $i$-th realization of $q_{j}$ and $q_{j}$
is the $j$-th element of $\boldsymbol{q}_{1}^{p_{n}}$. We write
$\hat{f}_{\boldsymbol{Z}i},\hat{f}_{V\boldsymbol{Z}i},f_{\boldsymbol{Z}i},f_{V\boldsymbol{Z}i}$
for short for $\hat{f}\left(\boldsymbol{z}_{1i}^{\tilde{p}}\right),\hat{f}\left(v_{i},\boldsymbol{z}_{1i}^{\tilde{p}}\right),f\left(\boldsymbol{z}_{1i}^{p^{\ast}}\right),$
and $f\left(v_{i},\boldsymbol{z}_{1i}^{p^{\ast}}\right),$ respectively.
We note that $\hat{f}_{\boldsymbol{Z}i}$ and $f_{\boldsymbol{Z}i}$
may contain different elements of $Z$, but we still use those notations
if no confusion arises.

\begin{lemma} \label{LE:expansion}Suppose Assumption \ref{A:main},
\ref{A:continuousf_kernelK}, \ref{A:specialR} and \ref{A:GMM} hold.
In additional, we adopt $2r$-th order kernel to construct $\hat{f}\left(V\left\vert \boldsymbol{Z}_{1}^{\tilde{p}}\right.\right)$
for $\widehat{\tilde{y}}_{i}$. Then for $1\leq j\leq s^{\ast}$ 
\begin{align*}
 & \frac{1}{\sqrt{n}}\sum_{i=1}^{n}z_{ji}\left(\widehat{\tilde{y}}_{i}-\boldsymbol{x}_{i}^{\prime}\boldsymbol{\beta}^{\ast}\right)=\frac{1}{\sqrt{n}}\sum_{i=1}^{n}z_{ji}\left(\tilde{y}_{i}-\boldsymbol{x}_{i}^{\prime}\boldsymbol{\beta}^{\ast}\right)\\
 & +\frac{1}{\sqrt{n}}\sum_{i=1}^{n}z_{ji}\left(y_{i}-\boldsymbol{1}\left(v_{i}>0\right)\right)\left[\frac{\hat{f}_{\boldsymbol{Z}i}-\mathbb{E}\left(\hat{f}_{\boldsymbol{Z}i}\right)}{\mathbb{E}\left(\hat{f}_{V\boldsymbol{Z}i}\right)}+\frac{\mathbb{E}\left(\hat{f}_{\boldsymbol{Z}i}\right)\left(\mathbb{E}\left(\hat{f}_{V\boldsymbol{Z}i}\right)-\hat{f}_{V\boldsymbol{Z}i}\right)}{\left[\mathbb{E}\left(\hat{f}_{V\boldsymbol{Z}i}\right)\right]^{2}}\right]+o_{P}\left(1\right)\\
 & =\frac{1}{\sqrt{n}}\sum_{i=1}^{n}q_{ji}+o_{P}\left(1\right),
\end{align*}
and for $j\in A\cup B,$ 
\[
\frac{1}{\sqrt{n}}\sum_{i=1}^{n}\left[z_{ji}\left(\widehat{\tilde{y}}_{i}-\boldsymbol{x}_{i}^{\prime}\boldsymbol{\beta}^{\ast}\right)-\underline{\eta}_{j}^{\ast}\right]=\frac{1}{\sqrt{n}}\sum_{i=1}^{n}q_{ji}+o_{P}\left(1\right).
\]

\end{lemma}

\begin{lemma} \label{LE:uniformrate}Suppose Assumption \ref{A:main},
\ref{A:continuousf_kernelK}, \ref{A:specialR} and \ref{A:GMM} hold,
then 
\[
\left\Vert \overline{\boldsymbol{\hat{m}}}_{n}\left(\boldsymbol{\theta}\right)-\mathbb{E}\left[\boldsymbol{m}\left(\boldsymbol{\theta}\right)\right]\right\Vert =O_{P}\left(\sqrt{\left.p_{n}\right/n}\right),
\]
uniformly over a compact set of $\boldsymbol{\theta}$. \end{lemma}

\begin{lemma} \label{LE:matrix_conv}Suppose Assumption \ref{A:GMM}
holds. Then 
\[
\left\Vert \frac{\partial\overline{\boldsymbol{\hat{m}}}_{n}}{\partial\boldsymbol{\theta}_{B}^{\prime}}-\Gamma_{\boldsymbol{\theta}_{B}}\right\Vert =o_{P}\left(1\right).
\]

\end{lemma}

\begin{lemma} \label{LE:consistency}Suppose Assumption \ref{A:main},
\ref{A:continuousf_kernelK}, \ref{A:specialR} and \ref{A:GMM} hold,
then $\boldsymbol{\hat{\theta}}-\boldsymbol{\theta}^{\ast}=O_{P}\left(\sqrt{\left.p_{n}\right/n}\right).$
\end{lemma}

\begin{lemma} \label{LE:selection}Suppose Assumption \ref{A:main},
\ref{A:continuousf_kernelK}, \ref{A:specialR} and \ref{A:GMM} hold,
then 
\[
\Pr\left(\underline{\hat{\eta}}_{j}=0,\text{ for all }j\in A\right)\rightarrow1,\text{ and }\Pr\left(\min_{j\in B}\left\vert \underline{\hat{\eta}}_{j}\right\vert >0\right)\rightarrow1.
\]

\end{lemma}

\begin{lemma} \label{LE:asymptotics}Suppose Assumption \ref{A:main},
\ref{A:continuousf_kernelK}, \ref{A:specialR} and \ref{A:GMM} hold.
Then, 
\[
\boldsymbol{\hat{\theta}}_{B}\boldsymbol{-\theta}_{B}^{\ast}=\left[\frac{\partial\overline{\boldsymbol{\hat{m}}}_{n}}{\partial\boldsymbol{\theta}_{B}^{\prime}}^{\prime}\mathbf{W}_{n}\frac{\partial\overline{\boldsymbol{\hat{m}}}_{n}}{\partial\boldsymbol{\theta}_{B}^{\prime}}\right]^{-1}\left[\frac{\partial\overline{\boldsymbol{\hat{m}}}_{n}}{\partial\boldsymbol{\hat{\theta}}_{B}^{\prime}}^{\prime}\mathbf{W}_{n}\overline{\boldsymbol{\hat{m}}}_{n}\left(\boldsymbol{\theta}^{\ast}\right)\right]+o_{P}\left(n^{-1/2}\right).
\]

\end{lemma}

\begin{lemma} \label{LE:asym_step0}Suppose Assumption \ref{A:main},
\ref{A:continuousf_kernelK}, \ref{A:specialR} and \ref{A:GMM} hold.
For any $p_{n}\times1$ vector $\boldsymbol{e}$ such that $\left\Vert \boldsymbol{e}\right\Vert =1,$
the following holds 
\[
\sqrt{n}\boldsymbol{e}^{\prime}\Omega_{n}^{-1/2}\overline{\boldsymbol{\hat{m}}}_{n}\left(\boldsymbol{\theta}^{\ast}\right)\overset{d}{\rightarrow}N\left(0,1\right).
\]

\end{lemma}

\begin{proof}[Proof of Theorem \ref{TH:GMM}]The first part is
proved in Lemma \ref{LE:selection}.

We turn to the second part. Note from Lemma \ref{LE:asymptotics},
\begin{align*}
\sqrt{n}\boldsymbol{e}^{\prime}\Sigma_{n}^{-1/2}\left(\boldsymbol{\hat{\theta}}_{B}\boldsymbol{-\theta}_{B}^{\ast}\right) & =\sqrt{n}\boldsymbol{e}^{\prime}\Sigma_{n}^{-1/2}\left(\frac{\partial\overline{\boldsymbol{\hat{m}}}_{n}}{\partial\boldsymbol{\theta}_{B}^{\prime}}^{\prime}\mathbf{W}_{n}\frac{\partial\overline{\boldsymbol{\hat{m}}}_{n}}{\partial\boldsymbol{\theta}_{B}^{\prime}}\right)^{-1}\left(\frac{\partial\overline{\boldsymbol{\hat{m}}}_{n}}{\partial\boldsymbol{\hat{\theta}}_{B}^{\prime}}^{\prime}\mathbf{W}_{n}\overline{\boldsymbol{\hat{m}}}_{n}\left(\boldsymbol{\theta}^{\ast}\right)\right)+o_{P}\left(1\right)\\
 & =\sqrt{n}\left[\boldsymbol{e}^{\prime}\Sigma_{n}^{-1/2}\left(\frac{\partial\overline{\boldsymbol{\hat{m}}}_{n}}{\partial\boldsymbol{\theta}_{B}^{\prime}}^{\prime}\mathbf{W}_{n}\frac{\partial\overline{\boldsymbol{\hat{m}}}_{n}}{\partial\boldsymbol{\theta}_{B}^{\prime}}\right)^{-1}\frac{\partial\overline{\boldsymbol{\hat{m}}}_{n}}{\partial\boldsymbol{\hat{\theta}}_{B}^{\prime}}^{\prime}\mathbf{W}_{n}\Omega_{n}^{1/2}\right]\Omega_{n}^{-1/2}\overline{\boldsymbol{\hat{m}}}_{n}\left(\boldsymbol{\theta}^{\ast}\right)+o_{P}\left(1\right)\\
 & \equiv\sqrt{n}\boldsymbol{e}^{\ast\prime}\Omega_{n}^{-1/2}\overline{\boldsymbol{\hat{m}}}_{n}\left(\boldsymbol{\theta}^{\ast}\right)+o_{P}\left(1\right),
\end{align*}
where $\boldsymbol{e}^{\ast\prime}\equiv\boldsymbol{e}^{\prime}\Sigma_{n}^{-1/2}\left(\frac{\partial\overline{\boldsymbol{\hat{m}}}_{n}}{\partial\boldsymbol{\theta}_{B}^{\prime}}^{\prime}\mathbf{W}_{n}\frac{\partial\overline{\boldsymbol{\hat{m}}}_{n}}{\partial\boldsymbol{\theta}_{B}^{\prime}}\right)^{-1}\frac{\partial\overline{\boldsymbol{\hat{m}}}_{n}}{\partial\boldsymbol{\hat{\theta}}_{B}^{\prime}}^{\prime}\mathbf{W}_{n}\Omega_{n}^{1/2}$.
Obviously $\boldsymbol{e}^{\ast}$ is a $p_{n}\times1$ vector$.$
If we can verify that $\boldsymbol{e}^{\ast\prime}\boldsymbol{e}^{\ast}=1+o_{P}\left(1\right),$
applying Lemma \ref{LE:asym_step0} yields the desired result: 
\[
\sqrt{n}\boldsymbol{e}^{\ast\prime}\Omega_{n}^{-1/2}\overline{\boldsymbol{\hat{m}}}_{n}\left(\boldsymbol{\theta}^{\ast}\right)\overset{d}{\rightarrow}N\left(0,1\right).
\]
So the remaining task it to show $\boldsymbol{e}^{\ast\prime}\boldsymbol{e}^{\ast}=1+o_{P}\left(1\right).$
To this end, we first note the result from Lemma \ref{LE:matrix_conv}
that $\left\Vert \frac{\partial\overline{\boldsymbol{\hat{m}}}_{n}}{\partial\boldsymbol{\theta}_{B}^{\prime}}-\Gamma_{\boldsymbol{\theta}_{B}}\right\Vert =o_{P}\left(1\right)$.
Using this result, we show it by 
\begin{align*}
\boldsymbol{e}^{\ast\prime}\boldsymbol{e}^{\ast} & =\boldsymbol{e}^{\prime}\Sigma_{n}^{-1/2}\left(\Gamma_{\boldsymbol{\theta}_{B}}^{\prime}\mathbf{W}_{n}\Gamma_{\boldsymbol{\theta}_{B}}\right)^{-1}\Gamma_{\boldsymbol{\theta}_{B}}^{\prime}\mathbf{W}_{n}\Omega_{n}^{1/2}\Omega_{n}^{1/2\prime}\mathbf{W}_{n}\Gamma_{\boldsymbol{\theta}_{B}}\left(\Gamma_{\boldsymbol{\theta}_{B}}^{\prime}\mathbf{W}_{n}\Gamma_{\boldsymbol{\theta}_{B}}\right)^{-1}\Sigma_{n}^{-1/2\prime}\boldsymbol{e+}o_{P}\left(1\right)\\
 & =\boldsymbol{e}^{\prime}\Sigma_{n}^{-1/2}\Sigma_{n}\Sigma_{n}^{-1/2\prime}\boldsymbol{e+}o_{P}\left(1\right)=\boldsymbol{e}^{\prime}\boldsymbol{e+}o_{P}\left(1\right)=1+o_{P}\left(1\right).
\end{align*}
\end{proof}

\section{Proofs of Technical Lemmas in Appendices \ref{APP:A}--\ref{APP:C}\label{APP:lemmas}}

We present the proofs of the technical lemmas in Appendices \ref{APP:A}--\ref{APP:C}. Recall
that we assume $H_{jkm}$ is \textit{nonnegative} without loss of
generality.

\begin{proof}[Proof of Lemma \ref{LE:3rdU_NoM}] Lemma \ref{LE:3rdU_NoM}
is a direct result of the second part of Theorem 5.6.1A in \citet{Serfling1980}.
To see that, We could set $t=n^{-1/2}\varsigma_{n},$ $m=3,$ $b=M,$
using the fact that $\sigma^{2}=$Var$\left(H_{jkm,1}\right)\leq\mathbb{E}(H_{jkm,1}^{2})\leq\mathbb{E}(H_{jkm}^{2})\leq B_{H2},$
and finally put $2$ before the bound because Theorem 5.6.1A was only
on one side, and we control the probability on both sides (note that
$t,m,b$ and $\sigma^{2}$ are the notations in Theorem 5.6.1A in
\citet{Serfling1980}). Note this result acts like Bennett's inequality
(see, e.g., Corollary 2.11 in \citet{Boucheron_L_M}) for U-statistics.)
\end{proof}

\begin{proof}[Proof of Lemma \ref{LE:3rdU_M}] \textbf{Preparations.}
First, the finiteness of $\left(2+\delta\right)$-th moment of conditional
expectation holds by Jensen's inequality and the laws of the iterated
means: 
\[
\mathbb{E}\left[\mathbb{E}\left(H\left(X_{j},X_{k},X_{m}\right)|X_{j},X_{k}\right)^{2+\delta}\right],\mathbb{E}\left[\mathbb{E}\left(H\left(X_{j},X_{k},X_{m}\right)|X_{j}\right)^{2+\delta}\right]\leq\mathbb{E}\left[H\left(X_{j},X_{k},X_{m}\right)^{2+\delta}\right]<\infty.
\]
In addition, the following result is useful for deriving the probability
bound: 
\begin{align}
\mathbb{E}\left(H_{jkm,2}^{2}\right) & =\mathbb{E}\left[H_{jkm}^{2}\boldsymbol{1}\left(H_{jkm}>M\right)\right]\leq\mathbb{E}\left[H_{jkm}^{2}\frac{H_{jkm}^{\delta}}{M^{\delta}}\boldsymbol{1}\left(H_{jkm}>M\right)\right]\nonumber \\
 & \leq\mathbb{E}\left[H_{jkm}^{2}\frac{H_{jkm}^{\delta}}{M^{\delta}}\right]\leq\frac{B_{H\delta}}{M^{\delta}}.\label{EQ:hDineq}
\end{align}
\smallskip{}

\noindent \textbf{Decomposition.} By the symmetry of $H$ in $X$s
and the i.i.d. of $\left\{ X_{j}\right\} _{j=1}^{n}$, we are able
to decompose the third-order U-statistics as follows: 
\begin{align*}
 & \frac{1}{n\left(n-1\right)\left(n-2\right)}\sum_{j,k,m=1,j\neq k\neq m}^{n}H_{jkm,2}-\mathbb{E}\left(H_{jkm,2}\right)\\
= & \frac{1}{n\left(n-1\right)\left(n-2\right)}\sum_{j,k,m=1,j\neq k\neq m}^{n}\left\{ H_{jkm,2}-\mathbb{E}\left(H_{jkm,2}|X_{j},X_{k}\right)-\mathbb{E}\left(H_{jkm,2}|X_{j},X_{m}\right)-\mathbb{E}\left(H_{jkm,2}|X_{m},X_{k}\right)\right.\\
 & \left.+\mathbb{E}\left(H_{jkm,2}|X_{j}\right)+\mathbb{E}\left(H_{jkm,2}|X_{k}\right)+\mathbb{E}\left(H_{jkm,2}|X_{m}\right)-\mathbb{E}\left(H_{jkm,2}\right)\right\} \\
 & +\frac{3}{n\left(n-1\right)}\sum_{j,k=1,j\neq k}^{n}\left\{ \mathbb{E}\left(H_{jkm,2}|X_{j},X_{k}\right)-\mathbb{E}\left(H_{jkm,2}|X_{j}\right)-\mathbb{E}\left(H_{jkm,2}|X_{k}\right)+\mathbb{E}\left(H_{jkm,2}\right)\right\} \\
 & +\frac{3}{n}\sum_{j=1}^{n}\left\{ \mathbb{E}\left(H_{jkm,2}|X_{j}\right)-\mathbb{E}\left(H_{jkm,2}\right)\right\} \\
\equiv & R_{n1}+R_{n2}+T_{n1}.
\end{align*}
Thus, 
\begin{align}
 & \Pr\left(\left\vert \frac{\sqrt{n}}{n\left(n-1\right)\left(n-2\right)}\sum_{j,k,m=1,j\neq k\neq m}^{n}\left[H_{jkm,2}-\mathbb{E}\left(H_{jkm,2}\right)\right]\right\vert \geq\varsigma_{n}\right)\nonumber \\
= & \Pr\left(\sqrt{n}\left\vert R_{n1}+R_{n2}+T_{n1}\right\vert \geq\varsigma_{n}\right)\nonumber \\
\leq & \Pr\left(\sqrt{n}\left\vert R_{n1}\right\vert \geq\frac{\varsigma_{n}}{3}\right)+\Pr\left(\sqrt{n}\left\vert R_{n2}\right\vert \geq\frac{\varsigma_{n}}{3}\right)+\Pr\left(\sqrt{n}\left\vert T_{n1}\right\vert \geq\frac{\varsigma_{n}}{3}\right),\label{EQ:3rdUdecomp}
\end{align}
where the last inequality holds by the fact that 
\[
\left\{ \left\vert X_{1}+X_{2}+X_{3}\right\vert \geq C\right\} \subseteq\left\{ \left\vert X_{1}\right\vert \geq\frac{C}{3}\right\} \cup\left\{ \left\vert X_{2}\right\vert \geq\frac{C}{3}\right\} \cup\left\{ \left\vert X_{3}\right\vert \geq\frac{C}{3}\right\} 
\]
for any random variables $X_{1},X_{2},$ and $X_{3}$ and positive
$C.$\smallskip{}

\noindent \textbf{Dealing with }$R_{n1}$\textbf{. }For $R_{n1},$
some simple calculation implies that the expectations of all the cross-products
in the summation is zero. As a result, 
\begin{equation}
\mathbb{E}\left(R_{n1}^{2}\right)\leq\frac{\mathbb{E}\left(H_{jkm,2}^{2}\right)}{n\left(n-1\right)\left(n-2\right)}\leq\frac{B_{H\delta}}{M^{\delta}n\left(n-1\right)\left(n-2\right)},\label{EQ:ERn12}
\end{equation}
by (\ref{EQ:hDineq}). We apply Markov's inequality and obtain 
\begin{equation}
\Pr\left(\sqrt{n}\left\vert R_{n1}\right\vert \geq\frac{\varsigma_{n}}{3}\right)\leq9\varsigma_{n}^{-2}\mathbb{E}\left[nR_{n1}^{2}\right]\leq9B_{H\delta}M^{-\delta}n^{-2}\varsigma_{n}^{-2}\equiv C_{1}M^{-\delta}n^{-2}\varsigma_{n}^{-2},\label{EQ:3rdUdecomp1}
\end{equation}
after substituting (\ref{EQ:ERn12}) in and letting $C_{1}\equiv9B_{H\delta}.$\smallskip{}

\noindent \textbf{Dealing with }$R_{n2}$\textbf{. }Similarly for
$R_{n2},$ the expectations of all the cross products in the summation
are zero. Consequently, 
\begin{equation}
\mathbb{E}\left(R_{n2}^{2}\right)\leq\frac{9\mathbb{E}\left(H_{jkm,2}^{2}\right)}{n\left(n-1\right)}\leq\frac{9B_{H\delta}}{M^{\delta}n\left(n-1\right)},\label{EQ:Rn22}
\end{equation}
by (\ref{EQ:hDineq}). Apply the Markov's inequality again,
we obtain 
\begin{equation}
\Pr\left(\sqrt{n}\left\vert R_{n2}\right\vert \geq\frac{\varsigma_{n}}{3}\right)\leq9\varsigma_{n}^{-2}\mathbb{E}\left[nR_{n2}^{2}\right]=81B_{H\delta}n^{-1}M^{-\delta}\varsigma_{n}^{-2}\equiv C_{2}M^{-\delta}n^{-1}\varsigma_{n}^{-2},\label{EQ:3rdUdecomp2}
\end{equation}
after substituting (\ref{EQ:Rn22}) in and letting $C_{2}\equiv81B_{H\delta}.$\smallskip{}

\noindent \textbf{Dealing with }$T_{n1}$\textbf{. }For $T_{n1},$
we first define some constants 
\begin{equation}
D_{1}=\left(1+\frac{2}{2+\delta}\right)^{2+\delta}\text{ and }D_{2}=2\left(4+\delta\right)^{-2}e^{-\left(2+\delta\right)}.\label{EQ:Ds}
\end{equation}
Since the series in $T_{n1}$ is i.i.d., we are able to apply the
moderate deviation theory for i.i.d. series developed in \citet{Nagaev}
on $T_{n1}$. Specifically, applying Corollary 1.8 in \citet{Nagaev}
on $\Pr\left(\sqrt{n}\left\vert T_{n1}\right\vert \geq\frac{\varsigma_{n}}{3}\right)$
yields 
\begin{align}
\Pr\left(\sqrt{n}\left\vert T_{n1}\right\vert \geq\frac{\varsigma_{n}}{3}\right) & \leq D_{1}\mathbb{E}\left(H_{jkm,2}^{2}\right)n^{-\delta/2}\varsigma_{n}^{-\left(2+\delta\right)}+2\exp\left(-D_{2}\left[\mathbb{E}\left(H_{jkm,2}^{2}\right)\right]^{-1}\varsigma_{n}^{2}\right)\nonumber \\
 & \leq D_{1}B_{H\delta}M^{-\delta}n^{-\delta/2}\left(\varsigma_{n}/3\right)^{-\left(2+\delta\right)}+2\exp\left(-D_{2}B_{H\delta}^{-1}M^{\delta}\left(\varsigma_{n}/3\right)^{2}\right),\label{EQ:3rdUdecomp3}
\end{align}
where the second line holds by (\ref{EQ:hDineq}).\smallskip{}

\noindent \textbf{Final Result. }Finally, we substitute results
(\ref{EQ:3rdUdecomp1}), (\ref{EQ:3rdUdecomp2}), and (\ref{EQ:3rdUdecomp3})
into (\ref{EQ:3rdUdecomp}) to obtain 
\begin{align*}
 & \Pr\left(\left\vert \frac{\sqrt{n}}{n\left(n-1\right)\left(n-2\right)}\sum_{j,k,m=1,j\neq k\neq m}^{n}\left[H_{jkm,2}-\mathbb{E}\left(H_{jkm,2}\right)\right]\right\vert \geq\varsigma_{n}\right)\\
\leq & C_{1}M^{-\delta}n^{-1}\varsigma_{n}^{-2}+C_{2}M^{-\delta}n^{-\delta/2}\varsigma_{n}^{-\left(2+\delta\right)}+2\exp\left(-C_{3}M^{\delta}\varsigma_{n}^{2}\right),
\end{align*}
where without loss of generality we re-use the constants $C_{1}$,
$C_{2}$ and $C_{3}$ and subsume some constants, including $B_{H\delta},$
into $C,$ we only keep the dominating terms, and the inequality continues
to hold by making $C_{1}$ and $C_{2}$ large enough and $C_{3}$
small enough. \end{proof}

\begin{proof}[Proof of Lemma \ref{LE:3rdU_-1}] \textbf{Decomposition.}
We first conduct the following decomposition:
\begin{align*}
 & \frac{1}{n^{3}}\sum_{j,k=1}^{n}\left[H_{jkk}-\mathbb{E}\left(H_{jkk}\right)\right]\\
= & \frac{1}{n^{3}}\sum_{j,k=1}^{n}\left[H_{jkk,1}-\mathbb{E}\left(H_{jkk,1}\right)\right]+\frac{1}{n^{3}}\sum_{j,k=1}^{n}\left[H_{jkk,2}-\mathbb{E}\left(H_{jkk,2}\right)\right]\\
\equiv & R_{n1}+R_{n2}.
\end{align*}
Then 
\begin{align}
 & \Pr\left(\left\vert \frac{\sqrt{n}}{n^{3}}\sum_{j,k=1}^{n}\left[H_{jkk}-\mathbb{E}\left(H_{jkm}\right)\right]\right\vert \geq\varsigma_{n}\right)\nonumber \\
\leq & \Pr\left(\sqrt{n}\left\vert R_{n1}\right\vert \geq\frac{\varsigma_{n}}{2}\right)+\Pr\left(\sqrt{n}\left\vert R_{n2}\right\vert \geq\frac{\varsigma_{n}}{2}\right),\label{EQ:Pr_hjkk}
\end{align}
where the last inequality holds by the fact that 
\[
\left\{ \left\vert X_{1}+X_{2}\right\vert \geq C\right\} \subseteq\left\{ \left\vert X_{1}\right\vert \geq\frac{C}{2}\right\} \cup\left\{ \left\vert X_{2}\right\vert \geq\frac{C}{2}\right\} 
\]
for any random variables $X_{1},X_{2}$ and positive $C.$ With this
decomposition, we obtain the probability bound by dealing with the
above terms one by one.\smallskip{}

\noindent\textbf{Dealing with }$R_{n1}$\textbf{. }We deal with the
first term in (\ref{EQ:Pr_hjkk}) first. Note
\[
\frac{\sqrt{n}}{n^{3}}\sum_{j,k=1}^{n}\left|H_{jkk,1}-\mathbb{E}\left(H_{jkk,1}\right)\right|\leq\frac{n^{2}\sqrt{n}}{n^{3}}2M<\frac{\varsigma_{n}}{3},
\]
due to the definition of $H_{jkk,1}$ and the condition $M<n^{1/2}\varsigma_{n}/5.$
As a result, 
\begin{equation}
\Pr\left(\sqrt{n}\left\vert R_{n1}\right\vert \geq\frac{\varsigma_{n}}{2}\right)=0.\label{EQ:Pr_hjkk1}
\end{equation}

\noindent\textbf{Dealing with }$R_{n2}$\textbf{. }We turn to $R_{n2}.$
Note $\mathbb{E}\left(\left[H_{jkk,2}-\mathbb{E}\left(H_{jkk,2}\right)\right]\left[H_{j^{\prime}k^{\prime}k^{\prime},2}-\mathbb{E}\left(H_{j^{\prime}k^{\prime}k^{\prime},2}\right)\right]\right)=0$
for $j\neq j^{\prime},k^{\prime},$ and $k\neq j^{\prime},k^{\prime}.$
A total of $n\left(n-1\right)\left(n^{2}-3n+3\right)$ terms after
expanding the summation in $\mathbb{E}\left(\sum_{j,k=1}^{n}\left[H_{jkk,2}-\mathbb{E}\left(H_{jkk,2}\right)\right]\right){}^{2}$
is in this category\footnote{There exist two cases for making those terms zero. Case 1:$\ j=k,$
$j^{\prime},k^{\prime}\neq j,$ $n\left(n-1\right)^{2}$ terms. Case
2: $j\neq k,$ $j^{\prime},k^{\prime}\neq j$ or $k,$ $n\left(n-1\right)\left(n-2\right)^{2}$
terms.}, and so we have $6n^{3}-6n^{2}+3n$ non-zero terms left. Note that
\begin{align}
\mathbb{E}\left(H_{jkk,2}^{2}\right) & =\mathbb{E}\left[H_{jkk}^{2}\boldsymbol{1}\left(H_{jkk}>M\right)\right]\leq\mathbb{E}\left[H_{jkk}^{2}\frac{H_{jkk}^{\delta}}{M^{\delta}}\boldsymbol{1}\left(H_{jkk}>M\right)\right]\nonumber \\
 & \leq\mathbb{E}\left[H_{jkk}^{2}\frac{H_{jkk}^{\delta}}{M^{\delta}}\right]\leq\frac{B_{H\delta}}{M^{\delta}},\label{EQ:hjkk_2delta}
\end{align}
and 
\begin{equation}
\mathbb{E}\left(\left[H_{jkk,2}-\mathbb{E}\left(H_{jkk,2}\right)\right]\left[H_{j^{\prime}k^{\prime}k^{\prime},2}-\mathbb{E}\left(H_{j^{\prime}k^{\prime}k^{\prime},2}\right)\right]\right)\leq\sqrt{\mathbb{E}\left(H_{jkk,2}^{2}\right)\mathbb{E}\left(H_{j^{\prime}k^{\prime}k^{\prime},2}^{2}\right)}\leq\frac{B_{H\delta}}{M^{\delta}},\label{EQ:hjkk_2bound1}
\end{equation}
by Cauchy-Schwarz inequality and using (\ref{EQ:hjkk_2delta}). Thus,
\begin{align}
 & \mathbb{E}\left\{ \left(\frac{\sqrt{n}}{n^{3}}\sum_{j,k=1}^{n}\left[H_{jkk,2}-\mathbb{E}\left(H_{jkk,2}\right)\right]\right)^{2}\right\} \nonumber \\
= & \mathbb{E}\left\{ \frac{1}{n^{5}}\sum_{j,k=1}^{n}\sum_{j^{\prime},k^{\prime}=1}^{n}\mathbb{E}\left(\left[H_{jkk,2}-\mathbb{E}\left(H_{jkk,2}\right)\right]\left[H_{j^{\prime}k^{\prime}k^{\prime},2}-\mathbb{E}\left(H_{j^{\prime}k^{\prime}k^{\prime},2}\right)\right]\right)\right\} \nonumber \\
\leq & \frac{6n^{3}-6n^{2}+3n}{n^{5}}\frac{B_{H\delta}}{M^{\delta}},\label{EQ:hjkk_2fb}
\end{align}
since only $6n^{3}-6n^{2}+3n$ terms are non-zero and each term is
bounded by $\frac{B_{H\delta}}{M^{\delta}}$ by (\ref{EQ:hjkk_2bound1}).

The probability bound can be obtained as 
\begin{align}
\Pr\left(\sqrt{n}\left\vert R_{n2}\right\vert \geq\frac{\varsigma_{n}}{2}\right) & =\Pr\left(\left\vert \frac{\sqrt{n}}{n^{3}}\sum_{j,k=1}^{n}\left[H_{jkk,2}-\mathbb{E}\left(H_{jkk,2}\right)\right]\right\vert \geq\frac{\varsigma_{n}}{2}\right)\nonumber \\
 & \leq4\varsigma_{n}^{-2}\mathbb{E}\left\{ \left(\frac{\sqrt{n}}{n^{3}}\sum_{j,k=1}^{n}\left[H_{jkk,2}-\mathbb{E}\left(H_{jkk,2}\right)\right]\right)^{2}\right\} \nonumber \\
 & \leq4\varsigma_{n}^{-2}\frac{6n^{3}-6n^{2}+3n}{n^{5}}\frac{B_{H\delta}}{M^{\delta}}\leq Cn^{-2}M^{-\delta}\varsigma_{n}^{-2},\label{EQ:Pr_hjkk2}
\end{align}
for some positive $C,$ where the second line holds by Markov inequality
and the last line holds by substituting (\ref{EQ:hjkk_2fb}) in.\smallskip{}

\noindent\textbf{Final result. }Substitute results (\ref{EQ:Pr_hjkk1})
and (\ref{EQ:Pr_hjkk2}) back into (\ref{EQ:Pr_hjkk}) to obtain the
desired result. \end{proof}

\begin{proof}[Proof of Lemma \ref{LE:3rdU_-2}] $H_{jjj}-\mathbb{E}\left(H_{jjj}\right)$
is i.i.d. across $j.$ We apply the moderate deviation theory for
i.i.d. series developed in \citet{Nagaev} again. We continue to use
the constants $D_{1}$ and $D_{2}$ defined in equation (\ref{EQ:Ds})
for the proof here. After applying Corollary 1.8 in \citet{Nagaev},
we obtain 
\begin{align*}
 & \Pr\left(\left\vert \frac{\sqrt{n}}{n^{3}}\sum_{j=1}^{n}\left[H_{jjj}-\mathbb{E}\left(H_{jjj}\right)\right]\right\vert \geq\varsigma_{n}\right)\leq D_{1}B_{H\delta}n^{-\delta/2}\left(n^{2}\varsigma_{n}\right)^{-\left(2+\delta\right)}+2\exp\left(-D_{2}B_{H\delta}^{-1}\left(n^{2}\varsigma_{n}\right)^{2}\right)\\
\leq & C_{1}n^{-4-5\delta/2}\varsigma_{n}^{-2-\delta}+2\exp\left(-C_{2}n^{4}\varsigma_{n}^{2}\right),
\end{align*}
for some positive $C_{1}$ and $C_{2}$, as desired. \end{proof}

\begin{proof}[Proof of Lemma \ref{LE:3rdU}] \textbf{First case.}
We deal with the first part of the lemma first. The conclusion is
a result of a combination of the results in Lemmas \ref{LE:3rdU_NoM},
\ref{LE:3rdU_M}, \ref{LE:3rdU_-1}, and \ref{LE:3rdU_-2}. This can
be seen from the following decomposition, 
\begin{align*}
 & \frac{\sqrt{n}}{n^{3}}\sum_{j,k,m=1}^{n}\left[H_{jkm}-\mathbb{E}\left(H_{jkm}\right)\right]\\
 & =\frac{\sqrt{n}}{n^{3}}\sum_{j,k,m=1,j\neq k\neq m}^{n}\left[H_{jkm,1}-\mathbb{E}\left(H_{jkm,1}\right)\right]+\frac{\sqrt{n}}{n^{3}}\sum_{j,k,m=1,j\neq k\neq m}^{n}\left[H_{jkm,2}-\mathbb{E}\left(H_{jkm,2}\right)\right]\\
 & +\frac{3\sqrt{n}}{n^{3}}\sum_{j,k=1}^{n}\left[H_{jkk}-\mathbb{E}\left(H_{jkm}\right)\right]-\frac{2\sqrt{n}}{n^{3}}\sum_{j=1}^{n}\left[H_{jjj}-\mathbb{E}\left(H_{jjj}\right)\right],
\end{align*}
and thus 
\begin{align}
 & \Pr\left(\left\vert \frac{\sqrt{n}}{n^{3}}\sum_{j,k,m=1}^{n}\left[H_{jkm}-\mathbb{E}\left(H_{jkm}\right)\right]\right\vert \geq\varsigma_{n}\right)\nonumber \\
\leq & \Pr\left(\left\vert \frac{\sqrt{n}}{n^{3}}\sum_{j,k,m=1,j\neq k\neq m}^{n}\left[H_{jkm,1}-\mathbb{E}\left(H_{jkm,1}\right)\right]\right\vert \geq\frac{\varsigma_{n}}{4}\right)\nonumber \\
 & +\Pr\left(\left\vert \frac{\sqrt{n}}{n^{3}}\sum_{j,k,m=1,j\neq k\neq m}^{n}\left[H_{jkm,2}-\mathbb{E}\left(H_{jkm,2}\right)\right]\right\vert \geq\frac{\varsigma_{n}}{4}\right)\nonumber \\
 & +\Pr\left(\left\vert \frac{3\sqrt{n}}{n^{3}}\sum_{j,k=1}^{n}\left[H_{jkk}-\mathbb{E}\left(H_{jkm}\right)\right]\right\vert \geq\frac{\varsigma_{n}}{4}\right)\nonumber \\
 & +\Pr\left(\left\vert \frac{2\sqrt{n}}{n^{3}}\sum_{j=1}^{n}\left[H_{jjj}-\mathbb{E}\left(H_{jjj}\right)\right]\right\vert \geq\frac{\varsigma_{n}}{4}\right),\label{EQ:3rdU_f}
\end{align}
where each of the above terms can be handled using Lemmas \ref{LE:3rdU_NoM},
\ref{LE:3rdU_M}, \ref{LE:3rdU_-1}, and \ref{LE:3rdU_-2}, respectively.
We ignore the effects of $\left(n-1\right)^{-1}$ or $\left(n-2\right)^{-1}$
in the denominators when applying Lemmas \ref{LE:3rdU_NoM} and \ref{LE:3rdU_M}
here.

We set $M=n^{1/2}\varsigma_{n}^{-1}.$ Obviously, $\varsigma_{n}$
and $M$ satisfy the conditions required in Lemma \ref{LE:3rdU_-1}.
Then 
\[
\exp\left(-\frac{\left\lfloor n/3\right\rfloor \varsigma_{n}^{2}}{2n\left(B_{H2}+Mn^{-1/2}\varsigma_{n}/3\right)}\right)=\exp\left(-\frac{\left\lfloor n/3\right\rfloor \kappa_{n}\log\left(n\right)}{2n\left(B_{H2}+1/3\right)}\right)\lesssim n^{-M^{\ast}},
\]
for any finite $M^{\ast}$ since $\kappa_{n}\rightarrow\infty$.\ Using
the above, applying Lemmas \ref{LE:3rdU_NoM}, \ref{LE:3rdU_M}, \ref{LE:3rdU_-1},
and \ref{LE:3rdU_-2} on (\ref{EQ:3rdU_f}), and keeping only the
dominant terms, we obtain 
\begin{align*}
 & \Pr\left(\left\vert \frac{\sqrt{n}}{n^{3}}\sum_{j,k,m=1}^{n}\left[H_{jkm}-\mathbb{E}\left(H_{jkm}\right)\right]\right\vert \geq\varsigma_{n}\right)\\
\leq & C_{1}M^{-\delta}n^{-1}\varsigma_{n}^{-2}+C_{2}M^{-\delta}n^{-\delta/2}\varsigma_{n}^{-\left(2+\delta\right)}\leq C_{1}n^{-1-\delta/2}\varsigma_{n}^{-2+\delta}+C_{2}n^{-\delta}\varsigma_{n}^{-2},
\end{align*}
for some positive $C_{1}$ and $C_{2}$, and the inequality can hold by making $C_{1}$ and $C_{2}$ sufficiently large.

\noindent \textbf{Second case. }We turn to the second case. The result
is straightforward, given the way we show the result in the first
part. Using the fact that $\varsigma_{n}\propto n^{C}$ for some positive
$C,$ we are able to set $M=n^{1/2}\varsigma_{n}^{1/2}$ so that the
probability bound in Lemma \ref{LE:3rdU_NoM} decays exponentially
and is negligible. We conduct similar decomposition as we do in the
first case, and substitute $M=n^{1/2}\varsigma_{n}^{1/2}$\ into
the probability bounds obtained from Lemmas \ref{LE:3rdU_NoM}, \ref{LE:3rdU_M},
\ref{LE:3rdU_-1}, and \ref{LE:3rdU_-2}. After keeping only the dominating
terms, we obtain 
\begin{align*}
 & \Pr\left(\left\vert \frac{\sqrt{n}}{n^{3}}\sum_{j,k,m=1}^{n}\left[H_{jkm}-\mathbb{E}\left(H_{jkm}\right)\right]\right\vert \geq\varsigma_{n}\right)\\
\leq & C_{1}M^{-\delta}n^{-1}\varsigma_{n}^{-2}+C_{2}M^{-\delta}n^{-\delta/2}\varsigma_{n}^{-\left(2+\delta\right)}\leq C_{1}n^{-1-\delta/2}\varsigma_{n}^{-2-\delta/2}+C_{2}n^{-\delta}\varsigma_{n}^{-2-3\delta/2},
\end{align*}
as desired. \end{proof}

\begin{proof}[Proof of Lemma \ref{LE:2ndU}] The proof of this
lemma is straightforward given Lemma \ref{LE:3rdU}. We omit the proof
here due to the similarity. \end{proof}

\begin{proof}[Proof of Lemma \ref{LE:Vn_deviation}] We first
define 
\[
S_{n2,1}\left(V\right)=\frac{1}{n^{2}}\sum_{j,k=1}^{n}\left\vert v_{j}-v_{k}\right\vert \text{ and }S_{n2,2}\left(Z_{l}\right)=\frac{1}{n^{2}}\sum_{j,k=1}^{n}\left\vert z_{lj}-z_{lk}\right\vert .
\]
Then 
\[
S_{n2}\left(V,Z_{l}\right)=S_{n2,1}\left(V\right)S_{n2,2}\left(Z_{l}\right).
\]
We rewrite $S_{n3}\left(V,Z_{l}\right)$ as 
\begin{align*}
S_{n3}\left(V,Z_{l}\right)= & \frac{1}{n^{3}}\sum_{j=1}^{n}\sum_{k,m=1}^{n}\left\vert v_{j}-v_{k}\right\vert \left\vert z_{lj}-z_{lm}\right\vert \\
= & \frac{1}{n^{3}}\sum_{j,k,m=1}^{n}\frac{1}{6}\left\{ \left\vert v_{j}-v_{k}\right\vert \left\vert z_{lj}-z_{lm}\right\vert +\left\vert v_{j}-v_{m}\right\vert \left\vert z_{lj}-z_{lk}\right\vert +\left\vert v_{k}-v_{j}\right\vert \left\vert z_{lk}-z_{lm}\right\vert \right.\\
 & \left.+\left\vert v_{k}-v_{m}\right\vert \left\vert z_{lk}-z_{lj}\right\vert +\left\vert v_{m}-v_{j}\right\vert \left\vert z_{lm}-z_{lk}\right\vert +\left\vert v_{m}-v_{k}\right\vert \left\vert z_{lm}-z_{lj}\right\vert \right\} .
\end{align*}
After this transformation, $S_{n3}\left(V,Z_{l}\right)$ contains
terms that are symmetric in $\left(j,k,m\right)$. As a result, we
can apply the lemmas we developed for the U-statistics for $S_{n1}\left(V,Z_{l}\right),S_{n2,1}\left(V,Z_{l}\right),$
$S_{n2,2}\left(V,Z_{l}\right),$ and $S_{n3}\left(V,Z_{l}\right).$

Let $(\tilde{V},\tilde{Z}_{l})$ be an independent copy of $(V,Z_{l})$.
We define the following population mean: 
\begin{align*}
S_{1}\left(V,Z_{l}\right) & =\mathbb{E}(|V-\tilde{V}||Z_{l}-\tilde{Z}_{l}|),\text{ }S_{2,1}\left(V\right)=\mathbb{E}(|V-\tilde{V}|),\text{ }\\
S_{2,2}(Z_{l}) & =\mathbb{E}(|Z_{l}-\tilde{Z}_{l}|)\text{ and }S_{3}\left(V,Z_{l}\right)=\mathbb{E}[\mathbb{E}(|V-\tilde{V}|\text{ }|V)\mathbb{E}(|Z_{l}-\tilde{Z}_{l}|\text{ }|Z_{l})].
\end{align*}
It is not hard to see that $S_{1}\left(V,Z_{l}\right),S_{2,1}\left(V\right),$
$S_{2,2}\left(Z_{l}\right)$ and $S_{3}\left(V,Z_{l}\right)$ are
the limit of $S_{n1}\left(V,Z_{l}\right),S_{n2,1}\left(V\right),$
$S_{n2,2}\left(Z_{l}\right),$ and $S_{n3}\left(V,Z_{l}\right),$
respectively, and 
\[
\mathcal{V}^{2}\left(V,Z_{l}\right)=S_{1}\left(V,Z_{l}\right)+S_{2,1}\left(V\right)S_{2,2}\left(Z_{l}\right)-2S_{3}\left(V,Z_{l}\right).
\]
Moreover, using the moment conditions imposed in Assumption \ref{A:main},
\[
0<S_{2,1}\left(V\right),S_{2,2}\left(Z_{l}\right)\leq C_{S}
\]
uniformly for all $l,$ for a positive $C_{S}.$

We conduct the following decomposition: 
\begin{align}
 & \Pr\left(\sqrt{n}\left\vert \mathcal{V}_{n}^{2}\left(V,Z_{l}\right)-\mathcal{V}^{2}\left(V,Z_{l}\right)\right\vert \geq\varsigma_{n}\right)\nonumber \\
= & \Pr\left(\sqrt{n}\left\vert S_{n1}\left(V,Z_{l}\right)+S_{n2,1}\left(V\right)S_{n2,2}\left(Z_{l}\right)-2S_{n3}\left(V,Z_{l}\right)-\mathcal{V}^{2}\left(V,Z_{l}\right)\right\vert \geq\varsigma_{n}\right)\nonumber \\
\leq & \Pr\left(\sqrt{n}\left\vert S_{n1}\left(V,Z_{l}\right)-S_{1}\left(V,Z_{l}\right)\right\vert \geq\frac{\varsigma_{n}}{3}\right)+\Pr\left(\sqrt{n}\left\vert S_{n2,1}\left(V\right)S_{n2,2}\left(Z_{l}\right)-S_{2,1}\left(V\right)S_{2,2}\left(Z_{l}\right)\right\vert \geq\frac{\varsigma_{n}}{3}\right)\nonumber \\
 & +\Pr\left(\sqrt{n}\left\vert S_{n3}\left(V,Z_{l}\right)-S_{3}\left(V,Z_{l}\right)\right\vert \geq\frac{\varsigma_{n}}{6}\right)\nonumber \\
\equiv & T_{n1}+T_{n2}+T_{n3}.\label{EQ:Vn_decomp}
\end{align}
For $T_{n2},$ 
\begin{align}
T_{n2}= & \Pr\left(\sqrt{n}\left\vert S_{n2,1}\left(V\right)S_{n2,2}\left(Z_{l}\right)-S_{2,1}\left(V\right)S_{2,2}\left(Z_{l}\right)\right\vert \geq\frac{\varsigma_{n}}{3}\right)\nonumber \\
= & \Pr\left(\sqrt{n}\left\vert S_{n2,1}\left(V\right)\left[S_{n2,2}\left(Z_{l}\right)-S_{2,2}\left(Z_{l}\right)\right]+S_{2,2}\left(Z_{l}\right)\left[S_{n2,1}\left(V\right)-S_{2,1}\left(V\right)\right]\right\vert \geq\frac{\varsigma_{n}}{3}\right)\nonumber \\
\leq & \Pr\left(\sqrt{n}\left\vert S_{n2,1}\left(V\right)\left[S_{n2,2}\left(Z_{l}\right)-S_{2,2}\left(Z_{l}\right)\right]\right\vert \geq\frac{\varsigma_{n}}{6}\right)\nonumber \\
 & +\Pr\left(\sqrt{n}\left\vert S_{2,2}\left(Z_{l}\right)\left[S_{n2,1}\left(V\right)-S_{2,1}\left(V\right)\right]\right\vert \geq\frac{\varsigma_{n}}{6}\right)\nonumber \\
\equiv & T_{n2,1}+T_{n2,2}.\label{EQ:Vn_decomp1}
\end{align}
We turn to $T_{n2,1}:$ 
\begin{align}
T_{n2,1} & =\Pr\left(\sqrt{n}\left\vert S_{n2,1}\left(V\right)\left[S_{n2,2}\left(Z_{l}\right)-S_{2,2}\left(Z_{l}\right)\right]\right\vert \geq\frac{\varsigma_{n}}{6}\right)\nonumber \\
 & \leq\Pr\left(\sqrt{n}\left\vert S_{n2,2}\left(Z_{l}\right)-S_{2,2}\left(Z_{l}\right)\right\vert \geq\frac{\varsigma_{n}}{6\left(C_{S}+\varsigma_{n}/\sqrt{n}\right)}\right)+\Pr\left(\left\vert S_{n2,1}\left(V\right)\right\vert \geq C_{S}+\varsigma_{n}/\sqrt{n}\right)\nonumber \\
 & \leq\Pr\left(\sqrt{n}\left\vert S_{n2,2}\left(Z_{l}\right)-S_{2,2}\left(Z_{l}\right)\right\vert \geq\frac{\varsigma_{n}}{6C}\right)+\Pr\left(\sqrt{n}\left\vert S_{n2,1}\left(V\right)-S_{2,1}\left(V\right)\right\vert \geq\varsigma_{n}\right)\nonumber \\
 & \equiv T_{n2,1,1}+T_{n2,1,2},\label{EQ:Vn_decomp2}
\end{align}
for some $C>0,$ where the third line holds by $\varsigma_{n}\ll\sqrt{n}$
and thus $C_{S}+\varsigma_{n}/\sqrt{n}\leq C$ for some $C$. For
$T_{n2,2},$ 
\begin{align}
T_{n2,2} & =\Pr\left(\sqrt{n}\left\vert S_{2,2}\left(Z_{l}\right)\left[S_{n2,1}\left(V\right)-S_{2,1}\left(V\right)\right]\right\vert \geq\frac{\varsigma_{n}}{6}\right)\nonumber \\
 & \leq\Pr\left(\sqrt{n}\left\vert S_{n2,1}\left(V\right)-S_{2,1}\left(V\right)\right\vert \geq\frac{\varsigma_{n}}{6C_{S}}\right),\label{EQ:Vn_decomp3}
\end{align}
because $S_{2,2}\left(Z_{l}\right)\leq C_{S}.$

Finally, combining (\ref{EQ:Vn_decomp}), (\ref{EQ:Vn_decomp1}),
(\ref{EQ:Vn_decomp2}), and (\ref{EQ:Vn_decomp3}) yields 
\[
\Pr\left(\sqrt{n}\left\vert \mathcal{V}_{n}^{2}\left(V,Z_{l}\right)-\mathcal{V}^{2}\left(V,Z_{l}\right)\right\vert \geq\varsigma_{n}\right)\leq T_{n1}+T_{n3}+T_{n2,1,1}+T_{n2,1,2}+T_{n2,2},
\]
and each of the above terms can be handled by either the first part
of Lemma \ref{LE:3rdU} or the first part of Lemma \ref{LE:2ndU}.
With Assumption \ref{A:main}, after applying Lemmas \ref{LE:3rdU}
and \ref{LE:2ndU} on $T_{n1},T_{n3},T_{n2,1,1},T_{n2,1,2},$ and
$T_{n2,2},$ we obtain 
\[
\Pr\left(\sqrt{n}\left\vert \mathcal{V}_{n}^{2}\left(V,Z_{l}\right)-\mathcal{V}^{2}\left(V,Z_{l}\right)\right\vert \geq\varsigma_{n}\right)\leq C_{1}n^{-1-\delta/2}\varsigma_{n}^{-2+\delta}+C_{2}n^{-\delta}\varsigma_{n}^{-2},
\]
for some positive $C_{1}$ and $C_{2},$\ after some large $n,$
because we can make $C_{1}$ and $C_{2}$ large enough to the make the inequality holds.

The second part of the Lemma is just a by-product of the previous
proof. It is not hard to see that 
\[
\Pr\left(\left\vert S_{n2}\left(V,Z_{l}\right)-S_{2}\left(V,Z_{l}\right)\right\vert \geq\frac{D_{1}}{2}\right)=\Pr\left(\left\vert S_{n2}\left(V,Z_{l}\right)-\mathbb{E}\left(\left\vert V-\tilde{V}\right\vert \right)\mathbb{E}\left(\left\vert Z_{l}-\tilde{Z}_{l}\right\vert \right)\right\vert \geq\frac{D_{1}}{2}\right)
\]
is the same as $T_{n2}$ in equation (\ref{EQ:Vn_decomp1}) if we
set $\varsigma_{n}=\frac{3D_{1}\sqrt{n}}{2}.$ Note $T_{n2}=T_{n2,1,1}+T_{n2,1,2}+T_{n2,2},$
we apply the second part of Lemmas \ref{LE:3rdU} and \ref{LE:2ndU}
to obtain 
\[
\Pr\left(\left\vert S_{n2}\left(V,Z_{l}\right)-\mathbb{E}\left(\left\vert V-\tilde{V}\right\vert \right)\mathbb{E}\left(\left\vert Z_{l}-\tilde{Z}_{l}\right\vert \right)\right\vert \geq\frac{D_{1}}{2}\right)\leq C_{3}n^{-2-3\delta/4}+C_{4}n^{-1-7\delta/4},
\]
for some positive $C_{3}$ and $C_{4}.$ \end{proof}

\begin{proof}[Proof of Lemma \ref{LE:sign_consistent}] This
Lemma is Lemma 3.1 from \citet{LinEtal2009}. We refer readers to
their paper for the proof.
\end{proof}

\begin{proof} [Proof of Lemma \ref{LE:matrixXX'}]To prove this
lemma, we first present two inequalities for generic $s^{\ast}\times s^{\ast}$
symmetric matrices $\mathbf{A}$ and $\mathbf{B}$: 
\[
\max\left\{ \left\vert \rho_{\min}\left(\mathbf{A}\right)\right\vert ,\left\vert \rho_{\min}\left(\mathbf{B}\right)\right\vert \right\} \leq s^{\ast}\max\left\{ \left\Vert \mathbf{A}\right\Vert _{\infty},\left\Vert \mathbf{B}\right\Vert _{\infty}\right\} 
\]
and 
\[
\left\vert \rho_{\min}\left(\mathbf{A}\right)-\rho_{\min}\left(\mathbf{B}\right)\right\vert \leq\max\left\{ \left\vert \rho_{\min}\left(\mathbf{A-B}\right)\right\vert ,\left\vert \rho_{\min}\left(\mathbf{B-A}\right)\right\vert \right\} .
\]
A proof of the above results can be found at the proof of Lemma A.4
in \citet{Su_et_al2022}.

Applying a similar process as for the proof of Lemma \ref{LE:3rdU}
to the summation of i.i.d. series $X_{j}X_{l},$ we get 
\begin{align}
 & \Pr\left(\left\vert \frac{1}{n}\sum_{i=1}^{n}x_{ji}x_{li}-\mathbb{E}\left(X_{j}X_{l}\right)\right\vert \geq\frac{D_{3}}{2s^{\ast}}\right)\nonumber \\
 & =\Pr\left(\left\vert \frac{1}{\sqrt{n}}\sum_{i=1}^{n}\left(x_{ji}x_{li}-\mathbb{E}\left(X_{j}X_{l}\right)\right)\right\vert \geq\frac{\sqrt{n}D_{3}}{2s^{\ast}}\right)\nonumber \\
 & \leq C_{1}n^{-5/4-3\delta/8}s^{\ast3/2+\delta/4}+C_{2}n^{3/4-7\delta/8}s^{\ast1/2+3\delta/4},\label{EQ:xjxl_diff}
\end{align}
where we use the moment condition $\mathbb{E}\left(\left\vert X_{j}X_{l}\right\vert ^{1+\delta/2}\right)=\mathbb{E}\left(\left\vert X_{j}X_{l}\right\vert ^{2+\left(\delta/2-1\right)}\right)<\infty,$
the condition $\delta>2,$ and the results in Lemma \ref{LE:3rdU}
to i.i.d. series by replacing $\delta$ with $\delta/2-1$ and $\varsigma_{n}$
with $n^{1/2}/s^{\ast}.$ Now, using the two inequalities presented
at the beginning of the proof, we can obtain 
\begin{align*}
 & \Pr\left(\rho_{\min}\left(\mathbf{C}_{n}^{\left(1,1\right)}\right)<\frac{D_{3}}{2}\right)\\
 & \leq\Pr\left(\left\vert \rho_{\min}\left(\mathbf{C}_{n}^{\left(1,1\right)}\right)-\rho_{\min}\left(\mathbb{E}\left(\boldsymbol{X}_{1}^{s^{\ast}}\boldsymbol{X}_{1}^{s^{\ast}\prime}\right)\right)\right\vert >\frac{D_{3}}{2}\right)\\
 & \leq\Pr\left(s^{\ast}\left\Vert \mathbf{C}_{n}^{\left(1,1\right)}-\mathbb{E}\left(\boldsymbol{X}_{1}^{s^{\ast}}\boldsymbol{X}_{1}^{s^{\ast}\prime}\right)\right\Vert _{\infty}>\frac{D_{3}}{2}\right)\\
 & =\Pr\left(\max_{j,l=1,2,...,s^{\ast}}\left\vert \frac{1}{n}\sum_{i=1}^{n}x_{ji}x_{li}-\mathbb{E}\left(X_{j}X_{l}\right)\right\vert >\frac{D_{3}}{2s^{\ast}}\right)\\
 & \leq\sum_{j,l=1}^{s^{\ast}}\Pr\left(\left\vert \frac{1}{n}\sum_{i=1}^{n}x_{ji}x_{li}-\mathbb{E}\left(X_{j}X_{l}\right)\right\vert >\frac{D_{3}}{2s^{\ast}}\right)\\
 & =C_{1}n^{-5/4-3\delta/8}s^{\ast7/2+\delta/4}+C_{2}n^{3/4-7\delta/8}s^{\ast5/2+3\delta/4}=o\left(1\right)
\end{align*}
where the second line holds because $\rho_{\min}\left(\mathbb{E}\left(\boldsymbol{X}_{1}^{s^{\ast}}\boldsymbol{X}_{1}^{s^{\ast}\prime}\right)\right)>D_{3}$
by assumption, the third line holds by applying the first two inequalities
at the beginning of the proof, the fifth line holds by applying the
union bound, and the last line holds \textbf{$C_{s}<1/2$} and $\delta>4$.
The result of the lemma holds by 
\[
\Pr\left(\rho_{\min}\left(\mathbf{C}_{n}^{\left(1,1\right)}\right)\geq\frac{D_{3}}{2}\right)=1-\Pr\left(\rho_{\min}\left(\mathbf{C}_{n}^{\left(1,1\right)}\right)<\frac{D_{3}}{2}\right)=1-o\left(1\right).
\]
\end{proof}

\begin{proof} [Proof of Lemma \ref{LE:matrixXX'C21}]
\begin{align*}
 & \Pr\left(\max_{j=s^{*}+1,...,p_{n}}\left\Vert \frac{1}{n}\mathbf{X}_{jn}^{\prime}\mathbf{X}_{n}^{\left(1\right)}\right\Vert \geq2C_{X2}\sqrt{s^{*}}\right)\\
\leq & \sum_{j=s^{\ast}+1}^{p_{n}}\Pr\left(\left\Vert \frac{1}{n}\mathbf{X}_{jn}^{\prime}\mathbf{X}_{n}^{\left(1\right)}\right\Vert \geq2C_{X2}\sqrt{s^{*}}\right)\\
\leq & \sum_{j=s^{\ast}+1}^{p_{n}}\Pr\left(\left[\sum_{l=1}^{s^{\ast}}\left(\frac{1}{n}\sum_{i=1}^{n}x_{ji}x_{li}\right)^{2}\right]^{1/2}\geq2C_{X2}\sqrt{s^{*}}\right)\\
\leq & \sum_{j=s^{\ast}+1}^{p_{n}}\sum_{l=1}^{s^{\ast}}\Pr\left(\left\vert \frac{1}{n}\sum_{i=1}^{n}x_{ji}x_{li}-\mathbb{E}\left(x_{ji}x_{li}\right)\right\vert \geq C_{X2}\right)\\
\leq & p_{n}s^{\ast}\left(C_{1}n^{-5/4-3\delta/8}+C_{2}n^{3/4-7\delta/8}\right),
\end{align*}
where the last line holds using (\ref{EQ:xjxl_diff}) by removing
terms related to $s^{\ast}.$\textbf{ }The above is $o\left(1\right),$due
to $C_{p}<1,C_{s}<1/2,$ and $\delta>4$.
\end{proof}

\begin{proof}[Proof of Lemma \ref{LE:prepare}]We show the results
of the lemma by conditioning on the event $\hat{E}^{*},$ that we
have selected all the relevant $X$ for $V$ in the first stage, and
optimal bandwidths are chosen for $\hat{f}\left(V|\boldsymbol{X}\right)$.
Since we demonstrate in Section \ref{SEC:f(v,x)reduce}\ that $\Pr\left(\hat{E}^{*}\right)=1-o\left(1\right),$
the results of the Lemma without conditioning on $\hat{E}^{*}$ follow
by the definition of $o_{P}\left(1\right)$ and the fact that 
\begin{align*}
\Pr\left(A\right) & =\Pr\left(A|\hat{E}^{*}\right)\Pr\left(\hat{E}^{*}\right)+\Pr\left(A\left\vert \left(\hat{E}^{*}\right)^{c}\right.\right)\Pr\left(\left(\hat{E}^{*}\right)^{c}\right)\\
 & \leq\Pr\left(A|\hat{E}^{*}\right)+\Pr\left(\left(\hat{E}^{*}\right)^{c}\right)=o\left(1\right)
\end{align*}
if $\Pr\left(A|\hat{E}^{*}\right)=o\left(1\right)$ for any event
$A.$

\textbf{Part 1.} We show the first result in this part of the proof.
The proof is based on assuming that $\hat{E}^{*}$ holds. We write
$\hat{f}_{\boldsymbol{X}i},\hat{f}_{V\boldsymbol{X}i},f_{\boldsymbol{X}i},f_{V\boldsymbol{X}i}$
for short for $\hat{f}\left(\boldsymbol{x}_{1i}^{\tilde{p}}\right),\hat{f}\left(v_{i},\boldsymbol{x}_{1i}^{\tilde{p}}\right),f\left(\boldsymbol{x}_{1i}^{p^{\ast}}\right),$
and $f\left(v_{i},\boldsymbol{x}_{1i}^{p^{\ast}}\right),$ respectively.
We note that $\hat{f}_{\boldsymbol{X}}$ and $f_{\boldsymbol{X}}$
may contain different elements of $X$, but we still use those notations
if no confusion arises.

We decompose $\frac{\hat{f}_{\boldsymbol{X}i}}{\hat{f}_{V\boldsymbol{X}i}}-\frac{f_{\boldsymbol{X}i}}{f_{V\boldsymbol{X}i}}$
as 
\begin{align}
 & \frac{\hat{f}_{\boldsymbol{X}i}}{\hat{f}_{V\boldsymbol{X}i}}-\frac{f_{\boldsymbol{X}i}}{f_{V\boldsymbol{X}i}}=\frac{\hat{f}_{\boldsymbol{X}i}-\mathbb{E}\left(\hat{f}_{\boldsymbol{X}i}\right)}{\mathbb{E}\left(\hat{f}_{V\boldsymbol{X}i}\right)}+\frac{\mathbb{E}\left(\hat{f}_{\boldsymbol{X}i}\right)\left(\mathbb{E}\left(\hat{f}_{V\boldsymbol{X}i}\right)-\hat{f}_{V\boldsymbol{X}i}\right)}{\left[\mathbb{E}\left(\hat{f}_{V\boldsymbol{X}i}\right)\right]^{2}}+\frac{\mathbb{E}\left(\hat{f}_{\boldsymbol{X}i}\right)}{\mathbb{E}\left(\hat{f}_{V\boldsymbol{X}i}\right)}-\frac{f_{\boldsymbol{X}i}}{f_{V\boldsymbol{X}i}}\nonumber \\
 & +\frac{\left[\hat{f}_{\boldsymbol{X}i}-\mathbb{E}\left(\hat{f}_{\boldsymbol{X}i}\right)\right]\left[\mathbb{E}\left(\hat{f}_{V\boldsymbol{X}i}\right)-\hat{f}_{V\boldsymbol{X}i}\right]}{\hat{f}_{V\boldsymbol{X}i}\mathbb{E}\left(\hat{f}_{V\boldsymbol{X}i}\right)}+\frac{\mathbb{E}\left(\hat{f}_{\boldsymbol{X}i}\right)\left[\mathbb{E}\left(\hat{f}_{V\boldsymbol{X}i}\right)-\hat{f}_{V\boldsymbol{X}i}\right]^{2}}{\hat{f}_{V\boldsymbol{X}i}\left[\mathbb{E}\left(\hat{f}_{V\boldsymbol{X}i}\right)\right]^{2}}\nonumber \\
 & \equiv\frac{\hat{f}_{\boldsymbol{X}i}-\mathbb{E}\left(\hat{f}_{\boldsymbol{X}i}\right)}{\mathbb{E}\left(\hat{f}_{V\boldsymbol{X}i}\right)}+\frac{\mathbb{E}\left(\hat{f}_{\boldsymbol{X}i}\right)\left(\mathbb{E}\left(\hat{f}_{V\boldsymbol{X}i}\right)-\hat{f}_{V\boldsymbol{X}i}\right)}{\left[\mathbb{E}\left(\hat{f}_{V\boldsymbol{X}i}\right)\right]^{2}}+R_{n1,i}+R_{n2,i}.\label{EQ:fvx-fvx^}
\end{align}
Conditional on $\hat{E}^{*},$ 
\begin{align*}
\frac{\mathbb{E}\left(\hat{f}_{\boldsymbol{X}i}\right)}{\mathbb{E}\left(\hat{f}_{V\boldsymbol{X}i}\right)} & =\frac{\mathbb{E}\left[\Pi_{l=1}^{p^{\ast}}K_{h_{l}}\left(x_{ji}-x_{li}\right)\Pi_{l=p^{\ast}+1}^{\tilde{p}}K_{h_{l}}\left(x_{ji}-x_{li}\right)\right]}{\mathbb{E}\left[K_{h_{V}}\left(v\right)\Pi_{l=1}^{p^{\ast}}K_{h_{l}}\left(x_{ji}-x_{li}\right)\Pi_{l=p^{\ast}+1}^{\tilde{p}}K_{h_{l}}\left(x_{ji}-x_{li}\right)\right]}\\
 & =\frac{\mathbb{E}\left[\Pi_{l=1}^{p^{\ast}}K_{h_{l}}\left(x_{ji}-x_{li}\right)\right]\mathbb{E}\left[\Pi_{l=p^{\ast}+1}^{\tilde{p}}K_{h_{l}}\left(x_{ji}-x_{li}\right)\right]}{\mathbb{E}\left[K_{h_{V}}\left(v\right)\Pi_{l=1}^{p^{\ast}}K_{h_{l}}\left(x_{ji}-x_{li}\right)\right]\mathbb{E}\left[\Pi_{l=p^{\ast}+1}^{\tilde{p}}K_{h_{l}}\left(x_{ji}-x_{li}\right)\right]}\\
 & =\frac{\mathbb{E}\left[\Pi_{l=1}^{p^{\ast}}K_{h_{l}}\left(x_{ji}-x_{li}\right)\right]}{\mathbb{E}\left[K_{h_{V}}\left(v\right)\Pi_{l=1}^{p^{\ast}}K_{h_{l}}\left(x_{ji}-x_{li}\right)\right]},
\end{align*}
where the second line holds by the independence assumption in Assumption
\ref{A:main}. With that, it is not hard to see that 
\begin{align*}
R_{n1,i} & =\frac{\mathbb{E}\left(\hat{f}_{\boldsymbol{X}i}\right)}{\mathbb{E}\left(\hat{f}_{V\boldsymbol{X}i}\right)}-\frac{f_{\boldsymbol{X}i}}{f_{V\boldsymbol{X}i}}=\frac{\mathbb{E}\left[\Pi_{l=1}^{p^{\ast}}K_{h_{l}}\left(x_{ji}-x_{li}\right)\right]}{\mathbb{E}\left[K_{h_{V}}\left(v\right)\Pi_{l=1}^{p^{\ast}}K_{h_{l}}\left(x_{ji}-x_{li}\right)\right]}-\frac{f\left(x_{1i},...,x_{p^{\ast}i}\right)}{f\left(v_{i},x_{1i},...,x_{p^{\ast}i}\right)}\\
 & =O\left(h_{1}^{2r}+...+h_{p^{\ast}}^{2r}\right)=o\left(n^{-1/2}\right),
\end{align*}
due to the fact that $K\left(\cdot\right)$ is an $2r$-th order kernel
and $h_{1}^{r},...,h_{p^{\ast}}^{r}$ is $o\left(n^{-1/4}\right)$
conditional on $\hat{E}^{*}$.

Following the proof of Theorem 1.4 in \citet{LiRacine}, we can show
that $\hat{f}_{\boldsymbol{X}i}-\mathbb{E}\left(\hat{f}_{\boldsymbol{X}i}\right)=o_{P}\left(n^{-1/4}\right)$
$\hat{f}_{V\boldsymbol{X}i}-\mathbb{E}\left(\hat{f}_{V\boldsymbol{X}i}\right)=o_{P}\left(n^{-1/4}\right)$
uniformly over the compact support of $V$ and $X$. Thus $R_{n2,i}=o_{P}\left(n^{-1/2}\right)$
uniformly over $i=1,2,...,n.$

Summarize the results so far in this part, 
\[
\frac{1}{\sqrt{n}}\sum_{i=1}^{n}x_{ji}\left(y_{i}-\boldsymbol{1}\left(v_{i}>0\right)\right)\left[R_{n1,i}+R_{n2,i}\right]=o_{P}\left(1\right).
\]
As a result, 
\begin{align*}
 & \frac{1}{\sqrt{n}}\sum_{i=1}^{n}x_{ji}\left(\widehat{\tilde{y}}_{i}-\boldsymbol{x}_{i}^{\prime}\boldsymbol{\beta}^{\ast}\right)\\
 & =\frac{1}{\sqrt{n}}\sum_{i=1}^{n}x_{ji}\left(\tilde{y}_{i}-\boldsymbol{x}_{i}^{\prime}\boldsymbol{\beta}^{\ast}\right)+\frac{1}{\sqrt{n}}\sum_{i=1}^{n}x_{ji}\left(\widehat{\tilde{y}}_{i}-\tilde{y}_{i}\right)\\
 & =\frac{1}{\sqrt{n}}\sum_{i=1}^{n}x_{ji}\left(\tilde{y}_{i}-\boldsymbol{x}_{i}^{\prime}\boldsymbol{\beta}^{\ast}\right)+\frac{1}{\sqrt{n}}\sum_{i=1}^{n}x_{ji}\left(y_{i}-\boldsymbol{1}\left(v_{i}>0\right)\right)\left[\frac{\hat{f}_{\boldsymbol{X}i}}{\hat{f}_{V\boldsymbol{X}i}}-\frac{f_{\boldsymbol{X}i}}{f_{V\boldsymbol{X}i}}\right]\\
 & =\frac{1}{\sqrt{n}}\sum_{i=1}^{n}x_{ji}\left(\tilde{y}_{i}-\boldsymbol{x}_{i}^{\prime}\boldsymbol{\beta}^{\ast}\right)+\frac{1}{\sqrt{n}}\sum_{i=1}^{n}x_{ji}\left(y_{i}-\boldsymbol{1}\left(v_{i}>0\right)\right)\left[\frac{\hat{f}_{\boldsymbol{X}i}-\mathbb{E}\left(\hat{f}_{\boldsymbol{X}i}\right)}{\mathbb{E}\left(\hat{f}_{V\boldsymbol{X}i}\right)}+\frac{\mathbb{E}\left(\hat{f}_{\boldsymbol{X}i}\right)\left(\mathbb{E}\left(\hat{f}_{V\boldsymbol{X}i}\right)-\hat{f}_{V\boldsymbol{X}i}\right)}{\left[\mathbb{E}\left(\hat{f}_{V\boldsymbol{X}i}\right)\right]^{2}}\right]\\
 & +\frac{1}{\sqrt{n}}\sum_{i=1}^{n}x_{ji}\left(y_{i}-\boldsymbol{1}\left(v_{i}>0\right)\right)\left[R_{n1,i}+R_{n2,i}\right]\\
 & =\frac{1}{\sqrt{n}}\sum_{i=1}^{n}x_{ji}\left(\tilde{y}_{i}-\boldsymbol{x}_{i}^{\prime}\boldsymbol{\beta}^{\ast}\right)+\frac{1}{\sqrt{n}}\sum_{i=1}^{n}x_{ji}\left(y_{i}-\boldsymbol{1}\left(v_{i}>0\right)\right)\left[\frac{\hat{f}_{\boldsymbol{X}i}-\mathbb{E}\left(\hat{f}_{\boldsymbol{X}i}\right)}{\mathbb{E}\left(\hat{f}_{V\boldsymbol{X}i}\right)}+\frac{\mathbb{E}\left(\hat{f}_{\boldsymbol{X}i}\right)\left(\mathbb{E}\left(\hat{f}_{V\boldsymbol{X}i}\right)-\hat{f}_{V\boldsymbol{X}i}\right)}{\left[\mathbb{E}\left(\hat{f}_{V\boldsymbol{X}i}\right)\right]^{2}}\right]\\
 & +o_{P}\left(1\right).
\end{align*}

\textbf{Part 2.} We assume $\hat{E}^{*}$ holds for this part of proof,
and we do the following decomposition first. 
\begin{align}
 & \Pr\left(\max_{j=1,...,p_{n}}\left\vert \frac{1}{\sqrt{n}}\sum_{i=1}^{n}x_{ji}\left(\widehat{\tilde{y}}_{i}-\boldsymbol{x}_{i}^{\prime}\boldsymbol{\beta}^{\ast}\right)\right\vert \geq\varsigma_{n}\text{ }\right)\nonumber \\
 & \leq\Pr\left(\max_{j=1,...,p_{n}}\left\vert \frac{1}{\sqrt{n}}\sum_{i=1}^{n}x_{ji}\left(\tilde{y}_{i}-\boldsymbol{x}_{i}^{\prime}\boldsymbol{\beta}^{\ast}\right)\right\vert \geq\frac{\varsigma_{n}}{3}\right)\nonumber \\
 & +\Pr\left(\max_{j=1,...,p_{n}}\left\vert \frac{1}{\sqrt{n}}\sum_{i=1}^{n}x_{ji}\left(y_{i}-\boldsymbol{1}\left(v_{i}>0\right)\right)\left[\frac{\hat{f}_{\boldsymbol{X}i}-\mathbb{E}\left(\hat{f}_{\boldsymbol{X}i}\right)}{\mathbb{E}\left(\hat{f}_{V\boldsymbol{X}i}\right)}+\frac{\mathbb{E}\left(\hat{f}_{\boldsymbol{X}i}\right)\left(\mathbb{E}\left(\hat{f}_{V\boldsymbol{X}i}\right)-\hat{f}_{V\boldsymbol{X}i}\right)}{\left[\mathbb{E}\left(\hat{f}_{V\boldsymbol{X}i}\right)\right]^{2}}\right]\right\vert \geq\frac{\varsigma_{n}}{3}\text{ }\right)\nonumber \\
 & +\Pr\left(\max_{j=1,...,p_{n}}\left\vert \frac{1}{\sqrt{n}}\sum_{i=1}^{n}x_{ji}\left(y_{i}-\boldsymbol{1}\left(v_{i}>0\right)\right)\left[R_{n1,i}+R_{n2,i}\right]\right\vert \geq\frac{\varsigma_{n}}{3}\text{ }\right)\nonumber \\
 & \equiv T_{n1}+T_{n2}+T_{n3}\label{EQ:in_decomp}
\end{align}
holds by the fact that $\Pr\left(|X_{1}+X_{2}+X_{3}|\geq\varsigma_{n}\right)\leq\Pr\left(|X_{1}|\geq\frac{\varsigma_{n}}{3}\right)+\Pr\left(|X_{2}|\geq\frac{\varsigma_{n}}{3}\right)+\Pr\left(|X_{3}|\geq\frac{\varsigma_{n}}{3}\right)$
for any random variables $X_{1},X_{2},$ and $X_{3}.$

Using the moment conditions and the proof in other lemmas, it is easy
to see that 
\[
\Pr\left(\max_{j=1,...,p_{n}}\left\{ \frac{1}{n}\sum_{i=1}^{n}x_{ji}\left(y_{i}-\boldsymbol{1}\left(v_{i}>0\right)\right)\right\} \leq M\right)=1-o\left(1\right)
\]
for a large constant $M.$ Since $\varsigma_{n}\rightarrow\infty,$
the results on $R_{n1,i}$ and $R_{n2,i}$ in Part 1 imply that 
\begin{equation}
T_{n3}=\Pr\left(\max_{j=1,...,p_{n}}\left\vert \frac{1}{\sqrt{n}}\sum_{i=1}^{n}x_{ji}\left(y_{i}-\boldsymbol{1}\left(v_{i}>0\right)\right)\left[R_{n1,i}+R_{n2,i}\right]\right\vert \geq\frac{\varsigma_{n}}{3}\right)=o\left(1\right).\label{EQ:Tn3}
\end{equation}

We can go through a much simpler process as in Lemma \ref{LE:3rdU}
and show that 
\[
\Pr\left(\left\vert \frac{1}{\sqrt{n}}\sum_{i=1}^{n}x_{ji}\left(\tilde{y}_{i}-\boldsymbol{x}_{i}^{\prime}\boldsymbol{\beta}^{\ast}\right)\right\vert \geq\frac{\varsigma_{n}}{3}\text{ }\right)\leq C_{1}n^{-1-\delta/2}\varsigma_{n}^{-2-\delta/2}+C_{2}n^{-\delta}\varsigma_{n}^{-2-3\delta/2},
\]
for some positive $C_{1}$ and $C_{2}$ because $x_{ji}\left(\tilde{y}_{i}-\boldsymbol{x}_{i}^{\prime}\boldsymbol{\beta}^{\ast}\right)$
is mean 0 by the identification result and is i.i.d. across $i.$
As a result, 
\begin{align}
T_{n1} & =\Pr\left(\max_{j=1,...,p_{n}}\left\vert \frac{1}{\sqrt{n}}\sum_{i=1}^{n}x_{ji}\left(\tilde{y}_{i}-\boldsymbol{x}_{i}^{\prime}\boldsymbol{\beta}^{\ast}\right)\right\vert \geq\frac{\varsigma_{n}}{3}\right)\nonumber \\
 & \leq\sum_{j=1}^{p_{n}}\Pr\left(\left\vert \frac{1}{\sqrt{n}}\sum_{i=1}^{n}x_{ji}\left(\tilde{y}_{i}-\boldsymbol{x}_{i}^{\prime}\boldsymbol{\beta}^{\ast}\right)\right\vert \geq\frac{\varsigma_{n}}{3}\right)\nonumber \\
 & \leq p_{n}\left(C_{1}n^{-1-\delta/2}\varsigma_{n}^{-2-\delta/2}+C_{2}n^{-\delta}\varsigma_{n}^{-2-3\delta/2}\right)=o\left(1\right),\label{EQ:Tn1}
\end{align}
where the last inequality holds by $C_{p}<1$ and $\delta>4$.

For $T_{n2},$ the term inside the probability function is a standard
second order U-statistics and is the $R_{n3,j}$ defined in (\ref{eq:rn3j}),
for details, see \citet{Lewbel2000}. Since $p_{n}\ll n$, applying
the result on $R_{n3,j}$ in Lemma \ref{LE:rn3j} yields 
\begin{equation}
T_{n2}\leq o\left(1\right).\label{EQ:Tn2}
\end{equation}

Substitute equations (\ref{EQ:Tn3}), (\ref{EQ:Tn1}), and (\ref{EQ:Tn2})
into equation (\ref{EQ:in_decomp}), we obtain the desired results:
\[
\Pr\left(\max_{j=1,...,p_{n}}\left\vert \frac{1}{\sqrt{n}}\sum_{i=1}^{n}x_{ji}\left(\widehat{\tilde{y}}_{i}-\boldsymbol{x}_{i}^{\prime}\boldsymbol{\beta}^{\ast}\right)\right\vert \geq\varsigma_{n}\text{ }\right)\leq o\left(1\right).
\]
\end{proof}

\begin{proof}[Proof of Lemma \ref{LE:AB}]By the definition of
$A,$ $A^{c}$ is a subset of 
\[
\left\{ \left\Vert \left(\mathbf{C}_{n}^{\left(1,1\right)}\right)^{-1}\hat{\boldsymbol{Q}}^{\left(1\right)}\right\Vert _{\infty}>\sqrt{n}\left[\min_{j=1,...,s^{*}}\left|\beta_{j}\right|-\frac{\lambda_{n}}{2n}\left\Vert \left(\mathbf{C}_{n}^{\left(1,1\right)}\right)^{-1}\boldsymbol{S}^{\left(1\right)}\right\Vert _{\infty}\right]\right\} .
\]
We analyze the three terms in the above one by one.

The first term 
\begin{align*}
\left\Vert \left(\mathbf{C}_{n}^{\left(1,1\right)}\right)^{-1}\hat{\boldsymbol{Q}}^{\left(1\right)}\right\Vert _{\infty} & \leq\left\Vert \left(\mathbf{C}_{n}^{\left(1,1\right)}\right)^{-1}\hat{\boldsymbol{Q}}^{\left(1\right)}\right\Vert \leq\rho_{\min}^{-1}\left\{ \mathbf{C}_{n}^{\left(1,1\right)}\right\} \left\Vert \hat{\boldsymbol{Q}}^{\left(1\right)}\right\Vert \\
 & \leq\rho_{\min}^{-1}\left\{ \mathbf{C}_{n}^{\left(1,1\right)}\right\} \sqrt{s^{*}}\left\Vert \hat{\boldsymbol{Q}}^{\left(1\right)}\right\Vert _{\infty}.
\end{align*}

The third term in the above satisfies 
\begin{align*}
\frac{\lambda_{n}}{2n}\left\Vert \left(\mathbf{C}_{n}^{\left(1,1\right)}\right)^{-1}\boldsymbol{S}^{\left(1\right)}\right\Vert _{\infty} & \leq\frac{\lambda_{n}}{2n}\left\Vert \left(\mathbf{C}_{n}^{\left(1,1\right)}\right)^{-1}\boldsymbol{S}^{\left(1\right)}\right\Vert \\
 & \leq\frac{\lambda_{n}}{2n}\rho_{\min}^{-1}\left\{ \mathbf{C}_{n}^{\left(1,1\right)}\right\} \left\Vert \boldsymbol{S}^{\left(1\right)}\right\Vert \\
 & \leq\frac{\lambda_{n}}{2n}\rho_{\min}^{-1}\left\{ \mathbf{C}_{n}^{\left(1,1\right)}\right\} \frac{\sqrt{s^{*}}}{\min_{j=1,...,s^{*}}\left|\tilde{\beta}_{j}^{-\gamma}\right|}\\
 & =O_{P}\left(\lambda_{n}n^{C_{s}/2+\alpha\gamma-1}\right).
\end{align*}

The second term by assumption satisfies $\min_{j=1,...,s^{*}}\left|\beta_{j}\right|>n^{-\alpha}$.
In addition, 
\[
\lambda_{n}n^{C_{s}/2+\alpha\gamma-1}=n^{C_{\lambda}+C_{s}/2+\alpha\gamma-1}\ll n^{-\alpha}
\]
due to $C_{\lambda}<1/2-\alpha\gamma-C_{s}/2<1-\alpha-\alpha\gamma-C_{s}/2$
and $\alpha<1/2$ in Assumption \ref{A:adaptiveLasso}. Therefore,
$\min_{j=1,...,s^{*}}\left|\beta_{j}\right|$ is the dominant term
on the right with very high probability.

The above analysis imply that 
\begin{align*}
\Pr\left(A^{c}\right) & \leq\Pr\left(\rho_{\min}^{-1}\left\{ \mathbf{C}_{n}^{\left(1,1\right)}\right\} \sqrt{s^{*}}\left\Vert \hat{\boldsymbol{Q}}^{\left(1\right)}\right\Vert _{\infty}>\frac{\sqrt{n}}{2}\min_{j=1,...,s^{*}}\left|\beta_{j}\right|\right)+o\left(1\right)\\
 & =\Pr\left(\left\Vert \hat{\boldsymbol{Q}}^{\left(1\right)}\right\Vert _{\infty}>\rho_{\min}\left\{ \mathbf{C}_{n}^{\left(1,1\right)}\right\} \frac{\sqrt{n/s^{*}}}{2}\min_{j=1,...,s^{*}}\left|\beta_{j}\right|\right)+o\left(1\right)\\
 & =o\left(1\right),
\end{align*}
where the last line we uses the results from Lemmas \ref{LE:matrixXX'}
($\rho_{\min}\left\{ \mathbf{C}_{n}^{\left(1,1\right)}\right\} $
is bounded away from zero with very high probability), Lemma \ref{LE:prepare},
and $\frac{\sqrt{n/s^{*}}}{2}\min_{j=1,...,s^{*}}\left|\beta_{j}\right|\apprge n^{1/2-C_{s}/2-\alpha}$
with $1/2-C_{s}/2-\alpha>0$ imposed in Assumption \ref{A:adaptiveLasso}.
This is the desired result.

We now show the result for $B.$ Similarly to the above proof, we
try to bound the probability of $B^{c}$ and $B^{c}$ satisfies

\begin{align*}
B^{c} & \subseteq\left\{ \left\Vert \left|\mathbf{C}_{n}^{\left(2,1\right)}\left(\mathbf{C}_{n}^{\left(1,1\right)}\right)^{-1}\hat{\boldsymbol{Q}}^{\left(1\right)}-\hat{\boldsymbol{Q}}^{\left(2\right)}\right|+\frac{\lambda_{n}}{2\sqrt{n}}\left|\mathbf{C}_{n}^{\left(2,1\right)}\left(\mathbf{C}_{n}^{\left(1,1\right)}\right)^{-1}\boldsymbol{S}^{\left(1\right)}\right|\right\Vert _{\infty}>\frac{\lambda_{n}}{2\sqrt{n}}\min_{j=s^{*}+1,...,p_{n}}\varpi_{j},\right\} \\
 & =\left\{ \max_{j=s^{*}+1,...,p_{n}}\left[\left|\frac{1}{n}\mathbf{X}_{jn}^{\prime}\mathbf{X}_{n}^{\left(1\right)}\left(\mathbf{C}_{n}^{\left(1,1\right)}\right)^{-1}\hat{\boldsymbol{Q}}^{\left(1\right)}-\hat{Q}_{j}\right|+\frac{\lambda_{n}}{2\sqrt{n}}\left|\frac{1}{n}\mathbf{X}_{jn}^{\prime}\mathbf{X}_{n}^{\left(1\right)}\left(\mathbf{C}_{n}^{\left(1,1\right)}\right)^{-1}\boldsymbol{S}^{\left(1\right)}\right|\right]\right.\\
 & \;\;\;\;>\left.\frac{\lambda_{n}}{2\sqrt{n}}\min_{j=s^{*}+1,...,p_{n}}\varpi_{j}\right\} .
\end{align*}
Further, by the definition of $\varpi_{j}$ and the rate on $\tilde{\boldsymbol{\beta}}$,
\[
\frac{\lambda_{n}}{2\sqrt{n}}\min_{j=s^{*}+1,...,p_{n}}\varpi_{j}\apprge_{P}\lambda_{n}n^{\varrho\gamma-1/2}.
\]
Thus there exist a positive $C$ such that with very high probability
\[
\frac{\lambda_{n}}{2\sqrt{n}}\min_{j=s^{*}+1,...,p_{n}}\varpi_{j}\geq C\lambda_{n}n^{\varrho\gamma-1/2}.
\]
Then 
\begin{align}
\Pr\left(B^{c}\right) & \leq\Pr\left(\max_{j=s^{*}+1,...,p_{n}}\left|\frac{1}{n}\mathbf{X}_{jn}^{\prime}\mathbf{X}_{n}^{\left(1\right)}\left(\mathbf{C}_{n}^{\left(1,1\right)}\right)^{-1}\hat{\boldsymbol{Q}}^{\left(1\right)}\right|>\frac{C\lambda_{n}n^{\varrho\gamma-1/2}}{3}\right)\nonumber \\
 & +\Pr\left(\max_{j=s^{*}+1,...,p_{n}}\left|\hat{Q}_{j}\right|>\frac{C\lambda_{n}n^{\varrho\gamma-1/2}}{3}\right)\nonumber \\
 & +\Pr\left(\max_{j=s^{*}+1,...,p_{n}}\frac{\lambda_{n}}{2\sqrt{n}}\left|\frac{1}{n}\mathbf{X}_{jn}^{\prime}\mathbf{X}_{n}^{\left(1\right)}\left(\mathbf{C}_{n}^{\left(1,1\right)}\right)^{-1}\boldsymbol{S}^{\left(1\right)}\right|>\frac{C\lambda_{n}n^{\varrho\gamma-1/2}}{3}\right)+o\left(1\right)\nonumber \\
 & \equiv T_{n1}+T_{n2}+T_{n3}.\label{eq:PrBc}
\end{align}

We show the bound for the terms in the above one by one. 
\begin{align}
T_{n1} & \leq\sum_{j=s^{*}+1}^{n}\Pr\left(\left|\frac{1}{n}\mathbf{X}_{jn}^{\prime}\mathbf{X}_{n}^{\left(1\right)}\left(\mathbf{C}_{n}^{\left(1,1\right)}\right)^{-1}\hat{\boldsymbol{Q}}^{\left(1\right)}\right|>\frac{C\lambda_{n}n^{\varrho\gamma-1/2}}{3}\right)\nonumber \\
 & \leq\sum_{j=s^{*}+1}^{n}\Pr\left(\left\Vert \frac{1}{n}\mathbf{X}_{jn}^{\prime}\mathbf{X}_{n}^{\left(1\right)}\right\Vert \geq2C_{X2}\sqrt{s^{*}}\right)+\Pr\left(\rho_{\min}^{-1}\left(\mathbf{C}_{n}^{\left(1,1\right)}\right)>\frac{2}{D_{3}}\right)\nonumber \\
 & +\Pr\left(\left\Vert \hat{\boldsymbol{Q}}^{\left(1\right)}\right\Vert >\frac{CD_{3}}{12C_{X2}}\frac{\lambda_{n}n^{\varrho\gamma-1/2}}{\sqrt{s^{*}}}\right)\nonumber \\
 & \leq\sum_{j=s^{*}+1}^{n}\Pr\left(\left\Vert \frac{1}{n}\mathbf{X}_{jn}^{\prime}\mathbf{X}_{n}^{\left(1\right)}\right\Vert \geq2C_{X2}\sqrt{s^{*}}\right)+\Pr\left(\rho_{\min}^{-1}\left(\mathbf{C}_{n}^{\left(1,1\right)}\right)>\frac{2}{D_{3}}\right)\nonumber \\
 & +\Pr\left(\max_{j=s^{*}+1,...,p_{n}}\left|\hat{Q}_{j}\right|>\frac{CD_{3}}{12C_{X2}}\frac{\lambda_{n}n^{\varrho\gamma-1/2}}{s^{*}}\right)\nonumber \\
 & =o\left(1\right),\label{eq:Tn1_}
\end{align}
where we apply Lemmas \ref{LE:matrixXX'}, \ref{LE:matrixXX'C21},
and \ref{LE:prepare}, and use the condition that 
\[
\frac{\lambda_{n}n^{\varrho\gamma-1/2}}{s^{*}}=n^{C_{\lambda}+\varrho\gamma-1/2-C_{s}}\text{ and }C_{\lambda}+\varrho\gamma-1/2-C_{s}>0
\]
imposed in Assumption \ref{A:adaptiveLasso}.

For $T_{n2},$ the second last line in (\ref{eq:Tn1_}) implies that
\[
T_{n2}\leq T_{n1}=o\left(1\right).
\]

For $T_{n3},$ note that 
\[
\max_{j=s^{*}+1,...,p_{n}}\left|S_{j}\right|\apprle\max_{j=s^{*}+1,...,p_{n}}\left|\tilde{\beta}_{j}\right|^{-\gamma}=O_{P}\left(n^{\alpha\gamma}\right).
\]
As in (\ref{eq:Tn1_}), $T_{n3}$ can be similarly rewritten as 
\begin{align*}
T_{n3} & =\Pr\left(\max_{j=s^{*}+1,...,p_{n}}\left|\frac{1}{n}\mathbf{X}_{jn}^{\prime}\mathbf{X}_{n}^{\left(1\right)}\left(\mathbf{C}_{n}^{\left(1,1\right)}\right)^{-1}\boldsymbol{S}^{\left(1\right)}\right|>\frac{2Cn^{\varrho\gamma}}{3}\right)\\
 & \leq\sum_{j=s^{*}+1}^{n}\Pr\left(\left\Vert \frac{1}{n}\mathbf{X}_{jn}^{\prime}\mathbf{X}_{n}^{\left(1\right)}\right\Vert \geq2C_{X2}\sqrt{s^{*}}\right)+\Pr\left(\rho_{\min}^{-1}\left(\mathbf{C}_{n}^{\left(1,1\right)}\right)>\frac{2}{D_{3}}\right)\\
 & +\Pr\left(\max_{j=1,...,s^{*}}\left|S_{j}\right|>\frac{CD_{3}}{12C_{X2}}\frac{n^{\varrho\gamma}}{s^{*}}\right)\\
 & =o\left(1\right),
\end{align*}
where the last line holds by $n^{\alpha\gamma}\ll n^{\varrho\gamma-C_{s}}$
due to $\gamma>C_{s}/\left(\varrho-\alpha\right).$

Substitute the results on $T_{n1},T_{n2}$ and $T_{n3}$ back into
(\ref{eq:PrBc}), we obtain 
\[
\Pr\left(B^{c}\right)=o\left(1\right),
\]
as desired.
\end{proof}

\begin{proof}[Proof of Lemma \ref{LE:expansion}] Similar to
the proof of Lemma \ref{LE:prepare}, we show the results of the lemma
by conditioning on the event $\hat{E}^{*},$ that we have selected
all the relevant $X$ for $V$ in the first stage, and optimal bandwidths
are chosen for $\hat{f}\left(V|\boldsymbol{X}\right)$. The unconditional
results can be obtained using a similar argument as in the proof of
Lemma \ref{LE:prepare}.

We decompose $\frac{\hat{f}_{\boldsymbol{Z}i}}{\hat{f}_{V\boldsymbol{Z}i}}-\frac{f_{\boldsymbol{Z}i}}{f_{V\boldsymbol{Z}i}}$
as 
\begin{align}
 & \frac{\hat{f}_{\boldsymbol{Z}i}}{\hat{f}_{V\boldsymbol{Z}i}}-\frac{f_{\boldsymbol{Z}i}}{f_{V\boldsymbol{Z}i}}=\frac{\hat{f}_{\boldsymbol{Z}i}-\mathbb{E}\left(\hat{f}_{\boldsymbol{Z}i}\right)}{\mathbb{E}\left(\hat{f}_{V\boldsymbol{Z}i}\right)}+\frac{\mathbb{E}\left(\hat{f}_{\boldsymbol{Z}i}\right)\left(\mathbb{E}\left(\hat{f}_{V\boldsymbol{Z}i}\right)-\hat{f}_{V\boldsymbol{Z}i}\right)}{\left[\mathbb{E}\left(\hat{f}_{V\boldsymbol{Z}i}\right)\right]^{2}}+\frac{\mathbb{E}\left(\hat{f}_{\boldsymbol{Z}i}\right)}{\mathbb{E}\left(\hat{f}_{V\boldsymbol{Z}i}\right)}-\frac{f_{\boldsymbol{Z}i}}{f_{V\boldsymbol{Z}i}}\nonumber \\
 & +\frac{\left[\hat{f}_{\boldsymbol{Z}i}-\mathbb{E}\left(\hat{f}_{\boldsymbol{Z}i}\right)\right]\left[\mathbb{E}\left(\hat{f}_{V\boldsymbol{Z}i}\right)-\hat{f}_{V\boldsymbol{Z}i}\right]}{\hat{f}_{V\boldsymbol{Z}i}\mathbb{E}\left(\hat{f}_{V\boldsymbol{Z}i}\right)}+\frac{\mathbb{E}\left(\hat{f}_{\boldsymbol{Z}i}\right)\left[\mathbb{E}\left(\hat{f}_{V\boldsymbol{Z}i}\right)-\hat{f}_{V\boldsymbol{Z}i}\right]^{2}}{\hat{f}_{V\boldsymbol{Z}i}\left[\mathbb{E}\left(\hat{f}_{V\boldsymbol{Z}i}\right)\right]^{2}}\nonumber \\
 & \equiv\frac{\hat{f}_{\boldsymbol{Z}i}-\mathbb{E}\left(\hat{f}_{\boldsymbol{Z}i}\right)}{\mathbb{E}\left(\hat{f}_{V\boldsymbol{Z}i}\right)}+\frac{\mathbb{E}\left(\hat{f}_{\boldsymbol{Z}i}\right)\left(\mathbb{E}\left(\hat{f}_{V\boldsymbol{Z}i}\right)-\hat{f}_{V\boldsymbol{Z}i}\right)}{\left[\mathbb{E}\left(\hat{f}_{V\boldsymbol{Z}i}\right)\right]^{2}}+R_{n1,i}+R_{n2,i}.\label{EQ:fvz}
\end{align}
Conditional on $\hat{E}^{*},$ 
\begin{align*}
\frac{\mathbb{E}\left(\hat{f}_{\boldsymbol{Z}i}\right)}{\mathbb{E}\left(\hat{f}_{V\boldsymbol{Z}i}\right)} & =\frac{\mathbb{E}\left[\Pi_{l=1}^{p^{\ast}}K_{h_{l}}\left(z_{ji}-z_{li}\right)\Pi_{l=p^{\ast}+1}^{\tilde{p}}K_{h_{l}}\left(z_{ji}-z_{li}\right)\right]}{\mathbb{E}\left[K_{h_{V}}\left(v\right)\Pi_{l=1}^{p^{\ast}}K_{h_{l}}\left(z_{ji}-z_{li}\right)\Pi_{l=p^{\ast}+1}^{\tilde{p}}K_{h_{l}}\left(z_{ji}-z_{li}\right)\right]}\\
 & =\frac{\mathbb{E}\left[\Pi_{l=1}^{p^{\ast}}K_{h_{l}}\left(z_{ji}-z_{li}\right)\right]\mathbb{E}\left[\Pi_{l=p^{\ast}+1}^{\tilde{p}}K_{h_{l}}\left(z_{ji}-z_{li}\right)\right]}{\mathbb{E}\left[K_{h_{V}}\left(v\right)\Pi_{l=1}^{p^{\ast}}K_{h_{l}}\left(z_{ji}-x_{li}\right)\right]\mathbb{E}\left[\Pi_{l=p^{\ast}+1}^{\tilde{p}}K_{h_{l}}\left(z_{ji}-z_{li}\right)\right]}\\
 & =\frac{\mathbb{E}\left[\Pi_{l=1}^{p^{\ast}}K_{h_{l}}\left(z_{ji}-z_{li}\right)\right]}{\mathbb{E}\left[K_{h_{V}}\left(v\right)\Pi_{l=1}^{p^{\ast}}K_{h_{l}}\left(z_{ji}-z_{li}\right)\right]},
\end{align*}
where the second line holds by the independence assumption in Assumption
\ref{A:main}. With that, it is not hard to see that 
\begin{align*}
R_{n1,i} & =\frac{\mathbb{E}\left(\hat{f}_{\boldsymbol{Z}i}\right)}{\mathbb{E}\left(\hat{f}_{V\boldsymbol{Z}i}\right)}-\frac{f_{\boldsymbol{Z}i}}{f_{V\boldsymbol{Z}i}}=\frac{\mathbb{E}\left[\Pi_{l=1}^{p^{\ast}}K_{h_{l}}\left(z_{ji}-z_{li}\right)\right]}{\mathbb{E}\left[K_{h_{V}}\left(v\right)\Pi_{l=1}^{p^{\ast}}K_{h_{l}}\left(z_{ji}-z_{li}\right)\right]}-\frac{f\left(z_{1i},...,z_{p^{\ast}i}\right)}{f\left(v_{i},z_{1i},...,z_{p^{\ast}i}\right)}\\
 & =O\left(h_{1}^{2r}+...+h_{p^{\ast}}^{2r}\right)=o\left(n^{-1/2}\right),
\end{align*}
due to the fact that $K\left(\cdot\right)$ is an $2r$-th order kernel
and $h_{1}^{r},...,h_{p^{\ast}}^{r}$ is $o\left(n^{-1/4}\right)$
conditional on $\hat{E}^{*}$.

Following the proof of Theorem 1.4 in \citet{LiRacine}, we can show
that $\hat{f}_{\boldsymbol{Z}i}-\mathbb{E}\left(\hat{f}_{\boldsymbol{Z}i}\right)=o_{P}\left(n^{-1/4}\right),$
$\hat{f}_{V\boldsymbol{Z}i}-\mathbb{E}\left(\hat{f}_{V\boldsymbol{Z}i}\right)=o_{P}\left(n^{-1/4}\right)$
uniformly over the compact support of $V$ and $\boldsymbol{Z}$.
Thus $R_{n2,i}=o_{P}\left(n^{-1/2}\right)$ uniformly over $i=1,2,...,n.$

Summarize the results, for any $j,$ 
\[
\frac{1}{\sqrt{n}}\sum_{i=1}^{n}z_{ji}\left(y_{i}-\boldsymbol{1}\left(v_{i}>0\right)\right)\left[R_{n1,i}+R_{n2,i}\right]=o_{P}\left(1\right).
\]
As a result, 
\begin{align*}
 & \frac{1}{\sqrt{n}}\sum_{i=1}^{n}z_{ji}\left(\widehat{\tilde{y}}_{i}-\boldsymbol{x}_{i}^{\prime}\boldsymbol{\beta}^{\ast}\right)\\
 & =\frac{1}{\sqrt{n}}\sum_{i=1}^{n}z_{ji}\left(\tilde{y}_{i}-\boldsymbol{x}_{i}^{\prime}\boldsymbol{\beta}^{\ast}\right)+\frac{1}{\sqrt{n}}\sum_{i=1}^{n}z_{ji}\left(\widehat{\tilde{y}}_{i}-\tilde{y}_{i}\right)\\
 & =\frac{1}{\sqrt{n}}\sum_{i=1}^{n}x_{ji}\left(\tilde{y}_{i}-\boldsymbol{x}_{i}^{\prime}\boldsymbol{\beta}^{\ast}\right)+\frac{1}{\sqrt{n}}\sum_{i=1}^{n}z_{ji}\left(y_{i}-\boldsymbol{1}\left(v_{i}>0\right)\right)\left[\frac{\hat{f}_{\boldsymbol{Z}i}}{\hat{f}_{V\boldsymbol{Z}i}}-\frac{f_{\boldsymbol{Z}i}}{f_{V\boldsymbol{Z}i}}\right]\\
 & =\frac{1}{\sqrt{n}}\sum_{i=1}^{n}x_{ji}\left(\tilde{y}_{i}-\boldsymbol{x}_{i}^{\prime}\boldsymbol{\beta}^{\ast}\right)+\frac{1}{\sqrt{n}}\sum_{i=1}^{n}z_{ji}\left(y_{i}-\boldsymbol{1}\left(v_{i}>0\right)\right)\left[\frac{\hat{f}_{\boldsymbol{Z}i}-\mathbb{E}\left(\hat{f}_{\boldsymbol{Z}i}\right)}{\mathbb{E}\left(\hat{f}_{V\boldsymbol{Z}i}\right)}+\frac{\mathbb{E}\left(\hat{f}_{\boldsymbol{Z}i}\right)\left(\mathbb{E}\left(\hat{f}_{V\boldsymbol{Z}i}\right)-\hat{f}_{V\boldsymbol{Z}i}\right)}{\left[\mathbb{E}\left(\hat{f}_{V\boldsymbol{Z}i}\right)\right]^{2}}\right]\\
 & +\frac{1}{\sqrt{n}}\sum_{i=1}^{n}z_{ji}\left(y_{i}-\boldsymbol{1}\left(v_{i}>0\right)\right)\left[R_{n1,i}+R_{n2,i}\right]\\
 & =\frac{1}{\sqrt{n}}\sum_{i=1}^{n}z_{ji}\left(\tilde{y}_{i}-\boldsymbol{x}_{i}^{\prime}\boldsymbol{\beta}^{\ast}\right)+\frac{1}{\sqrt{n}}\sum_{i=1}^{n}z_{ji}\left(y_{i}-\boldsymbol{1}\left(v_{i}>0\right)\right)\left[\frac{\hat{f}_{\boldsymbol{Z}i}-\mathbb{E}\left(\hat{f}_{\boldsymbol{Z}i}\right)}{\mathbb{E}\left(\hat{f}_{V\boldsymbol{Z}i}\right)}+\frac{\mathbb{E}\left(\hat{f}_{\boldsymbol{Z}i}\right)\left(\mathbb{E}\left(\hat{f}_{V\boldsymbol{Z}i}\right)-\hat{f}_{V\boldsymbol{Z}i}\right)}{\left[\mathbb{E}\left(\hat{f}_{V\boldsymbol{Z}i}\right)\right]^{2}}\right]\\
 & +o_{P}\left(1\right).
\end{align*}
Note that $\frac{1}{\sqrt{n}}\sum_{i=1}^{n}z_{ji}\left(y_{i}-\boldsymbol{1}\left(v_{i}>0\right)\right)\left[\frac{\hat{f}_{\boldsymbol{Z}i}-\mathbb{E}\left(\hat{f}_{\boldsymbol{Z}i}\right)}{\mathbb{E}\left(\hat{f}_{V\boldsymbol{Z}i}\right)}+\frac{\mathbb{E}\left(\hat{f}_{\boldsymbol{Z}i}\right)\left(\mathbb{E}\left(\hat{f}_{V\boldsymbol{Z}i}\right)-\hat{f}_{V\boldsymbol{Z}i}\right)}{\left[\mathbb{E}\left(\hat{f}_{V\boldsymbol{Z}i}\right)\right]^{2}}\right]$
is a U-statistic, and 
\[
\frac{1}{\sqrt{n}}\sum_{i=1}^{n}z_{ji}\left(\widehat{\tilde{y}}_{i}-\boldsymbol{x}_{i}^{\prime}\boldsymbol{\beta}^{\ast}\right)=\frac{1}{\sqrt{n}}\sum_{i=1}^{n}q_{ji}+o_{P}\left(1\right),
\]
for $1\leq j\leq s^{\ast}$, for details, see \citet{Lewbel2000}.
The results for $j\in A\cup B$ similarly hold. \end{proof}

\begin{proof}[Proof of Lemma \ref{LE:uniformrate}]
$\boldsymbol{m}\left(\boldsymbol{\theta}\right)$
is $p_{n}\times1$ vector, collecting all the moment conditions. At
the true value of parameters, by definition $\mathbb{E}\left[\boldsymbol{m}\left(\boldsymbol{\theta}^{\ast}\right)\right]=\mathbb{E}\left[\boldsymbol{Z}_{1}^{p_{n}}\left(\tilde{y}_{i}-\boldsymbol{x}_{i}^{\prime}\boldsymbol{\beta}^{\ast}\right)\right]=\boldsymbol{0,}$
thus $\mathbb{E}\left[\boldsymbol{m}\left(\boldsymbol{\theta}\right)\right]=\mathbb{E}\left[\boldsymbol{m}\left(\boldsymbol{\theta}\right)-\boldsymbol{m}\left(\boldsymbol{\theta}^{\ast}\right)\right].$
We turn to the object of interest. The $j$-th element of $\overline{\boldsymbol{\hat{m}}}_{n}\left(\boldsymbol{\theta}\right)-\mathbb{E}\left[\boldsymbol{m}\left(\boldsymbol{\theta}\right)\right]$
is 
\begin{align*}
 & n^{-1}\sum_{i=1}^{n}z_{ji}\left(\widehat{\tilde{y}}_{i}-\boldsymbol{x}_{i}^{\prime}\boldsymbol{\beta}\right)-\mathbb{E}\left[z_{ji}\left(\tilde{y}_{i}-\boldsymbol{x}_{i}^{\prime}\boldsymbol{\beta}\right)\right]\\
 & =n^{-1}\sum_{i=1}^{n}z_{ji}\left(\widehat{\tilde{y}}_{i}-\boldsymbol{x}_{i}^{\prime}\boldsymbol{\beta}^{\ast}\right)-n^{-1}\sum_{i=1}^{n}z_{ji}\left(\boldsymbol{x}_{i}^{\prime}\left(\boldsymbol{\beta-\beta}^{\ast}\right)\right)-\mathbb{E}\left[z_{ji}\left(\tilde{y}_{i}-\boldsymbol{x}_{i}^{\prime}\boldsymbol{\beta}^{\ast}\right)\right]+\mathbb{E}\left[z_{ji}\left(\boldsymbol{x}_{i}^{\prime}\left(\boldsymbol{\beta-\beta}^{\ast}\right)\right)\right]\\
 & =n^{-1}\sum_{i=1}^{n}z_{ji}\left(\widehat{\tilde{y}}_{i}-\boldsymbol{x}_{i}^{\prime}\boldsymbol{\beta}^{\ast}\right)-n^{-1}\sum_{i=1}^{n}\left\{ z_{ji}\left(\boldsymbol{x}_{i}^{\prime}\left(\boldsymbol{\beta-\beta}^{\ast}\right)\right)-\mathbb{E}\left[z_{ji}\left(\boldsymbol{x}_{i}^{\prime}\left(\boldsymbol{\beta-\beta}^{\ast}\right)\right)\right]\right\} \\
 & =O_{P}\left(n^{-1/2}\right),
\end{align*}
due to Lemma \ref{LE:expansion}, i.i.d. across $i,$ and the finite
fourth moments of $Z$ and $X.$ The above holds uniformly over a
compact set of $\boldsymbol{\theta}$ because $\boldsymbol{m}\left(\boldsymbol{\theta}\right)$
is linear in $\boldsymbol{\theta.}$

An immediate result is that 
\begin{align*}
\left\Vert \overline{\boldsymbol{\hat{m}}}_{n}\left(\boldsymbol{\theta}\right)-\mathbb{E}\left[\boldsymbol{m}\left(\boldsymbol{\theta}\right)\right]\right\Vert  & =\sqrt{\sum_{j=1}^{p_{n}}\left[n^{-1}\sum_{i=1}^{n}z_{ji}\left(\widehat{\tilde{y}}_{i}-\boldsymbol{x}_{i}^{\prime}\boldsymbol{\beta}\right)-\mathbb{E}\left[z_{ji}\left(\tilde{y}_{i}-\boldsymbol{x}_{i}^{\prime}\boldsymbol{\beta}\right)\right]\right]^{2}}\\
 & =O_{P}\left(\sqrt{\left.p_{n}\right/n}\right),
\end{align*}
holds uniformly over a compact set of $\boldsymbol{\theta}$. \end{proof}

\begin{proof} [Proof of Lemma \ref{LE:matrix_conv}] The key conditions
we need from Assumption \ref{A:GMM} are $p_{n}^{3}/n\rightarrow0$
and the finite fourth moment of $X$ and $Z.$ By some elementary
calculation, 
\[
\left\Vert \frac{\partial\overline{\boldsymbol{\hat{m}}}_{n}}{\partial\boldsymbol{\theta}_{B}^{\prime}}-\Gamma_{\boldsymbol{\theta}_{B}}\right\Vert \leq p_{n}k^{\ast}\left\Vert \frac{\partial\overline{\boldsymbol{\hat{m}}}_{n}}{\partial\boldsymbol{\theta}_{B}^{\prime}}-\Gamma_{\boldsymbol{\theta}_{B}}\right\Vert _{\infty}.
\]
With it, for any small positive $\epsilon,$ 
\begin{align*}
\Pr\left(\left\Vert \frac{\partial\overline{\boldsymbol{\hat{m}}}_{n}}{\partial\boldsymbol{\theta}_{B}^{\prime}}-\Gamma_{\boldsymbol{\theta}_{B}}\right\Vert >\epsilon\right) & \leq\Pr\left(p_{n}k^{\ast}\left\Vert \frac{\partial\overline{\boldsymbol{\hat{m}}}_{n}}{\partial\boldsymbol{\theta}_{B}^{\prime}}-\Gamma_{\boldsymbol{\theta}_{B}}\right\Vert _{\infty}>\epsilon\right)\\
 & =\Pr\left(\max_{j=1,2,..,p_{n},l=1,2,...,k^{\ast}}\left\vert n^{-1}\sum_{i=1}^{n}z_{ji}x_{li}-\mathbb{E}\left(z_{ji}x_{li}\right)\right\vert >\frac{\epsilon}{p_{n}k^{\ast}}\right)\\
 & \leq\sum_{j=1}^{p_{n}}\sum_{l=1}^{k^{\ast}}\Pr\left(\left\vert n^{-1}\sum_{i=1}^{n}z_{ji}x_{li}-\mathbb{E}\left(z_{ji}x_{li}\right)\right\vert >\frac{\epsilon}{p_{n}k^{\ast}}\right)\\
 & \leq\sum_{j=1}^{p_{n}}\sum_{l=1}^{k^{\ast}}\frac{p_{n}^{2}k^{\ast2}}{\epsilon^{2}}\mathbb{E}\left[\left(n^{-1}\sum_{i=1}^{n}z_{ji}x_{li}-\mathbb{E}\left(z_{ji}x_{li}\right)\right)^{2}\right]\\
 & \leq\frac{Cp_{n}^{3}k^{\ast3}}{n\epsilon^{2}}\rightarrow0,
\end{align*}
for some positive $C,$ where the fourth line holds by the Markov
inequality, and the last line holds by the finite fourth moment of
$Z$ and $X$, and the assumption that $p_{n}^{3}/n\rightarrow0.$
\end{proof}

\begin{proof}[Proof of Lemma \ref{LE:consistency}] We present
some useful results first. We let $\mathbb{E}_{n}$ denotes the expectation
by treating $\hat{\boldsymbol{\theta}}$ as a fixed constant. Due
to $\mathbb{E}\left[\boldsymbol{m}\left(\boldsymbol{\theta}^{\ast}\right)\right]=0$
and Lemma \ref{LE:uniformrate}, 
\begin{equation}
\overline{\boldsymbol{\hat{m}}}_{n}\left(\boldsymbol{\theta}^{\ast}\right)^{\prime}\mathbf{W}_{n}\overline{\boldsymbol{\hat{m}}}_{n}\left(\boldsymbol{\theta}^{\ast}\right)\leq C\left\Vert \overline{\boldsymbol{\hat{m}}}_{n}\left(\boldsymbol{\theta}^{\ast}\right)\right\Vert ^{2}=O_{P}\left(p_{n}/n\right).\label{eq:m^1}
\end{equation}
Recall that $\underline{\eta}_{j}^{*}=0$ for $j=1,...,d_{A},$ thus
\begin{equation}
\lambda_{n}\sum_{j=1}^{d_{A}+d_{B}}\varpi_{j}\left|\underline{\eta}_{j}^{*}\right|-\lambda_{n}\sum_{j=1}^{d_{A}+d_{B}}\varpi_{j}\left|\underline{\hat{\eta}}_{j}\right|\leq\lambda_{n}\sum_{j=d_{A}+1}^{d_{A}+d_{B}}\varpi_{j}\left(\left|\underline{\eta}_{j}^{*}\right|-\left|\underline{\hat{\eta}}_{j}\right|\right)\leq\lambda_{n}\left\Vert \boldsymbol{\varpi}_{B}\right\Vert \left\Vert \underline{\hat{\boldsymbol{\eta}}}_{B}^{*}-\underline{\boldsymbol{\eta}}_{B}^{*}\right\Vert .\label{eq:weta^}
\end{equation}
Again by $\mathbb{E}\left[\boldsymbol{m}\left(\boldsymbol{\theta}^{\ast}\right)\right]=0$
and the linearity of $\boldsymbol{m},$ 
\begin{equation}
\mathbb{E}_{n}\left[\boldsymbol{m}\left(\boldsymbol{\hat{\theta}}\right)\right]=\mathbb{E}_{n}\left[\boldsymbol{m}\left(\boldsymbol{\hat{\theta}}\right)\right]-\mathbb{E}\left[\boldsymbol{m}\left(\boldsymbol{\theta}^{\ast}\right)\right]=C_{1}\left\Vert \boldsymbol{\hat{\theta}}-\boldsymbol{\theta}^{\ast}\right\Vert ,\label{eq:mtheta^}
\end{equation}
for some positive $C_{1}$. By the uniform convergence result in Lemma
\ref{LE:uniformrate} and using (\ref{eq:mtheta^}), 
\begin{align}
\overline{\boldsymbol{\hat{m}}}_{n}\left(\boldsymbol{\hat{\theta}}\right)^{\prime}\mathbf{W}_{n}\overline{\boldsymbol{\hat{m}}}_{n}\left(\boldsymbol{\hat{\theta}}\right) & =\mathbb{E}_{n}\left[\boldsymbol{m}\left(\boldsymbol{\hat{\theta}}\right)\right]'\mathbf{W}_{n}\mathbb{E}_{n}\left[\boldsymbol{m}\left(\boldsymbol{\hat{\theta}}\right)\right]+2\mathbb{E}_{n}\left[\boldsymbol{m}\left(\boldsymbol{\hat{\theta}}\right)\right]'\mathbf{W}_{n}\left\{ \overline{\boldsymbol{\hat{m}}}_{n}\left(\boldsymbol{\hat{\theta}}\right)-\mathbb{E}_{n}\left[\boldsymbol{m}\left(\boldsymbol{\hat{\theta}}\right)\right]\right\} \nonumber \\
 & +\left\{ \overline{\boldsymbol{\hat{m}}}_{n}\left(\boldsymbol{\hat{\theta}}\right)-\mathbb{E}_{n}\left[\boldsymbol{m}\left(\boldsymbol{\hat{\theta}}\right)\right]\right\} '\mathbf{W}_{n}\left\{ \overline{\boldsymbol{\hat{m}}}_{n}\left(\boldsymbol{\hat{\theta}}\right)-\mathbb{E}_{n}\left[\boldsymbol{m}\left(\boldsymbol{\hat{\theta}}\right)\right]\right\} \nonumber \\
 & \geq C_{2}\left\Vert \boldsymbol{\hat{\theta}}-\boldsymbol{\theta}^{\ast}\right\Vert ^{2}+\left\Vert \boldsymbol{\hat{\theta}}-\boldsymbol{\theta}^{\ast}\right\Vert \cdot O_{P}\left(\sqrt{p_{n}/n}\right)+O_{P}\left(p_{n}/n\right),\label{eq:m^2}
\end{align}
for some positive $C_{2}.$

Let $\mathcal{L}_{n}\left(\boldsymbol{\theta}\right)$\ denote the
objective function 
\[
\mathcal{L}_{n}\left(\boldsymbol{\theta}\right)\equiv\overline{\boldsymbol{\hat{m}}}_{n}\left(\boldsymbol{\theta}\right)^{\prime}\mathbf{W}_{n}\overline{\boldsymbol{\hat{m}}}_{n}\left(\boldsymbol{\theta}\right)+\lambda_{n}\sum_{j=1}^{d_{A}+d_{B}}\varpi_{j}\left|\underline{\eta}_{j}\right|.
\]
Then 
\[
\mathcal{L}_{n}\left(\boldsymbol{\hat{\theta}}\right)\leq\mathcal{L}_{n}\left(\boldsymbol{\theta}^{\ast}\right),
\]
which implies 
\[
\overline{\boldsymbol{\hat{m}}}_{n}\left(\boldsymbol{\hat{\theta}}\right)^{\prime}\mathbf{W}_{n}\overline{\boldsymbol{\hat{m}}}_{n}\left(\boldsymbol{\hat{\theta}}\right)\leq\overline{\boldsymbol{\hat{m}}}_{n}\left(\boldsymbol{\theta}^{\ast}\right)^{\prime}\mathbf{W}_{n}\overline{\boldsymbol{\hat{m}}}_{n}\left(\boldsymbol{\theta}^{\ast}\right)+\lambda_{n}\sum_{j=1}^{d_{A}+d_{B}}\varpi_{j}\left|\underline{\eta}_{j}^{*}\right|-\lambda_{n}\sum_{j=1}^{d_{A}+d_{B}}\varpi_{j}\left|\underline{\hat{\eta}}_{j}\right|.
\]
Substituting results of (\ref{eq:m^1}), (\ref{eq:weta^}), and (\ref{eq:m^2})
into the above yields 
\begin{equation}
C_{2}\left\Vert \boldsymbol{\hat{\theta}}-\boldsymbol{\theta}^{\ast}\right\Vert ^{2}+\left\Vert \boldsymbol{\hat{\theta}}-\boldsymbol{\theta}^{\ast}\right\Vert \cdot O_{P}\left(\sqrt{p_{n}/n}\right)+O_{P}\left(p_{n}/n\right)\leq O_{P}\left(p_{n}/n\right)+\lambda_{n}\left\Vert \boldsymbol{\varpi}_{B}\right\Vert \left\Vert \underline{\hat{\boldsymbol{\eta}}}_{B}^{*}-\underline{\boldsymbol{\eta}}_{B}^{*}\right\Vert .\label{eq:theta_bound}
\end{equation}
Since $\left\Vert \boldsymbol{\hat{\theta}}-\boldsymbol{\theta}^{\ast}\right\Vert ^{2}=\left\Vert \hat{\boldsymbol{\beta}}-\boldsymbol{\beta}\right\Vert ^{2}+\left\Vert \underline{\hat{\boldsymbol{\eta}}}_{A}^{*}-\underline{\boldsymbol{\eta}}_{A}^{*}\right\Vert ^{2}+\left\Vert \underline{\hat{\boldsymbol{\eta}}}_{B}^{*}-\underline{\boldsymbol{\eta}}_{B}^{*}\right\Vert ^{2},$
the above first implies that 
\[
\left\Vert \underline{\hat{\boldsymbol{\eta}}}_{B}^{*}-\underline{\boldsymbol{\eta}}_{B}^{*}\right\Vert =O_{P}\left(\sqrt{p_{n}/n}+\lambda_{n}\left\Vert \boldsymbol{\varpi}_{B}\right\Vert \right).
\]
Note $\left\Vert \boldsymbol{\varpi}_{B}\right\Vert =O_{P}\left(a_{n}^{-\gamma}\sqrt{p_{n}}\right),$the
rate in Assumption \ref{A:adaptiveLasso}(7) implies $\lambda_{n}\left\Vert \boldsymbol{\varpi}_{B}\right\Vert =O_{P}\left(\sqrt{p_{n}/n}\right).$
Therefore, we obtain 
\[
\left\Vert \underline{\hat{\boldsymbol{\eta}}}_{B}^{*}-\underline{\boldsymbol{\eta}}_{B}^{*}\right\Vert =O_{P}\left(\sqrt{p_{n}/n}\right).
\]
Now with the above rate, (\ref{eq:theta_bound}) is ready to imply
that 
\[
\left\Vert \boldsymbol{\hat{\theta}}-\boldsymbol{\theta}^{\ast}\right\Vert =O_{P}\left(\sqrt{p_{n}/n}\right),
\]
as desired.
\end{proof}

\begin{proof}[Proof of Lemma \ref{LE:selection}] We begin the
discussion on the first part. Let $\boldsymbol{e}_{j}$ be a $p_{n}\times1$
vector with the $j$th element being 1 and others being 0. By the
Karush-Kuhn-Tucker condition, $\underline{\hat{\eta}}_{j}=0$ if 
\[
\left\vert \boldsymbol{e}_{j}^{\prime}\mathbf{W}_{n}\overline{\boldsymbol{\hat{m}}}_{n}\left(\boldsymbol{\hat{\theta}}\right)\right\vert <\varpi_{j}\frac{\lambda_{n}}{2}.
\]
Thus, it is equivalent to showing 
\begin{equation}
\Pr\left(\max_{j\in A}\left\vert \boldsymbol{e}_{j}^{\prime}\mathbf{W}_{n}\overline{\boldsymbol{\hat{m}}}_{n}\left(\boldsymbol{\hat{\theta}}\right)\right\vert <\varpi_{j}\frac{\lambda_{n}}{2}\right)\rightarrow1.\label{EQ:p_KKT}
\end{equation}
To this end, we bound $\overline{\boldsymbol{\hat{m}}}_{n}\left(\boldsymbol{\hat{\theta}}\right)$
as follows. We use $\mathbb{E}_{n}$ denote the expecation by treating
$\boldsymbol{\hat{\theta}}$ as a constant. Then, 
\begin{align*}
\left\Vert \overline{\boldsymbol{\hat{m}}}_{n}\left(\boldsymbol{\hat{\theta}}\right)\right\Vert  & \leq\left\Vert \overline{\boldsymbol{\hat{m}}}_{n}\left(\boldsymbol{\hat{\theta}}\right)-\mathbb{E}_{n}\left[\boldsymbol{m}\left(\boldsymbol{\hat{\theta}}\right)\right]\right\Vert +\left\vert \mathbb{E}_{n}\left[\boldsymbol{m}\left(\boldsymbol{\hat{\theta}}\right)\right]-\mathbb{E}\left[\boldsymbol{m}\left(\boldsymbol{\theta}^{\ast}\right)\right]\right\vert \\
 & \leq O_{P}\left(\sqrt{\left.p_{n}\right/n}\right)+C\left\Vert \boldsymbol{\hat{\theta}-\theta}^{\ast}\right\Vert =O_{P}\left(\sqrt{\left.p_{n}\right/n}\right),
\end{align*}
where we apply the results in Lemmas \ref{LE:uniformrate} and \ref{LE:matrix_conv}.
Note $\mathbf{W}_{n}$ is assumed to be full rank and with uniformly
bounded eigenvalues. So 
\[
\max_{j\in A}\left\Vert \boldsymbol{e}_{j}^{\prime}\mathbf{W}_{n}\right\Vert \leq C.
\]
Combining the above two results yields 
\[
\max_{j\in A}\left\vert \boldsymbol{e}_{j}^{\prime}\mathbf{W}_{n}\overline{\boldsymbol{\hat{m}}}_{n}\left(\boldsymbol{\hat{\theta}}\right)\right\vert \leq\max_{j\in A}\left\Vert \boldsymbol{e}_{j}^{\prime}\mathbf{W}_{n}\right\Vert \left\Vert \overline{\boldsymbol{\hat{m}}}_{n}\left(\boldsymbol{\hat{\theta}}\right)\right\Vert \leq O_{P}\left(\sqrt{\left.p_{n}\right/n}\right).
\]
Further, by the convergence rate \ref{A:GMM} (6), $\min_{j=1,...,d_{A}}\varpi_{j}\lambda_{n}\gg_{P}\sqrt{\left.p_{n}\right/n}$.
Thus equation (\ref{EQ:p_KKT}) holds, as desired.

We turn to the second part. 
\begin{align*}
\Pr\left(\min_{j\in B}\left\vert \underline{\hat{\eta}}_{j}\right\vert >0\right) & \geq\Pr\left(\min_{j\in B}\left[\left\vert \underline{\eta}_{j}^{\ast}\right\vert -\left\vert \underline{\hat{\eta}}_{j}-\underline{\eta}_{j}^{\ast}\right\vert \right]>0\right)\\
 & \geq\Pr\left(\min_{j\in B}\left\vert \underline{\eta}_{j}^{\ast}\right\vert -\max_{j\in B}\left\vert \underline{\hat{\eta}}_{j}-\underline{\eta}_{j}^{\ast}\right\vert >0\right)\\
 & \geq\Pr\left(\min_{j\in B}\left\vert \underline{\eta}_{j}^{\ast}\right\vert -\left\Vert \boldsymbol{\hat{\theta}}-\boldsymbol{\theta}^{\ast}\right\Vert >0\right)\rightarrow1,
\end{align*}
because Assumption \ref{A:GMM} implies $\min_{j\in B}\left\{ \underline{\eta}_{j}^{\ast}\right\} \gg\sqrt{p_{n}/n},$
and $\left\Vert \boldsymbol{\hat{\theta}}-\boldsymbol{\theta}^{\ast}\right\Vert =O_{P}\left(\sqrt{p_{n}/n}\right)$
as shown in Lemma \ref{LE:consistency}$.$
\end{proof}

\begin{proof}[Proof of Lemma \ref{LE:asymptotics}] As shown
in Lemma \ref{LE:selection}, $\Pr\left(\underline{\hat{\eta}}_{j}=0,\text{ for all }j\in A\right)\rightarrow1$
and $\Pr\left(\min_{j\in B}\left\vert \underline{\hat{\eta}}_{j}\right\vert >0\right)\rightarrow1.$
For the asymptotics, we assume that these two events hold, that is
$\underline{\hat{\eta}}_{j}=0,$ for all $j\in A$ and $\min_{j\in B}\left\vert \underline{\hat{\eta}}_{j}\right\vert >0$.
We now show the limiting distribution of $\boldsymbol{\hat{\theta}}_{B}$=$\left(\hat{\boldsymbol{\beta}};\underline{\boldsymbol{\hat{\eta}}}_{B}^{*}\right)$.

Without loss of generality, we assume $\underline{\boldsymbol{\hat{\eta}}}_{B}^{*}$
are all positive to simplify notation. Using the notation in (\ref{EQ:pd_linear}),
$\boldsymbol{\hat{\theta}}_{B}$ and $\hat{\boldsymbol{\theta}}$
satisfy the following first-order condition: 
\[
\frac{\partial\overline{\boldsymbol{\hat{m}}}_{n}}{\partial\boldsymbol{\hat{\theta}}_{B}^{\prime}}^{\prime}\mathbf{W}_{n}\overline{\boldsymbol{\hat{m}}}_{n}\left(\boldsymbol{\hat{\theta}}\right)+\left(\begin{array}{c}
\boldsymbol{0}_{k^{*}\times1}\\
\lambda_{n}\boldsymbol{\varpi}_{B}
\end{array}\right)=\boldsymbol{0}_{\left(k^{*}+d_{B}\right)\times1},
\]
where $\boldsymbol{\varpi}_{B}$ is a $d_{B}\times1$ vector collecting
$\varpi_{j}$ for $j\in B$. Since $\overline{\boldsymbol{\hat{m}}}_{n}\left(\boldsymbol{\hat{\theta}}\right)$
is linear in $\boldsymbol{\theta}$ and $\underline{\hat{\eta}}_{j}=0,$
for all $j\in A\boldsymbol{,}$ 
\[
\overline{\boldsymbol{\hat{m}}}_{n}\left(\boldsymbol{\hat{\theta}}\right)-\overline{\boldsymbol{\hat{m}}}_{n}\left(\boldsymbol{\theta}^{\ast}\right)=\frac{\partial\overline{\boldsymbol{\hat{m}}}_{n}}{\partial\boldsymbol{\theta}_{B}^{\prime}}\left(\boldsymbol{\hat{\theta}}_{B}\boldsymbol{-\theta}_{B}^{\ast}\right).
\]
Substitute it back into the above, we obtain 
\[
\frac{\partial\overline{\boldsymbol{\hat{m}}}_{n}\left(\boldsymbol{\hat{\theta}}\right)}{\partial\boldsymbol{\hat{\theta}}_{B}^{\prime}}^{\prime}\mathbf{W}_{n}\overline{\boldsymbol{\hat{m}}}_{n}\left(\boldsymbol{\theta}^{\ast}\right)+\frac{\partial\overline{\boldsymbol{\hat{m}}}_{n}\left(\boldsymbol{\hat{\theta}}\right)}{\partial\boldsymbol{\theta}_{B}^{\prime}}^{\prime}\mathbf{W}_{n}\frac{\partial\overline{\boldsymbol{\hat{m}}}_{n}\left(\boldsymbol{\tilde{\theta}}\right)}{\partial\boldsymbol{\theta}_{B}^{\prime}}\left(\boldsymbol{\hat{\theta}}_{B}\boldsymbol{-\theta}_{B}^{\ast}\right)+\left(\begin{array}{c}
\boldsymbol{0}_{k^{*}\times1}\\
\lambda_{n}\boldsymbol{\varpi}_{B}
\end{array}\right)=\boldsymbol{0.}
\]
Due to the linearity, $\frac{\partial\overline{\boldsymbol{\hat{m}}}_{n}\left(\boldsymbol{\hat{\theta}}\right)}{\partial\boldsymbol{\theta}_{B}^{\prime}}=\frac{\partial\overline{\boldsymbol{\hat{m}}}_{n}\left(\boldsymbol{\tilde{\theta}}\right)}{\partial\boldsymbol{\theta}_{B}^{\prime}}=\frac{\partial\overline{\boldsymbol{\hat{m}}}_{n}}{\partial\boldsymbol{\hat{\theta}}_{B}^{\prime}}$.
Therefore, 
\begin{align*}
\boldsymbol{\hat{\theta}}_{B}\boldsymbol{-\theta}_{B}^{\ast} & =-\left(\frac{\partial\overline{\boldsymbol{\hat{m}}}_{n}}{\partial\boldsymbol{\theta}_{B}^{\prime}}^{\prime}\mathbf{W}_{n}\frac{\partial\overline{\boldsymbol{\hat{m}}}_{n}}{\partial\boldsymbol{\theta}_{B}^{\prime}}\right)^{-1}\left[\left(\frac{\partial\overline{\boldsymbol{\hat{m}}}_{n}}{\partial\boldsymbol{\hat{\theta}}_{B}^{\prime}}^{\prime}\mathbf{W}_{n}\overline{\boldsymbol{\hat{m}}}_{n}\left(\boldsymbol{\theta}^{\ast}\right)\right)+\left(\begin{array}{c}
\boldsymbol{0}_{k^{*}\times1}\\
\lambda_{n}\boldsymbol{\varpi}_{B}
\end{array}\right)\right]\\
 & =-\left(\frac{\partial\overline{\boldsymbol{\hat{m}}}_{n}}{\partial\boldsymbol{\theta}_{B}^{\prime}}^{\prime}\mathbf{W}_{n}\frac{\partial\overline{\boldsymbol{\hat{m}}}_{n}}{\partial\boldsymbol{\theta}_{B}^{\prime}}\right)^{-1}\left(\frac{\partial\overline{\boldsymbol{\hat{m}}}_{n}}{\partial\boldsymbol{\hat{\theta}}_{B}^{\prime}}^{\prime}\mathbf{W}_{n}\overline{\boldsymbol{\hat{m}}}_{n}\left(\boldsymbol{\theta}^{\ast}\right)\right)+o_{P}\left(n^{-1/2}\right),
\end{align*}
where the last line holds due to 
\[
\left\Vert \begin{array}{c}
\boldsymbol{0}_{k^{*}\times1}\\
\lambda_{n}\boldsymbol{\varpi}_{B}
\end{array}\right\Vert =O_{P}\left(\sqrt{p_{n}\times\frac{\lambda_{n}^{2}}{a_{n}^{2\gamma}}}\right)=o_{P}\left(n^{-1/2}\right).
\]
The above uses each element in $\boldsymbol{\varpi}_{B}$ being $O_{P}\left(a_{n}^{-\gamma}\right),\text{ and }a_{n}^{-\gamma}\lambda_{n}\ll1/\sqrt{np_{n}}$.
\end{proof}

\begin{proof}[Proof of Lemma \ref{LE:asym_step0}] We have shown
the influence term of each element in $\overline{\boldsymbol{\hat{m}}}_{n}\left(\boldsymbol{\theta}^{\ast}\right)\ $in
Lemma \ref{LE:expansion}. Thus, by definition, $\Omega_{n}$ is limit
of $n\cdot\overline{\boldsymbol{\hat{m}}}_{n}\left(\boldsymbol{\theta}^{\ast}\right)\overline{\boldsymbol{\hat{m}}}_{n}\left(\boldsymbol{\theta}^{\ast}\right)^{\prime}$
element by element. The uniform convergence of $n\cdot\overline{\boldsymbol{\hat{m}}}_{n}\left(\boldsymbol{\theta}^{\ast}\right)\overline{\boldsymbol{\hat{m}}}_{n}\left(\boldsymbol{\theta}^{\ast}\right)^{\prime}$\ to
$\Omega_{n}$ can be verified if $p_{n}$ is not too large. Specifically,
one sufficient condition of the convergence of $n\cdot\overline{\boldsymbol{\hat{m}}}_{n}\left(\boldsymbol{\theta}^{\ast}\right)\overline{\boldsymbol{\hat{m}}}_{n}\left(\boldsymbol{\theta}^{\ast}\right)^{\prime}$
to $\Omega_{n}$\ and the limiting distribution of$\sqrt{n}\boldsymbol{e}^{\prime}\Omega_{n}^{-1/2}\overline{\boldsymbol{\hat{m}}}_{n}\left(\boldsymbol{\theta}^{\ast}\right)$
is the Lindeberg's condition for triangular arrays. We can follow
the proof of Lemma 4.2 in \citet{ChengLiao2015} to show that $p_{n}^{3}/n\rightarrow0$
and the finite fourth moment of $X$ and $Z$ (imposed in Assumption
\ref{A:GMM}) is sufficient for Lindeberg's condition. We omit the
details due to the similarity. Applying the Lindeberg Central Limit
Theorem yields the desired result. \end{proof}

\section{Error Rate of the Lasso Estimator for the Binary Choice Model\label{APP:lasso}}

In this section, we examine the error rates of the Lasso estimator,
aiming to replicate the results from \citet{Belloni_et_al2012} (the
Lasso part). We assume $\boldsymbol{X}$ are exogenous, and we try
to find the rate of the difference between the estimated model and
the true model.

As this problem differs from those discussed in Section \ref{SEC:theoryapply},
the notation used here may have different meanings compared to other
sections, even if some of it might appear the same.

\subsection{The Setup}

Following the set up in \citet{Belloni_et_al2012}, we revise the
choice model in (\ref{EQ:model}) to 
\begin{align}
Y & =\boldsymbol{1}\left(V+\boldsymbol{X}\boldsymbol{\beta}^{\ast}+R\left(\boldsymbol{X}\right)+\varepsilon>0\right),\label{eq:model_approximate}
\end{align}
where 
\begin{equation}
\left\Vert \boldsymbol{\beta}^{\ast}\right\Vert _{0}=s^{*}=o\left(n\right),\text{ and }R\left(\boldsymbol{X}\right)<C_{R}\apprle\sqrt{s^{\ast}/n}.\label{eq:ModelApproximate}
\end{equation}
We define 
\[
\Gamma=\text{support}\left(\boldsymbol{\beta}^{\ast}\right).
\]
Of course, this set is unknown. For any $p_{n}\times1$ vector $\boldsymbol{a}$,
$\boldsymbol{a}_{\Gamma}$ means that $a_{\Gamma,j}=a_{j}$ for $j\in\Gamma$,
and $a_{\Gamma,j}=0$ for $j\in\Gamma^{c}$. We adopt this convention
for other index sets. Without loss of generality, we assume the first
$s^{*}$ elements of $\boldsymbol{\beta}^{\ast}$ are nonzero. $\boldsymbol{X}$
are exogenous, and thus they play the role of $\boldsymbol{Z}_{1}^{p_{n}}$
in Assumption \ref{A:specialR}. We use the notation $X$ instead
of $Z$ in this section to avoid confusion. As noted in Section \ref{SEC:VariableSelection},
signals (with nonzero $\beta^{*}$) may or may not overlap with the
$X$s relevant to $V$, because the two issues (namely, ``true signals
in the binary choice model'', and ``relevant $X$s to $V$'') are
independent in general. However, we do not reorder $X$ to distinguish
these two issues in this section for notational convenience.

The role of $R\left(\boldsymbol{X}\right)$ is to suggest that the
linear function inside the indicator function is an approximation
of the true model, although the error from this approximation, $R\left(\boldsymbol{X}\right),$
remains small. We require $R\left(\boldsymbol{X}\right)$ to be uniformly
bounded. This strengthens the corresponding condition in \citet{Belloni_et_al2012},
where it is only required that the above holds in probability. This
modification is necessary to ensure that the large support condition
in Assumption \ref{A:specialR} (3) still applies to the modified
model.

Suppose Assumption \ref{A:specialR} holds, then the result in \citet{Lewbel2000}
implies 
\[
\tilde{Y}=\frac{Y-\boldsymbol{1}\left(V>0\right)}{f\left(V|\boldsymbol{X}\right)}
\]
satisfies 
\[
\tilde{Y}=\boldsymbol{X}'\boldsymbol{\beta}^{\ast}+R\left(\boldsymbol{X}\right)+u,\text{ with }\mathbb{E}\left(u|\boldsymbol{X}\right)=0.
\]
The feasible $\tilde{y}_{i}$ for observation $i$ is denoted as $\widehat{\tilde{y}}_{i}$,
and 
\[
\widehat{\tilde{y}}_{i}=\frac{y_{i}-\boldsymbol{1}\left(v_{i}>0\right)}{\hat{f}(v_{i}|\boldsymbol{x}_{1i}^{\tilde{p}})},
\]
the same as the $\widehat{\tilde{y}}_{i}$ in (\ref{eq:feasibley}).
The task here is to estimate the approximate model $\boldsymbol{X}'\boldsymbol{\beta}^{\ast}$
and find the rate of the error $\boldsymbol{X}^{\prime}\boldsymbol{\hat{\beta}}-\boldsymbol{X}\boldsymbol{\beta}^{\ast}$,
where $\boldsymbol{\hat{\beta}}$ is the Lasso estimator defined below.
Specifically, $\boldsymbol{\hat{\beta}}$ is obtained from 
\[
\boldsymbol{\hat{\beta}}=\arg\min_{\boldsymbol{\beta}\in\mathbb{R}^{p_{n}}}\widehat{Q}\left(\boldsymbol{\beta}\right)+\frac{\lambda}{n}\left\Vert \hat{\Upsilon}\boldsymbol{\beta}\right\Vert _{1},
\]
where 
\[
\widehat{Q}\left(\boldsymbol{\beta}\right)=\frac{1}{n}\sum_{i=1}^{n}\left(\widehat{\tilde{y}}_{i}-\boldsymbol{x}_{i}'\boldsymbol{\beta}\right)^{2},
\]
and $\hat{\Upsilon}$ is a $p_{n}\times p_{n}$ diagonal matrix specifying
penalty loadings, a feasible proxy (discussed in Appendix \ref{APP:ChoiceOfLoadings})
for the infeasible loadings: 
\[
\hat{\Upsilon}^{0}=\text{diag}\left(\hat{\varUpsilon}_{1}^{0},\ldots,\hat{\varUpsilon}_{p_{n}}^{0}\right),\hat{\varUpsilon}_{j}^{0}=\sqrt{\frac{1}{n}\sum_{i=1}^{n}q_{ji}^{2}},j=1,2,\ldots,p_{n},
\]
where 
\[
q_{ji}=u_{i}x_{ji}+\mathbb{E}\left(\left.u_{i}x_{ji}\right|\boldsymbol{x}_{1i}^{p^{*}}\right)-\mathbb{E}\left(\left.u_{i}x_{ji}\right|\boldsymbol{x}_{1i}^{p^{*}},v_{i}\right).
\]
Similar to what has been discussed in Appendix \ref{APP:C}, $q_{ji}$
is the influence term in the error term, $\widehat{\tilde{y}}_{i}-\boldsymbol{x}_{i}'\boldsymbol{\beta^{*}},$
due to the fact that $\widehat{\tilde{y}}_{i}$ is not directly observed
and estimated from data. With the above loadings, one can obtain
the probability bound of 
\[
\sqrt{n}\max_{j=1,\ldots,p_{n}}\left|\left(\hat{\varUpsilon}_{j}^{0}\right)^{-1}\frac{1}{n}\sum_{i=1}^{n}q_{ji}\right|,
\]
using moderate deviation theory of \citet{JingShaoWang}. For this
innovation, we refer readers to Section 2.3 and Appendix B of \citet{Belloni_et_al2012}
for details.

Let $T$ denote a set of indices, $T\subseteq\{1,2,\ldots,p_{n}\}$,
and let $|T|$ denote the number of elements in $T$. As in \citet{BickelEtal2009}
and \citet{Belloni_et_al2012}, we define a constant related to the
restricted eigenvalue condition: 
\[
\kappa_{C}\equiv\min_{\substack{\boldsymbol{\delta}\in\mathbb{R}^{p_{n}}:\left\Vert \boldsymbol{\delta}_{T^{c}}\right\Vert _{1}\leq C\left\Vert \boldsymbol{\delta}_{T}\right\Vert _{1},\\
\boldsymbol{\delta}\neq0,|T|\leq s^{*}
}
}\frac{\sqrt{s^{*}\boldsymbol{\delta}'\left(n^{-1}\sum_{i=1}^{n}\boldsymbol{x}_{i}\boldsymbol{x}_{i}'\right)\boldsymbol{\delta}}}{\left\Vert \boldsymbol{\delta}_{T}\right\Vert _{1}}.
\]
The above quantity depends on $n$, but we suppress the dependence.

\subsection{Assumptions and Results}

We impose the following assumptions. They are essentially the same
as Conditions AS, RE, and RF and restrictions on tuning parameters
in \citet{Belloni_et_al2012}. However, we do need a much stronger
rate on $p_{n}$.

\begin{assumptionp}{6}\label{A:lasso_approximate} 
\begin{enumerate}
\item[(1)] Model (\ref{eq:model_approximate}) holds along with the conditions
in (\ref{eq:ModelApproximate}). 
\item[(2)] Uniformly for any $C>0,$ there exists a $C_{\text{RE}}>0,$ such
that $\kappa_{C}\geq C_{\text{RE}}$ with high probability as $n\rightarrow\infty.$ 
\item[(3)] $\mathbb{E}\left(\left|u_{i}\right|^{8+\delta}\right)<\infty$ and
$\max_{j=1,\ldots,p_{n}}\mathbb{E}\left(\left|x_{ji}\right|^{8+\delta}\right)<\infty,$
for a $\delta>8/7.$ 
\item[(4)] $s^{*}\ll n,$ and $p_{n}\apprle n^{3/2}\log n$. 
\item[(5)] Let $\Phi$ denote the cumulative distribution function of the standard
normal. Set 
\[
\lambda=4c\sqrt{n}\Phi^{-1}\left(1-\frac{\gamma}{2p_{n}}\right)
\]
for some $c>0,$ where $\gamma\rightarrow0$ and $\log\left(\frac{1}{\gamma}\right)\lesssim\log\left(p_{n}\vee n\right).$ 
\end{enumerate}
\end{assumptionp}

Since we essentially quote conditions from \citet{Belloni_et_al2012},
we will only briefly explain their roles and focus on the differences.
We refer readers back to the original paper for detailed explanations and justifications of their validity. Assumption \ref{A:lasso_approximate}
(1) states the approximate model. Importantly, we do not impose the
``beta-min'' condition here. (2) outlines the restricted eigenvalue
condition, standard in the Lasso literature. (3) defines
the moment conditions to ensure that $\max_{j=1,\ldots,p_{n}}\mathbb{E}\left(q_{ji}^{4}\right)<\infty$.
(4) restricts the growth rates of $p_{n}$, allowing $s^{*}$ and
$p_{n}$ to diverge at a certain rate. However, the primary restriction
on $s^{*}$ comes from (2). We can only permit $p_{n}\apprle n^{3/2}\log n$,
due to an error term from the first-step estimation, specifically,
$R_{n3,j}$ in (\ref{eq:rn3j}), which is considered a small order
term from a U-statistic. The error bound on $R_{n3,j}$ in Lemma \ref{LE:rn3j}
is not sharp because it only uses Markov's inequality. If the error
bound could be improved, we could relax the restriction on $p_{n}$.
(5) is to ensure that 
\begin{equation}
\Pr\left(\frac{\lambda}{n}\geq4c\max_{j=1,\ldots,p_{n}}\left|\left(\hat{\varUpsilon}_{j}^{0}\right)^{-1}\frac{1}{n}\sum_{i=1}^{n}q_{ji}\right|\right)\geq1-\gamma+o\left(1\right)\rightarrow1,\label{eq:new_maximumineq}
\end{equation}
by the moderate deviation theory of \citet{JingShaoWang}. We set
$\lambda$ twice as large as in \citet{Belloni_et_al2012} to control
the additional error term $R_{n3,j}$ from the first-step estimation.

As in (3.2) in \citet{Belloni_et_al2012}, we note that the $\hat{\Upsilon}$
we propose in Appendix \ref{APP:ChoiceOfLoadings} is asymptotically
valid, and it satisfies 
\begin{equation}
\ell\hat{\Upsilon}^{0}\leq\hat{\Upsilon}\leq\upsilon\hat{\Upsilon}^{0},\label{eq:claim_loadings}
\end{equation}
with $0<\ell\leq1\leq\upsilon$ such that $\ell\stackrel{P}{\rightarrow}1$
and $\upsilon\stackrel{P}{\rightarrow}\upsilon'\geq1.$

We present the following theorem which replicates Theorem 1 (the most
important result on Lasso) in \citet{Belloni_et_al2012} for our case.

\begin{theorem} \label{TH:LASSO} Suppose Assumptions \ref{A:main},
\ref{A:continuousf_kernelK}, \ref{A:specialR}, and \ref{A:lasso_approximate}
hold. Further, $r>\left(p^{\ast}+1\right)/2$ and we adopt $2r$-th
order kernel and the $\hat{\boldsymbol{h}}$ obtained from (\ref{EQ:bandwidth})
to construct $\hat{f}(V|\boldsymbol{Z}_{1}^{\tilde{p}})$ for $\widehat{\tilde{y}}_{i}$,
where the $r$ is defined in Assumption \ref{A:continuousf_kernelK}.
Then the Lasso estimator $\boldsymbol{\hat{\beta}}$ and the Lasso
fit $\boldsymbol{X}^{\prime}\boldsymbol{\hat{\beta}}$ satisfy

\[
\sqrt{\frac{1}{n}\sum_{i=1}^{n}\left(\boldsymbol{x}_{i}'\boldsymbol{\hat{\beta}}-\boldsymbol{x}_{i}'\boldsymbol{\beta^{*}}\right)^{2}}=O_{P}\left(\frac{1}{\kappa_{\bar{C}}}\sqrt{\frac{s^{*}\log\left(\left.p_{n}\right/\gamma\right)}{n}}\right),
\]
and

\[
\left\Vert \boldsymbol{\hat{\beta}}-\boldsymbol{\beta^{*}}\right\Vert _{1}=O_{P}\left(\frac{1}{\left(\kappa_{2\bar{C}}\right)^{2}}\sqrt{\frac{s^{*2}\log\left(\left.p_{n}\right/\gamma\right)}{n}}\right),
\]
where $\bar{C}=\left\Vert \hat{\Upsilon}^{0}\right\Vert _{\infty}\left\Vert \left(\hat{\Upsilon}^{0}\right)^{-1}\right\Vert _{\infty}\left(\upsilon c+1\right)/\left(\ell c-1\right),$
and $\ell c>1$.

\end{theorem}

The post-Lasso estimation result of Theorem 2 in \citet{Belloni_et_al2012}
is straightforward given the above result.

\subsection{Proof of Theorem \ref{TH:LASSO} }

\textbf{Proof.} If we observe $\tilde{y}_{i}$ and use it for estimation,
our framework exactly fits the one in \citet{Belloni_et_al2012},
and the results follow accordingly. Thus, the key aspect of the proof
here is to demonstrate the impact of the approximation error, $\widehat{\tilde{y}}_{i}-\tilde{y}_{i}$,
on the results. Moreover, as the selection error in the first stage
(relevant $X$ to $V$) is very small, we assume that we achieve the
correct selection for the proof that follows. The results still hold
with the selection (see the argument in the proof of Lemma \ref{LE:expansion})
and the fact that we only demonstrate the results hold in probability.

We show the results in five steps. In Step 0, we present some results
of $\widehat{\tilde{y}}_{i}-\tilde{y}_{i}$. Steps 1 to 4 essentially
follow Steps 1 to 4 in the proof of Theorem 1 in \citet{Belloni_et_al2012},
but taking into account the result from Step 0.

We introduce a new $p_{n}\times1$ vector, 
\[
\boldsymbol{S}=2\left(\hat{\Upsilon}^{0}\right)^{-1}\frac{1}{n}\sum_{i=1}^{n}\boldsymbol{q}_{i},
\]
with 
\[
\boldsymbol{q}=u\boldsymbol{x}+\mathbb{E}\left(u\boldsymbol{x}\mid\boldsymbol{X}_{1}^{p^{\ast}}\right)-\mathbb{E}\left(u\boldsymbol{x}\mid\boldsymbol{X}_{1}^{p^{\ast}},V\right),
\]
and the difference 
\[
\hat{\boldsymbol{\delta}}=\hat{\boldsymbol{\beta}}-\boldsymbol{\beta}^{*}.
\]

\noindent \textbf{Step 0.} Using a similar argument to Lemma \ref{LE:expansion}---by
substituting $Z$ with $X$ and incorporating $R\left(\boldsymbol{X}\right)$
from (\ref{eq:model_approximate})---we obtain: 
\begin{align}
 & \frac{1}{n}\sum_{i=1}^{n}x_{ji}\left(\widehat{\tilde{y}}_{i}-\boldsymbol{x}_{i}^{\prime}\boldsymbol{\beta}^{\ast}\right)\nonumber \\
\stackrel{(\text{i})}{=} & \frac{1}{n}\sum_{i=1}^{n}x_{ji}u_{i}+\frac{1}{n}\sum_{i=1}^{n}x_{ji}R\left(\boldsymbol{x}_{i}\right)+\nonumber \\
 & \frac{1}{n}\sum_{i=1}^{n}x_{ji}\left(y_{i}-\boldsymbol{1}\left(v_{i}>0\right)\right)\left[\frac{\hat{f}_{\boldsymbol{X}i}-\mathbb{E}\left(\hat{f}_{\boldsymbol{X}i}\right)}{\mathbb{E}\left(\hat{f}_{V\boldsymbol{X}i}\right)}+\frac{\mathbb{E}\left(\hat{f}_{\boldsymbol{X}i}\right)\left(\mathbb{E}\left(\hat{f}_{V\boldsymbol{X}i}\right)-\hat{f}_{V\boldsymbol{X}i}\right)}{\left[\mathbb{E}\left(\hat{f}_{V\boldsymbol{X}i}\right)\right]^{2}}\right]+\frac{1}{n}\sum_{i=1}^{n}R_{1,i}\nonumber \\
\stackrel{(\text{ii})}{=} & \frac{1}{n}\sum_{i=1}^{n}q_{ji}+\frac{1}{n}\sum_{i=1}^{n}x_{ji}\left[R\left(\boldsymbol{x}_{i}\right)+R_{1,i}+R_{2,i}\right]+R_{n3,j}\nonumber \\
\equiv & \frac{1}{n}\sum_{i=1^{n}}q_{ji}+\frac{1}{n}\sum_{i=1}^{n}x_{ji}\tilde{R}_{i}+R_{n3,j}.\label{eq:xy_decomp}
\end{align}
Since this applies the same argument as in Lemma \ref{LE:expansion},
we only briefly explain the process. For (i), note that $\tilde{y}_{i}=\boldsymbol{x}_{i}'\boldsymbol{\beta}^{\ast}+R\left(\boldsymbol{x}_{i}\right)+u_{i}$;
the difference between $\widehat{\tilde{y}}_{i}$ and $\tilde{y}_{i}$
includes the terms on the third line, and $R_{1,i}=o_{P}\left(n^{-1/2}\right)$
uniformly for all $i$ (due to the uniform convergence of $\hat{f}$).
Step (ii) involves standard calculations and recognizes the first
term on the third line as a U-statistic with an asymptotically negligible
bias term. In addition, $\frac{1}{n}\sum_{i=1}^{n}q_{ji}$ denotes
the leading term of the U-statistic, $R_{2,i}$ represents the bias
term from the nonparametric estimation, and $R_{n3,j}$ is the small
order term of the U-statistic (details below). Due to the order of
the kernel we use, the bias term $R_{2,i}=o\left(n^{-1/2}\right)$
uniformly for all $i$. According to standard results, e.g., in \citet{PowellEtal1989}
and \citet{Lewbel2000}, $R_{n3,j}$ is a second-order U-statistic
and can be written as: 
\begin{equation}
R_{n3,j}=\frac{1}{n^{2}}\sum_{l,l'=1}^{n}\left\{ H\left(\boldsymbol{\xi}_{jl},\boldsymbol{\xi}_{jl'}\right)-\mathbb{E}\left[\left.H\left(\boldsymbol{\xi}_{jl},\boldsymbol{\xi}_{jl'}\right)\right|\boldsymbol{\xi}_{jl}\right]-\mathbb{E}\left[\left.H\left(\boldsymbol{\xi}_{jl},\boldsymbol{\xi}_{jl'}\right)\right|\boldsymbol{\xi}_{jl'}\right]+\mathbb{E}\left[H\left(\boldsymbol{\xi}_{jl},\boldsymbol{\xi}_{jl'}\right)\right]\right\} ,\label{eq:rn3j}
\end{equation}
where $\boldsymbol{\xi}$ denotes all involved variables. The function
$H\left(\boldsymbol{\xi}_{jl},\boldsymbol{\xi}_{jl'}\right)$ is given
by: 
\[
H\left(\boldsymbol{\xi}_{jl},\boldsymbol{\xi}_{jl'}\right)=x_{jl}\tilde{y}_{l}\Pi_{k=1}^{p^{\ast}}K_{h}\left(x_{kl}-x_{kl'}\right)+x_{jl}\tilde{y}_{l}K_{h}\left(v_{l}-v_{l'}\right)\Pi_{k=1}^{p^{\ast}}K_{h}\left(x_{kl}-x_{kl'}\right),
\]
assuming the same $h$ for all covariates for simplicity. The expectation
of the squared $R_{n3,j}$ is proportional to $1/\left(n^{2}h^{p^{*}+1}\right)$,
as the expectation of cross-products for $l\neq l'$ is zero, and
the second moment of $H$ is of rate $h^{-\left(p^{*}+1\right)}$.
Furthermore, Lemma \ref{LE:rn3j} shows that, $\max_{j=1,\ldots,p_{n}}\left|R_{n3,j}\right|<\sqrt{\frac{\log n}{n}}$
with probability approaching one. In this last line, $\tilde{R}_{i}\equiv R\left(\boldsymbol{x}_{i}\right)+R_{1,i}+R_{2,i}$.
To summarize: 
\[
\max_{i=1,\ldots,n}\left|\tilde{R}_{i}\right|=O_{P}\left(\sqrt{s^{\ast}/n}\right)\text{ and }\max_{j=1,\ldots,p_{n}}\left|R_{n3,j}\right|=O_{P}\left(\sqrt{\frac{\log n}{n}}\right).
\]

\noindent \textbf{Step 1.} The proof of this step basically repeats
Step 1 in the proof of Theorem 1 of \citet{Belloni_et_al2012}, but
we include the treatment of the new small order terms, $\tilde{R}_{i}$
and $R_{n3,j}$, in the analysis. We define 
\[
\tilde{\kappa}_{C}=\min_{\substack{\boldsymbol{\delta}\in\mathbb{R}^{p_{n}}:\\
\left\Vert \hat{\Upsilon}^{0}\boldsymbol{\delta}_{\Gamma^{c}}\right\Vert _{1}\leq C\left\Vert \hat{\Upsilon}^{0}\boldsymbol{\delta}_{\Gamma}\right\Vert _{1},\\
\boldsymbol{\delta}\neq0
}
}\frac{\sqrt{s^{*}\boldsymbol{\delta}'\left(n^{-1}\sum_{i=1}^{n}\boldsymbol{x}_{i}\boldsymbol{x}_{i}'\right)\boldsymbol{\delta}}}{\left\Vert \hat{\Upsilon}^{0}\boldsymbol{\delta}_{\Gamma}\right\Vert _{1}},
\]
and denote 
\[
a=\min_{j=1,\ldots,p_{n}}\hat{\varUpsilon}_{j}^{0}\leq\max_{j=1,\ldots,p_{n}}\hat{\varUpsilon}_{j}^{0}=b.
\]
For the same reason as in \citet{Belloni_et_al2012}, $a$ and $b$
are bounded and bounded away from zero with probability approaching
1, and 
\begin{equation}
\tilde{\kappa}_{c_{0}}\geq\frac{1}{b}\kappa_{\bar{C}},\label{eq:k_relation}
\end{equation}
with 
\[
c_{0}=\frac{\upsilon c+1}{\ell c-1},\bar{C}=\left\Vert \hat{\Upsilon}^{0}\right\Vert _{\infty}\left\Vert \left(\hat{\Upsilon}^{0}\right)^{-1}\right\Vert _{\infty}\frac{\upsilon c+1}{\ell c-1},\text{ and }c\text{ satisfies }\ell c>1.
\]

Using (\ref{eq:xy_decomp}), we obtain

\begin{align}
 & Q\left(\boldsymbol{\hat{\beta}}\right)-Q\left(\boldsymbol{\beta}^{*}\right)\nonumber \\
= & \frac{1}{n}\sum_{i=1}^{n}\left(\boldsymbol{x}_{i}'\hat{\boldsymbol{\delta}}\right)^{2}-2\left[\frac{1}{n}\sum_{i=1}^{n}\left(\widehat{\tilde{y}}_{i}-\boldsymbol{x}_{i}'\boldsymbol{\beta}^{*}\right)\boldsymbol{x}_{i}\right]'\hat{\boldsymbol{\delta}}\nonumber \\
= & \frac{1}{n}\sum_{i=1}^{n}\left(\boldsymbol{x}_{i}'\hat{\boldsymbol{\delta}}\right)^{2}-2\left(\frac{1}{n}\sum_{i=1}^{n}\boldsymbol{q}_{i}\right)'\hat{\boldsymbol{\delta}}-2\left(\frac{1}{n}\sum_{i=1}^{n}\tilde{R}_{i}\boldsymbol{x}_{i}\right)'\hat{\boldsymbol{\delta}}-2\boldsymbol{R}_{n3}'\hat{\boldsymbol{\delta}},\label{eq:Q_decomp}
\end{align}
where $\boldsymbol{R}_{n3}$ is a $p_{n}\times1$ vector and collects
$R_{n3,j}$, $j=1,...,p_{n}$. By optimality of $\boldsymbol{\hat{\beta}}$
\begin{equation}
Q\left(\boldsymbol{\hat{\beta}}\right)-Q\left(\boldsymbol{\beta}^{*}\right)\leq\frac{\lambda}{n}\left(\left\Vert \hat{\Upsilon}\boldsymbol{\beta}^{*}\right\Vert _{1}-\left\Vert \hat{\Upsilon}\boldsymbol{\hat{\beta}}\right\Vert _{1}\right).\label{eq:optimalb}
\end{equation}
The by the decomposition in (\ref{eq:Q_decomp}), 
\begin{align}
\left|Q\left(\boldsymbol{\hat{\beta}}\right)-Q\left(\boldsymbol{\beta}^{*}\right)-\frac{1}{n}\sum_{i=1}^{n}\left(\boldsymbol{x}_{i}'\hat{\boldsymbol{\delta}}\right)^{2}\right| & =\left|2\left(\frac{1}{n}\sum_{i=1}^{n}\boldsymbol{q}_{i}\right)'\hat{\boldsymbol{\delta}}+2\frac{1}{n}\sum_{i=1}^{n}\tilde{R}_{i}\boldsymbol{x}_{i}'\hat{\boldsymbol{\delta}}+2\boldsymbol{R}_{n3}'\hat{\boldsymbol{\delta}}\right|\nonumber \\
 & \leq\left\Vert \boldsymbol{S}\right\Vert _{\infty}\left\Vert \hat{\Upsilon}^{0}\hat{\boldsymbol{\delta}}\right\Vert _{1}+2\max_{i=1,...,n}\left|\tilde{R}_{i}\right|\sqrt{\frac{1}{n}\sum_{i=1}^{n}\left(\boldsymbol{x}_{i}'\hat{\boldsymbol{\delta}}\right)^{2}}+2\left\Vert \boldsymbol{R}_{n3}\right\Vert _{\infty}\left\Vert \hat{\boldsymbol{\delta}}\right\Vert _{1}.\label{eq:Q_diff}
\end{align}
By the restriction on $\lambda$ in Assumption \ref{A:lasso_approximate}
(4) and (5) and that result (\ref{eq:new_maximumineq}) (to be shown
in Step 2) holds with very high probability, 
\begin{equation}
\left\Vert \boldsymbol{S}\right\Vert _{\infty}\leq\frac{1}{2c}\cdot\frac{\lambda}{n},\text{ and }C\sqrt{\frac{\log n}{n}}\leq\frac{\lambda}{n},\text{ for some positive \ensuremath{C.}}\label{eq:on_lambda}
\end{equation}
Then 
\begin{equation}
2\left\Vert \boldsymbol{R}_{n3}\right\Vert _{\infty}\left\Vert \hat{\boldsymbol{\delta}}\right\Vert _{1}\leq2b^{-1}\left\Vert \boldsymbol{R}_{n3}\right\Vert _{\infty}\left\Vert \hat{\Upsilon}^{0}\hat{\boldsymbol{\delta}}\right\Vert _{1}\stackrel{(\text{i})}{\leq}\frac{C}{2c}\sqrt{\frac{\log n}{n}}\left\Vert \hat{\Upsilon}^{0}\hat{\boldsymbol{\delta}}\right\Vert _{1}\leq\frac{1}{2c}\cdot\frac{\lambda}{n}\left\Vert \hat{\Upsilon}^{0}\hat{\boldsymbol{\delta}}\right\Vert _{1}\label{eq:onR}
\end{equation}
with probability approaching one, where (i) applies Lemma \ref{LE:rn3j}.

(\ref{eq:optimalb}), (\ref{eq:Q_diff}), (\ref{eq:on_lambda}) and
(\ref{eq:onR}) imply that 
\begin{align*}
\frac{1}{n}\sum_{i=1}^{n}\left(\boldsymbol{x}_{i}'\hat{\boldsymbol{\delta}}\right)^{2} & \leq\frac{\lambda}{n}\left(\left\Vert \hat{\Upsilon}\boldsymbol{\beta}^{*}\right\Vert _{1}-\left\Vert \hat{\Upsilon}\boldsymbol{\hat{\beta}}\right\Vert _{1}\right)+\frac{1}{c}\cdot\frac{\lambda}{n}\left\Vert \hat{\Upsilon}^{0}\hat{\boldsymbol{\delta}}\right\Vert _{1}+2\max_{i=1,...,n}\left|\tilde{R}_{i}\right|\sqrt{\frac{1}{n}\sum_{i=1}^{n}\left(\boldsymbol{x}_{i}'\hat{\boldsymbol{\delta}}\right)}\\
 & =\frac{\lambda}{n}\left(\left\Vert \hat{\Upsilon}\hat{\boldsymbol{\delta}}_{\Gamma}\right\Vert _{1}-\left\Vert \hat{\Upsilon}\hat{\boldsymbol{\delta}}_{\Gamma^{c}}\right\Vert _{1}\right)+\frac{1}{c}\cdot\frac{\lambda}{n}\left\Vert \hat{\Upsilon}^{0}\hat{\boldsymbol{\delta}}\right\Vert _{1}+2\max_{i=1,...,n}\left|\tilde{R}_{i}\right|\sqrt{\frac{1}{n}\sum_{i=1}^{n}\left(\boldsymbol{x}_{i}'\hat{\boldsymbol{\delta}}\right)}\\
 & \leq\left(\upsilon+\frac{1}{c}\right)\frac{\lambda}{n}\left\Vert \hat{\Upsilon}^{0}\hat{\boldsymbol{\delta}}_{\Gamma}\right\Vert _{1}-\left(\ell-\frac{1}{c}\right)\frac{\lambda}{n}\left\Vert \hat{\Upsilon}^{0}\hat{\boldsymbol{\delta}}_{\Gamma^{c}}\right\Vert _{1}+2\max_{i=1,...,n}\left|\tilde{R}_{i}\right|\sqrt{\frac{1}{n}\sum_{i=1}^{n}\left(\boldsymbol{x}_{i}'\hat{\boldsymbol{\delta}}\right)},
\end{align*}
where the last line uses (\ref{eq:claim_loadings}) and the definition
of $\Gamma$. Now we reach the same equation as in (C.3) in \citet{Belloni_et_al2012}.
Continue their line of analysis, we can obtain the corresponding part
of Lemma 6 
\begin{equation}
\sqrt{\frac{1}{n}\sum_{i=1}^{n}\left[\boldsymbol{x}_{i}'\left(\boldsymbol{\hat{\beta}}-\boldsymbol{\beta}^{*}\right)\right]^{2}}=O_{P}\left(\frac{\lambda\sqrt{s^{*}}}{n\tilde{\kappa}_{c_{0}}}+\sqrt{\frac{s^{*}}{n}}\right)\label{eq:xb-b}
\end{equation}
and 
\begin{equation}
\left\Vert \hat{\Upsilon}^{0}\left(\boldsymbol{\hat{\beta}}-\boldsymbol{\beta}^{*}\right)\right\Vert _{1}=O_{P}\left(\frac{\sqrt{s^{*}}}{\tilde{\kappa}_{2c_{0}}}\left(\frac{\lambda\sqrt{s^{*}}}{n\tilde{\kappa}_{c_{0}}}+\sqrt{\frac{s^{*}}{n}}\right)+\frac{s^{*}}{\lambda}\right).\label{eq:rb-b}
\end{equation}
with $c_{0}=\frac{\upsilon c+1}{\ell c-1}.$ Note, we are only able
to obtain the asymptotic bound instead of the non-asymptotic bound
in their paper because we apply the asymptotic result in Lemma \ref{LE:rn3j}.

\noindent \textbf{Step 2.} We need to show in this step that for $\lambda=4c\sqrt{n}\Phi^{-1}\left(1-\gamma\left/\left(2p_{n}\right)\right.\right)$
\[
\Pr\left(2cn\left\Vert \boldsymbol{S}\right\Vert _{\infty}>\lambda\right)=o(1).
\]
As in Lemma 5 in \citet{Belloni_et_al2012}, it requires that there
exists some $b_{n}\rightarrow\infty$ such that 
\[
2\Phi^{-1}\left(1-\gamma\left/\left(2p_{n}\right)\right.\right)\leq\frac{n^{1/6}}{b_{n}}\min_{j=1,...,p_{n}}\frac{\left[\mathbb{E}\left(q_{ji}^{2}\right)\right]^{1/2}}{\left[\mathbb{E}\left(\left|q_{ji}\right|^{3}\right)\right]^{1/3}}.
\]
This clearly is the case because $2\Phi^{-1}\left(1-\gamma\left/\left(2p_{n}\right)\right.\right)\apprle\log\left(n\right).$

\noindent \textbf{Step 3.} Define the expectation of the infeasible
loadings 
\[
\Upsilon^{0}\equiv\text{diag}\left(\sqrt{\mathbb{E}\left(q_{1i}^{2}\right)},...,\sqrt{\mathbb{E}\left(q_{1p_{n}}^{2}\right)}\right).
\]
Lemma \ref{LE:moment_uniform} shows the uniform convergence: 
\[
\max_{j=1,...,p_{n}}\left|\hat{\varUpsilon}_{j}^{0}-\sqrt{\mathbb{E}\left(q_{ji}^{2}\right)}\right|\stackrel{P}{\rightarrow}0.
\]

\noindent \textbf{Step 4.} Given that $\lambda=4c\sqrt{n}\Phi^{-1}\left(1-\gamma\left/\left(2p_{n}\right)\right.\right)$$\apprle\sqrt{n\log\left(\left.p_{n}\right/\gamma\right)}$,
and asymptotic valid loading $\hat{\Upsilon},$ (\ref{eq:xb-b}) implies
that 
\begin{align*}
\sqrt{\frac{1}{n}\sum_{i=1}^{n}\left[\boldsymbol{x}_{i}'\left(\boldsymbol{\hat{\beta}}-\boldsymbol{\beta}^{*}\right)\right]^{2}} & =O_{P}\left(\frac{1}{\tilde{\kappa}_{c_{0}}}\sqrt{\frac{s^{*}\log\left(\left.p_{n}\right/\gamma\right)}{n}}+\sqrt{\frac{s^{*}}{n}}\right)\\
 & =O_{P}\left(\frac{1}{\tilde{\kappa}_{c_{0}}}\sqrt{\frac{s^{*}\log\left(\left.p_{n}\right/\gamma\right)}{n}}\right)\\
 & =O_{P}\left(\frac{1}{\kappa_{\bar{C}}}\sqrt{\frac{s^{*}\log\left(\left.p_{n}\right/\gamma\right)}{n}}\right),
\end{align*}
by (\ref{eq:k_relation}). From (\ref{eq:rb-b}), the second part
of the result can be obtained as 
\begin{align*}
\left\Vert \boldsymbol{\hat{\beta}}-\boldsymbol{\beta}^{*}\right\Vert _{1} & \leq\left\Vert \left(\hat{\Upsilon}^{0}\right)^{-1}\right\Vert _{\infty}\left\Vert \hat{\Upsilon}^{0}\left(\boldsymbol{\hat{\beta}}-\boldsymbol{\beta}^{*}\right)\right\Vert _{1}\\
 & =O_{P}\left(\left\Vert \left(\hat{\Upsilon}^{0}\right)^{-1}\right\Vert _{\infty}\frac{\sqrt{s^{*}}}{\tilde{\kappa}_{2c_{0}}}\left(\frac{\lambda\sqrt{s^{*}}}{n\tilde{\kappa}_{c_{0}}}+\sqrt{\frac{s^{*}}{n}}\right)+\frac{s^{*}}{\lambda}\right)\\
 & =O_{P}\left(\frac{1}{\left(\tilde{\kappa}_{2c_{0}}\right)^{2}}\sqrt{\frac{s^{*2}\log\left(\left.p_{n}\right/\gamma\right)}{n}}\right)\\
 & =O_{P}\left(\frac{1}{\left(\kappa_{2\bar{C}}\right)^{2}}\sqrt{\frac{s^{*2}\log\left(\left.p_{n}\right/\gamma\right)}{n}}\right),
\end{align*}
by (\ref{eq:k_relation}). 
\begin{flushleft}
▗ 
\par\end{flushleft}

We impose Assumption \ref{A:lasso_approximate} for the following
Lemmas.

\begin{lemma}\label{LE:rn3j}For $R_{n3,j}$ defined in (\ref{eq:rn3j}),
there exists a $C_{2}>0,$ uniformly for $j=1,2,...,p_{n}$, any $C_{1}>0,$
\[
\Pr\left(\left|R_{n3,j}\right|>C_{1}\sqrt{\frac{\log n}{n}}\right)\ll\frac{C_{2}}{n^{3/2}\log n}
\]
If $p_{n}\apprle n^{3/2}\log n$, 
\[
\Pr\left(\max_{j=1,...,p_{n}}\left|R_{n3,j}\right|>C_{1}\sqrt{\frac{\log n}{n}}\right)\rightarrow0.
\]

\end{lemma}

\begin{proof}[Proof of Lemma \ref{LE:rn3j}] By the moment condition
in Assumption \ref{A:lasso_approximate} (4), the fourth moment of
$H\left(\boldsymbol{\xi}_{jl},\boldsymbol{\xi}_{jl'}\right)$ exits.
We write 
\[
H_{ll'}=H\left(\boldsymbol{\xi}_{jl},\boldsymbol{\xi}_{jl'}\right)-\mathbb{E}\left[\left.H\left(\boldsymbol{\xi}_{jl},\boldsymbol{\xi}_{jl'}\right)\right|\boldsymbol{\xi}_{jl}\right]-\mathbb{E}\left[\left.H\left(\boldsymbol{\xi}_{jl},\boldsymbol{\xi}_{jl'}\right)\right|\boldsymbol{\xi}_{jl'}\right]+\mathbb{E}\left[H\left(\boldsymbol{\xi}_{jl},\boldsymbol{\xi}_{jl'}\right)\right].
\]
for short, where we suppress its dependence on $j$. Then, 
\begin{equation}
\mathbb{E}\left(R_{n3,j}^{4}\right)=\frac{1}{n^{8}}\sum_{l_{1},l_{2}=1}^{n}\sum_{l_{3},l_{4}=1}^{n}\sum_{l_{5},l_{6}=1}^{n}\sum_{l_{7},l_{8}=1}^{n}\mathbb{E}\left(H_{l_{1}l_{2}}H_{l_{3}l_{4}}H_{l_{5}l_{6}}H_{l_{7}l_{8}}\right).\label{eq:Rn3^4}
\end{equation}
If one element of those $\left(l_{2j-1},l_{2j}\right),$ $j=1,...,4$,
is different any of the rest elements $l$, the expectation is zero.
To see that, suppose $l_{8}\neq l_{1},...,l_{7},$ then 
\[
\mathbb{E}\left(H_{l_{1}l_{2}}H_{l_{3}l_{4}}H_{l_{5}l_{6}}H_{l_{7}l_{8}}\right)=\mathbb{E}\left[H_{l_{1}l_{2}}H_{l_{3}l_{4}}H_{l_{5}l_{6}}\mathbb{E}\left(H_{l_{7}l_{8}}\left|\boldsymbol{\xi}_{\mathrm{1}}\mathrm{,...,}\boldsymbol{\xi}_{\mathrm{7}}\right.\right).\right]
\]
By the independence of observations across $l$, 
\begin{align*}
 & \mathbb{E}\left(H_{l_{7}l_{8}}\left|\boldsymbol{\xi}_{1},...,\boldsymbol{\xi}_{7}\right.\right)\\
= & \mathbb{E}\left\{ \left.H\left(\boldsymbol{\xi}_{l_{7}},\boldsymbol{\xi}_{l_{8}}\right)-\mathbb{E}\left[\left.H\left(\boldsymbol{\xi}_{l_{7}},\boldsymbol{\xi}_{l_{8}}\right)\right|\boldsymbol{\xi}_{l_{7}}\right]-\mathbb{E}\left[\left.H\left(\boldsymbol{\xi}_{l_{7}},\boldsymbol{\xi}_{l_{8}}\right)\right|\boldsymbol{\xi}_{l_{8}}\right]+\mathbb{E}\left[H\left(\boldsymbol{\xi}_{l_{7}},\boldsymbol{\xi}_{l_{8}}\right)\right]\right|\boldsymbol{\xi}_{1},...,\boldsymbol{\xi}_{7}\right\} \\
= & \mathbb{E}\left\{ -\mathbb{E}\left[\left.H\left(\boldsymbol{\xi}_{l_{7}},\boldsymbol{\xi}_{l_{8}}\right)\right|\boldsymbol{\xi}_{l_{8}}\right]+\mathbb{E}\left[H\left(\boldsymbol{\xi}_{l_{7}},\boldsymbol{\xi}_{l_{8}}\right)\right]\right\} \\
= & 0.
\end{align*}
Or $l_{8}=l_{7}\neq l_{1},...,l_{6},$ 
\begin{align}
 & \mathbb{E}\left(H_{l_{7}l_{8}}\left|\boldsymbol{\xi}_{1},...,\boldsymbol{\xi}_{6}\right.\right)=\mathbb{E}\left(H_{l_{7}l_{8}}\right)=0.\label{eq:l7l8}
\end{align}
Since $l_{1},...,l_{8}$ can take maximum 4 different values if no
$l$ is different any of the rest $l$, the number of nonzero terms
in (\ref{eq:Rn3^4}) is proportional to $n^{4}$. Therefore \footnote{Note we can integrate one $h^{p^{*}+1}$ out for the expectation when
$l\neq l'$. When $l=l'$, the number of nonzero terms are much smaller
than $n^{4}$, e.g., see the point in (\ref{eq:l7l8}). } 
\[
\mathbb{E}\left(R_{n3,j}^{4}\right)\leq\frac{\mathbb{E}\left[H\left(\boldsymbol{\xi}_{jl},\boldsymbol{\xi}_{jl'}\right)^{4}\right]}{n^{4}}\leq\frac{C}{n^{4}h^{3\left(p^{*}+1\right)}}.
\]
Using the above and applying Markov's inequality, 
\[
\Pr\left(\left|R_{n3,j}\right|>C_{1}\sqrt{\frac{\log n}{n}}\right)\leq\frac{n\mathbb{E}\left(R_{n3,j}^{4}\right)}{C_{1}^{2}\log n}\leq\frac{C}{n^{4}h^{3\left(p^{*}+1\right)}}\frac{n}{C_{1}^{2}\log n}\leq\frac{C/C_{1}^{2}}{\left(nh^{p^{*}+1}\right)^{3}\log n}\ll\frac{C_{2}}{n^{3/2}\log n},
\]
where the last line holds by the rate in Theorem \ref{TH:main1} ($\sqrt{nh^{p^{*}+1}}\gg n^{1/4}$)
and by letting $C_{2}=C/C_{1}^{2}$. The second part is a direct result
from the above: 
\begin{align*}
\Pr\left(\max_{j=1,\ldots,p_{n}}\left|R_{n3,j}\right|>\frac{\log(n)}{\sqrt{n}}\right) & \leq\sum_{j=1}^{p_{n}}\Pr\left(\left|R_{n3,j}\right|>\frac{\log(n)}{\sqrt{n}}\right)\\
 & \ll\frac{p_{n}C_{2}}{n^{3/2}\log n}.
\end{align*}
If $p_{n}\apprle n^{3/2}\log n,$ 
\[
\Pr\left(\max_{j=1,\ldots,p_{n}}\left|R_{n3,j}\right|>\frac{\log(n)}{\sqrt{n}}\right)\ll1,
\]
as desired. \end{proof}

\begin{lemma}\label{LE:moment_uniform}For any small $\epsilon>0$
and $p_{n}\apprle n^{3/2}\log n$, 
\[
\Pr\left(\max_{j=1,...,p_{n}}\left|\frac{1}{n}\sum_{i=1}^{n}q_{ji}^{2}-\mathbb{E}\left(q_{ji}^{2}\right)\right|>\epsilon\right)\rightarrow0.
\]
\end{lemma}

\begin{proof}[Proof of Lemma \ref{LE:moment_uniform}] By the
moment conditions in Assumption \ref{A:lasso_approximate}(4), $\mathbb{E}\left[\left(q_{ji}^{2}\right)^{2+\delta/4}\right]$
exists. Similar to the second part of Lemma \ref{LE:2ndU} by setting
$\varsigma_{n}=n^{1/2}\epsilon$ (note the result in this Lemma also
applies to independent sample means), 
\begin{align*}
 & \Pr\left(\max_{j=1,...,p_{n}}\left|\frac{1}{n}\sum_{i=1}^{n}q_{ji}^{2}-\mathbb{E}\left(q_{ji}^{2}\right)\right|>\epsilon\right)\\
\leq & p_{n}\max_{j=1,...,p_{n}}\Pr\left(\left|\frac{1}{n}\sum_{i=1}^{n}q_{ji}^{2}-\mathbb{E}\left(q_{ji}^{2}\right)\right|>\epsilon\right)\\
\leq & p_{n}C\left[n^{-1-\delta/8}n^{-1-\delta/8}+n^{-\delta/4}n^{-1-3\delta/16}\right]\\
\rightarrow & 0,
\end{align*}
due to $\delta>8/7,$ and $p_{n}\apprle n^{3/2}\log n$.
\end{proof}

\subsection{Choice of \textmd{\normalsize{}{}{}{}$\hat{\Upsilon}$\label{APP:ChoiceOfLoadings} }}

We propose to choose $\hat{\Upsilon}$ in the below, following Appendix
A in \citet{Belloni_et_al2012}. Recall that

\[
\hat{\Upsilon}^{0}=\text{diag}\left(\hat{\varUpsilon}_{1}^{0},...,\hat{\varUpsilon}_{p_{n}}^{0}\right),\hat{\varUpsilon}_{j}^{0}=\sqrt{\frac{1}{n}\sum_{i=1}^{n}q_{ji}^{2}}
\]
and 
\[
q_{ji}=u_{i}x_{ji}+\mathbb{E}\left(\left.u_{i}x_{ji}\right|\boldsymbol{x}_{1i}^{p^{*}}\right)-\mathbb{E}\left(\left.u_{i}x_{ji}\right|\boldsymbol{x}_{1i}^{p^{*}},v_{i}\right),j=1,2,...,p_{n}.
\]
Set $\lambda=2.2\sqrt{n}\Phi^{-1}\left(1-\frac{0.1}{p_{n}\log\left(n\right)}\right)$.

\noindent \textbf{Step 0.} For $\hat{\Upsilon},$ set 
\[
\hat{\varUpsilon}_{j}=\sqrt{\frac{1}{n}\sum_{i=1}^{n}x_{ji}^{2}\left(\hat{\tilde{y}}_{i}-\overline{\hat{\tilde{y}}}\right)^{2}},j=1,2,...,p_{n}.
\]
We then obtain $\boldsymbol{\hat{\beta}^{*}}.$

\noindent \textbf{Step 1.} Using $\boldsymbol{\hat{\beta}^{*}}$ from
previous estimation, we then refine 
\[
\hat{\varUpsilon}_{j}=\sqrt{\frac{1}{n}\sum_{i=1}^{n}\hat{q}_{ji}^{2}},
\]
with 
\[
\hat{q}_{ji}=\hat{\tilde{y}}_{i}-\boldsymbol{x}_{i}'\boldsymbol{\hat{\beta}^{*}}+\widehat{\mathbb{E}\left(\left.\hat{\tilde{y}}_{i}x_{ji}\right|\boldsymbol{x}_{1i}^{p^{*}}\right)}-\widehat{\mathbb{E}\left(\left.\hat{\tilde{y}}_{i}x_{ji}\right|\boldsymbol{x}_{1i}^{p^{*}},v_{i}\right)},
\]
where $\widehat{\cdot}$ denotes standard nonparametric regression.
We re-estimate the model and obtain an update of $\boldsymbol{\hat{\beta}^{*}}.$

\noindent \textbf{Step 2.} Repeat Step 1 $K$ (e.g., 15) times.

One note is that we use the information on the selected $\boldsymbol{x}_{1i}^{p^{*}}$
from the previous selection as in Section \ref{SEC:proceduref}.

\section{Tables\label{APP:tables}}

\subsection{Tables for Simulations}
\begin{center}
\begin{table}[H]
\global\long\def\thetable{1A}%
\centering \small \caption{Design 1, Performance of DC Screening and Variable Selection}
\label{tab:1A} %
\begin{tabular}{cccccccc}
    \toprule
          &       & \multicolumn{2}{c}{DC Variable Screening} &       & \multicolumn{3}{c}{Variable Selection} \\
\cmidrule{3-4}\cmidrule{6-8}          &       & $\Pr(\{1\} \in \tilde{\mathcal{A}})$ & $\Pr(\{2\} \in \tilde{\mathcal{A}})$ &       &       & Correct (\%) & Incorrect \\
    \midrule
    \multirow{9}[6]{*}{$p_n=14$} & \multirow{3}[2]{*}{$n=500$} &       &       &       & $\gamma = 2$ & 10.154 (84.61\%) & 0.006 \\
          &       & 100.00\% & 95.00\% &       & $\gamma = 3$ & 10.144 (84.53\%) & 0.006 \\
          &       &       &       &       & $\gamma = 4$ & 10.574 (88.12\%) & 0.008 \\
\cmidrule{2-8}          & \multirow{3}[2]{*}{$n=1000$} &       &       &       & $\gamma = 2$ & 10.260 (85.50\%) & 0.014 \\
          &       & 100.00\% & 99.40\% &       & $\gamma = 3$ & 10.862 (90.52\%) & 0.014 \\
          &       &       &       &       & $\gamma = 4$ & 11.502 (95.85\%) & 0.014 \\
\bottomrule
    \end{tabular}%
\end{table}

\begin{table}[H]
\global\long\def\thetable{1B}%
\centering \small \caption{Design 1, Performance of Adaptive Lasso Regression}
\label{tab:1B} %
    \begin{tabular}{cccccccccccc}
    \toprule
          &       &       & \multicolumn{4}{c}{$\beta_2$}   &       & \multicolumn{4}{c}{$\beta_3$} \\
\cmidrule{4-7}\cmidrule{9-12}          &       &       & MEANB & RMSE  & MEDB  & MAD   &       & MEANB & RMSE  & MEDB  & MAD \\
    \midrule
    \multirow{9}[6]{*}{$p_n=14$} & \multirow{3}[2]{*}{$n=500$} & $\gamma = 2$ & -0.071 & 0.249 & -0.085 & 0.194 &       & -0.016 & 0.249 & -0.008 & 0.191 \\
          &       & $\gamma = 3$ & -0.050 & 0.241 & -0.058 & 0.186 &       & 0.002 & 0.245 & 0.004 & 0.189 \\
          &       & $\gamma = 4$ & -0.046 & 0.240 & -0.054 & 0.184 &       & 0.006 & 0.243 & 0.006 & 0.187 \\
\cmidrule{2-12}          & \multirow{3}[2]{*}{$n=1000$} & $\gamma = 2$ & -0.016 & 0.197 & -0.018 & 0.137 &       & 0.045 & 0.180 & 0.041 & 0.136 \\
          &       & $\gamma = 3$ & -0.011 & 0.195 & -0.011 & 0.135 &       & 0.050 & 0.180 & 0.048 & 0.136 \\
          &       & $\gamma = 4$ & -0.011 & 0.195 & -0.012 & 0.135 &       & 0.050 & 0.180 & 0.048 & 0.136 \\
\bottomrule
    \end{tabular}%
\end{table}

\begin{table}[H]
\global\long\def\thetable{1C}%
\centering \small \caption{Design 1, Performance of Probit Regression}
\label{tab:1C} %
    \begin{tabular}{clccccccccc}
    \toprule
          &       & \multicolumn{4}{c}{$\beta_2/\rho$} &       & \multicolumn{4}{c}{$\beta_3/\rho$} \\
\cmidrule{3-6}\cmidrule{8-11}          &       & MEANB & RMSE  & MEDB  & MAD   &       & MEANB & RMSE  & MEDB  & MAD \\
    \midrule
    \multirow{3}[2]{*}{$p_n=14$} & $n=500$ & 0.008 & 0.167 & 0.001 & 0.114 &       & -0.003 & 0.158 & -0.010 & 0.102 \\
          & $n=1000$ & 0.000 & 0.114 & 0.008 & 0.078 &       & -0.002 & 0.114 & -0.013 & 0.078 \\
          & $n=2000$ & -0.005 & 0.083 & -0.008 & 0.057 &       & 0.006 & 0.078 & 0.006 & 0.056 \\
    \midrule
    \multirow{3}[2]{*}{$p_n=29$} & $n=500$ & 0.001 & 0.181 & -0.002 & 0.119 &       & 0.001 & 0.173 & 0.000 & 0.109 \\
          & $n=1000$ & 0.001 & 0.115 & 0.002 & 0.078 &       & 0.008 & 0.118 & 0.004 & 0.078 \\
          & $n=2000$ & 0.005 & 0.084 & 0.003 & 0.056 &       & 0.001 & 0.084 & 0.002 & 0.056 \\
    \midrule
    \multirow{3}[2]{*}{$p_n=49$} & $n=500$ & 0.010 & 0.186 & -0.002 & 0.129 &       & 0.002 & 0.188 & -0.006 & 0.124 \\
          & $n=1000$ & 0.006 & 0.116 & 0.001 & 0.076 &       & -0.002 & 0.115 & -0.005 & 0.076 \\
          & $n=2000$ & -0.008 & 0.085 & -0.010 & 0.061 &       & -0.005 & 0.079 & -0.008 & 0.054 \\
    \bottomrule
    \end{tabular}%
\end{table}

\begin{table}[H]
\global\long\def\thetable{2A}%
\centering \small \caption{Design 2, Performance of DC Screening and Variable Selection}
\label{tab:2A} %
    \begin{tabular}{cccccccc}
    \toprule
          &       & \multicolumn{2}{c}{DC Variable Screening} &       & \multicolumn{3}{c}{Variable Selection} \\
\cmidrule{3-4}\cmidrule{6-8}          &       & $\Pr(\{1\} \in \tilde{\mathcal{A}})$ & $\Pr(\{2\} \in \tilde{\mathcal{A}})$ &       &       & Correct (\%) & Incorrect \\
    \midrule
    \multirow{9}[6]{*}{$p_n=14$} & \multirow{3}[2]{*}{$n=500$} &       &       &       & $\gamma = 2$ & 10.042 (83.68\%) & 0.006 \\
          &       & 100.00\% & 95.00\% &       & $\gamma = 3$ & 10.260 (85.50\%) & 0.006 \\
          &       &       &       &       & $\gamma = 4$ & 10.620 (88.50\%) & 0.010 \\
\cmidrule{2-8}          & \multirow{3}[2]{*}{$n=1000$} &       &       &       & $\gamma = 2$ & 10.372 (86.43\%) & 0.012 \\
          &       & 100.00\% & 99.40\% &       & $\gamma = 3$ & 11.050 (92.08\%) & 0.012 \\
          &       &       &       &       & $\gamma = 4$ & 11.636 (96.97\%) & 0.014 \\
\bottomrule
    \end{tabular}%
\end{table}

\begin{table}[H]
\global\long\def\thetable{2B}%
\centering \small \caption{Design 2, Performance of Adaptive Lasso Regression}
\label{tab:2B} %
    \begin{tabular}{cccccccccccc}
    \toprule
          &       &       & \multicolumn{4}{c}{$\beta_2$}   &       & \multicolumn{4}{c}{$\beta_3$} \\
\cmidrule{4-7}\cmidrule{9-12}          &       &       & MEANB & RMSE  & MEDB  & MAD   &       & MEANB & RMSE  & MEDB  & MAD \\
    \midrule
    \multirow{9}[6]{*}{$p_n=14$} & \multirow{3}[2]{*}{$n=500$} & $\gamma = 2$ & -0.071 & 0.249 & -0.069 & 0.190 &       & -0.068 & 0.277 & -0.056 & 0.216 \\
          &       & $\gamma = 3$ & -0.055 & 0.238 & -0.056 & 0.182 &       & -0.053 & 0.272 & -0.043 & 0.211 \\
          &       & $\gamma = 4$ & -0.049 & 0.234 & -0.052 & 0.178 &       & -0.050 & 0.274 & -0.040 & 0.210 \\
\cmidrule{2-12}          & \multirow{3}[2]{*}{$n=1000$} & $\gamma = 2$ & -0.014 & 0.194 & -0.023 & 0.139 &       & -0.012 & 0.194 & -0.005 & 0.150 \\
          &       & $\gamma = 3$ & -0.009 & 0.192 & -0.014 & 0.137 &       & -0.006 & 0.190 & 0.000 & 0.147 \\
          &       & $\gamma = 4$ & -0.008 & 0.192 & -0.014 & 0.136 &       & -0.007 & 0.195 & -0.001 & 0.148 \\
\bottomrule
    \end{tabular}%
\end{table}

\begin{table}[H]
\global\long\def\thetable{2C}%
\centering \small \caption{Design 2, Performance of Probit Regression}
\label{tab:2C} %
    \begin{tabular}{clccccccccc}
    \toprule
          &       & \multicolumn{4}{c}{$\beta_2/\rho$} &       & \multicolumn{4}{c}{$\beta_3/\rho$} \\
\cmidrule{3-6}\cmidrule{8-11}          &       & MEANB & RMSE  & MEDB  & MAD   &       & MEANB & RMSE  & MEDB  & MAD \\
    \midrule
    \multirow{3}[2]{*}{$p_n=14$} & $n=500$ & 0.004 & 0.165 & -0.012 & 0.107 &       & -0.117 & 0.210 & -0.116 & 0.146 \\
          & $n=1000$ & 0.000 & 0.113 & 0.000 & 0.077 &       & -0.119 & 0.173 & -0.119 & 0.124 \\
          & $n=2000$ & -0.001 & 0.084 & 0.001 & 0.057 &       & -0.114 & 0.145 & -0.117 & 0.118 \\
    \midrule
    \multirow{3}[2]{*}{$p_n=29$} & $n=500$ & 0.008 & 0.176 & -0.001 & 0.120 &       & -0.102 & 0.215 & -0.112 & 0.154 \\
          & $n=1000$ & 0.000 & 0.117 & -0.001 & 0.078 &       & -0.117 & 0.173 & -0.114 & 0.120 \\
          & $n=2000$ & 0.002 & 0.081 & 0.001 & 0.055 &       & -0.116 & 0.152 & -0.117 & 0.121 \\
    \midrule
    \multirow{3}[2]{*}{$p_n=49$} & $n=500$ & 0.004 & 0.173 & 0.004 & 0.116 &       & -0.127 & 0.237 & -0.125 & 0.153 \\
          & $n=1000$ & 0.001 & 0.112 & 0.006 & 0.078 &       & -0.116 & 0.171 & -0.106 & 0.121 \\
          & $n=2000$ & -0.003 & 0.084 & -0.005 & 0.056 &       & -0.118 & 0.151 & -0.118 & 0.122 \\
    \bottomrule
    \end{tabular}%
\end{table}

\begin{table}[H]
\global\long\def\thetable{3A}%
\centering \caption{Design 3, Performance of DC Screening and Variable Selection}
\label{tab:3A} %
    \begin{tabular}{cccccccc}
    \toprule
          &       & \multicolumn{2}{c}{DC Variable Screening} &       & \multicolumn{3}{c}{Variable Selection} \\
\cmidrule{3-4}\cmidrule{6-8}          &       & $\Pr(\{1\} \in \tilde{\mathcal{A}})$ & $\Pr(\{2\} \in \tilde{\mathcal{A}})$ &       &       & Correct (\%) & Incorrect \\
    \midrule
    \multirow{9}[6]{*}{$p_n=14$} & \multirow{3}[2]{*}{$n=500$} &       &       &       & $\gamma = 2$ & 9.860 (82.17\%) & 0.054 \\
          &       & 100.00\% & 100.00\% &       & $\gamma = 3$ & 10.054 (83.78\%) & 0.060 \\
          &       &       &       &       & $\gamma = 4$ & 10.196 (84.97\%) & 0.100 \\
\cmidrule{2-8}          & \multirow{3}[2]{*}{$n=1000$} &       &       &       & $\gamma = 2$ & 9.948 (82.90\%) & 0.022 \\
          &       & 100.00\% & 100.00\% &       & $\gamma = 3$ & 10.284 (85.70\%) & 0.026 \\
          &       &       &       &       & $\gamma = 4$ & 10.958 (91.32\%) & 0.046 \\
\bottomrule
    \end{tabular}%
\end{table}

\begin{table}[H]
\global\long\def\thetable{3B}%
\centering \small \caption{Design 3, Performance of Adaptive Lasso Regression}
\label{tab:3B} %
    \begin{tabular}{cccccccccccc}
    \toprule
          &       &       & \multicolumn{4}{c}{$\beta_2$}   &       & \multicolumn{4}{c}{$\beta_3$} \\
\cmidrule{4-7}\cmidrule{9-12}          &       &       & MEANB & RMSE  & MEDB  & MAD   &       & MEANB & RMSE  & MEDB  & MAD \\
    \midrule
    \multirow{9}[6]{*}{$p_n=14$} & \multirow{3}[2]{*}{$n=500$} & $\gamma = 2$ & -0.125 & 0.336 & -0.103 & 0.257 &       & -0.131 & 0.383 & -0.137 & 0.298 \\
          &       & $\gamma = 3$ & -0.110 & 0.326 & -0.091 & 0.249 &       & -0.121 & 0.380 & -0.127 & 0.292 \\
          &       & $\gamma = 4$ & -0.111 & 0.348 & -0.083 & 0.255 &       & -0.110 & 0.388 & -0.115 & 0.294 \\
\cmidrule{2-12}          & \multirow{3}[2]{*}{$n=1000$} & $\gamma = 2$ & -0.058 & 0.246 & -0.045 & 0.179 &       & -0.083 & 0.270 & -0.072 & 0.205 \\
          &       & $\gamma = 3$ & -0.053 & 0.241 & -0.047 & 0.176 &       & -0.075 & 0.266 & -0.063 & 0.200 \\
          &       & $\gamma = 4$ & -0.059 & 0.249 & -0.056 & 0.180 &       & -0.081 & 0.285 & -0.067 & 0.206 \\
\bottomrule
    \end{tabular}%
\end{table}

\begin{table}[H]
\global\long\def\thetable{3C}%
\centering \small \caption{Design 3, Performance of Probit Regression}
\label{tab:3C} %
    \begin{tabular}{clccccccccc}
    \toprule
          &       & \multicolumn{4}{c}{$\beta_2/\rho$} &       & \multicolumn{4}{c}{$\beta_3/\rho$} \\
\cmidrule{3-6}\cmidrule{8-11}          &       & MEANB & RMSE  & MEDB  & MAD   &       & MEANB & RMSE  & MEDB  & MAD \\
    \midrule
    \multirow{3}[2]{*}{$p_n=14$} & $n=500$ & 0.015 & 0.208 & 0.015 & 0.143 &       & -0.185 & 0.306 & -0.185 & 0.214 \\
          & $n=1000$ & 0.010 & 0.138 & 0.006 & 0.093 &       & -0.181 & 0.243 & -0.186 & 0.194 \\
          & $n=2000$ & 0.018 & 0.105 & 0.016 & 0.070 &       & -0.192 & 0.225 & -0.188 & 0.188 \\
    \midrule
    \multirow{3}[2]{*}{$p_n=29$} & $n=500$ & 0.003 & 0.214 & -0.007 & 0.148 &       & -0.174 & 0.292 & -0.164 & 0.206 \\
          & $n=1000$ & 0.015 & 0.154 & 0.018 & 0.102 &       & -0.175 & 0.240 & -0.175 & 0.189 \\
          & $n=2000$ & 0.013 & 0.102 & 0.016 & 0.073 &       & -0.194 & 0.225 & -0.196 & 0.197 \\
    \midrule
    \multirow{3}[2]{*}{$p_n=49$} & $n=500$ & 0.002 & 0.225 & 0.001 & 0.143 &       & -0.192 & 0.311 & -0.202 & 0.232 \\
          & $n=1000$ & 0.016 & 0.152 & 0.014 & 0.104 &       & -0.198 & 0.259 & -0.189 & 0.193 \\
          & $n=2000$ & 0.016 & 0.106 & 0.005 & 0.069 &       & -0.195 & 0.227 & -0.189 & 0.189 \\
    \bottomrule
    \end{tabular}%
\end{table}

\begin{table}[H]
\global\long\def\thetable{4A}%
\centering \small \caption{Design 4, Performance of DC Screening and Moment Selection}
\label{tab:4A} %
    \begin{tabular}{cccccccc}
    \toprule
          &       & \multicolumn{2}{c}{DC Variable Screening} &       & \multicolumn{3}{c}{Moment Selection} \\
\cmidrule{3-4}\cmidrule{6-8}          &       & $\Pr(Z^{*} \in \tilde{\mathcal{Z}})$ & $\Pr(Z_1 \in \tilde{\mathcal{Z}})$ &       &       & Correct (\%) & Incorrect (\%) \\
    \midrule
    \multirow{9}[6]{*}{$(d_A,d_B)=(6,7)$} & \multirow{3}[2]{*}{$n=500$} &       &       &       & $\gamma = 1$ & 2.380 (39.37\%) & 0.640 (9.14\%) \\
          &       & 100.00\% & 100.00\% &       & $\gamma = 2$ & 4.500 (75.00\%) & 1.794 (25.63\%) \\
          &       &       &       &       & $\gamma = 3$ & 5.484 (91.40\%) & 3.440 (49.14\%) \\
\cmidrule{2-8}          & \multirow{3}[2]{*}{$n=1000$} &       &       &       & $\gamma = 1$ & 2.786 (46.43\%) & 0.178 (2.54\%) \\
          &       & 100.00\% & 100.00\% &       & $\gamma = 2$ & 4.942 (82.37\%) & 0.736 (10.51\%) \\
          &       &       &       &       & $\gamma = 3$ & 5.678 (94.63\%) & 2.306 (32.94\%) \\
\bottomrule
    \end{tabular}%
\end{table}

\begin{table}[H]
\global\long\def\thetable{4B}%
\centering \footnotesize \caption{Design 4, Performance of Penalized GMM Estimation}
\label{tab:4B} %
    \begin{tabular}{cccccccccccc}
    \toprule
          &       &       & \multicolumn{4}{c}{$\beta_0$}   &       & \multicolumn{4}{c}{$\beta_1$} \\
\cmidrule{4-7}\cmidrule{9-12}          &       &       & MEANB & RMSE  & MEDB  & MAD   &       & MEANB & RMSE  & MEDB  & MAD \\
    \midrule
    \multirow{9}[6]{*}{$(d_A,d_B)=(6,7)$} & \multirow{3}[2]{*}{$n=500$} & $\gamma = 1$ & -0.049 & 0.145 & -0.056 & 0.116 &       & -0.135 & 0.398 & -0.155 & 0.316 \\
          &       & $\gamma = 2$ & -0.052 & 0.147 & -0.060 & 0.118 &       & -0.155 & 0.399 & -0.171 & 0.321 \\
          &       & $\gamma = 3$ & -0.055 & 0.150 & -0.058 & 0.120 &       & -0.233 & 0.429 & -0.271 & 0.356 \\
\cmidrule{2-12}          & \multirow{3}[2]{*}{$n=1000$} & $\gamma = 1$ & -0.043 & 0.117 & -0.043 & 0.089 &       & -0.080 & 0.307 & -0.084 & 0.223 \\
          &       & $\gamma = 2$ & -0.042 & 0.117 & -0.042 & 0.089 &       & -0.074 & 0.293 & -0.055 & 0.207 \\
          &       & $\gamma = 3$ & -0.045 & 0.118 & -0.046 & 0.090 &       & -0.158 & 0.339 & -0.153 & 0.264 \\
\bottomrule
    \end{tabular}%
\end{table}

\begin{table}[H]
\global\long\def\thetable{5A}%
\centering \small \caption{Design 5, Performance of DC Screening and Moment Selection}
\label{tab:5A} %
    \begin{tabular}{cccccccc}
    \toprule
          &       & \multicolumn{2}{c}{DC Variable Screening} &       & \multicolumn{3}{c}{Moment Selection} \\
\cmidrule{3-4}\cmidrule{6-8}          &       & $\Pr(Z^{*} \in \tilde{\mathcal{Z}})$ & $\Pr(Z_1 \in \tilde{\mathcal{Z}})$ &       &       & Correct (\%) & Incorrect (\%) \\
    \midrule
    \multirow{9}[6]{*}{$(d_A,d_B)=(6,7)$} & \multirow{3}[2]{*}{$n=500$} &       &       &       & $\gamma = 1$ & 2.790 (46.50\%) & 0.630 (9.00\%) \\
          &       & 100.00\% & 100.00\% &       & $\gamma = 2$ & 4.922 (82.03\%) & 1.806 (25.80\%) \\
          &       &       &       &       & $\gamma = 3$ & 5.710 (95.17\%) & 3.460 (49.43\%) \\
\cmidrule{2-8}          & \multirow{3}[2]{*}{$n=1000$} &       &       &       & $\gamma = 1$ & 3.252 (54.20\%) & 0.152 (2.17\%) \\
          &       & 100.00\% & 100.00\% &       & $\gamma = 2$ & 5.392 (89.87\%) & 0.800 (11.43\%) \\
          &       &       &       &       & $\gamma = 3$ & 5.878 (97.97\%) & 2.432 (34.74\%) \\
\bottomrule
    \end{tabular}%
\end{table}

\begin{table}[H]
\global\long\def\thetable{5B}%
\centering \footnotesize \caption{Design 5, Performance of Penalized GMM Estimation}
\label{tab:5B} %
    \begin{tabular}{cccccccccccc}
    \toprule
          &       &       & \multicolumn{4}{c}{$\beta_0$}   &       & \multicolumn{4}{c}{$\beta_1$} \\
\cmidrule{4-7}\cmidrule{9-12}          &       &       & MEANB & RMSE  & MEDB  & MAD   &       & MEANB & RMSE  & MEDB  & MAD \\
    \midrule
    \multirow{9}[6]{*}{$(d_A,d_B)=(6,7)$} & \multirow{3}[2]{*}{$n=500$} & $\gamma = 1$ & -0.051 & 0.155 & -0.060 & 0.119 &       & -0.140 & 0.408 & -0.165 & 0.320 \\
          &       & $\gamma = 2$ & -0.053 & 0.158 & -0.066 & 0.122 &       & -0.167 & 0.423 & -0.181 & 0.332 \\
          &       & $\gamma = 3$ & -0.060 & 0.161 & -0.068 & 0.125 &       & -0.262 & 0.471 & -0.305 & 0.389 \\
\cmidrule{2-12}          & \multirow{3}[2]{*}{$n=1000$} & $\gamma = 1$ & -0.042 & 0.115 & -0.042 & 0.089 &       & -0.091 & 0.315 & -0.097 & 0.230 \\
          &       & $\gamma = 2$ & -0.043 & 0.115 & -0.040 & 0.089 &       & -0.101 & 0.316 & -0.098 & 0.234 \\
          &       & $\gamma = 3$ & -0.048 & 0.118 & -0.047 & 0.092 &       & -0.200 & 0.390 & -0.205 & 0.308 \\
\bottomrule
    \end{tabular}%
\end{table}

\begin{table}[H]
\global\long\def\thetable{6A}%
\centering \small \caption{Design 6, Performance of DC Screening and Moment Selection}
\label{tab:6A} %
    \begin{tabular}{cccccccc}
    \toprule
          &       & \multicolumn{2}{c}{DC Variable Screening} &       & \multicolumn{3}{c}{Moment Selection} \\
\cmidrule{3-4}\cmidrule{6-8}          &       & $\Pr(Z^{*} \in \tilde{\mathcal{Z}})$ & $\Pr(Z_1 \in \tilde{\mathcal{Z}})$ &       &       & Correct (\%) & Incorrect (\%) \\
    \midrule
    \multirow{9}[6]{*}{$(d_A,d_B)=(6,7)$} & \multirow{3}[2]{*}{$n=500$} &       &       &       & $\gamma = 1$ & 2.410 (40.17\%)) & 0.330 (4.71\%) \\
          &       & 100.00\% & 100.00\% &       & $\gamma = 2$ & 4.486 (74.77\%) & 0.914 (13.06\%) \\
          &       &       &       &       & $\gamma = 3$ & 5.454 (90.90\%) & 1.926 (27.51\%) \\
\cmidrule{2-8}          & \multirow{3}[2]{*}{$n=1000$} &       &       &       & $\gamma = 1$ & 2.808 (46.80\%) & 0.082 (1.17\%) \\
          &       & 100.00\% & 100.00\% &       & $\gamma = 2$ & 4.982 (83.03\%) & 0.228 (3.26\%) \\
          &       &       &       &       & $\gamma = 3$ & 5.676 (94.60\%)) & 0.758 (10.83\%) \\
\bottomrule
    \end{tabular}%
\end{table}

\begin{table}[H]
\global\long\def\thetable{6B}%
\centering \footnotesize \caption{Design 6, Performance of Penalized GMM Estimation}
\label{tab:6B} %
    \begin{tabular}{cccccccccccc}
    \toprule
          &       &       & \multicolumn{4}{c}{$\beta_0$}   &       & \multicolumn{4}{c}{$\beta_1$} \\
\cmidrule{4-7}\cmidrule{9-12}          &       &       & MEANB & RMSE  & MEDB  & MAD   &       & MEANB & RMSE  & MEDB  & MAD \\
    \midrule
    \multirow{9}[6]{*}{$(d_A,d_B)=(6,7)$} & \multirow{3}[2]{*}{$n=500$} & $\gamma = 1$ & -0.048 & 0.143 & -0.053 & 0.115 &       & -0.137 & 0.405 & -0.149 & 0.318 \\
          &       & $\gamma = 2$ & -0.049 & 0.144 & -0.055 & 0.116 &       & -0.139 & 0.405 & -0.115 & 0.315 \\
          &       & $\gamma = 3$ & -0.052 & 0.148 & -0.057 & 0.119 &       & -0.192 & 0.453 & -0.177 & 0.362 \\
\cmidrule{2-12}          & \multirow{3}[2]{*}{$n=1000$} & $\gamma = 1$ & -0.043 & 0.113 & -0.042 & 0.088 &       & -0.085 & 0.300 & -0.081 & 0.222 \\
          &       & $\gamma = 2$ & -0.042 & 0.113 & -0.042 & 0.089 &       & -0.049 & 0.272 & -0.029 & 0.189 \\
          &       & $\gamma = 3$ & -0.042 & 0.114 & -0.042 & 0.089 &       & -0.078 & 0.310 & -0.031 & 0.220 \\
\bottomrule
    \end{tabular}%
\end{table}
\end{center}

\subsection{Tables for the Application}

\begin{center}
\global\long\def\thetable{AP2}%
{\small \begin{longtable}[H]{ll}

\caption[Definitions of Variables]{Definitions of Variables} \label{tab:AP2} \\
\hline \hline
\endfirsthead
\multicolumn{2}{c} 
{{ \tablename\ \thetable{} -- continued from previous page}} \\
\hline
\endhead
\hline \multicolumn{2}{r}{{Continued on next page}} \\ \hline
\endfoot
\hline \hline
\endlastfoot
\multicolumn{2}{c}{Panel A: Dependent Variable}  \\\hline
$Y$	& Migration intention within a year \\\hline
\multicolumn{2}{c}{Panel B: The Special Regressor}  \\\hline
$V$	& Minus log precipitation \\\hline
\multicolumn{2}{c}{Panel C: Other Explanatory Variables}  \\\hline
nykid1	& Numbers of children aged from 0 to 5 \\
nykid2 & 	Numbers of children aged from 6 to 10 \\
nykid3	 & Numbers of children aged  from 10 and 15\\
\end{longtable}}

\global\long\def\thetable{AP3}%
{\small \begin{longtable}[p]{llllllll}

\caption[Summary Statistics]{Summary Statistics} \label{tab:AP3}\\
\hline \hline
Variable & MEAN & STD & MIN & LQ & MD & UQ & MAX  \\\hline
\endfirsthead
\multicolumn{8}{c}%
{{ \tablename\ \thetable{} -- continued from previous page}} \\\hline
Variable & MEAN & STD & MIN & LQ & MD & UQ & MAX  \\
\hline
\endhead
\hline \multicolumn{8}{r}{{Continued on next page}} \\ \hline
\endfoot
\hline \hline
\endlastfoot
\multicolumn{8}{c}{Panel A: Dependent Variable}  \\\hline
 $Y$       & 0.041 & 0.198 & 0.000 & 0.000 & 0.000 & 0.000 & 1.000 \\
\multicolumn{2}{c}{}  \\
\multicolumn{8}{c}{Panel B: The Special Regressor}  \\\hline
 $V$       & -0.078 & 0.168 & -0.548 & -0.177 & -0.047 & 0.044 & 0.307 \\
\multicolumn{2}{c}{}  \\
\multicolumn{8}{c}{Panel C:Other Explanatory Variables}  \\\hline
nykid1    & 0.140 & 0.368 & 0.000 & 0.000 & 0.000 & 0.000 & 3.000 \\
\end{longtable}Note: MEAN = sample mean, STD = standard deviation, MIN =
minimum, LQ = 25 percentile, MD = median, UQ = 75 percentile, MAX = maximum. }
\end{center}

\end{document}